# HEALTH INFORMATION STANDARDISATION AS A BASIS FOR LEARNING HEALTH SYSTEMS

**Scott McLachlan**

**DSysEng, GDInfSc, GDBus, GCertTTLP, MPhil (Sc), GDL, LLM**

Supervisor:

Professor Norman E Fenton

Submitted in fulfilment of the requirements for the degree of
Doctor of Philosophy

School of Electrical Engineering and Computer Science
Queen Mary, University of London
2019

# Keywords

# Abstract

Standardisation of healthcare has been the focus of hospital management and clinicians since the 1990's. Electronic health records were already intended to provide clinicians with real-time access to clinical knowledge and care plans while also recording and storing vast amounts of patient data. It took more than three decades for electronic health records to start to become ubiquitous in all aspects of healthcare. Learning health systems are the next stage in health information systems whose potential benefits have been promoted for more than a decade - yet few are seen in clinical practice. Clinical care process specifications are a primary form of clinical documentation used in all aspects of healthcare, but they lack standardisation. This thesis contends that this lack of standardisation was inherited by electronic health records and that this is a significant issue holding back the development and adoption of learning health systems. Standardisation of clinical documents is used to mitigate issues in electronic health records as a basis for enabling learning health systems. One type of clinical document, the caremap, is standardised in order to achieve an effective approach to containing resources and ensuring consistency and quality. This led not only to improved clinicians' comprehension and acceptance of the clinical document, but also to reduced time expended in developing complicated learning health systems built using the input of clinical experts.

# Table of Contents

# List of Figures

# List of Tables

# List of Abbreviations

| | |
|---|---|
| BN | Bayesian Network. |
| CA | Content Analysis |
| CCPS | Clinical Care Process Specifications |
| CPG | Clinical Practice Guidelines. |
| CDSS | Computerised Decision Support System |
| DP | Decision Point |
| DSS | Decision Support System |
| EBM | Evidence-Based Medicine |
| EHR | Electronic Health Records. |
| HI | Health Informatics. |
| HIE | Health Information Exchange |
| HIS | Health Information Systems |
| HIT | Health Information Technology |
| LHS | Learning Health Systems / Learning Healthcare Systems. |
| RS-EHR | Realistic Synthetic Electronic Health Records |
| TA | Thematic Analysis |

## LHS Taxonomy Abbreviations

| | |
|---|---|
| CI | Cohort Identification |
| PD | Positive Deviance |
| ND | Negative Deviance |
| PPRM | Predictive Patient Risk Modelling |
| PCROM | Predictive Care Risk and Outcome Models |
| CDSS | Clinical Decision Support Systems |
| CER | Comparative Effectiveness Research |
| IA | Intelligent Assistance |
| S | Surveillance |

# Statement of Original Authorship

I, *Scott McLachlan*, confirm that the research included within this thesis is my own work or that where it has been carried out in collaboration with, or supported by others, that this is duly acknowledged below and my contribution indicated. Previously published material is also acknowledged below.

I attest that I have exercised reasonable care to ensure that the work is original, and does not to the best of my knowledge break any UK law, infringe any third party's copyright or other Intellectual Property Right, or contain any confidential material.

I accept that the College has the right to use plagiarism detection software to check the electronic version of the thesis.

I confirm that this thesis has not been previously submitted for the award of a degree by this or any other university.

Signature:

Date: 10th February, 2020

London, UK.

# Acknowledgements

With thanks firstly to my supervisor, Norman Fenton. There were times while working with his group that I felt the pain of impostor syndrome. He ensured that I had more than enough things to do and when I kept achieving them, pointed out my achievements to remind me why I actually deserved to be there. Thanks also to my co-supervisors and academic advisors: Martin Neil - whose affable and inquiring nature made my thinking and the work I was doing more robust; Kuda Dube – who patiently reminded me of what I should already know, and guided me to read the right directions when I did not; and William Marsh – who acted much like the polishing abrasive that releases the precious stone from the rock in a jeweller's tumbler. I always work at my best with a little tension and under pressure. And finally, to Evangelia (Lina) Kyrimi. You are a brilliant researcher and collaborator who has quietly served as an example and guide to many of us on the PamBayesian team.

There is a long list of people who deserve some mention. The PamBayesian team I worked with: Eugene, Marianna, Chris, Jiali, Maggie, Ali, Hamit, Paul, Graham, Sarah, and Bridget. And my awesome external collaborators: Tom Gallagher, Henry Potts, Owen Johnson and Derek Buchanan.

I only hope that someday I can return to you as much time, energy and support to you as you gave me.

**Dedication:**
As with any thesis there are a lot of people you interact with and seemingly small events that happen in the background that all contribute to the success of the whole. For me this journey has been almost 25 years in the making. But through those years there are a small number of good-hearted people that, had our paths not crossed, I wouldn't be here receiving this doctorate degree.

Firstly, for those who were not fortunate enough to be here and see where I ended up. For my grandmother, Esma Baxter. She was a tough, no-nonsense pioneering country woman but when she hugged you, you knew you were exactly where you wanted to be. For my aunt, Laurel Ryan. A lady who accepted me as her nephew and always made me feel welcome and comfortable in her presence. During our Saturday morning visits to her house she started a life-long obsession with *iced vovo* and *monte carlo* biscuits that my much older waistline silently endures.

And for those still among us. For my aunt, Gwen Morris. You fed me when I needed feeding, treated me as a human being when others ignored me, and talked to me when no one else would. And along with Aunty Laurel, your sister and *partner in crime*, you took me places and helped me to experience the world so that I could learn how to act and be comfortable socially. That I finally stopped living down to the low expectations of others and pulled myself together enough to even have the title *doctor* bestowed upon me is due more to you and your influence than any other single person. Thank you.

To my brothers. It took the better part of three decades, but we finally have an open dialogue and accept each other's different perspectives of the world.

To my younger brother Craig. Don't give up. It's hard sometimes but *make time*. I know and believe that you too can do it. Never forget that you can always call on me for help.

Jen, you were right, and I was wrong. YingYing, it was a once in a lifetime experience. Lisa Ryan, thank you and best wishes.

For Danika, Thomas and Liam

For James

And to my nieces Chelsea, Annabella, Evie and Grace. Keep your dad honest and twisted as tightly as possible around your little fingers.

# Related Publications

The following primary publications which are further discussed in Appendix F relate to the work presented in this thesis:

**[1] McLachlan, S**., Kyrimi, E., Dube, K., Hitman, G.A., Simmonds, J., & Fenton, N.E. (2019). Towards standardisation of clinical care process specifications. *Manuscript accepted for publication in the Sage Health Informatics Journal (HIJ).*

**[2] McLachlan, S**., Kyrimi, E., Dube, K., & Fenton, N. (2019). Clinical Caremap Development: How can caremaps standardise care when they are not standardised? *Proceedings of the 12th International Joint Conference on Biomedical Systems and Technologies (BIOSTEC 2019)*, volume 5: HEALTHINF, pp 123-134. DOI: 10.5220/0007522601230134

**[3] McLachlan, S**., Dube, K., Buchanan, D., Lean, S., Johnson, O., Potts, H.W.W., Gallagher, T., Marsh, W., & Fenton, N. (2018) Learning Health Systems: The research community awareness challenge. *Journal of Innovation in Health Informatics*, 25(1).

**[4] McLachlan, S**., Potts, H.W.W., Dube, K., Buchanan, D., Lean, S., Gallagher, T., Johnson, O., Daley, B., Marsh, W., & Fenton, N. (2018) The Heimdall Framework for supporting characterisation of Learning Health Systems. *Journal of Innovation in Health Informatics*, 25(2).

**[5] McLachlan, S**., Johnson, O., Dube, K., Buchanan, D., Potts, H.W.W., Gallagher, T., & Fenton, N. (2019). A framework for analysing learning health systems: Are we removing the most impactful barriers? *Journal of Learning Health Systems*, DOI: 10.1002/lrh2.10189

**[6] McLachlan, S**., Dube, K., Kyrimi, E., & Fenton, N. (2019). LAGOS: Learning Health Systems and how they can integrate with patient care. *BMJ Health and Care Informatics (BMJHCI)* https://dx.doi.org/10.1136/bmjhci-2019-100037.

**[7] McLachlan, S**., Kyrimi, E., Daley, B., Dube, K., & Fenton, N. (2019). TaSC: Incorporating clinical decisions into the standardised Caremap. *Manuscript submitted to IEEE ICHI 2020.*

The following secondary publications relate to the work presented in this thesis:

**[8]** Daley, B., Hitman, G., Fenton, N., & **McLachlan, S**. (2018). Systematic review of the quality of Clinical Practice Guidelines on the Identification and Management of Diabetes in Pregnancy. *BMJ Open*. 9(e027285).

**[9]** Kyrimi, E., Neves, M., Neil, M., Marsh, W., **McLachlan, S**., & Fenton, N. (2019). Medical Idioms: Reasoning patterns to develop medical Bayesian Networks. *Manuscript submitted to BMJ Health and Care Informatics (BMJHCI)*

**[10]** **McLachlan, S.,** Kyrimi, E., Dube, K., & Fenton, N. (2020) Standardising Clinical Caremaps: Model, Method and Graphical Notation for Caremap Specification. *Manuscript accepted for publication in Communications in Computer and Information Science*, Springer Nature, DE.

# Chapter 1: Introduction

In 2016, the *per person* spend in US dollars on healthcare for the United States of America (USA) was $9,892, Germany $4,876, Australia $4,708, France $4,600, United Kingdom (UK) $4,192, and New Zealand (NZ) $3,590 [1]. Including government, prepaid private and out-of-pocket expenses, the annual global cost of healthcare was calculated as falling between US$7.7 and US$7.83 trillion for 2014 [2, 3], being the last year in which this total amount could be located[1]. This amount is forecast to rise to US$8.7 trillion by 2020 [4], and more than double to over US$18 trillion annually by 2033 [2]. In the UK, around 70% of the National Health Service (NHS) budget is currently spent on treating slightly more than 30%[2] of the population who have been diagnosed with a chronic condition [5, 6]. Hospital overspend has left the NHS with debts over £12bn [7], and NHS trusts like that which operates Barts and The London hospital struggle under the weight of Private Finance Initiative (PFI) contracts that will see it repay more than £7bn on a 43-year loan of only £1.1bn [8].

Electronic health records (EHR) were intended to standardise health information and the documentation of interactions between clinicians and patients, enable execution of standardised models of clinical and guideline knowledge, and allow use of coding systems like SNOMED that were designed to standardise terminology [9, 10]. They were also seen as a vehicle to standardise business information, communication between healthcare providers, and the recording process and format of patient data [11, 10, 12]. Some hospitals believed they would realise substantial returns on investment from implementing EHR; in some cases, in the range of tens of millions of dollars [13]. EHR were envisioned to be a singular, or at least significant, solution for increasing healthcare quality and safety, reducing costs and increasing efficiency [14, 15]. However, a large number of barriers have meant that EHR are yet to deliver on many of these promises and goals.

---

[1] More recent figures published by the WHO and the OECD for 2016 provide global health as: (i) cost per capita; and (ii) percentage of GDP, but have not as yet provided the global total expenditure amount.
[2] In 2014 chronic disease affected 23% of the English population, and 46% of the Scottish population (source: [5, 6]).

In the early 1990's the focus of hospital management and clinical literature shifted from their prior reliance on project management and total quality management tools more common to industry, to promoting *standardisation* as being of paramount importance to the future of quality and efficiency in healthcare [16-19]. More recently researchers and clinicians have expanded this focus even further, joining together to describe the multitude of inefficiencies and voice the need for current healthcare services to transition to a system that continuously learns and improves: termed the *Learning Health System* (LHS) [20-22]. LHS are a recent concept, with the majority of works in the domain having been published only since 2011 [23]. Although they are not well understood, they embody the relationship between care practice, research and knowledge translation and are recognised to be one of the major potential computing technological advances in healthcare [24-26].

This introductory chapter describes the background for standardisation and learning health systems (section 1.1) and outlines the thesis research problem, hypothesis and objectives (section 1.2). Section 1.3 describes the thesis significance while section 1.5 summarises its novelty. Finally, section 1.6 provides an outline of the remaining chapters of the thesis.

## 1.1 BACKGROUND

### 1.1.1 Standardisation

Researchers, politicians and hospital managers have sought to achieve standardisation of such things as clinical decisions, diagnostic and therapeutic models, evidence-based guidelines, approaches to patient care, practice standards and even the clinical information recorded in health records [27-30, 18, 31, 32]. Standardisation in the name of quality care and outcomes has become the single-minded national focus of healthcare service delivery for entire countries [33, 29, 32]. This type of standardisation has been passionately promoted such that multiple teams within the same country, or even the same organisation, can be seen developing standardisation frameworks with some degree of similarity and overlap [34-36]. However, a review of the literature and current working practices in hospitals shows that the drive towards standardisation has had little effect on the definition, development and structure of clinical documentation. As shown in Appendix D, authors cannot seem to agree as to whether some types of clinical care process specifications (CCPS) represent distinct clinical documents with different content and purpose or are simply different names for the same clinical document. As we approached the end of the first full decade of standardisation, calls for standardised clinical documents continued to increase [29, 37-42], and an unmet need was also identified in calls to resolve poorly standardised taxonomy and nomenclatures being used in developing and cataloguing some clinical documentation [43].

The UK's NHS has promoted *efficiency* as their model for continued viability over the next five years [44]. However, health service providers everywhere are finding this goal difficult to achieve because, as this thesis contends, it, and successful implementation and integration of LHS in clinical practice, are restrained by a lack of standardisation. Even though standardisation of healthcare practice, procedure and tools is seen to improve healthcare in terms of quality, outcomes and accountability, a lack of development and significant resistance persists such that standardisation has remained a key issue [45, 39, 46]. NHS hospitals, like health service providers of many countries, struggle to find financial resources sufficient to employ adequate staff to meet patient demand [47, 48]. This makes the potential benefits from standardisation and the targeted treatment identification of LHS especially important in the struggle to contain budgets. Standardisation of care processes can significantly aid providers in their efforts to optimise quality *and* efficiency [49, 50] yet in many cases, and especially in the management of chronic diseases like diabetes, standardisation is lacking and as a result, care remains costly and suboptimal [51-53].

Traditionally, standards and innovation have been seen in juxtaposition: standardisation obstructs innovation while innovation disrupts the orderly nature of standards [54-56]. While it is widely accepted that standards are important, they are not well understood [57]. There is however, a growing understanding which this thesis will explore in chapter 2, that engaging with standards can support the goals and activities of innovation [54, 58, 56, 57].

### 1.1.2 Learning Health Systems

In July 2006 the Institute of Medicine (IoM)[3] convened the *Roundtable on Evidence-Based Medicine* committee with the intention *to help transform the way evidence on clinical effectiveness is generated and used to improve health and health care* [59]. Their overarching goal was that by 2020 more than 90 percent of clinical decisions will be supported with near real-time clinical data [59]. It was this committee that introduced the concept of LHS, which they promoted as necessary to meeting ever-growing demands for evidence to improve healthcare delivery, health outcomes and ensure the ongoing economic viability of healthcare services [59]. Committee members either alone or in small groups developed papers that were presented as chapter sections in the IoM *Learning Healthcare System: Workshop Summary* proceedings [59], each describing how a variety of elements that existed in healthcare environments of the day were necessary or could be used to help create LHS, including: large system databases; current evidence creation and gathering approaches like randomised control trials (RCTs) and medical device evaluation trials; mathematical modelling; pharmacogenetics; quality improvement processes from industry; the internet; clinical education processes; and most importantly, the Electronic Health Record (EHR).

The IoM defined the Learning Health System (LHS) as one *in which progress in science, informatics and care culture converges to continuously create new knowledge as a natural by-product of a care process in which best practice is applied for continuous improvement* [60].

In 2011 the IoM renamed the committee as the *Roundtable on Value and Science-Driven Health Care* and convened their second workshop, titled the *Digital Infrastructure for the Learning Health System: The Foundation for Continuous Improvement in Health and Health Care* [60]. With the same charter and vision statement and comprised of many of the same members, the proceedings of this second workshop focused more specifically on the role of digital health data systems and their ability to be the information backbone for LHS [60]. While they acknowledge their task as *building on previous efforts*, their introductions carefully skirted a significant issue: that very little progress had been made in the years since their original 2006 workshop [60].

---

[3] The IoM was established in 1970 under the USA's National Academy of Sciences charter to specifically study and address policy matters of medicine, healthcare conduct, and the health of the public. Annually, the IoM convene as many as 50 workshops where members with a range of multidisciplinary expertise are brought together for what are described as roundtable focused on a particular domain or issue of contemporary interest.

LHS remain largely an academic proposition; the best-known systems that could be classified as LHS solutions like IBM's Watson Health and Google's DeepMind Health, either fail to recognise they represent an LHS solution or avoid entirely identifying themselves within the wider LHS domain. Further, while a literature search for LHS finds many recent publications, the literature returned is not sufficiently representative to describe the entire domain and remains sparse when it comes to resolving why the precision-medicine approach of LHS [61, 62], a concept considered to be the next evolution of evidence-based medicine [63], remains an elusive goal. As publicly-funded health systems like those of the United Kingdom, Australia and New Zealand struggle to meet ever-increasing demand with shrinking budgets, we may come to appreciate the IoM's efforts to establish LHS as the natural evolution of evidence-based medicine: one capable of improving public health while reducing healthcare costs.

## 1.2 RESEARCH PROBLEM, HYPOTHESIS AND OBJECTIVES

### 1.2.1 Research problem

In spite of agreement on the importance of standardisation there has been little progress on standardisation of clinical care process specification documents. Standardisation of the building blocks, the underlying documentation and the data they collect, is an important and necessary step [64]. We want to know what effect this lack of standardisation has had on health information systems like EHR and LHS, and whether solutions to standardise clinical documents can mitigate the barriers that have prevented their adoption and use in clinical settings.

### 1.2.2 Thesis research hypothesis

The research hypothesis is that by understanding the ongoing barriers that have inhibited health IT implementation, it is possible to develop models to achieve the standardisation necessary for development and implementation of LHS with a new approach to standardising CCPS.

### 1.2.3 Objectives

The research objectives are:

1. Investigate and identify the lack of standardisation in CCPS;
2. Investigate the issues currently constraining implementation and clinical adoption of LHS, including:
    a. Explore and apply an approach for analysing barriers and benefits of health information technology (HIT) implementation;
    b. Identify approaches that have mitigated other HIT adoption barriers, and whether these may improve the success of LHS implementations.
3. Explore: (a) the relationship between CCPS, EHR and LHS; (b) the role of standardisation in that relationship; (c) their application in clinical practice; and, (d) a unifying model that integrates and focuses each on their application to clinical practice and treatment of the individual patient;
4. Investigate and evaluate an approach to standardising CCPS.

## 1.3 SIGNIFICANCE

Presently, 53% of all publications self-identifying in the LHS domain propose a potential solution that is yet to be built and tested [23]. Of the remaining, those that have been constructed in some form were most often built only to test the theory, almost none are in clinical use, and many could be categorised as primarily for identifying patient cohorts for use in other research. Of those systems in use which could be classified as LHS, such as the much-hyped IBM Watson Health, we first heard the unbelievable promises [65, 66] and then watched as tens of millions of dollars and half a decade delivered a system with narrow applicability that made unsafe treatment recommendations [67] and failed in its mission *to cure cancer* [68]. Such public failings should not be considered a failure of LHS generally. Rather, it represents a lesson we had to learn and which we should embrace on the path towards efficiency, quality and precision medicine. Identification of the issues that currently make LHS expensive and difficult to adopt is an important first step to resolving these problems and setting about the implementation of solutions that make cost effective, accessible and accurate precision medicine available to all healthcare service consumers.

## 1.4 RELATIONSHIP BETWEEN OUTPUTS AND THESIS OBJECTIVES AND OVERALL NARRATIVE

This section relates the research objectives described in Section 1.2.3 with the novel contributions listed in Section 1.5.3 and overall narrative of this thesis.

Contributions 1 and 2 focus on resolving requirements for Objective 1. Contribution 1, presented in Chapter 5, provides a stable base from which to define and describe CCPS and in doing so, identifies that the lack of standardisation extends beyond structure and content to even include the names used for particular clinical documents. Contribution 2, also in Chapter 5, exemplifies the proposed taxonomy in a case study describing CCPS for the care of those with Diabetes Mellitus in the United Kingdom.

Contributions 3 and 4 apply directly to Objective 4 in that they develop, apply and evaluate an approach termed *TaSC* to standardisation of one type of CCPS, the caremap. Contribution 3, which describes the standardised approach to the development, structure and presentation of caremaps is presented in Chapter 8. Contribution 4 is the subject of Chapter 9, and presents both quantitative resource assessment and qualitative survey evaluation of the standardised caremap.

Contributions 5 and 6 occur within the scope of Objective 2 and are presented collectively as components of *the Heimdall Framework* in Chapter 6. Contributions 5 and 6 describe and validate a complete taxonomy of LHS solutions, while Contribution 7 presents the results of literature reviews of EHR and LHS literature into their barriers, benefits and facilitators.

Contributions 8 and 9 apply to Objective 3. Contribution 8 comparatively analyses the barriers, benefits and facilitators for EHR and LHS using a new extension to ITPOSMO that is described in Section 3.3.2 and applied in Section 6.4 under the title *ITPOSMO-BBF*. Contribution 9 resolved the *LAGOS framework* which is presented in Chapter 7.

## 1.5 NOVELTY

The thesis makes several novel research contributions in the areas of *standardising clinical documentation* and in identifying and mitigating *the research community awareness challenge* that has restrained LHS. We summarise these contributions here and link them to the seven publications (listed on page ix) that are direct outputs of the thesis.

### 1.5.1 Standardising Clinical Documentation

There is extensive literature on approaches and benefits that can be realised from standardisation of healthcare processes and procedures. However, while there has been an attempt to standardise the definition and identify the structure of a single clinical care process specification (CCPS), the clinical pathway, confusion continues with respect to the broad range of CCPS that are currently in use. This thesis reviews the literature and provides a new taxonomy defining and describing all relevant CCPS and demonstrates application of that taxonomy in the context of a case study in documentation for Type 2 Diabetes. This research is the focus of Publication 1 listed on page ix.

Further, this thesis presents a thorough examination of the history, development and current state for another type of CCPS, the caremap, and presents one solution for standardising caremap structure, content and development that is distilled directly from analysis of a large pool of literature and clinical expertise. This research is the focus of Publication 2.

While validating caremaps with clinicians, the presence of latent clinical decisions within activity nodes at points where a caremap path diverges, and which guide selection of the appropriate path for a given patient, were discovered. Indication of clinical decisions necessary to identification of the patient's progression through the caremap gave caremaps greater utility, increased their accuracy and made them more approachable for clinicians. This research is the focus of Publication 7.

### 1.5.2 The LHS Research Community Awareness Challenge

Review of proceedings from two health informatics conferences highlighted that most authors at both failed to appropriately identify the proposed solutions of their papers as LHS. Discussions with some of the authors at one of those conferences highlighted that many were unaware of LHS, and for those that were, they were unaware of an identifiable comprehensive taxonomy that could be used to rule solutions as LHS, or indeed as a particular type of LHS. This issue is described in this thesis as *the research community awareness challenge* and was the focus of Publication 3.

While a review of the literature identified three papers that had attempted to produce descriptions for the different LHS types, there were distinct differences between all three such that none provided a complete picture of the domain. This thesis presents a taxonomy describing nine novel types of LHS and validates that taxonomy against a collection of 240 published solutions that self-identified as LHS. This research is the focus of Publication 4.

It was clear from the literature that in contrast to EHR, LHS as a research domain was still in its infancy. While extensive research had looked at the benefits, barriers and facilitating factors for successful implementation of EHR, only anecdotal observations existed in the literature for LHS. This research collected and analysed those observations and contrasted them with those described in systematic reviews of the EHR literature, finding that LHS are inheriting many of the same challenges from EHR, and that the majority of facilitating action is directed towards promoting or resolving human factors issues, and very little facilitating effort is directed towards resolving barriers that lead to improved efficiency, patient safety or health outcomes. This research was the focus of Publication 5.

Discussions with clinicians during the conduct of this research identified that there was a lack of understanding regarding where the different LHS and CCPS applied in clinical care, and how they were related to the levels of medical application: population medicine, evidence-based medicine and precision medicine. This thesis presents a framework which integrates and describes LHS and CCPS in the context of their relationship to the levels of medical application. This research was the focus of Publication 6.

### 1.5.3 Summary of New Contributions

Nine contributions are presented in this thesis:

1. A taxonomy for CCPS;
2. A case study demonstrating application and validation of the CCPS taxonomy;
3. An approach to standardising one type of CCPS, the caremap;
4. Case studies demonstrating application, and a convenience survey of clinicians to assess the usability and accuracy of the standardised caremap model;
5. A taxonomy for LHS;
6. Validation of the LHS taxonomy against all research self-identifying as LHS;
7. A systematic review to identify the Benefits, Barriers and Facilitating factors (BBF) for EHR and LHS;
8. Comparative analysis of the BBF for EHR and LHS;
9. A framework to guide integration of new healthcare tools.

## 1.6 THESIS OUTLINE

Figure 1 illustrates the structure and organisation of this thesis. This thesis consists of five major parts. **Part 1** consists of chapters 1 and 2. It describes the problem under investigation and sets the context by illuminating the background and history of standardisation in healthcare. **Part 2** presents the overall methodology and approach for the conduct of this research. A number of analysis methods are also described in detail. **Part 3** begins with chapter 4 which presents literature reviews into CCPS, EHR and LHS that are used in chapter 5 to demonstrate the lack of standardisation in clinical documentation. The results of these literature reviews are used to support development and presentation of a taxonomy and characterisation for clinical documentation. **Part 4** consists of chapters 6 to 9 which present the frameworks, standardisation approach, evaluation and results of this thesis. It begins in chapter 6 by presenting a taxonomy, characterisation and unifying framework for LHS, followed by identification and analysis of the barriers, benefits and facilitating factors for EHR and LHS implementations. Chapter 7 presents a framework to guide integration of new healthcare tools. Lastly, chapter's 8 and 9 present and evaluate a new standardisation approach for one type of clinical document, the caremap. **Part 5** consists only of chapter 10, and presents a review of the thesis, its contributions, and a conclusion.

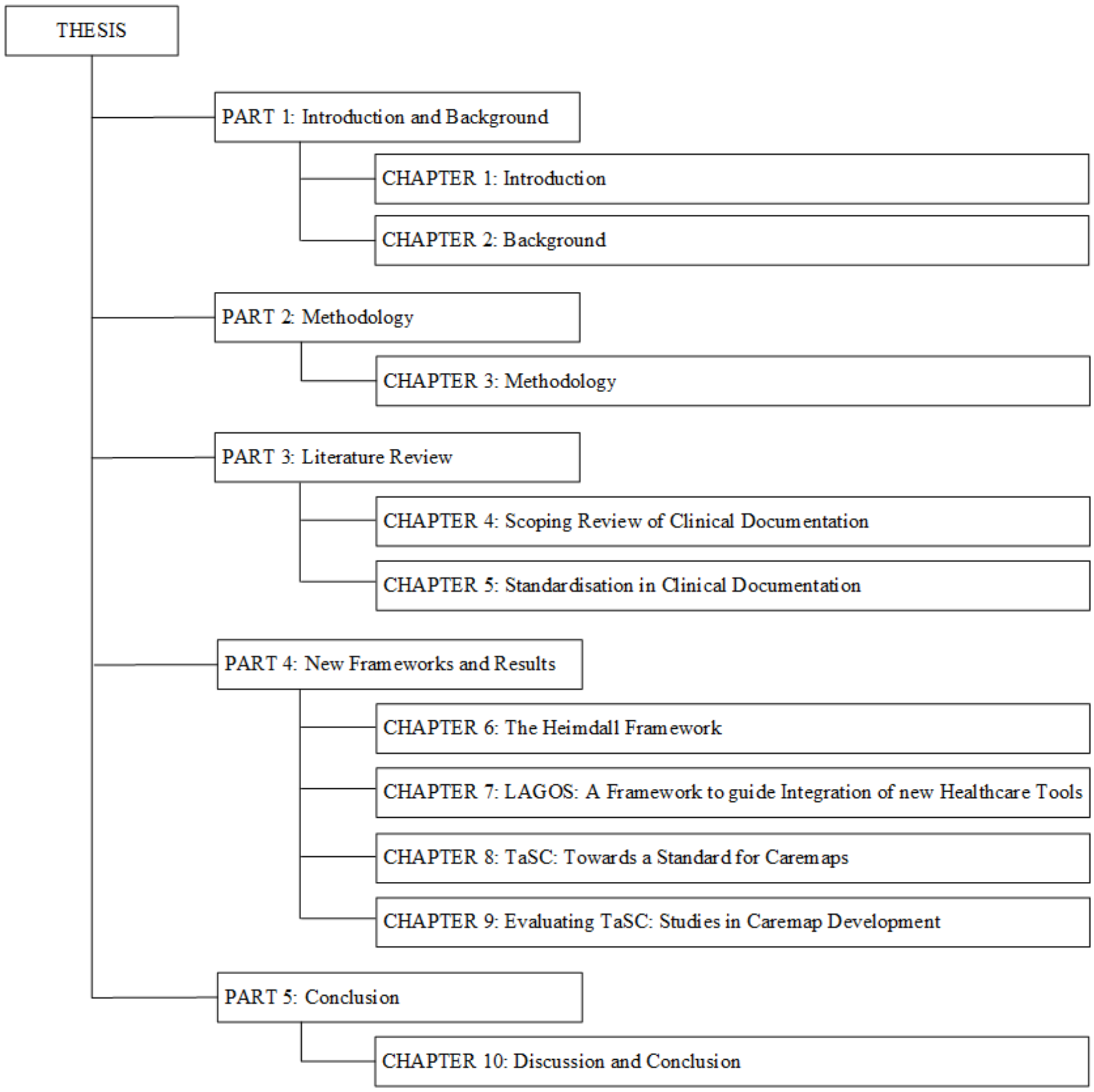

Figure 1: Thesis Structure

# Chapter 2: Background and Motivation

This chapter begins with discussion of the background and motivation for this research (section 0), followed by discussion of the interplay between standardisation and innovation (section 2.2).

## 2.1 THE DRIVE TOWARDS STANDARDISATION

Florence Nightingale transformed healthcare through the introduction of formal and descriptive documentation that would become a vital component of effective care delivery [69]. Clinical documentation, also known as clinical care process specifications (CCPS), were introduced to healthcare to streamline, record and improve healthcare delivery. CCPS developed from the notation and reporting methods of Nightingale's early nursing practice as a structured method to record the treatments performed and observations concerning the response of the patient to that treatment. Since then CCPS have continued to develop extensively, sometimes in ways that have had no direct relationship to patient care [69, 70]. It is suggested that clinical documentation represents only one essential component of the patient record [71]. That the practice of documenting care: (i) influences information collection and reasoning strategies; (ii) increases diagnostic accuracy; and (iii) can prompt clinicians. In doing these things it may also mitigate deficiencies in clinical practice [72, 73].

Unfortunately, CCPS have been hampered by a lack of standardisation in nomenclature, document structure, content, and in some cases, development. Clinical documentation gives form and structure to the processes that underpin population and evidence-based medicine (EBM). Electronic health records (EHR) are a health information system (HIS) created to replace paper and give digital form to clinical documentation [74-77]. Learning health systems (LHS) are a more recent evolution of HIS capable of identifying new knowledge for application in precision medicine from data-mining large collections of EHR [61, 62, 78].

Some CCPS present as templates to guide evidence-based care, while others provide templates onto which individual patient needs, metrics or treatments can be charted. CCPS serve many purposes, the most important of which is to ensure continuity and quality of patient care [79, 69]. EHR are CCPS and the patient care data normally captured on them in digital form [80, 81]. They capture into an electronic record the patient histories, allergies, acute health problems, test results, observations, measurements, treatments, prescription records and other information traditionally recorded in separate CCPS in the patient file [82, 83, 81]. Learning Health Systems (LHS) require EHR. They operate on the acquisition and use of knowledge data mined from collections of aggregated, filtered and analysed EHR [84]. LHS can be as simple as a rule-based monitor to identify and notify health regulators when a patient is diagnosed with a communicable disease or suffers an adverse event (surveillance) [85-87], or as complicated as a precision medicine system that identifies prospective cohorts of *like patients* from which to predict in near real-time treatment decisions that have a higher probability of being curative for the individual patient (cohort identification) [86, 87]. LHS could have as great an effect on patient care as Nightingale's advances in sanitation and note-taking during the Crimean War. However, as we come to the end of their first decade, LHS are yet to begin realising their true potential.

Standardised approaches to documentation, paper or digital, can ensure that each time a clinician approaches a particular CCPS, the content and format meet expectation, can be read more quickly, and are better retained; all of which improves patient safety and outcomes [88, 89]. Standardisation is a key mitigant for clinical challenges and enables clinicians to concentrate efforts on accurately identifying disease or developing appropriate treatments [90, 91]. Standardisation of clinical documentation improves the availability of information for subsequent care providers, reduces documentation errors, and improves patient safety and outcomes; yet efforts towards standardising clinical documentation are often resisted [92, 45, 93]. The call to standardise in healthcare is one which is repeated often in the literature; calling for standardisation of clinical terminology [91, 94], treatment approaches [95, 96], consistency in clinical documentation [97], or provision of uniform methods for the conduct of clinical investigations (tests) [98, 99] and units of value for reporting of measurands[4] [100, 101]. It is difficult to comprehend how many of these CCPS can be intended to standardise different aspects of the conduct, management or reporting of clinical care, when the nomenclature, content and method of construction for CCPS are themselves not standardised. This research is motivated both by a wish to see what LHS can become, and a belief, that whilst initially anecdotally based on more than a decade of experience in health informatics has been observed by others [102], that the failure to adopt LHS has possibly resulted from their inheritance of the unresolved familiar challenges of EHR and CCPS.

## 2.2 STANDARDISATION VERSUS INNOVATION

Standardisation and the use of established standards are ubiquitous in our daily lives [58]. Examples include the USA's CAFÉ and similar international fuel economy standards used to govern efficiency and emissions of new motor vehicles offered for sale [103]; standards instituted for terminology and language, especially for mission-critical applications like satellite and aeronautical navigation systems [104] and air traffic control [105]; and standards used to ensure safe development, testing, production, prescription and administration of medicines [106-108]. Standardisation has been described as *the activity of establishing and recording a limited set of solutions to actual or potential matching problems directed at benefits for the party or parties involved in balancing their needs and intending and expecting that these solutions will be repeatedly or continuously used during a certain period by a substantial number of the parties for whom they are meant* [109]. Standards generally consist of rules, guidelines, templates or characteristics for activities, or their results, that are provided for common and repeated use [110].

***Innovation*** involves the development and implementation of a new or significantly improved product, service or process, and includes all scientific, technological, organisational, financial and commercial steps which are, or are intended, to lead to its implementation [54, 111]. Innovation in technology and strategy is both a catalyst for modern economic growth [112, 54], and standardisation [113, 114]; yet standardisation, especially that which is unofficial or voluntary, is believed to be something that innately inhibits innovation [54, 115]. It is this last belief which has benefited from contemporary reconsideration: growing insight into the role standardisation plays in enabling innovation [54, 58, 55-57].

---

[4] One simple example is that of $HbA_{1c}$, which is used to monitor glycaemic control in diabetic patients. While there have been calls from as early as 2007 [98] for the reporting unit of measurement to be standardised. $HbA_{1c}$ is reported by some laboratories as: millimoles per litre (mmol); a percentage (%); or an estimation of average glucose (eAG).

A variety of approaches have demonstrated the beneficial role standards can have in supporting innovation. *Interoperability standards* describe how different components in an ecosystem work together - for example, the hardware and software in ICT systems [57]. *Anticipatory standards* describe the operation and interoperation of components of future systems not yet in operation [116]. *Formal standards* are high-quality but have a considerable development lead time as they are carefully deliberated by standards-writing organisations, such as the International Standards Organisation (ISO) and International Engineering Task Force (IETF) [117]. *De facto standards* autonomously stem from processes and interactions within the ecosystem, such as the dominance of Microsoft's operating system in personal computing or resilience of the QWERTY keyboard layout which while being originally designed to mitigate adjacent keys jamming on early mechanical typewriters, is still seen on devices like touchscreens which have no moving parts [117, 57]. Standards can also be described in terms of their *particularisation* or *extent to which they are standardised*: whether the organisation, service or approach is, for example, *wholly* or only *largely* standardised [118].

Motor vehicle production and use is constrained by a great many standards: directing safety, materials application, pollution, operation and maintenance. The same standards governing fuel efficiency and emissions discussed earlier, and which have removed many vehicles with inefficient large-bore engines from sale, actually stimulated innovation. This innovation includes the recently released homogenous charge gasoline compression engines using a system described as Spark Controlled Compression Ignition (SCCI). SCCI is claimed to reduce fuel consumption by as much as 20% [119, 120]. The standards also produced competition in innovation with another major vehicle manufacturer also releasing new technology this year, the Variable Compression Turbocharged (VC-T) engine [121]. There were also innovations that delivered the fully electric vehicle (fEV) by Tesla - a product that sits in a market space that must continue to innovate in order to meet anticipated standards requiring all passenger/commuter vehicles to become electric [122, 123]. While the standards discussed operate to ensure that motor vehicles marketed today cause less pollution, they do not, for example, inhibit a manufacturer's choice of colour, luxury options or the model name that might adorn your next vehicle. And as we have seen, far from inhibiting innovation, standards can beneficially support novel innovations.

While standardisation does not have to inhibit innovation (and can in principle lead to greater innovation), in much of the medical profession standardisation is seen as an inhibitor to professional expertise and innovation. In part, this is due to clinician resistance: with approaches at care standardisation derided as *cookbook* or *cookie-cutter medicine* that some say can only be effective after they have set aside the unique needs of individual patients [124-127]. In spite of current issues with clinical overuse and the financial crisis pervading healthcare service delivery, standardisation of key documentation could help healthcare service providers deliver managed care which is seen to reduce incidence of inappropriate care, resource consumption and overall cost [128, 129]. Yet healthcare remains one of the slowest industries to adopt process standardisation or to demonstrate that it has positive effects on patient safety and outcomes [130, 131, 127].

The British Medical Association and Royal College of Nursing developed a joint guidance stating that use of standardised forms is beneficial in reducing variation in healthcare practice [132]. The use of different versions of the same clinical care specification in different units within the same care facility, and between different care facilities, may not previously have been seen as such an important issue. However, in this increasingly digital healthcare environment we are seeing greater amounts of data being generated and captured daily, including from diagnostic devices used, or sensors worn, by the patient while in the community. Any differences in the documentation approach or data recording method results in fragmented data, complicates the integration of data about the same patient from different sources, and inhibits health information exchange (HIE) [133, 134].

When it comes to the practice of medicine, a large array of standards applies to almost every action a clinician undertakes. Built on a base of clinical practice guidelines, evidence-based medicine is perhaps the most broadly applied and well-known standardisation in medicine [135, 136]. Current efforts for standardisation in health informatics often focus on some element of: (i) how the clinician interacts with the system; or, (ii) data entry or composition. Our search for and review of systematic reviews of EHR implementations described later at Section 4.2 was unable to identify any mention of consideration of the potential for underlying issues to have arisen when non-standardised clinical documentation was digitised in the creation of EHR solutions, which if mentioned would have been in the context of being a barrier to their adoption. While this thesis focuses on standardisation of clinical documentation and the benefit this can bring for those seeking to develop and implement LHS, clinical documentation should be more thoroughly investigated as a potential source for a range of barriers that previously constrained EHR adoption, and which continue to prevent LHS from gaining a foothold in clinical practice.

## 2.3 THE RELATIONSHIP BETWEEN STANDARDISATION AND LHS

Our underlying assumption that standardisation of CCPS is a necessary and beneficial task is certainly not new. Many authors working on CCPS have been making the call for standardisation for several decades, as demonstrated by the proposal presented at the Conference on Guideline Standardisation (COGS) in 2002 [135]. Standardisation of CCPS can have direct implications on how that CCPS document could record information capable of supporting LHS. For example, consider the recommendation made by authors in the domain of orthopaedics [137, 138]. They suggest a call to standardise the methodology, classification, terminology and structure of documentation for reporting clinical complications, adverse clinical events and deviations from the expected post-treatment healing process would improve the accuracy of communication of complication events between clinicians. Moreover, where the proposed CCPS standardisation is then inherited into the formal structure and data definitions used to construct the EHR, this standardisation approach would also make available improved and consistent data capable of reuse in LHS [139]. This would enable multiple LHS uses, for example, that could: (i) investigate instances where the care provided may have been sub-optimal; or (ii) monitor for, and identify, instances of contaminated medicine, faulty or poorly manufactured medical devices or highlight clusters of post-surgical infections indicative of a site where sterilisation procedures may have broken down (a type of LHS defined as *Surveillance*).

We will show (in Chapter 6) that issues of data integrity, integration and interoperability are known to be significant barriers to LHS adoption. Standardisation of CCPS may represent a significant facilitating factor for mitigating these data-related issues. While this thesis is not the first work to propose a relationship between the lack of standardisation in CCPS and the heterogeneity in data representation in EHR [139], it is the first to draw the direct line from CCPS all the way to LHS and propose that resolving the first may be a strong factor in development and adoption of the last.

# Chapter 3: Methodology

Selection of the research approach, or methodology, is an important step for guiding research development, progression and outcomes, and enables the researcher to clearly describe both research and procedure [140, 141]. Many schools of research have previously coached students in the tenets of the *incompatibility thesis*: that qualitative and quantitative methodologies were in stark competition and must never be mixed in the same research [142, 143]. However, more contemporary proponents encourage a mixed approach as the path to becoming a more pragmatic researcher [144]. This chapter describes the methodology adopted to achieve the aims and objectives of the research. Section 3.1 provides the research design while also defining and describing the qualitative and quantitative research methodologies identified in the research approach; Section 3.2 discusses the overall research approach used and the stages in which the methodologies were implemented; Section 3.3 lists the instruments used or created in the study and justifies their use; and, section 3.4 discusses the ethical considerations of the research as well as its problems and limitations.

## 3.1 RESEARCH DESIGN

This research uses a number of established analytical methods:

### 3.1.1 Scoping Review

The scoping review synthesises and analyses a wide range of literature in order to extract evidence and bring greater clarity to a topic of inquiry [145]. The scoping study is often used to guide more focused lines of research, such as the systematic review [145]. By contrast, the systematic review uses a relatively narrow range of quality-assessed studies focused on answering well-defined questions where appropriate study designs are identified in advance. The scoping review is less interested in the quality and design of included studies that are used to address broader, less-specific topics [146]. The scoping review was used in the initial stages of this work in the opening literature review which identified many of the works that present LHS solutions do not identify within the LHS domain: the community awareness challenge described in Section 1.4.2. Knowledge gathered during the scoping review informed the search terms and initial concepts at the commencement of the systematic review.

### 3.1.2 Meta-narrative Review

For research with a complex and diverse collection of literature it can be difficult to understand the problem as a whole and requires exposure of tensions, mapping of diversity, and communication of complexity [147]. Meta-narrative review (MNR) seeks to explain seemingly disparate and conflicting data encountered in literature [148]. MNR attempts to contextualise terms, interpret relationships between different concepts, and recognise the impact of assumptions made by authors [148]. Meta-narrative review was used during the initial reading of EHR systematic reviews in order to understand and align the often heterogenous review and reporting methods used in each of the 23 works to discuss what was ostensibly the same topic.

### 3.1.3 Content Analysis and Thematic Analysis

Content analysis and thematic analysis are separate but interrelated qualitative approaches for descriptive data analysis with low levels of interpretation [149]. Content analysis (CA) is an accepted systematic coding and categorisation method for investigating texts and resolving a quantitative description of the features [150, 149]. CA establishes categories and then records the instances in which that category is evident or can be inferred from within the collected texts being analysed [150]. Thematic analysis (TA) is a more qualitative method used to identify, analyse and report patterns, or themes, that emerge as being important within the material being analysed [151, 150, 149]. TA provides the systematic element characteristic of CA, while additionally affording the ability to combine analysis of frequency with analysis of *in context* meaning, therefore providing a more truly qualitative analysis [150]. CA and TA are established methods regularly used in clinical, nursing and other healthcare contexts [151, 150, 152, 149].

Authors often represented the same general idea in a variety of ways. While CA resolved many common concepts, it is only once TA was applied that these concepts could be described by their underlying contextual themes. The following example analyses how authors discussed one of the most often-mentioned concepts, *cost*.

**EHR:**

1. Reduce RCT cost [153]
2. Offer low-cost, high-volume nuanced answers [153]
3. The initial price of the system should not be the overriding consideration as the organisation should be willing to avoid purely cost-oriented vendors as costs will soon mount [154]
4. EHR systems come with huge investment costs [154]
5. HIT improve quality and care coordination and reduce costs [155]
6. Physicians practices are slow to adopt EHR for a variety of reasons including high costs [155]

**LHS:**

7. Reduce cost of service delivery [85]
8. Cost-effective, secure health information exchange represents an important interoperability sub-challenge [156]
9. There were multiple additional systems-based considerations (e.g. startup costs) [157]
10. Integrating [our LHS] into this clinic costs approximately $50,000 [157]
11. Reduce costs by providing more efficient care [158]
12. Initial cost of implementing the registry was high [157]
13. Reduce the cost of conducting clinical trials [159]

While CA finds that many authors identify the concept *cost*, TA analyses its usage, in this case resolving two contrasting contexts.

Hence, implementing EHR and LHS:

1. can reduce the cost of healthcare delivery (1,2,5,7,11,13)
2. carries a high initial cost burden (3,4,6,8,9,10,12)

The first contextual theme presents as a *benefit* that comes from engaging the health information system, while the second clearly represents a *barrier* inhibiting adoption of these systems. In this way, common benefits and barriers to the adoption and use of EHR and LHS were resolved and analysed across the two collections of literature.

### 3.1.4 Case Studies

The case study (CS) is an ideal methodology used in many domains including information sciences and comes with a well-developed history and robust qualitative procedures for investigation and process validation [160-162]. They are a method for conducting and presenting comparative research into subject areas that include qualitative and mixed-mode information science inquiry [163]. CS are a valid methodology where the rigid approach of experimental research cannot or does not apply [160, 162, 164]. Each CS presents with a unique use case and reporting structure, and a range of case study types exist, including: exploratory, explanatory, descriptive, intrinsic, instrumental and collective [162, 164].

Computer science, information systems and information science are domains characterised by constant change and innovation [165]. All too often, IS researchers learn by studying the innovations developed and implemented by practitioners, rather than by providing the initial wisdom to underpin these new ideas [165]. Case studies allow researchers to capture the knowledge of practitioners, and using a broad variety of data sources to ensure the knowledge is looked at through multiple lenses, generalise this knowledge in the development and evaluation of new theories and approaches learned from it [166, 165].

The case study examples used in this research draw on knowledge and evidence developed from and with the assistance of clinicians in the domains of obstetrics and midwifery. Clinical practice guidelines were initially systematically evaluated [167]. With the input of clinical experts, novel clinical document standards for development, structure and content are evaluated through the conduct of case studies to produce compliant clinical documents.

## 3.2 RESEARCH APPROACH

This research is divided into the following stages:

**Stage 1:** Scoping and meta-narrative review with meta-analysis to:

1. Identify literature on each clinical documentation type, seeking articles that will help:
    a. Describe conception and intended purpose for the document type;
    b. Describe the structure, content and development process for the document, and;
    c. Demonstrate application of the document in a clinical context from inception to today.
2. Identify literature that presents scoping or systematic reviews of EHR implementations on which this research will;
    a. Use Concept Analysis (CA) and Thematic Analysis (TA).
3. Identify literature that proposes or presents an LHS solution, from which this research will;
    a. Use CA and TA to develop a classification and descriptive taxonomy that will allow those designing and implementing health informatics solutions to identify their work within the LHS domain.
    b. Use CA and TA to identify: (1) proposed benefits of LHS use for clinicians, healthcare organisations and consumers, and; (2) barriers that have inhibited adoption of LHS specifically.

**Stage 2:** Develop and present:

i. Standardised definitions and nomenclature for each identified type of clinical documentation;

ii. A methodology for standardised development, structure and content for those clinical documents where the literature demonstrates absence or discordance;

iii. A taxonomy describing current and foreseeable types of LHS;

iv. Quantitative presentation of the identified *barriers* and *benefits* for both EHR and LHS, along with the *facilitators* identified from EHR literature.

v. A unifying framework that effectively demonstrates the application of LHS within the multiple layers of a clinical learning organisation.

**Stage 3:** Evaluation:

i. Case study evaluation applying the clinical document standardisation method to creating artefacts for clinical examples in midwifery and obstetrics.

ii. Quantification of LHS solutions by classification of all those proposed or presented in the LHS literature review.

iii. Contrast, comparison and analysis of the barriers and benefits described for EHR and LHS to identify similarity and consistency and identify whether LHS present with similar issues and challenges as has been observed for EHR, and whether these issues and challenges may have been inherited from EHR.

iv. Quantitative and qualitative assessment of the application of facilitators in the domain of EHR, including evaluation of whether the type and approach to facilitating EHR may be applied successfully to increase the success of LHS implementation.

The above research approach was applied to each of the clinical documentation types identified in Section 5.1. For example, for caremaps (which we cover in detail in Chapter 8), the approach is summarised in Figure 2.

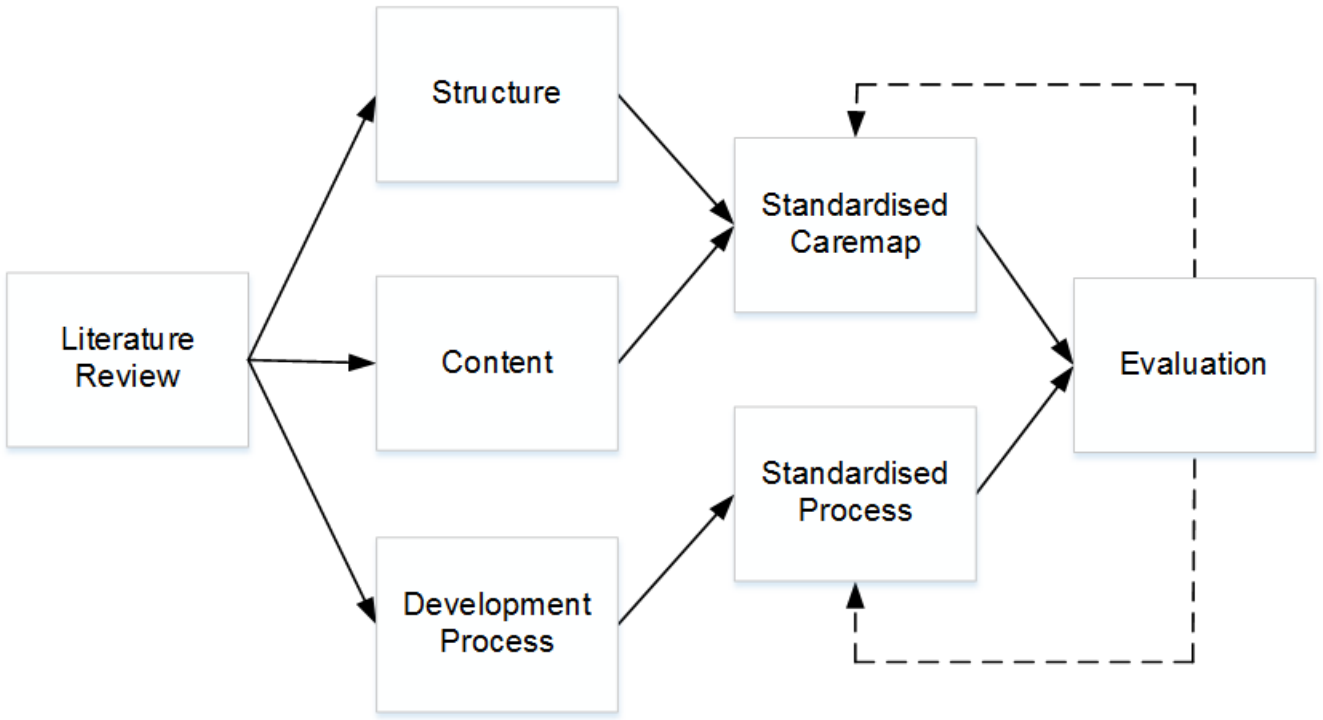

Figure 2: Research Process - Consensus formation and evaluation for the standardised caremap

In this example the research focused on the three components listed in Table 1 whose characteristics would come out of the TA and make up the review framework:

Table 1: Clinical Documentation Components for Standardisation

| | |
|---|---|
| Structure | What is the representational structure and notation for expressing the contemporary clinical document? |
| Content | What content types are consistently seen in the contemporary clinical document? |
| Development | What are the steps followed by authors to develop this contemporary clinical document type? |

The literature review is a ground-level consensus-forming function allowing researchers to identify implementation techniques and assess the degree of agreement within a domain [168, 169]. In this way, the literature pool of the caremaps study was used to identify common definition, structure and content elements, and a standard for the overall look and feel of caremaps. In addition, process steps that were consistently described led to a standardised caremap development process.

## 3.3 INSTRUMENTS

### 3.3.1 Formal Concept Analysis (FCA) Spreadsheets

The literature review process produced spreadsheets with the details and, for LHS papers, a brief description of each paper. CA then data mined the literature to inductively populate the spreadsheets initially in the form of formal concept analysis charts [170-172]. TA extended these with context, making each concept of interest identified from the literature more generic.

### 3.3.2 ITPOSMO-BBF

While many approaches exist for evaluating implementations of new systems, services, or the commission of evidence into practice, e.g. [173-175], one was sought which had received sufficient attention in the literature to demonstrate robustness and flexibility in application. ITPOSMO was originally proposed and developed for evaluating the gap between expectation and reality in government IT projects, but has also been used to evaluate EHR projects and help explain why health information systems have succeeded or failed [176]. In the information systems literature ITPOSMO is a widely-used framework for evaluating implementation challenges: for understanding critical factors that affected the success or failure of technology projects [177]. The ITPOSMO framework identifies seven dimensions for exploring the gap between a system's design and the reality of its implementation: (1) Information; (2) Technology; (3) Processes; (4) Objectives; (5) Staffing; (6) Management, and; (7) Other factors [178]. The seven dimensions are arranged into four aspects, each composed of related dimensions: IT-PO-SM-O.

Precedent exists for expanding ITPOSMO to enable additional scope and functionality, including: service quality analysis [179], survey-based study of consumer and public perceptions [180] and Socratic analysis of local e-government [181]. This work in Chapter 6 adapts ITPOSMO to identify barriers that have impeded implementations of health information technology, and the actions, described as facilitating factors, that authors have sought to use in their mitigation.

## 3.4 ETHICS AND LIMITATIONS

### 3.4.1 Ethics

The conduct of this research does not depend on or require the primary participation of human subjects, nor the secondary use of their personal health information.

### 3.4.2 Limitations

One key limitation of this work is that it is largely theoretical. While the clinical documents that are produced as part of this work can be evaluated and assessed by clinicians, the results presented that relate to EHR and LHS can only be properly evaluated in the future, and with some benefit of hindsight. This is because while it is a relatively simple matter to redevelop CCPS it may be some time before the standardisation changes propagate into the next generation of EHR, and before we might begin to see any effect on the adoption of LHS solutions in clinical practice.

## 3.5 SUMMARY

This chapter has introduced the methods applied generally across the research reported in this thesis. The methodologies were selected with close attention to the aims and objectives of this research. The research approach was described and two of the primary instrument types created during this research were introduced. The ethics requirements were then discussed before the chapter closed with discussion of the primary limitation arising from the study.

# Chapter 4: Scoping Review of Clinical Documentation

While literature on CCPS, EHR and LHS presents each as its own separate and isolated domain, we identify the existence of interdependency and, in many ways, inheritance between each. In this chapter we apply the methods described in section 3.1 to undertake a systematic review of each domain. Sections 4.1, 4.2 and 4.3 review the current state of CCPS, EHR and LHS respectively. Section 4.4 looks at whether other authors have attempted to develop a framework to unify CCPS, EHR and LHS. The chapter concludes in section 4.5 with a summary.

## 4.1 CCPS

A search was conducted using Scopia, Science Direct, PubMed, EBSCOhost, DOAJ and Elsevier for literature discussing the definition, development or use of a range of clinical document types including: caremaps (*caremap*, *care map* and *CareMap*), pathways (*care*, *clinical* and *critical pathways*), guidelines (*clinical practice guidelines*), plans (*care plans*), rules (*clinical decision rules*), and protocols (*care protocol*). In each case literature was sought where authors had attempted to: (a) define the document type; (b) detail the inputs used in development (evidence, literature, team composition, etc.); (c) identify the intended patients, conditions and audience; (d) detail the aim, goals and intended outcomes, and; (e) highlight potential benefits and barriers encountered during development and operationalisation of the document. To identify the scope of author positions on whether clinical care documents were the same or distinct, we collected data on how authors described and discussed them in this context as shown in the table presented in Appendix D. Where authors described that a document type was also known by or the same as others, these were indicated in green. Where they provided narrative that differentiated or distinguished different types, this is indicated in red. Separate rows are used to record subsequent instances identified in the same text being reviewed. In our primary spreadsheet we also collected any definitions authors provided for each document type, as well as descriptions of input materials, goals, uses, and whether they provided an exemplar document for the reader.

Many of the CCPS reviewed in this research were developed and refined during the 1980's and 90's in response to a number of key needs, including: (a) to control costs; and, (b) improve the quality of patient care [182, 32, 183]. Using project management (PM) and total quality management (TQM) tools more common to industry, hospital managers sought to reengineer the processes of patient care with the aim of reducing clinical resource overuse and error rates, along with concomitant improvements in patient outcomes [16, 17, 183, 19]. In spite of potential for these reductions, clinical costs have continued to increase and error rates persist with distressing frequency [30].

CCPS, also described as care process modelling, arose from a need to evaluate the quality of healthcare delivery [184, 185]. Documentation of care processes brings healthcare quality improvement as it helps those involved in patient care to develop a shared understanding of care to be provided for the patient, and a reference for identification of areas for future improvement [186-188]. The intention is to ensure that services provided to patients result in the desired health outcomes, consistent with current clinical knowledge [185]. Deviation from quality standards established in the CCPS is described as variation, and the presence of variation is the driving force necessitating adaptation of clinical change described in revised CCPS [189].

A range of CCPS are in use in clinical practice. However, there is vast disagreement in the literature regarding their titles, standard content and construction methods. Many terms are used for what are describe collectively in this research as CCPS, each of whose remit could generally be described as *the documentation of care processes for improving healthcare quality and outcomes*. These terms include: *clinical practice guidelines* (CPG) [190, 191] which are also known as *consensus-based guidelines* (CBG) [192] or *local operating procedures* (LOP) [193, 194]; *clinical decision rules* (CDR) [195]; *clinical pathways* [196]; *care plans* [197]; *treatment protocols* [198]; and *caremaps* [39]. Some authors describe *care pathways*, *clinical pathways*, *critical pathways*, *care plans* and *caremaps* as synonymous and interchangeable terms [199-202], while others consider them to be representative of separate documents [203, 204, 19]. This terminology confusion is further epitomised when we observe diagrams that internally describe themselves as one CCPS type, in this example [205, 206], *caremaps*, but which are captioned as one or more other types by the author.

Standardisation of definitions, presentations and development processes for most types of clinical documentation is lacking, and while for some a standard has been attempted, these attempts are either incomplete or have only further added to the confusion [207, 208, 39, 135]. Electronic health record (EHR) software products are notorious for their lack of standardisation; in general [209, 210], in process and presentation [211], in terminology [212], in structure [213], and in content, format and nomenclature [210]. Inconsistency, changes in format between like documents and poorly designed materials only increase ambiguity and create greater confusion for clinicians [29, 89, 214]. Where clinicians must perform processes and tasks differently it is difficult to achieve consistency, and it is only once those processes and tasks become better documented and understood that standardisation can occur [215]. It is disturbing to note a trend in the literature that recognises that implementation of EHR has seen many healthcare organisations neglect the quality of their paper-based records: extinguishing all efforts at business process re-engineering that would standardise or improve the structure, format and information captured by the clinical documentation that their EHR is based on [216, 217]. As one author succinctly puts it:

> "...computerising medical records in their current state will create more problems than it solves: *a mess computerised is a computerised mess*." [216]

Caremaps, clinical and critical pathways, clinical flow diagrams and nursing care plans are observed with vastly different content and appearance: within the same journal, from hospital to hospital, and even from ward to ward within the same hospital. While there is much literature presenting examples of each type of clinical document, and texts exist describing the development of some traditional and text-based types, a gap exists in the literature with regards to describing and standardising the development, structure and content for contemporary forms of some documents. Even though standardisation of healthcare practice, procedure and tools is seen to improve healthcare in terms of quality, outcomes and accountability, significant resistance persists such that standardisation remains an issue [45, 39, 46]. This should not be surprising when the issue of confusion in distinguishing these clinical tools persists [39].

## 4.2 ELECTRONIC HEALTH RECORDS

An additional literature search was conducted using the terms ("electronic healthcare record" along with "barriers" and "benefits" and "facilitating factors"). This was followed up by a second search replacing "healthcare" with "health". Articles were included where they presented as a scoping or systematic review of EHR implementations and provided analysis and discussion of all three elements: barriers, benefits and facilitators. Those not meeting these requirements were rejected. A further requirement was to limit the EHR review to articles published during the ten years since the IoM's initial LHS report, in order to align the EHR literature with that contemporaneously published for LHS. Figure 3 presents a PRISMA diagram showing the outcome of the EHR literature search, which resolved a pool of 26 articles for this review.

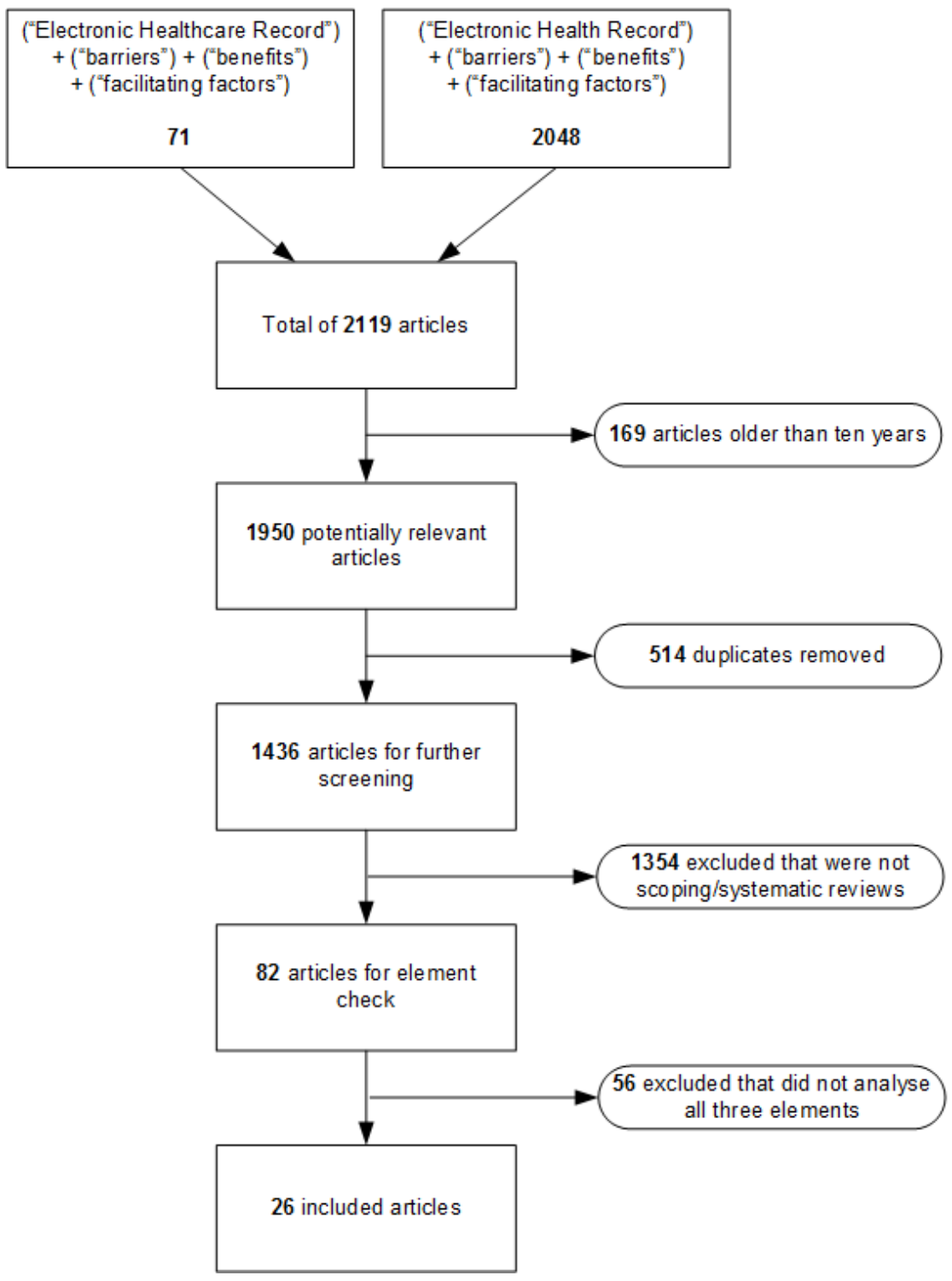

Figure 3: EHR Literature Selection (PRISMA)

EHR have become ubiquitous in healthcare, yet many hospitals and clinics still employ a mix of both digital and paper-based recordkeeping [218-221]. Among those EHR implementations in the hospital setting, many offer only limited functionality and occur as isolated islands of information, with separate EHRs tied to a particular ward, medical specialty or care pathway [222]. There have been some spectacular failures to realise the initial promise of EHRs, as seen with the UK's National Program for IT wherein the National Health Service (NHS) failed to deliver effective national hospital EHRs [223]. Health services seeking to implement EHR have long complained of slow adoption and limited implementation success rates [156, 86, 224, 225].

The design of many EHR applications, particularly those from the USA, come as an extension of health service billing software and often conflict with the needs of clinicians, leading to both workflow and information presentation challenges in clinical use [226]. The data contained in medical records typically exists within a specific context and there can be limitations on applying it for use in another context, or significant work needed to support that application [227]. Many commercial EHR solutions have proprietary, technological and cost barriers to integration with other systems and sources of health data relevant to the individual patient [228, 229]. With the best of intentions, many healthcare organisations have self-inflicted these issues by layering inflexible new technology over existing processes and procedures in the belief that EHR implementation meant simply replacing paper records with digital systems [230, 219]. Underlying this has been a lack of understanding about how clinicians interact with computers, and general disagreement as to whether such interaction enables or inhibits patient-centred care [231]. Successful implementation of new EHR requires both clinician- and user-led processes that re-evaluate practices and procedures, with a requisite period of adaptation and training for those who will use the resulting combination of new IT systems, documentation procedures and clinical workflows [232, 233].

Numerous studies have found that EHR increased efficiency while also improving quality of care, accuracy of clinical documentation and access to patient information [234-239]. Well implemented EHR also reduced healthcare delivery costs, resource use, and clinical errors [240, 234, 235]. Some studies reported improved communication between clinicians which acted to increase patient safety [240, 236-238], while a small number reported EHR increasing or enhancing patient confidentiality [241, 238].

Chapter 6 reports the review of the literature collection identified herein, finding that almost every author whose work presents an EHR implementation spoke of barriers or challenges. Overcoming these barriers has been a significant challenge to implementing EHR [242]. Organisations have invested heavily in efforts to identify and engage facilitators they consider could lead to successful implementations [234, 242, 243]. Some barriers are seen as endemic of, or specific to, the IT industry [243], while others are thought to require change in health policy and legislation [241-243].

## 4.3 LEARNING HEALTH SYSTEMS

A literature search was performed using the search terms ("LHS" and "learning healthcare system"), seeking works that proposed or presented an LHS solution. An initial read of abstracts was used to reject duplicates and papers not related to the central topic, including those that used the search term in context of learning in the academic or education sense [244, 245]. Conclusions and methods were then reviewed, seeking to remove papers that did not present or propose an LHS, for example: those exploring the medicolegal, ethical or societal aspects of LHS [246-248] Figure 4 shows that a total of 230 articles were identified for inclusion in the LHS part of this study.

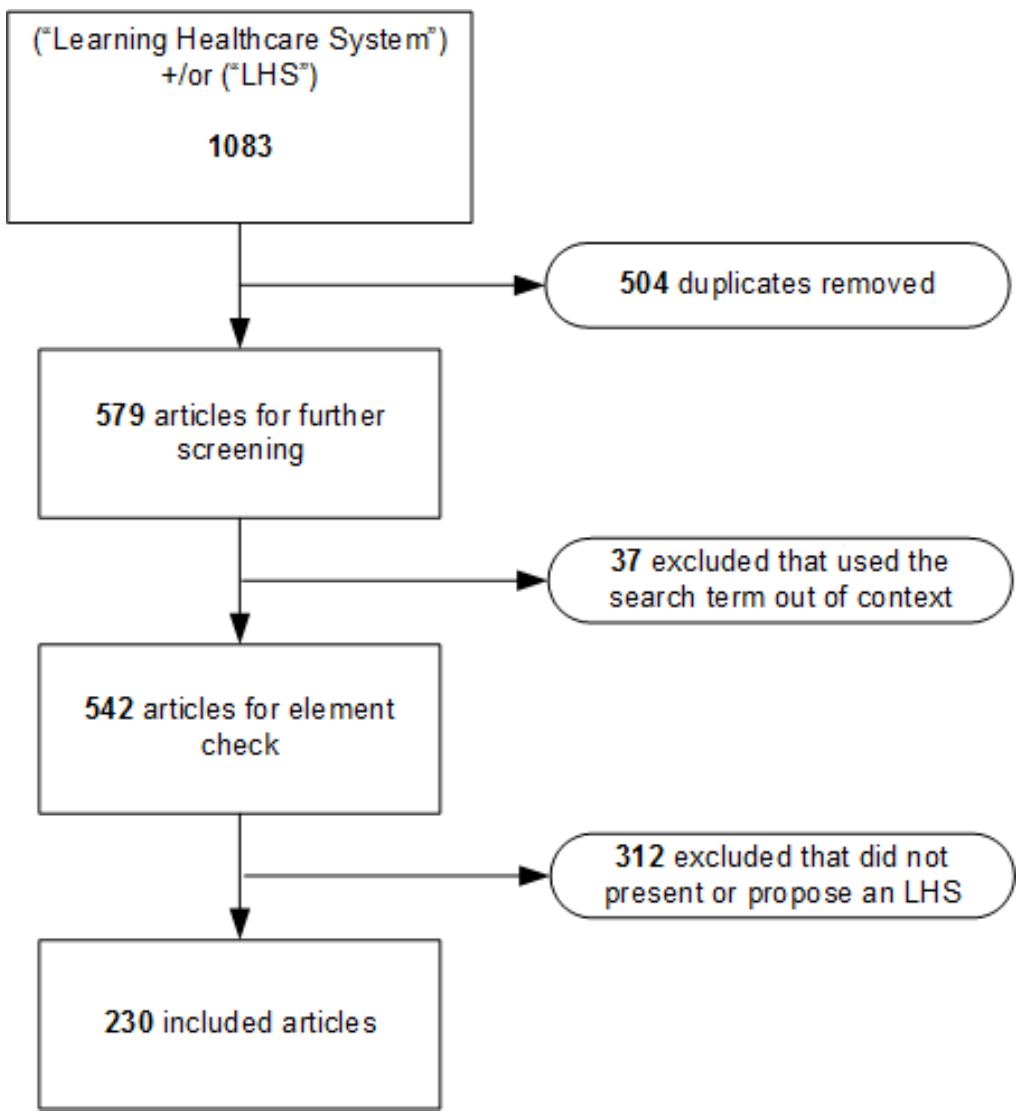

Figure 4: LHS literature selection (PRISMA)

Advancement of evidence-based medicine (EBM) was promoted as the primary driver for LHS [59] yet the IoM definition for the concept found in section 1.1.2 fails to describe attributes that would contribute to quality, safety, efficiency and efficacy of patient care [86, 23]. The four fundamental attributes listed in Table 2 provide significantly more tangible metrics for conceptually comprehending the LHS.

Table 2: Four Elements of an LHS (adapted from [249])

| | |
|---|---|
| 1 | An organisational architecture that facilitates formation of communities of patients, families, front-line clinicians, researchers, and health system leaders who collaborate to produce and use big data; |
| 2 | Large electronic health record and routinely collected healthcare datasets (big data); |
| 3 | Quality improvement for each patient at the point of care brought about by the integration of relevant new knowledge generated through research, and; |
| 4 | Observational research and clinical trials done in routine clinical care settings. |

Figure 5 shows that references to LHS in literature were initially quite limited. While some much earlier texts including one authored by Florence Nightingale in 1863 provide methodological references for health outcome monitoring and benchmarking suggestive of what would later become LHS [250, 251], the first true use of the term comes in 2007 around the time of the IoM's initial LHS report. However, it wasn't until the IoM's second LHS report in 2011 [60] that we begin to see a significant uptake in publications referring to or about LHS. The primary purpose for their second LHS report was to summarise presentations and discussions that had occurred during a series of roundtable workshops to identify opportunities, challenges and priorities for the development and application of new health information systems (HIS). HIS that would constitute one of the various types of LHS [60].

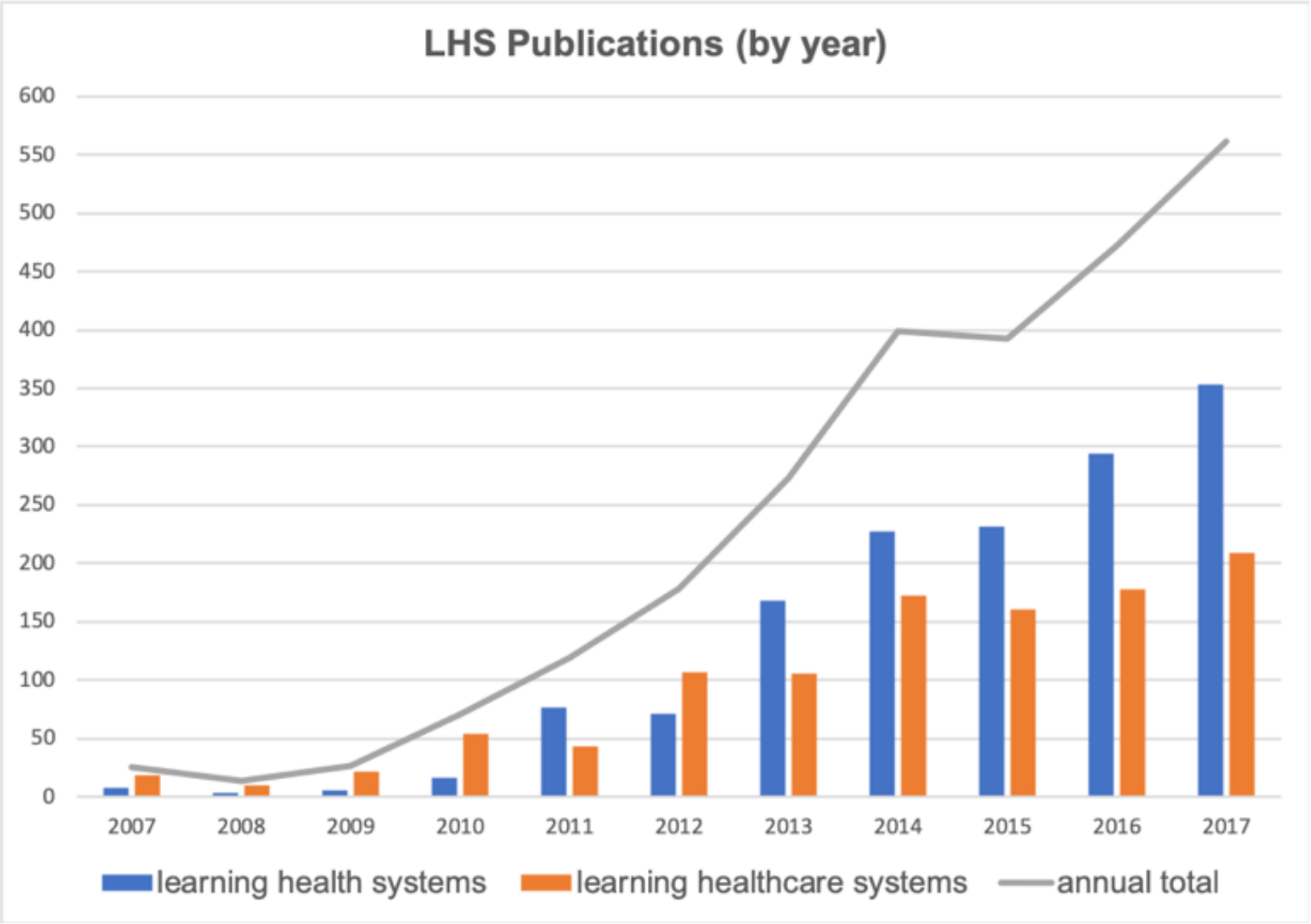

Figure 5: LHS Publications (by year)

Coincidentally, and with no reason given, the title for LHS was slightly amended in that report. *Healthcare* was shortened to just *health*. Figure 5 also shows that while use of *learning healthcare system* had been more prevalent in the years after their first LHS report and prior to the second, starting in 2011 and seemingly as a result of the title change in the second IoM report, *learning health systems* became the more commonly used nomenclature. It is notable however, that while the nomenclature has matured, definitions for LHS have not continued to develop and continue to be largely based on the descriptions provided by the two IoM reports [23].

LHS are seen to embody the core principles of 4P medicine, which are medical information systems that have the ability to be *predictive*, *preventative*, *personalised* and *participatory* [252, 253]. 4P approaches personalise medicine by identifying those at greater risk for complications for the purpose of targeting interventions [253, 59], but while many 4P approaches present in the literature as clinical decision support systems (CDSS) [254-256], CDSS are only one of a number of types of LHS [257, 86]. In parallel, maturation of technologies such as large datasets, machine learning, prediction and decision support, and ongoing increases in processing power will further enable development and use of LHS [252, 61, 84].

LHS are promoted as bringing numerous benefits for all stakeholders in healthcare. For clinicians, these include assessment of which laboratory or imaging investigations may be of greater diagnostic value given a particular patient's presentation and symptoms [258], reductions in the risk of causing patient harm from prescribing errors [248], and increased awareness of the potential and effects of pharmacogenetics [259]. The ability to use LHS technology to record, compare, contrast and present information in near real-time enhances the analysis and decision phases of the clinical care process. Patients benefit from clinicians having access to advanced knowledge developed from the experiences and outcomes of cohorts of past patients, the decrease in time taken to arrive at the most appropriate diagnosis, and reduced resource use and costs [260, 258, 261]. Others see the financial burden to implement and support health information technology (HIT) including LHS [234, 241, 237], coupled with the persistent need for data and systems standardisation [262, 237, 238, 263], interoperability [242, 264, 263] and integration [242, 243] as collective barriers that prevent broader LHS adoption.

LHS represent a vision to transform healthcare [265, 60]. It is suggested they could be the most significant development in healthcare since the advent of evidence based medicine (EBM) in the early 1990's [266, 23]. The LHS vision requires leveraging ongoing developments in EHR by developing new knowledge from the ever-increasing amounts of routine clinical data being accumulated [153, 265], innovating medicine from the slower population-based process of EBM towards rapid near real-time precision medicine [265, 267]. Realisation of some LHS also requires access to or development of clinical documentation, for example: development of some LHS requires clinical protocols, caremaps and guidelines [248, 268, 269, 39, 270]. While EHRs have become pervasive in all aspects of healthcare service delivery and contain ever-increasing amounts and types of health data [271, 272], they represent a rich and underutilised resource demonstrating that the transformational vision LHS promises is struggling to be realised [273].

Review of publications of a number of health informatics conferences held during 2017 showed that much of the published work that should naturally fall within the domain of LHS fails to be captured under that umbrella term [274]. Review of 28 papers from Proceedings of the 2017 International Conference on Health Informatics (ICHI'17) [275] resolved that 15 (54%) presented works that could be classified using the taxonomy presented in Section 6.2.2 as an LHS yet no single author identified their efforts as such. A similar review of 60 abstracts from the Medical Informatics Europe (MIE) 2017 Informatics for Health conference found that while 41 authors (68%) of that conference also present content classifiable as an LHS, only 2 authors (3%) actually described their approach thus. It is possible that this failure to identify work within the domain of LHS is due to a lack of awareness and understanding of the domain. Lacking awareness of the domain results in authors failing to recognise that their works are actually LHS. This limits the ability of other researchers to identify all relevant works from within the broader field of health informatics (HI) during LHS literature searches.

Presently, there are more than 7000 discoverable publications annually discussing topics concerning EHR, while LHS struggles to reach 700. Figure 6 shows this dramatic disparity in numbers between EHR and LHS publications when these terms are entered into a university library search engine. When considered in the light of authors who have described the lack of ontology or classification systems for LHS, it is clear that there is a lack of awareness around LHS which may explain why the LHS domain is avoided even by authors who are aware of its existence [276, 277, 274].

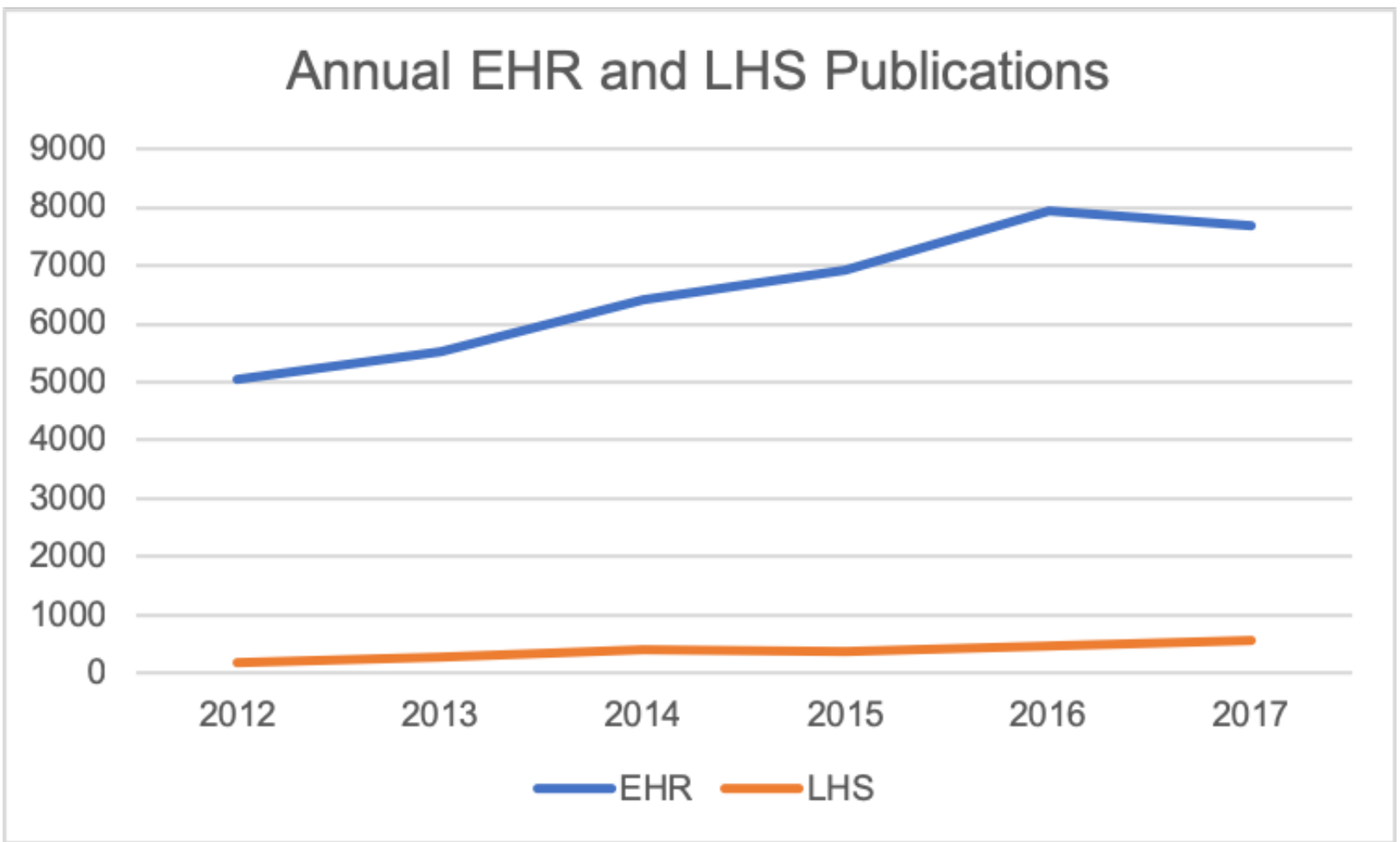

Figure 6: Annual LHS and EHR Publications

## 4.4 BRINGING CCPS, EHR AND LHS TOGETHER

We found no papers that had investigated or developed a practical or operational framework to unify LHS, the established levels of medicine practice described in Section 7.1.1, and common CCPS documentation. While other works have characterised subsets of LHS or particular CCPS, no work was located that characterised either in the context of medical application. And even though this thesis investigates all three, additional effort is required to integrate them into a single unified framework that can improve understanding and success in health informatics and LHS implementation.

## 4.5 SUMMARY AND IMPLICATIONS

In this chapter we have seen that confusion exists in the name, structure and content for clinical documents causing vast differences in their quality, applicability and usability. We also propose that clinical documentation suffer from a lack of standardisation, and that true standardisation in structure, content and development processes is beneficial if clinical documentation are to have a positive effect in the delivery of effective and safe patient care. This leads to the question: *How can clinical documentation standardise care, when the clinical documents themselves are not standardised?*

EHR were promoted to clinicians as a means to transition healthcare from paper to digital recordkeeping. It was asserted that they would deliver improved efficiency, accuracy, outcomes and patient safety because they provide easy access to important information, treatment orders and a complete, searchable record of all patient information. Given their beginnings as an enhancement of hospital billing information systems, significant barriers including clinician resistance have resulted in slow adoption of EHR. Multiple disparate EHR systems exist and even where two hospitals have contracted with the same supplier, variance can be seen in the presentation of patient and clinical information at each. Integration between systems is expensive, time consuming and has thus remained a major challenge to those seeking a truly universal EHR. Therefore, while EHR are now ubiquitous in healthcare settings, we must ask: *why is it that national health systems have not adopted a standard for EHR?* We should also consider *why it is that after almost four decades of EHR, their adoption still represents a significant challenge for health service providers?*

LHS are a natural evolution from EHR. Large collections of EHR are required in order to realise operational population-to-evidence-to-precision methods that LHS offer. Many LHS also require clinical documentation that is representative of current clinical best practice, clinical practice guidelines, and the processes and procedures of patient care delivery seen in caremaps and clinical workflows. While there are many examples in the literature of proposed LHS solutions, many others do not identify within the domain. Adoption rates for LHS in clinical settings remains quite low. This research further asks: *are LHS simply inheriting the challenges identified from research of their underlying antecedents, clinical documentation and EHR?* and, *is it possible to standardise those antecedents and identify facilitators previously employed in the mitigation of barriers to EHR that might be beneficial to implementation success rates for LHS?*

# Chapter 5: Standardisation in Clinical Documentation

Appropriate use of CCPS should see patient and clinician actively engage in a cycle of healthcare that can be represented as a learning circle as shown in Figure 7, in that CCPS in differing amounts and degrees: (1) guide, direct and inform patient care; (2) record the care that has been provided; and, (3) capture the present condition of the patient and their response to treatment. This last step informs the process that guides and directs clinical decisions about the next round of patient care, thus repeating the cycle as a continuous process until the patient is restored to their particular version of health.

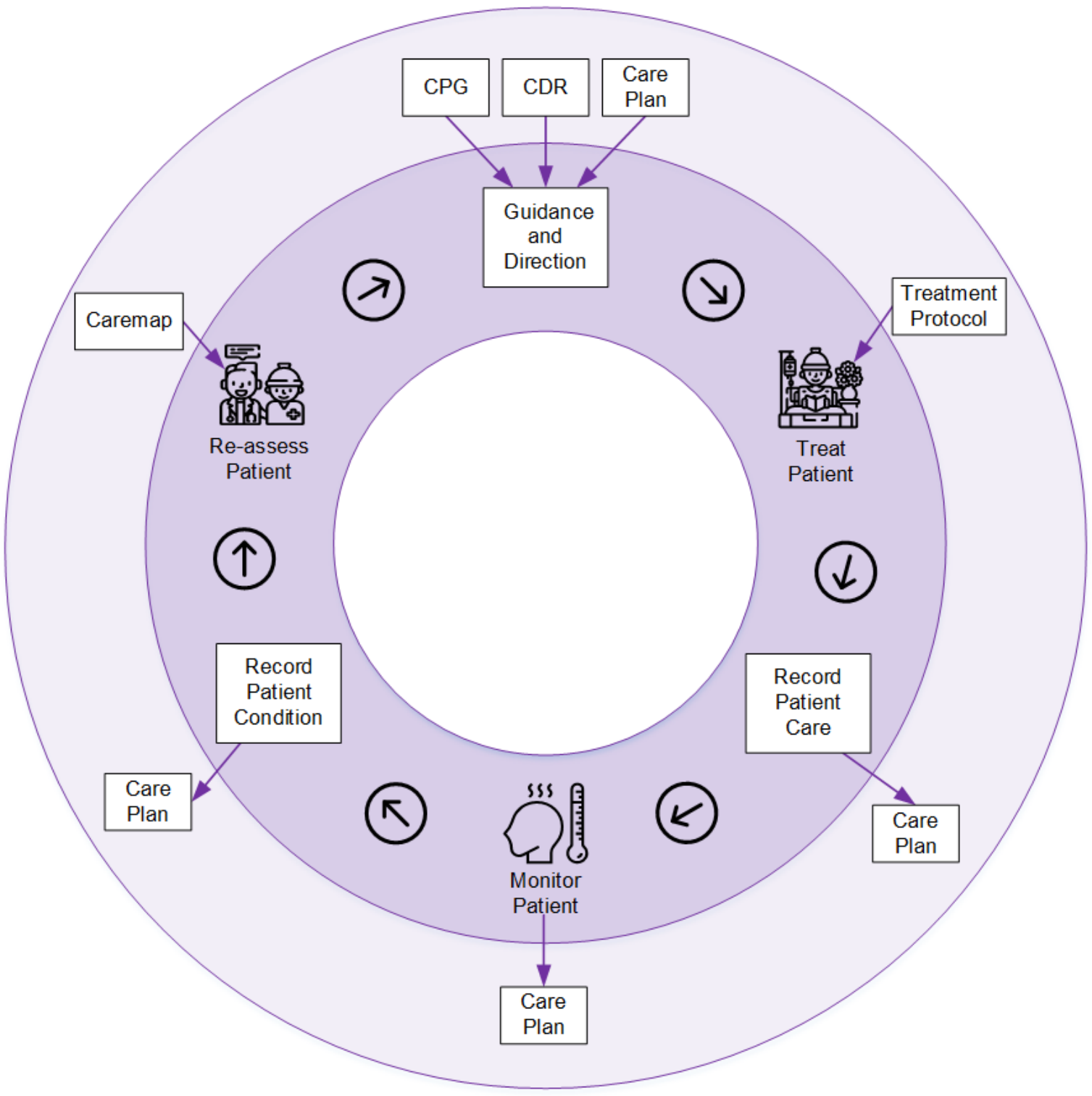

Figure 7: Clinical Documentation Learning Circle and interactions with CCPS

This chapter investigates the broad range of clinical documentation, seeking to resolve the aspects lacking standardisation identified in Section 1.1.1. This chapter asks: *Is it possible to resolve the nomenclature confusion and characterise the content and structure of these documents, and in doing so, create a standard for CCPS?*

Section 5.1 seeks to resolve the lack of standardisation in nomenclature for CCPS. Section 5.2 extends an existing approach developed to classify and characterise one type of CCPS, the *clinical pathway*, for application to the broad range of CCPS identified in section 4.1. This extended approach is used in development of a taxonomy (section 5.3) and characterisation (section 5.4) for CCPS. Section 5.5 demonstrates and evaluates this taxonomy and characterisation through application in a case study on CCPS for Type 2 Diabetes Mellitus in the United Kingdom. The chapter concludes with a discussion (section 0) and summary (section 5.8).

## 5.1 THE LACK OF STANDARDISATION IN CCPS

This section seeks to address the lack of standardisation in nomenclature for CCPS introduced in Section 1.1.1 and resolved in section 4.5 through a review of the literature collected by the search described in section 4.1.

### 5.1.1 Policy

Often seen as the remit of politics, policymaking can be driven or impeded by highly charged and competitive ideologies and value systems that mean the resulting policies are not always based on clear and convincing scientific evidence [278, 279]. International and national organisations can give rise to health policy [280-283] that can be applied at both a national and local level, with local policies often being representative of aspects of national or regional policy made relevant to specific local populations [284, 285, 283]. Policy can be directly determinant of a population's health and health outcomes, especially where it is used to regulate the distribution of health services in terms of human and material resource allocation, how healthcare services are funded, and the accessibility of services for patients [286, 283].

### 5.1.2 Clinical Practice Guidelines

It is generally accepted that clinical practice guidelines (CPG) are systematically developed statements based on critical assessment of scientific evidence, experience and consensus and they are intended to assist in decision making regarding appropriate care for patients in specific clinical circumstances [278, 287-291]. CPG seek to provide clinicians with a vital shortcut to identifying the underlying science that informs particular clinical decisions [288]. The hierarchy for CPG shown in Figure 8 was developed as part of this work. Although the hierarchical view has been adopted, this work also established potential for other possible views and perspectives, such as that of levels of customisation in which one guideline may exist at each level as a customised instance of a version at another level. Furthermore, in such a view a CPG may originate at any level in Figure 8 such that customisation may occur as a generalisation or specialisation process to suit the population or individual, respectively. Thus, a health facility-based CPG may become regionalised, nationalised or globalised while global CPG may be nationalised, regionalised or localised at a healthcare facility through customisation, i.e., generalisation or specialisation respectively. A key point to note here is that as you move from left to right in Figure 8, the patient population for which the CPG applies reduces and becomes less heterogeneous while the guideline becomes more specific as a direct result of customisation.

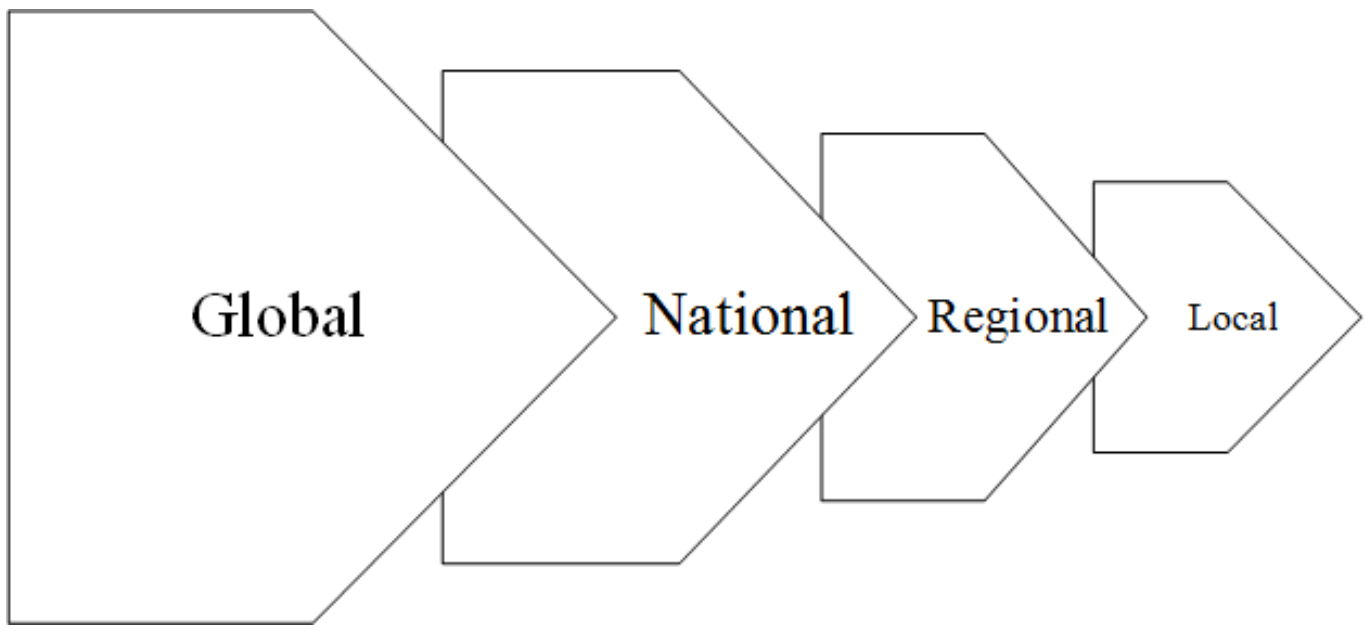

Figure 8: Hierarchy for Clinical Practice Guidelines

At the highest level are *global guidelines*: international guidelines published by organisations such as the World Health Organisation (WHO) [292, 293] and World Federation of Haemophilia (WFH) [294], and multinational guidelines from groups like the European Society of Cardiologists (ESC) [295]. International and multinational guidelines tend toward general or globally-applicable recommendations for the diagnosis or management of specific conditions or clinical presentations.

In combination with local synthesis of evidence, global guidelines are regularly used in the production of *national guidelines*, which are endorsed for and intended to inform consistent clinical care across entire countries and health sectors [296, 297].

*Regional guidelines* draw on health policy and use national guidelines in the production of a more operationalised form of CPG dealing with demographic population needs in specific states, counties, or municipalities [298].

*Local guidelines* is the term used herein to represent those CPGs developed by and for use in hospitals, general practice clinics and other patient-facing locations. They provide operationalised, instantiated evidence and clinical knowledge drawn from regional, national and global guidelines and are targeted to support and improve clinical decision-making and care delivered directly at the patient's side [298]. It is not uncommon for patients with the same condition to receive different treatment at different hospitals, and for different healthcare providers to disagree on what the best evidence may be in a given situation [299]. For this reason local guidelines may also reflect particular elements of organisational culture, and it is not unknown for hospitals to manipulate or tweak guideline diagnostic or treatment threshold values either: (a) where it may make the guideline more approachable to clinicians and increase adherence; (b) as a response to elements previously identified from clinical errors or; (c) where based on past experience, senior consultants feel it may aid in the avoidance of future poor patient outcomes and potential litigation [299-301, 298].

### 5.1.3 Clinical, Critical and Care Pathways

While CPGs and clinical pathways both provide a defined approach for a specific medical condition or treatment process, clinical pathways were a separate clinical care process specification (CCPS) developed in the 1980s and applied to healthcare practice with a distinct focus on: (a) the use of benchmarked outcomes; (b) cost containment through efficient resource use; and, (c) management of patient care within a structured hospitalisation period to produce a reduced length-of-stay (LOS) [302, 287, 303].

Clinical pathways was the term most frequently described as being the same or synonymous with others studied in this review, primarily caremaps. Clinical pathways was also the term most often distinguished as different from others, particularly CPGs. Confusion within authors' narratives as to whether clinical pathways, critical pathways and care pathways represented the same [304, 305, 200, 306] or distinct [278, 307] specifications was identified. An attempt at analysis of the definitions provided by authors was made but given broad variance in defining elements used across the literature pool this was not probative and had potential to increase this confusion. While some authors provided simple and direct statements, describing their particular pathway as a paper-based document or blueprint [16, 308, 309], others gave more comprehensive descriptions discussing systematic use of clinical evidence, the types of clinical staff and motivations driving the development and use of their pathways [304, 17, 310].

There was very limited evidence from which to infer the existence of minor operational differences between the three terms. In this analysis the term *critical pathways* was observed more often in situations where the care is short, intervention- or event-based or requiring of some urgency, i.e. in emergency medicine, surgical or intensive care domains [305]. The term *care pathways* was most often observed in the nursing domain describing the day-to-day routine of care provided for patients. And *clinical pathways* appeared to be more broadly applied and more holistic in that authors sought to address a wider range of patient issues than just focusing on this particular illness or clinical event. They were also observed across a range of care domains including: nursing, oncology, community-based care and mental health.

Despite these merely operational distinctions it would be disingenuous to identify pathways as three separate objects. [304] also investigated the nomenclature issue for pathways, determining that while *critical pathways* was the internationally accepted term, more than half of all authors in their study had shown a preference for *clinical pathway*. For all these reasons this thesis treats them as the single term, *clinical pathway*.

### 5.1.4 Caremaps

The term caremap refers to a graphical representation of the sequence of patient care activities to be performed for a specific medical condition suffered by either a patient or a cohort [311-313]. Caremaps have been in use, in one form or another, for around forty years [17, 314, 315]. The literature suggests they originated in the nursing domain, incorporating and extending the critical pathway method and bringing established project management methodologies into healthcare delivery [16, 17, 19]. Caremaps aim to standardise health care practice by organising and sequencing care workflow, ensuring standard of care, timely interventions and uniform outcomes using an appropriate level of resources [316, 207, 314, 311]. Caremaps also help track variance in clinical practice, as they provide a simple and effective visual method for identifying when care practice has deviated from the routine evidence-based pathway [317, 311].

Definitions drawn from the literature of the early-mid 1990's agree in principle that the caremap *presents as a graph or schedule of care activities described on a timeline and performed as part of the patient's treatment by a multidisciplinary team to produce identified outcomes* [316, 318, 314, 319, 311-313, 315].

Caremaps are observed in the literature using three different titles: (1) *caremaps*; (2) *CareMaps*, and; (3) *care maps*. The first appears to have been the original title prior to a Centre for Case Management (CCM) report and trademarking of the second in the early 1990's [316, 320]. In literature published after 1994 using *caremaps* it is not uncommon to see some mention of CCM or their trademark [321], although this is not always the case [322, 323]. It is possible that the use of *care maps* has come about as a defence to any potential issues that might arise from confusion with the trademark, as authors did not use this third title in context or reference to CCM [324, 311, 325].

Just as the literature presents three different titles for caremaps, it also describes three origins for caremaps, with distinct points of intersection between each that make it difficult to assess which may be the true history:

(1) Caremaps were an output of the Centre for Case Management (CCM) in 1991 [320]. CCM's CareMaps were similar in form and function to existing clinical pathways and were applied to specific patient populations that were commonly treated in many hospitals [320]. CCM went on to trademark the double-capitalised version CareMap but had not within the first decade undertaken any research to demonstrate the effectiveness of the concept whose invention they claimed [326].

(2) Caremaps naturally evolved as an expansion of earlier case management and care plans [315].

(3) Caremaps were developed during the 1980's at the New England Medical Centre (NEMC) [327, 313].

Caremaps arose in nursing where they incorporate and extend the critical pathways and bring established project management methodologies into healthcare delivery [202, 111, 328]. Indeed, from the early 1980's nurses were the primary users of caremaps [125].

Early caremaps were text-based and holistic. Rather than centring on only the immediate primary diagnosis or intervention, nurses developed these early caremaps to focus on the entire scope of care that might be necessary for the patient during hospitalisation. *Traditional caremaps* considered elements such as anxiety, rehabilitation, education, prevention and coping strategies and were intended to restore the patient to a normal quality of life [329, 330, 311-313].

In the second half of the 1990's clinicians began to realise that creating caremaps was easier for surgical procedures than other in-hospital care scenarios [331]. That caremaps would evolve in form and function had been expected as information technology and EBM developed [313]. From 1999 examples can be found of *transitional caremaps*: caremaps that whilst still being text-based, have reduced their focus to interventions that are part of the primary diagnosis [207, 208, 321].

Caremaps continued to evolve as graphical representations. *Contemporary caremaps* emerged as separate but complimentary components to clinical pathways and clinical practice guidelines (CPG) [320, 323]. Recent caremaps come linked to or provide graphical flow representation for the clinical practice guideline (CPG) or surgical event [332, 317, 325]. While retaining the purpose and flow and still conforming to the general principles of the caremap definition, many that are found annexed to CPGs have begun dropping the caremap title [333-336]. A summary and comparison of caremap evolution stages is shown in Table 3.

There were numerous examples of contemporary caremaps in the literature and annexed to hospital-based CPGs. Contemporary caremap literature tended towards focusing on establishing the clinical condition that justified creation of the caremap, such as: determination of incidence, risk factors and patient outcomes [332]; diagnosis and stabilisation of patients with an acute presentation [337], and; protocolising of ongoing treatment [325]. Presentation or discussion of the development process and elements for construction were rare, and more often had to be inferred from a thorough reading of each paper.

We found a single article written by a veterinarian and a lawyer which attempted a systematic description of the process for contemporary caremap development [338]. This article primarily focused on standardisation for the purpose of cost containment, and provides an example of mapping for a surgical procedure [338]. Given their focus and particular caremap construction model which, through their own exemplar application, only includes a temporally-ordered single-path representation of the gross steps of patient care, their paper was at best, merely formative. By their own admission, they deliberately limited relevant data analysed during the input design phase to what they felt was truly critical for identifying and understanding outliers. This results in a model lacking clinical applicability and a distinct lack of detail surrounding each care process. Their method requires significant work to adequately support true standardised clinical caremap development.

Table 3: Summary and comparison of caremap evolution stages

| | **Traditional**<br>(1980's to mid-1990's) * | **Transitional**<br>(Mid-1990's to mid 2000's) * | **Contemporary**<br>(2004 onwards) * |
|---|---|---|---|
| **Primary Author** | Nurses | Nurses and Doctors | Doctors |
| **Context** | Holistic | Primary condition | Single diagnostic, screening and/or intervention event. |
| **Foci** | Restoring the patient to normal life | Outcomes, cost and resource consumption | Efficiency of care delivery and outcomes, reduction of practice variation, bridge gap between evidence and practice |
| **Presentation** | Text-based | Text-based with some early flow examples | Flow diagram or graph |
| **Status** | Independent document | Independent or sometimes incorporated with CP document | Self-contained but often found appended to/contained in CPG |

CP = Clinical Pathway, CPG = Clinical Practice Guide
*All dates are approximate ranges

### 5.1.5 Clinical Decision Rule

Clinical Decision Rules (CDR) assist clinicians in their decision-making role by providing a link between published clinical evidence, best practice, and the clinical decision under consideration [278, 339, 340]. CDRs draw on evidence from guidelines to provide operationalised and efficient approaches to assessing probabilities for diagnostic, treatment, and prognostic decisions integrated into the clinical pathway [287, 339-341]. Rules guide clinicians in their clinical decision-making role, helping to establish pre-test probability, providing simple screening tests for common problems, and aiding in the estimation of risk [278, 342].

### 5.1.6 Care Plan

A Care plan is an organised multidisciplinary day-by-day list of care activities that healthcare providers will undertake to support identified patient problems along with intermediate outcome-based goals for assessing performance [343, 344]. While a ward or care unit may have a care plan template for patients with particular clinical presentations, such as for post-caesarean section [343], patients diagnosed with community-acquired pneumonia [345], or stroke [308]; being as care plans are specific to the patient they can only be defined once the nurse or clinician has assessed that patient's particular deficits and needs [310]. An implied relationship exists whereby the care plan may draw knowledge of the condition from the CPG however, in practice care plans include no supporting literature or knowledge gained from retrospective chart or patient record reviews that would justify the practices used as part of the plan [345]. Care plans are a constituent component of the clinical pathway [305, 346, 344], and due to their prescriptive and scheduled nature are credited as the component used to achieve reduced LOS [345].

### 5.1.7 Treatment Protocol

Known variously as clinical, care or treatment protocols (TP), these are standard descriptors of clinical care activities developed on the basis of guideline-based evidence, and usually found incorporated into clinical pathways and described against a timeline [16, 347]. Examples of TPs for severe acute respiratory syndrome [348], post resuscitation care [349], lung donor treatment [350] and rhabdomyolysis-based renal failure [347] show the structure of these to treatment steps arranged in sequence for a particular clinical presentation. Some also include outcome assessment factors for evaluating the success of protocol application or for consideration of whether to move the patient onto more aggressive treatment options [348, 349]. Given the structure and presentation of TPs, many have potential for description using the graphical formalisation of the caremap.

## 5.2 THE EXPANDED DE BLESER APPROACH

Clinical care process specification (CCPS) documents have been viewed from the perspective that they consist of elements that can be represented along continuums describing either their degree of prescriptiveness or input to healthcare quality improvement [304]. Both continuums and elements were applied in development of a characterisation framework, but in De Bleser et al's case, applied to *clinical pathways* only [304]. The continuum and elements were represented on two axes: the x-axis having elements ranking the pathway from descriptive to prescriptive, and the y-axis using elements that describe levels from current practice to healthcare quality improvement [304]. However, it was necessary to extend De Bleser's framework in two ways. The first adds an additional dimension to De Bleser's coordinate system approach by describing CCPS documents' evidential and practical applicability: that is, the addition of a z-axis describing the hierarchy of evidential strength contained within the CCPS document and how applicable or useable the document is in the delivery of clinical care to the individual patient. This is illustrated in Figure 9, where the newly applied axis is represented horizontally. The second extension applied the resulting extended framework to the hierarchy of six care process specifications and thus provides a standard for comparison and accurate characterisation of each clinical care process specification.

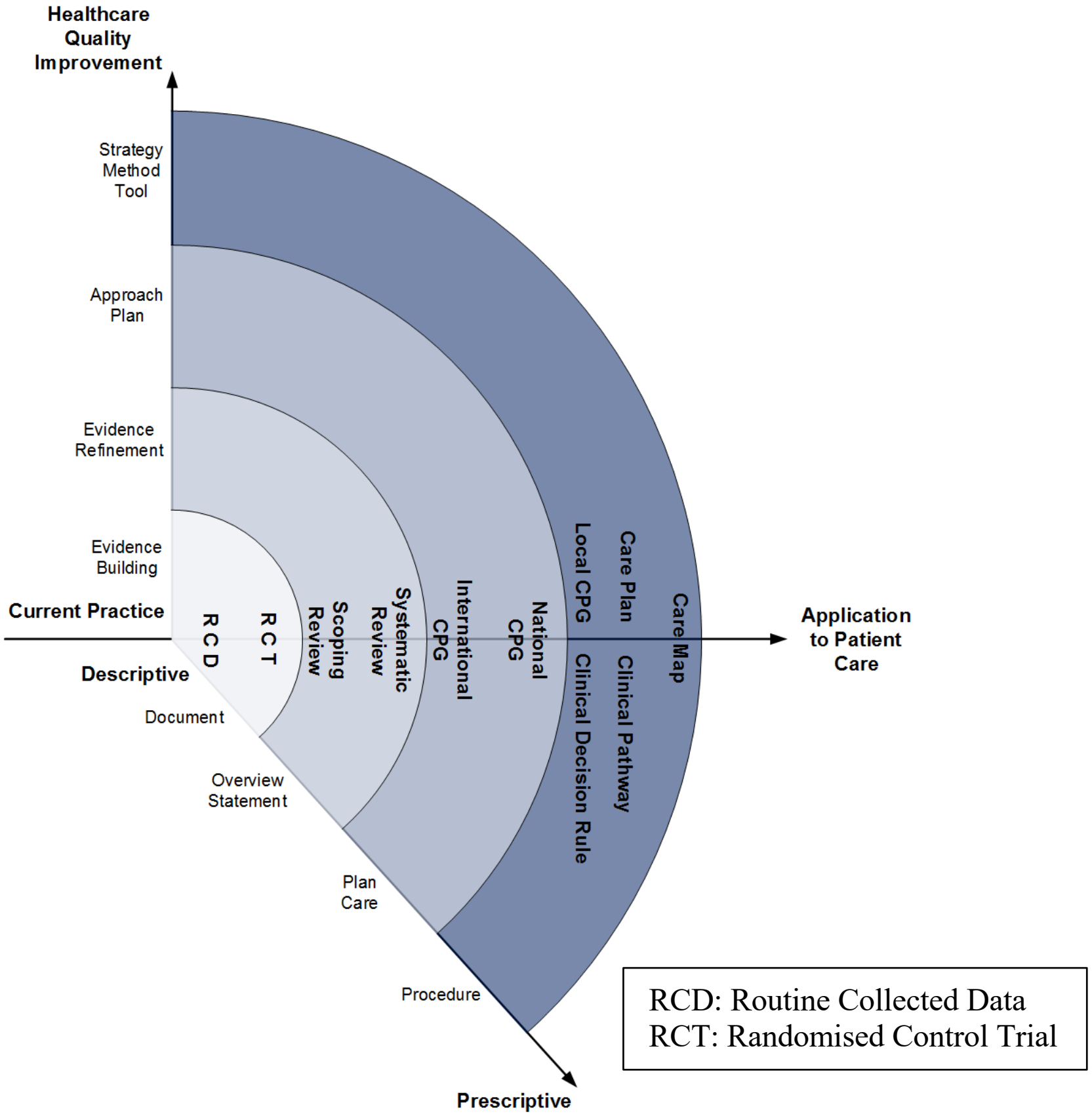

Figure 9: The expanded De Bleser approach

## 5.3 TAXONOMY FOR CLINICAL DOCUMENTATION

In developing the taxonomy shown in Figure 10, the literature collected in section 4.1 was reviewed using the CA, TA and meta-narrative methods described in section 3.1. All instances identified in the table in appendix D were considered, along with authors' definitions and descriptions of each document type and input from our clinical experts. The taxonomy presented identifies the categories of clinical documents we determined from our review as being unique or distinct, as well as a hierarchy for how authors described each as being derived from, dependant on, or acting as a component of another. A solid connecting line, or arc, represents where a strong and direct, almost parent-child relationship was resolved from the literature. A dashed connecting arc represents those instances where an indirect relationship was inferred from authors' narratives. Each document type and their described interrelationships are discussed in the following section.

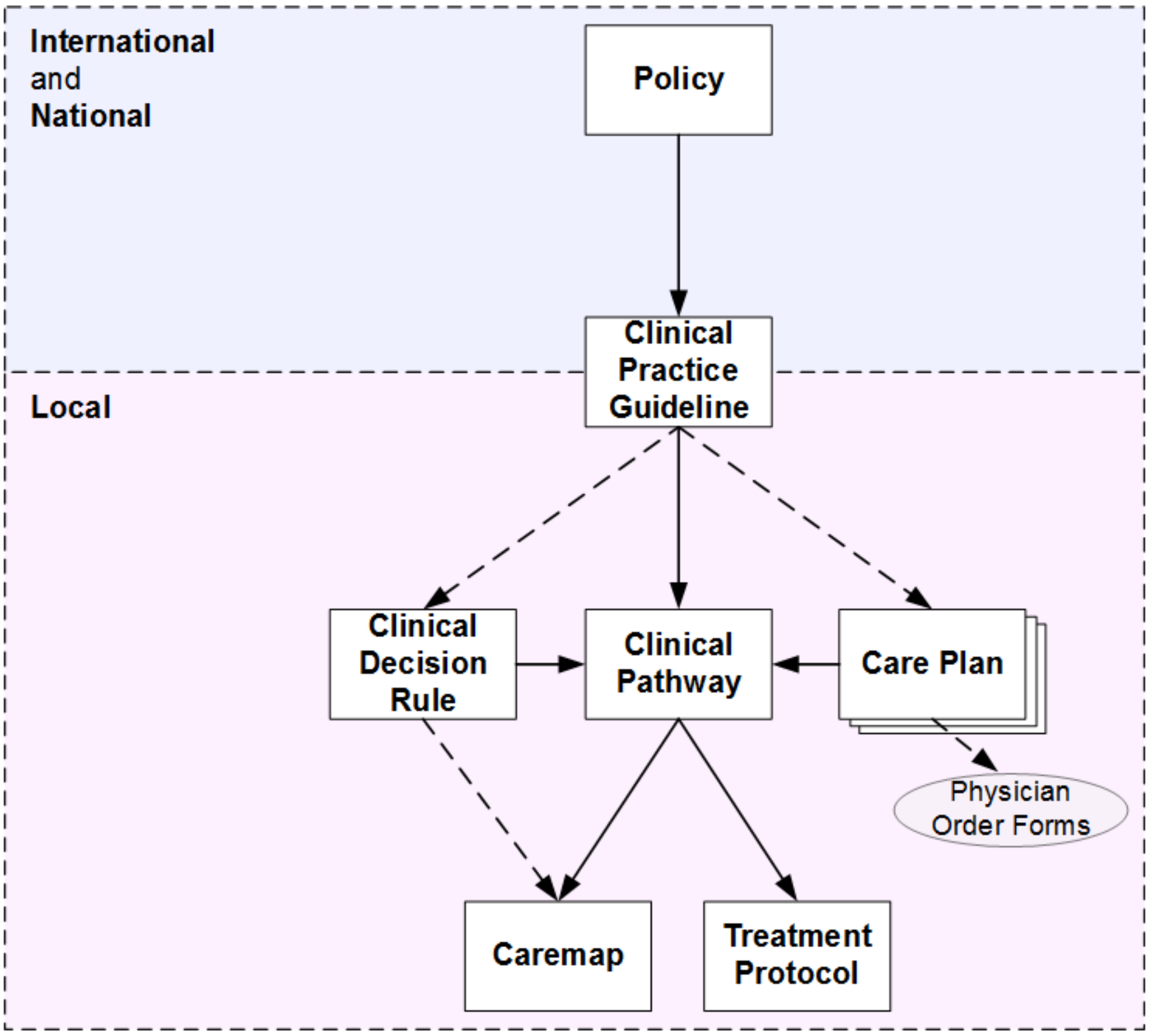

Figure 10: Taxonomy and hierarchy for clinical documentation

The relationship between this taxonomy and the expanded De Bleser model in Figure 9, especially in the hierarchy of systematically developed evidence described on the z-axis, is notable. In both cases the clinical care specifications naturally fall in order of clinical, or operational, usability.

## 5.4 CHARACTERISATION OF CLINICAL DOCUMENTATION

Table 4 provides a characterisation for CCPS based on their presentation, structure, the population they effect and their clinical intention. It also positions each document with reference to De-Bleser et al's continuums, or axes. The x-axis (descriptive / prescriptive) and the y-axis (healthcare quality improvement). And finally, it aligns the six clinical CCPS on a continuum in order of operational usability.

Table 4: Characterisation of Clinical Documentation

| | CPG | Clinical Pathway | Care Plan | CDR | Caremap | Treatment Protocol |
|---|---|---|---|---|---|---|
| **Presentation** | Primarily Free text | **Ensemble**<br>Free text<br>Structured text<br>Columnar (day-to-day)<br>Graphical description | Columnar<br>Temporal flow (day-to-day) | Structured text (If this, do that) | **Contemporary**<br>Graphical<br>Flow diagram<br>Cyclic<br><br>Temporal or sequential | Structured text<br><br>step-by-step instructions |
| **Application** | Cohort (condition/ intervention) Generic | Cohort (condition/intervention) Generic | Pt Specific | Condition investigation or intervention Specific | Event or condition specific | Intervention specific |
| **Grounding** | Evidence based<br>Graded evidence<br>Consensus based | Do not usually include discussion of evidence | Do not usually include discussion of evidence | Do not usually include discussion of evidence. May refer reader to CPG | Do not usually include discussion of evidence | Do not usually include discussion of evidence |
| **Contextual Domain** | All domains | More common in nursing | Nursing | Medicine | *Traditional:* nursing<br>*Transitional:* mostly nursing<br>*Contemporary:* more common in medicine, less common in nursing | All domains |
| **Aim/Goal** | Reduce variation from evidence-based practice | Define outcomes<br>Cost containment<br>Resource containment | Provide day-to-today evidence-based care schedule for a particular patient | Provide algorithm/rules to support clinical decision making | Reduce variation while expediting decisions by providing defined care and treatment paths | Operationalise particular treatment steps of the CPG |
| **Usability** | Least operational | | | ⇒ | | Most operational |
| **Descriptive/ Prescriptive** | Plan Care | Plan Care | Plan Care | Strategy/ Method/ Tool | Strategy/ Method/ Tool | Strategy/ Method/ Tool |
| **Healthcare Quality Improvement** | Approach/ Plan | Approach/ Plan | Approach/ Plan | Procedure | Procedure | Procedure |

This characterisation is robust in the sense that in addition to De Bleser's elements it also incorporates aspects that allow CCPS to be analysed more comprehensively in terms of their content, structure and target patient cohort. This table would be useful in tasks seeking to identify or design particular CCPS documentation and ensure that it covers all of the aspects incorporated in Table 4.

## 5.5 CCPS: CASE STUDY

This section applies the CCPS taxonomy and characterisations to clinical documentation from England on the topic of Type 2 Diabetes (T2D) and Diabetes in Pregnancy. In each of the following, any reference to a section identifies a section so labelled within the identified CCPS document being described. Each of the specifications cited includes a URL in the reference so that the reader may access and assess the source document.

### 5.5.1 Policy

Three related documents were identified within England that describe policy regarding diabetes, and more specifically Type 2 Diabetes (T2D). These include government, health and clinical policies and are listed in Table 5.

Table 5: Policy Documents for Type 2 Diabetes

| | |
|---|---|
| **Government Health Policy** | Health Matters: Preventing Type 2 Diabetes [351] |
| **National Health Policy** | National Strategy and policy to prevent Type 2 Diabetes [352] |
| **National Clinical Policy** | NHS England: Action for Diabetes (NHS, 2014) [353] |

Government policies generally open by expounding on the scope and scale of a public health issue. This information, which centres around the community and resource impact of the medical condition tends to be followed by pronouncements that are so simplified that they lose the proper meaning and efficacy they are intended to convey. In effect, they miss their goal and become more like promotional material intended for the media. An example from the Government Health Policy identified in Table 5 is shown in Figure 11, where, at the time of writing, the government set the goal of delivering weight loss, which was achieved in a very small subset of cherry-picked and highly motivated volunteers. As with all such experiments, caution should be used when claiming these results will be transferrable to the general population. Success in such situations is rarely achievable.

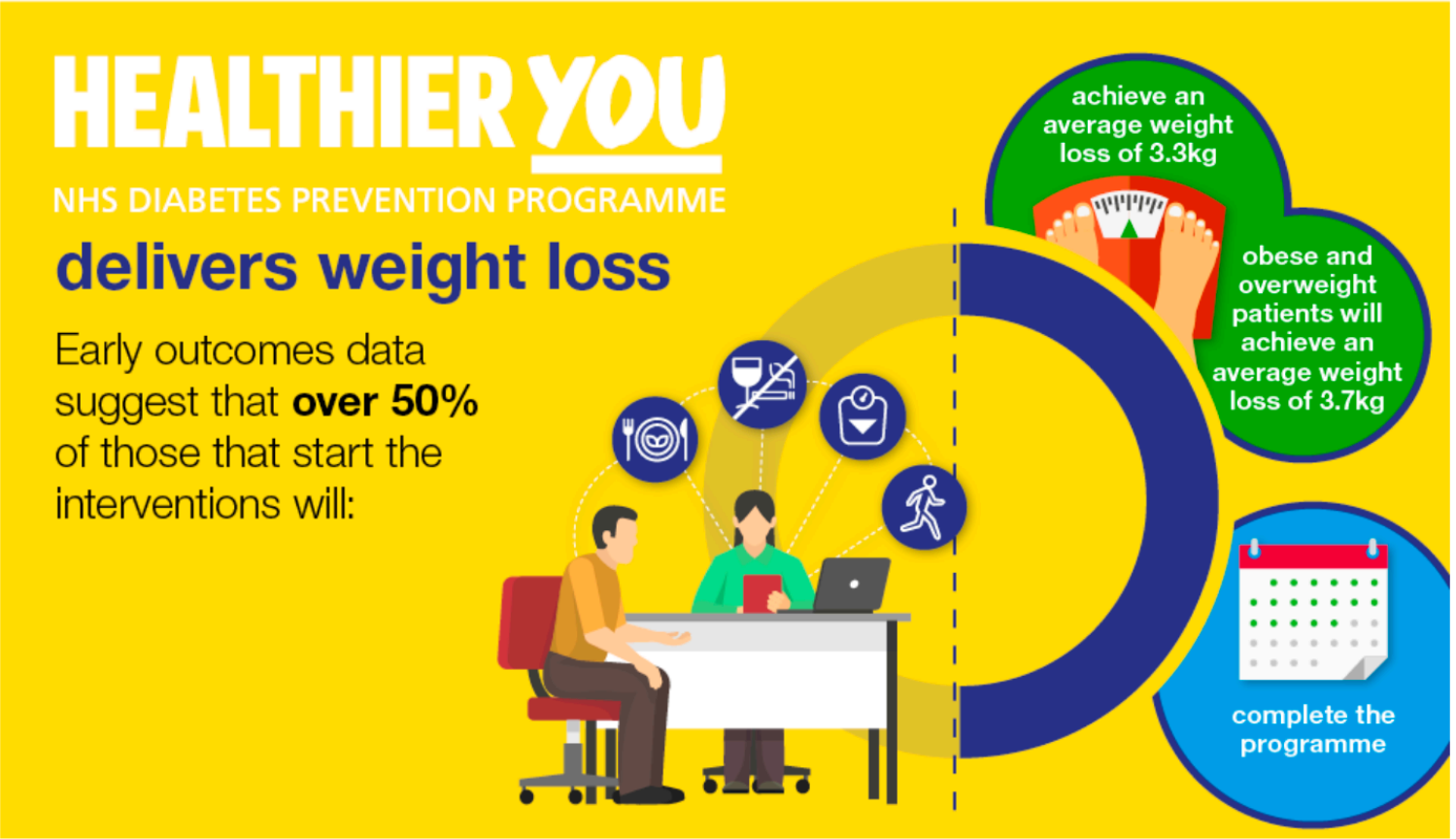

Figure 11: Clinical Practice Guidelines for T2D (UK Government, 2018) [351]

One component of this same policy which has become enshrined in UK taxation legislation, is the laudable challenge to remove 20% of the caloric content comprised by sugar from all products produced across the food and drinks industry. Whilst put in holistic context, this caloric-reduction policy carries over in section 4 of the identified National Health Policy. However, this policy approach has resulted in manufacturers finding it necessary to replace sugar primarily with artificial sweeteners like aspartame and saccharin. Both these non-nutritive sweeteners have received sufficient research to suggest that the health dangers of frequent or daily use may also be harmful [354-356]. This example demonstrates our claims regarding policy: (a) that it is politically motivated; (b) that it is often not based on established scientific evidence; and (c) that it tends towards having been constructed by politicians from many small *sound bites* seeking *air time* to expound the virtues of their policy and elicit support from the ordinary citizen.

### 5.5.2 Clinical Practice Guidelines

CPGs range from those that broadly present considered recommendations of best practice suitable internationally, through to hospital guidelines that are customised by consultants based on the characteristics they observe in the local population. A selection, again relevant to diabetes care in England, are shown in Table 6.

Table 6: Clinical Practice Guidelines for Type 2 Diabetes

| | |
|---|---|
| **International CPG** | IDF Clinical Practice recommendations for managing T2D in primary care [357]<br>Management of hyperglycaemia in Type 2 Diabetes (ADA & EASD) [358] |
| **National CPG** | Type 2 diabetes in adults: Management [359]<br>Pharmacological management of glycaemic control in people with Type 2 Diabetes [360] |
| **Local CPG** | Barts Health NHS Trust CPG: Diabetes in Pregnancy [361]<br>Type 2 Diabetes: Reducing hypoglycaemia [362] |

Similar to the government policy which claimed it had, or would, deliver weight loss, the IDF international guideline at section 1.2 recommends weight loss as the key tool for preventing T2D. It also presents more responsible with regards to its scientific and evidentiary basis than the government policy, in that it recommends avoidance of sugar and sugar-sweetened drinks rather than replacement with artificially sweetened versions.

The international guideline at section 1.3 also provides for a regimen of testing and thresholds to be used in diagnosing the diabetic patient. In this case the IDF guideline follows a diagnostic standard from another international guideline, that of the World Health Organisation (WHO). Similar diagnostic testing and threshold recommendations carry through in the first National CPG at section 1.6, and Barts Health Local CPG at section 4.3.

At each level the recommendations in CPGs are seen to become more capable of application to patient care. While the international and national guidelines are written from a best-practice perspective, hospital guidelines tend towards being more pragmatic and patient-centred. The international and national guidelines present the perfect scenario or perfect model for treatment, while local guidelines recognise that there are non-compliant patients and not everything will go according to plan. The contrast is as one might expect, and not dissimilar to that between fiction and reality.

### 5.5.3 Clinical Decision Rule

In many cases CDR can be identified and isolated from within the content of local CPGs. The diagnostic threshold rule for glycaemic control in pregnancy for diabetic mothers who were treated only with diet control in Table 7 is drawn from within the Barts Health NHS Trust Local CPG referenced in Table 6 comes:

Table 7: Clinical Decision Rule for Diabetes in Pregnancy

| | |
|---|---|
| 10.1 | Monitor blood glucose 4 hourly.<br><br>If blood glucose >7.0mmol/L on 2 consecutive occasions, start IV insulin/glucose infusion.<br><br>If the woman requires a Caesarean section and the blood glucose >7.0 mmol/L, start IV insulin/glucose regime. |

### 5.5.4 Clinical Pathway

Two English diabetes clinical pathways are referenced in Table 8. Each presents an approach to screening and patient assessment, care planning, patient support and education that includes lifestyle changes in diet, weight, exercise and insulin/glucose control. Each also makes reference to, and recommendations regarding, potential complications that may arise in future for these patients and advises the signs and symptoms that clinicians and patients should be mindful of. These pathways present as a generic blueprint for care of patients with diabetes.

Table 8: Clinical Pathway for Diabetes

| | |
|---|---|
| **Clinical Pathway** | Clinical Management: Optimal pathway [363] |
| **Clinical Pathway** | New Patient Clinic Pathway [364] |

### 5.5.5 Care Plan

Care plans are the first of the clinical care specifications that present with customisation for the immediate patient. Each of the example care plans in Table 9 presents with information and contact details for services the patient may require. More importantly however, each provides space and tables for recording the day-to-day condition, treatment and care recommendations specific to that patient.

Table 9: Care Plan for Diabetes

| | |
|---|---|
| **Care Plan** | My Personal Diabetes Health Care Plan and Record [365] |
| **Care Plan** | My Personal Diabetes handheld record and care plan [366] |

### 5.5.6 Caremaps

Caremaps provide a visual roadmap of care for a specific condition or treatment process. Caremaps can aid in treatment selection, understanding of disease progression, and can also provide the base knowledge that, together with clinical expertise, may be used in other forms of health research. The EPSRC-funded

PAMBAYESIAN project has demonstrated that caremaps can be used in the development of Bayesian-based LHS [367]. PAMBAYESIAN developed the caremaps listed in Table 10 for *diabetes in pregnancy* as part of research and development of clinical decision support LHS for use by both clinicians and patients.

Table 10: Caremaps for Diabetes in Pregnancy

| | |
|---|---|
| **Midwifery Booking Visit** | GDM Booking Visit [368] |
| **Diagnostic Caremap** | GDM Diagnostic Decisions 2.0 [369] |
| **Management Caremap** | GDM Management Decisions 2.0 [370] |

Caremaps can be further instructive with the addition of decision points. This addition provides detail at the points where clinicians are called to make diagnostic or treatment decisions for the instant patient. With the input of experienced clinicians, the PAMBAYESIAN caremaps include decision points and diagnostic criteria, as shown in Figure 12.

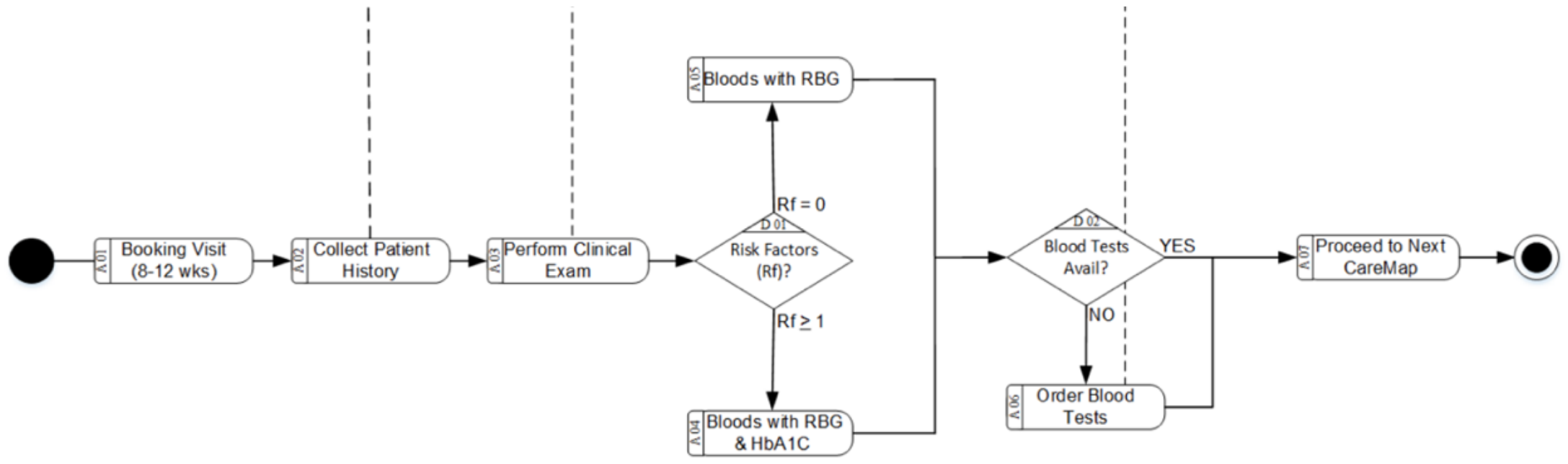

Figure 12: GDM Booking Visit Caremap with Decision Points

## 5.6 STANDARDISATION BEFORE COMPUTERISATION

**CCPS Computerisation:** Various types of CCPS such as caremaps, clinical practice guidelines and clinical protocols are increasingly becoming the subject of computerisation [371]. For computerisation to be easily achievable, CCPS need to be formalised in order to render them computer-interpretable [372]. Formalisation of CCPS refers to specification using precise and unambiguous terms and structures, usually using computational or mathematical formalisms, rendering the resulting CCPS as computer interpretable.

**Standardisation as a Prerequisite to Computerisation:** CCPS computerisation is enabled by CCPS formalisation which in turn is enabled by CCPS standardisation. Consequently, computerising CCPS that have not first been standardised creates more problems than it solves. While this research has referred to the need to standardise the structure and presentation of CCPS, there is also a need to standardise the content captured within the CCPS by standardising the underlying language. This is crucial to the challenge of formalisation to attain computer interpretability and hence computerisation.

**Role of Data and Information Standardisation in CCPS Computerisation:** Data standardisation is needed to resolve the *Data Triple-I Issues* of *integrity*, *integration* and *interoperability* identified in a 2019 literature review as significant barriers to adoption of health information technology [373]. The use of standardised clinical terminology and approaches to eliciting and reporting routine clinical data is necessary to improving clinical practice and communication and enabling accurate computerisation of health record data [374]. In our experience there are many existing healthcare provider projects in the United Kingdom, Australia and New Zealand that seek to standardise coding of data as it is captured. The issue observed is that they do this more often without considering whether the underlying CCPS was standardised first. Where standards are considered there tends to be disagreement as to which underlying methodology, terminology, data standard or data exchange protocol should prevail: SNOMED, ICD-10, METeOR, FHIR or some other.

**The Benefits of Standardisation in the Context of CCS Computerisation:** Standardisation of CCPS has been observed to improve data collection, data quality, and to enable automated coding and analysis [375, 376]. Standardisation of CCPS using approaches that include: (1) defined data definitions; and (2) attention to built-in computer interpretability [377] supports data standardisation, mitigating the Data Triple-I issues and increasing the chances for successful creation and adoption of LHS.

## 5.7 DISCUSSION

The confusion arising out of the absence of standardised definitions and nomenclature for clinical documents has been known since at least 2006 [378, 208, 304, 379]. Resulting from this lack of standard terminology and taxonomy have been successive and discordant attempts at defining and describing structures and development processes for some clinical documents [378, 304, 379]. These have only served to exacerbate the lack of standardisation. This is further intensified when we observe examples of authors who identified multiple document types as synonymous in name or structure [378, 208, 16, 343], only to differentiate them in the same research by describing one as a single or multi-component item of the other [16, 343], or a translation of the other represented in different structural form [378]. In [39] we reported examples where authors' described a clinical document in-text using one label listed in the CCPS taxonomy, only to find that the author's included clinical document titled or described itself using another e.g.: [205] and [206]. Such instances exemplify this issue and its potential to create lasting confusion.

While authors have attempted to define or present approaches to the development of different clinical care process specifications during the last two decades, significant variance in the complexity level, design approach, content and representational structures for the majority has persisted [207, 39]. As a result of this lack of standard and formal care process specifications, differences in communication and information transfer between clinicians providing care for the same patient can be substantial and ubiquitous, affecting the quality of care and introducing additional risk of harm for as many as 25% of all patients [380-382]. Error reporting documentation is similarly, and more often, not standardised. However, even where some effort has been made to standardise, clinical decision and treatment errors remain underreported. While clinicians and clinical researchers are encouraged to be honest and transparent in such situations, we can never be absolutely certain whether reported error figures quoted are representative of the entire scope of the problem [383-386].

Standardised approaches, especially in clinical documentation, ensure that each time a healthcare provider approaches that type of document, the format and content are consistent with expectation [39]. Standardised documentation ensures sufficient higher quality information is recorded and reported, enabling documents to be read quicker and content within to be better retained, all with the effect of improving overall patient safety and outcomes [380, 88, 381, 39, 89]. Standardisation of specifications also brings benefits that go beyond being operational or purely clinical. The data triple-I issue represents the biggest single barrier to learning health systems (LHS) [373, 387]. Standardisation of clinical care specifications can enable computer interpretability, which in turn supports data standardisation and increases the chances for successful LHS implementation.

## 5.8 SUMMARY

Since the early 2000's there has been a significant shift towards standardisation of healthcare and healthcare practice. Efforts to standardise healthcare can be held back by many things, one of which may be the lack of consistent nomenclature and definitions for CCPS documents, many of which are used every day in hospitals and general practice clinics. This research has addressed the challenge by presenting a 4-layer taxonomy for clinical care process specification documents that begins with those documents that are least able to be applied in clinical care and moves towards those which are most capable of use in the clinical setting, and evaluating this taxonomy through a case study looking at a range of clinical care process specification documents from the UK concerning diagnosis and treatment of patients with Type 2 Diabetes and diabetes in pregnancy. It is intended that the results and discussion should stimulate those in healthcare and healthcare service management to consider the potential benefits of standardising clinical documentation in the same way that evidence-based medicine aims to standardise healthcare. Standardising the way patient care is documented and reported will reduce significant data integration and interoperability issues as national health services continue to move towards implementing centralised electronic shared care records.

# Chapter 6: The Heimdall Framework

Using comprehensive analysis of literature for LHS collected by the search discussed in Chapter 4, this chapter initially seeks to characterise and present the first complete taxonomy for LHS, and position the identified types within the wider healthcare domain. Then, with addition of the EHR literature collection from Chapter 4, this chapter seeks to: (i) identify challenges that have inhibited adoption of health information systems for clinical practice; and (ii) ascertain whether there is a relationship between those factors that previously inhibited EHR adoption, and those that authors believe currently restrain acceptance of LHS.

Section 6.1 extends our general understanding of the issues that inhibited EHR through an analysis of the barriers and facilitating factors of EHR implementation. Section 6.2 identifies the concepts that define and characterise LHS, as well as presenting and validating a taxonomy for LHS. How LHS may integrate into the learning health organisation and draw efforts at EBM towards precision medicine is depicted in section 6.3 with presentation of a novel framework, the *Heimdall framework*. Section 6.4 extends the established ITPOSMO methodology, resulting in a new approach named ITPOSMO-BBF. ITPOSMO-BBF is then applied to analyse and compare the current state for EHR and LHS implementations and the results are discussed in section 6.5. Section 6.6 summarises the chapter.

## 6.1 EHR: BARRIERS AND FACILITATORS

The CA and TA methods discussed in section 3.1.3 were used on the literature collected at section 4.2 to identify barriers, benefits and facilitating factors, and examine the frequency and context within which authors described these themes. It was possible to categorise many identified themes within a key theme: a grouping of related themes. Key barrier and facilitator themes are listed in Table 11, which also lists the literature in which each theme was identified and whether the authors discussed it in the context of: (a) a barrier inhibiting implementation or use of EHR, or (b) if appropriately engaged, a potential facilitating factor increasing the likelihood of success for a new EHR implementation.

Table 11: EHR literature categorised into key themes by TA

| | Key Themes | Barriers | Facilitators |
|---|---|---|---|
| 1 | Willingness, interest or motivation to **adopt** new HIS and frameworks | [240, 235, 242, 237, 263] | [388] |
| 2 | **Training** and skills with computer systems and HIS | [389, 390] | [234, 243, 264, 237-239] |
| 3 | Data **standardisation**, **interoperability** and **integration** | [391, 392, 258, 393, 388, 394, 395, 224, 225, 396, 397] | [234, 84, 394, 238, 398] |
| 4 | Changes to **legislation**, **policy** and government-mandated financial factors (**incentives** or **penalties**) | [156, 258, 393, 394, 390, 399] | [388, 241, 400, 395, 224, 225, 401, 399, 402] |
| 5 | Capital investment, implementation, maintenance and support **costs** | [154, 240, 234, 393, 235, 241, 403, 242, 243, 262, 264, 390, 225, 237, 238, 398] | [154, 393, 235, 241, 262, 238, 263, 398] |
| 6 | Impact of EHR on health outcomes and patient-clinician encounter within the patient care **workflow** | [154, 240, 234, 241-243, 262, 237, 238, 263, 239, 398] | |
| 7 | **Privacy**, security, data integrity and accuracy | [404, 154, 240, 234, 235, 242, 264, 390, 225, 405, 237, 399, 238, 263] | [241, 238, 406] |
| 8 | Approvals and **ethics** oversight for use of digital health data | [236, 397] | |
| 9 | Organisational **culture**, management and clinician attitudes to change | [404, 154, 234, 242, 262, 237, 407] | [154, 234, 408] |
| 10 | Identifying and involving all relevant **stakeholders** | [409, 410, 263] | [154, 240, 237, 263, 397] |

## 6.2 A TAXONOMY AND CHARACTERISATION FOR LHS

By drawing on the literature collected in section 4.3, this section seeks to resolve the research community awareness challenge introduced in section 1.5.2 to develop a characterisation, taxonomy and unifying framework for LHS.

### 6.2.1 Characterisation and Conceptualisation of LHS

The IoM definition cited at section 1.1.2 which states that a Learning Health System (LHS) is one *in which progress in science, informatics and care culture converges to continuously create new knowledge as a natural by-product of a care process in which best practice is applied for continuous improvement* [60] is repeated and relied upon by most authors. However, that definition both *fails to contemplate what it means for something to be an LHS*, and *fails to describe attributes that would contribute to aspects of patient care that LHS systems should target*. Namely: quality, safety, efficiency and efficacy of patient care. The domain has seen little development of these aspects [274, 23]. Similarly, the IoM definition does little to develop an understanding of the attributes, or concepts, underpinning implementation and usage of LHS in clinical practice [23], with those that are observed, like those quoted in Table 2 of Chapter 4, being limited to the naive idea that an LHS is made up of an organisation forming communities who collaborate to learn new knowledge from big data for the purpose of quality improvement [411].

This work synthesises the unified conceptualisation and framework for systematically characterising LHS from various strands of investigations undertaken by the author. These strands are as follows:

- **Concept:** The problem of lack of understanding of the concept of LHS is described as the research community awareness challenge [274].
- **Taxonomy:** Development of a taxonomy describing the entire scope of LHS as observed in the current literature and showing how each type positions within the larger learning health organisation [23].
- **Framework:** Development of a unifying framework showing the role and context for LHS and its integration into the learning healthcare organisation [23].
- **Implementation:** Identification of the benefits, barriers and potential facilitators for LHS, and comparative analysis of how these may have arisen from, or be related to, those that have impacted EHR implementation during the preceding thirty years [387].
- **Realisation:** Demonstrating how the taxonomy, framework and factors for successful LHS implementation can be applied to a contemporary project to identify those aspects that constitute LHS, the type of LHS, the barriers to which facilitators may be applied and the benefits that may result.

The accumulation of this research found that the concepts that define or are associated with LHS (conceptual composition of LHS) as shown in Figure 13 are:

- **Patient data:** Collections of patient data in the form of EHR are seen as a near-inexhaustible source from which to learn new knowledge [61, 62].
- **Computing System:** Not limited to the computers used to access or store patient data, but also includes those which contain the programs and perform the machine learning, prediction and other computing necessary to learning and applying knowledge [248, 62].

- **Learning:** The concept of learning as it relates to LHS is the processes that analyse data to derive or generate new knowledge [248, 61, 102, 62].
- **New knowledge:** The new knowledge learned from patient data can advance our understanding of the underlying mechanisms of disease and patients' response to treatment [249, 62].
- **Near real-time:** The current driving ambition for LHS is to expedite the process, often described in terms of a seventeen-year lag, of getting knowledge to inform clinical decisions from scientific discovery to clinical use [249, 61, 87, 62].

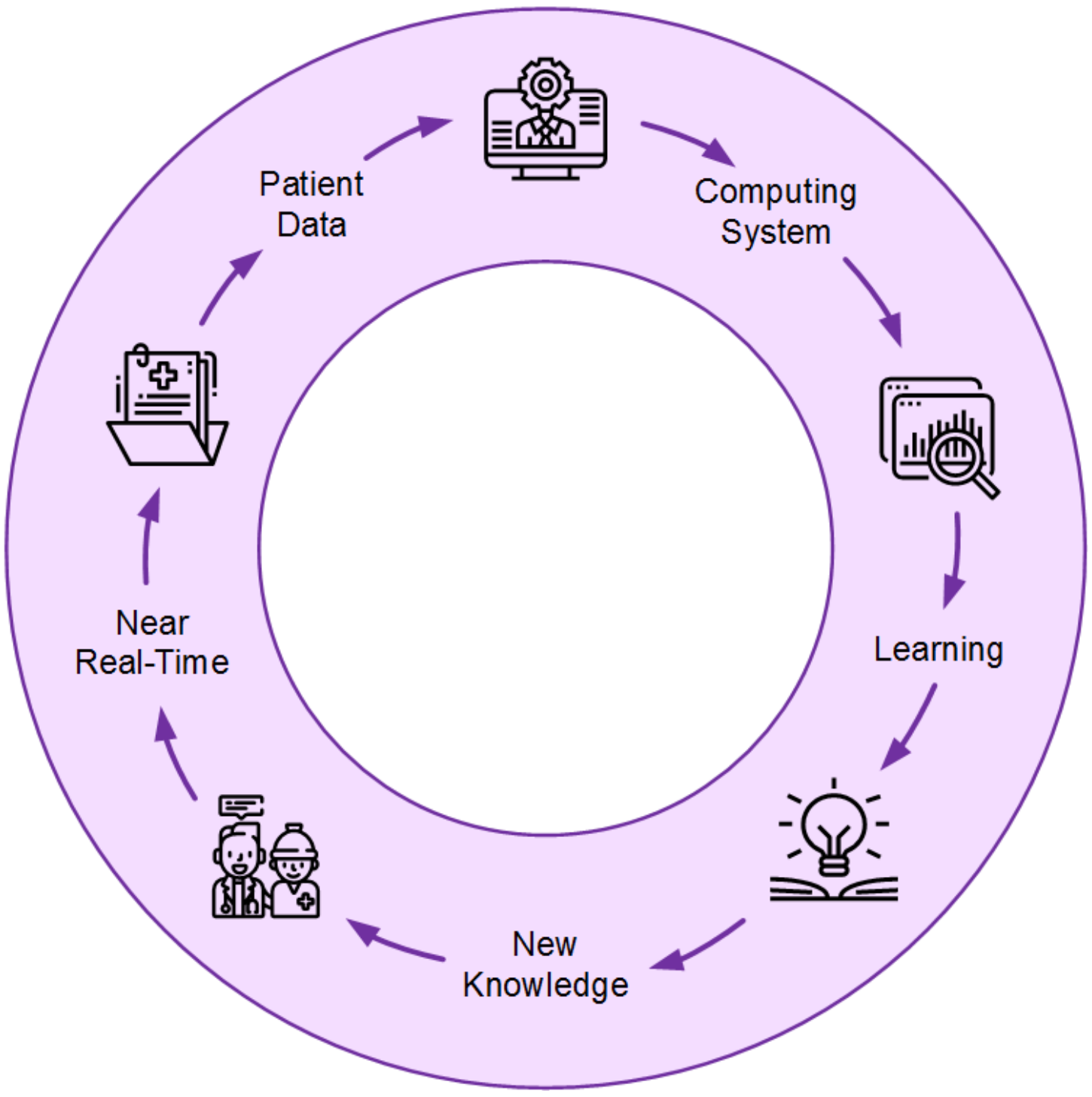

Figure 13: Five Concepts that define Learning Health Systems

### 6.2.2 Taxonomy for LHS

Only three papers could be identified that proposed classification systems for LHS [257, 86, 87]. Surveillance and Comparative Effectiveness Research were the only types common to all three. Figure 14 unifies the knowledge identified from all three papers into a taxonomy of nine LHS classification types, indicating abbreviations (initials) and the primary reference for each.

***Cohort Identification* (CI)** seeks patients with similar attributes, used to determine the feasibility of studies and quantify numbers of potential patients that may be helped [87]. CI is also the first operational step of most other LHS types.

***Positive Deviance* (PD)** uses outcome data to benchmark clinical care. PD identifies elements of safer, more effective, timely and patient-centred care, recognising beneficial behaviours for incorporation into another clinician's practice [412]. PD can also identify common traits of patients benefiting from a treatment, using these to identify others who may benefit from the same intervention [87].

***Negative Deviance* (ND)** identifies instances of sub-optimal care. ND presupposes some particular clinical behaviour negatively impacted patient care and the resulting outcome. The clinician critically evaluates the care provided, investigating causes for sub-optimal results [257].

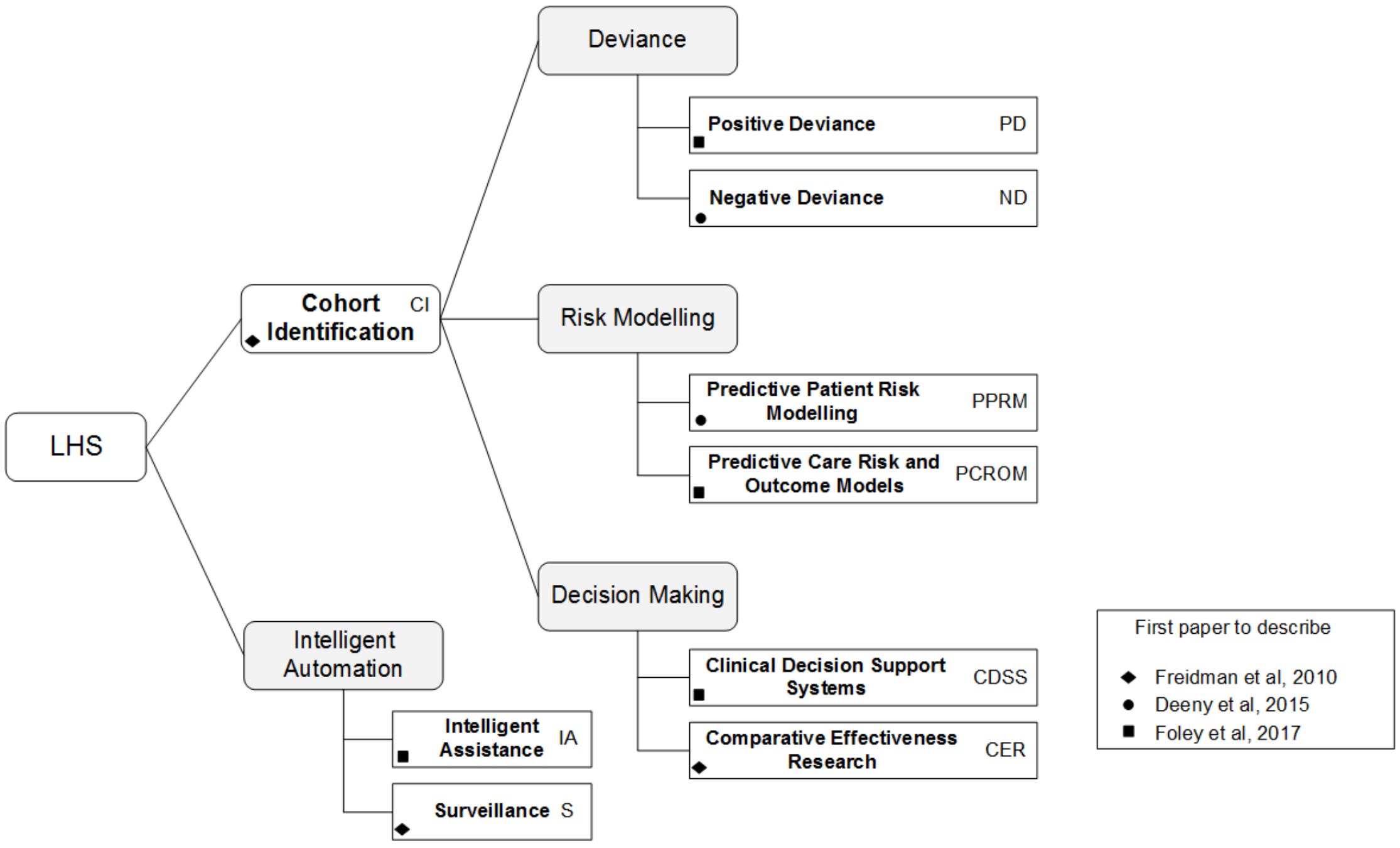

Figure 14: LHS Taxonomy

***Predictive Patient Risk Modelling* (PPRM)** uses patterns discovered in patient datasets to identify cohorts at higher risk for future adverse events. PPRM can also use routine health data to identify 'triple fail' events; where treatment fails to improve patient care experiences, population health, or lower healthcare costs [413].

***Predictive Care Risk and Outcome Models* (PCROM)** algorithms identify situations of high risk for 'unsafe,' 'delayed,' or 'inefficient' care, providing estimates of the effectiveness of different interventions [86]. Geisinger Health Systems have incorporated PCROM approaches into clinical software, identifying when spikes in hospital activity or patient non-attendance may occur [414].

***Clinical Decision Support Systems*** **(CDSS)** are active knowledge systems where two or more characteristics of the patient are matched to computerised knowledge bases with algorithms generating patient-specific treatment recommendations [415-417].

***Comparative Effectiveness Research*** **(CER)** compares interventions and outcomes within an EHR dataset to determine the most effective treatment; using a method considered more efficient than randomised control trials [86]. CER isolates patients with similar attributes to the current patient, returning knowledge on treatments that deliver optimum health outcomes [418].

***Intelligent Assistance*** **(IA)** uses data sources to automate routine processes such as prepopulating pathology orders and clinical notes, or summarising patient case notes prior to consultations [86].

***Surveillance*** **(S)** monitors EHR data for outbreaks of disease (e.g. measles) or treatment issues (e.g. contaminated medicines or increased frequency for post-surgical infections). Examples observed include Health and Demographic Surveillance Systems (HDSS) used in sub-Saharan Africa [419].

### 6.2.3 Validation of the LHS Taxonomy

To validate that the LHS taxonomy included all currently classifiable types of LHS, proposed or presented solutions from the LHS literature collection described in Chapter 4 were reviewed and classified using the taxonomic descriptions. Some proposed solutions were incompletely described; but it was observed that the intention of the authors was always clear from the information presented. All LHS solutions conformed easily to one of the identified taxonomic types. It is proposed that this validates the taxonomy as it has been presented. This validation of the taxonomy also found CER were the most prevalent type of LHS presented in the literature, as identified in Table 12. CER was followed at some distance by stand-alone CI, although it should be noted that this classification type had not been identified by either [257] or [86].

Table 12: Distribution of LHS Solutions (per 100 publications)

| | |
|---|---|
| Comparative Effectiveness Research (CER) | 44 |
| Cohort Identification (CI) | 14 |
| Clinical Decision Support System (CDSS) | 13 |
| Predictive Patient Risk Modelling (PPRM) | 10 |
| Positive Deviance (PD) | 8 |
| Intelligent Assistance (IA) | 3 |
| Negative Deviance (ND) | 3 |
| Surveillance (S) | 3 |
| Predictive Care Risk and Outcome Models (PCROM) | 2 |

## 6.3 HEIMDALL: THE INTEGRATED LHS FRAMEWORK

Just as the Norse God Heimdall was said to be the son of nine mothers, the Heimdall framework starts from the nine LHS classifications to develop the integrated LHS framework in Figure 15. The diagram's conical structure demonstrates how the use of technology (large datasets and processing systems) to record, store, index and present information, flowing into and improving the learning processes used in EBM, can focus clinical practice towards delivery of precision medicine (PM). This enables the learning healthcare organisation to engage in decisions individualised to match unique patient characteristics.

PM results from enhancing the generalised population health approach using attributes in the EHR to constrain analysis for diagnosis and treatment options to cohorts predominately matching the presenting patient's profile. As the clinician enters attributes about the current patient, the speed and accuracy of decisions increase as illustrated by arrows in Figure 15. LHS draw knowledge from a reducing cohort whose attributes predominately match the current patient as illustrated by the circular design. In examining a cross-section of Figure 15, the larger white circle represents the entire population used to select the most effective common treatment. The light grey circle reduces that population to those who share some basic attributes with the patient that would normally be identified in the slower learning organisation approach of EBM. The inner dark circle further reduces the population to a significant cohort with clinical, genetic and socioeconomic attributes predominately matching the patient at the centre. The interrelationship between the Heimdall framework and the LHS taxonomy is shown down the right side of the conical portion of the diagram. While LHS are technology solutions, the majority operate in the context of either the treatment provider's learning organisation, or the clinician's primary patient-facing role.

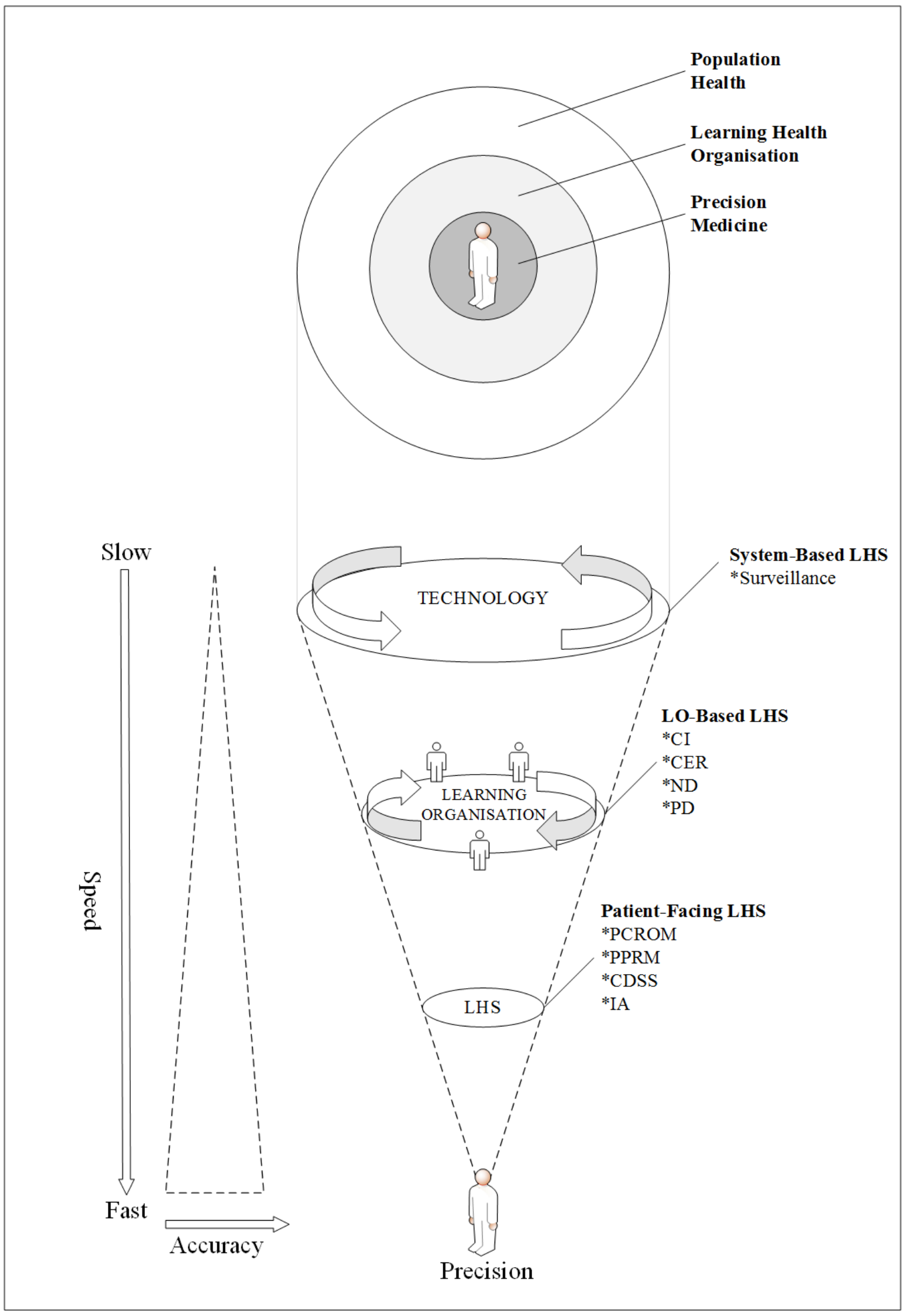

Figure 15: The Heimdall LHS Unifying Framework

Incorporated within this framework is the concept of the clinical lifecycle as shown in Figure 16 and adapted from multiple works in this review, including [418, 257, 414]. The right side of the diagram represents largely clinician driven aspects, while the left side identifies those aspects where LHS technology delivers improvements. The more challenging barriers to LHS regarding data quality, interoperability and standardisation all result from activities on the right side of the lifecycle.

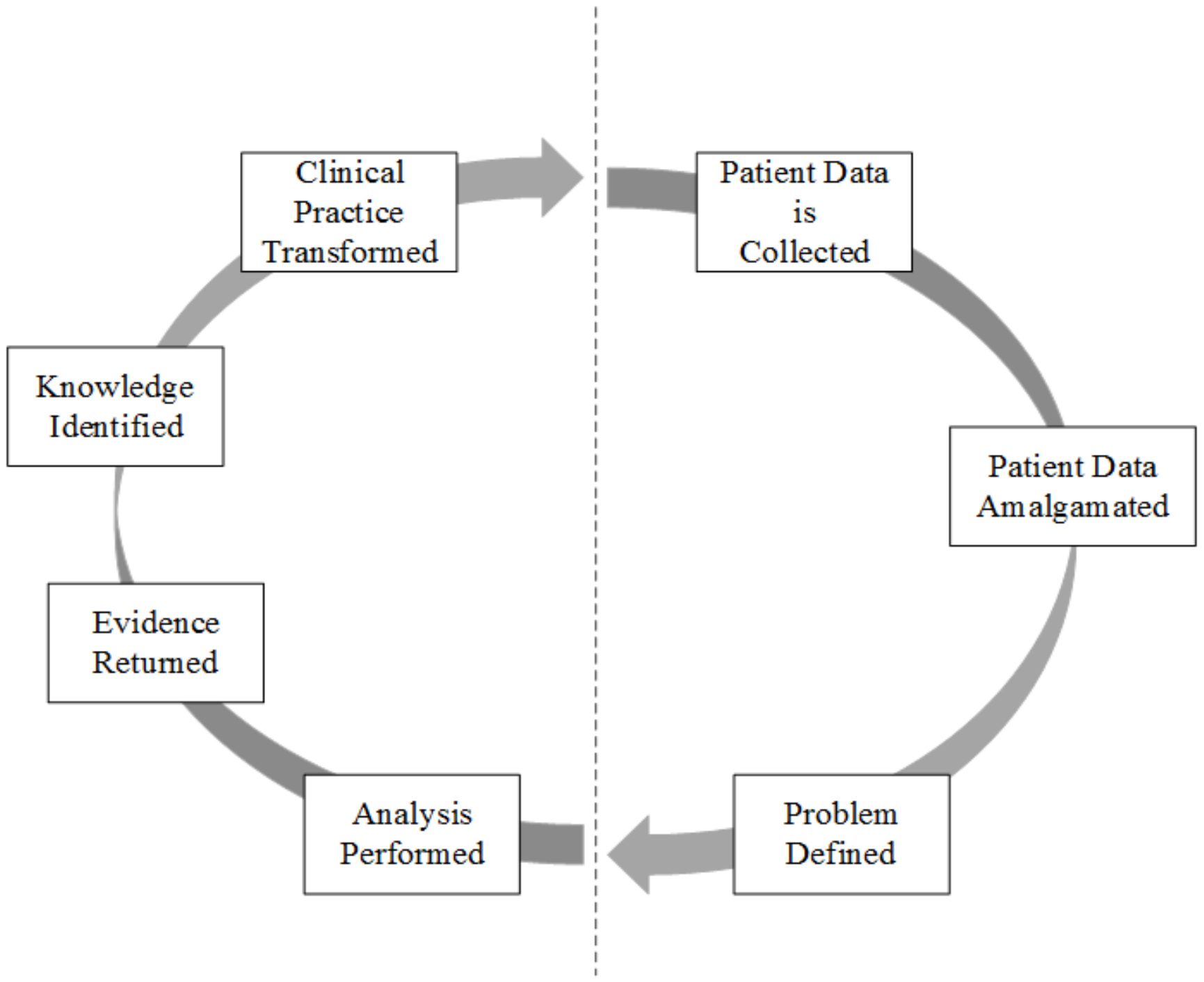

Figure 16: A generalisation of the clinical knowledge translation lifecycle

This cycle is repeated, both for surveilling the proposed transformation, and to seek further items of knowledge [418, 257, 414]. LHS engender a close relationship between care, research and knowledge translation, aimed at providing a platform for integrating various data to better understand patients [420]. LHS are commonly described using an iterative lifecycle similar to many other EBM processes where: (a) patient data is collected by clinicians, (b) aligned, transformed, and amalgamated into larger data sets; (c) a problem is defined and, (d) analysis performed; (e) with evidence data returned, and; (f) made into new knowledge, that is; (g) used to transform clinical practice.

The taxonomy and Heimdall framework together provide a toolkit for characterising LHS literature in terms of the following thematic and analytical aspects to assess whether the LHS demonstrates:

- *Taxonomic Consistency* – conforms to the taxonomy.
- *Patient Focused* – ensures personalised health care, known as PM.
- *Technology Usage* – uses health IT with big data, machine learning algorithms and automation.

- *Decision Support* – near real-time support to clinicians, bringing recent scientific advances, machine learning and EBM together at point-of-care.
- *Application of LHS* – goes beyond selecting 'most likely treatment for a population', to selecting the 'most applicable treatment for an individual'.
- *Barriers and Further Observations* – challenges limiting implementation.

As for EHR, CA and TA were used to identify barriers and facilitators for LHS. These were grouped using the same process and key themes and are shown in Table 13.

Table 13: LHS literature categorised into key themes using TA

| | **Key Themes** | **Barriers** | **Facilitators** |
|---|---|---|---|
| 1 | Willingness, interest or motivation to **adopt** new HIS and frameworks | [156, 260, 392, 421, 86, 393, 422, 62, 405] | [418, 156, 157, 423, 392, 87] |
| 2 | **Training** and skills with computer systems and HIS | [244, 246, 62, 396, 406] | [424, 425] |
| 3 | Data **standardisation**, **interoperability** and **integration** | [156, 426, 391, 427, 159, 421, 86, 428, 225, 405, 429, 430, 406, 397, 425] | [418, 158, 427, 159, 431, 428, 402] |
| 4 | Changes to **legislation**, **policy** and government-mandated financial factors (**incentives** or **penalties**) | [156, 246, 391, 392, 257, 431, 432, 421, 87, 433, 388, 434, 390, 435] | [158, 392, 432, 433, 388, 394, 436-439] |
| 5 | Capital investment, implementation, maintenance and support **costs** | [156, 157, 427, 421, 268, 440, 400, 26, 441, 442] | [153, 244, 158, 252, 86, 84, 389, 268, 396, 441] |
| 6 | Impact of LHS on health outcomes and patient-clinician encounter within the patient care **workflow** | [391, 392, 258, 388, 400, 390, 405, 397, 402] | [428, 405] |
| 7 | **Privacy**, security, data integrity and accuracy | [418, 156, 246, 391, 260, 392, 257, 248, 421, 84, 388, 62, 428, 400, 395, 401, 26, 429, 397, 402, 443] | [158, 159] |
| 8 | Approvals and **ethics** oversight for use of digital health data | [247, 248, 444, 434, 400, 445, 396, 270, 446, 399, 406] | [400, 399] |
| 9 | Organisational **culture**, management and clinician attitudes to change | [418, 158, 431, 86, 447, 433, 265, 62, 396, 430, 406] | [431, 84, 448, 430, 425] |
| 10 | Identifying and involving all relevant **stakeholders** | [418, 449, 424, 246, 158, 427, 260, 392, 431, 421, 422, 444, 436, 450, 448, 400, 401, 26, 441, 451, 452, 397] | [157, 431, 453, 84, 448, 440, 26, 446] |

## 6.4 ITPOSMO-BBF ANALYSIS

This research extended the ITPOSMO framework to support comparative analysis of barriers, benefits and facilitators that are identified from the literature for both EHR and LHS: naming the resulting new framework ITPOSMO-BBF. ITPOSMO-BBF adopts the aspects and dimensions of the original ITPOSMO framework [178], and combines these with the concept of a framework for analysing barriers and facilitators from [454]. The resulting new framework developed for this research is then used to identify related benefits that authors identify as resulting, whether directly from mitigating the barriers or simply from applying the identified facilitators. Figure 17 provides an example of the ITPOSMO-BBF diagrammatic structure that: (1) identifies barriers and facilitators to the implementation of EHR and LHS and the frequency with which they were discussed in the literature; (2) quantifies the relationships between facilitators and barriers, and; (3) identifies potential benefits authors expect from resolving the listed barriers.

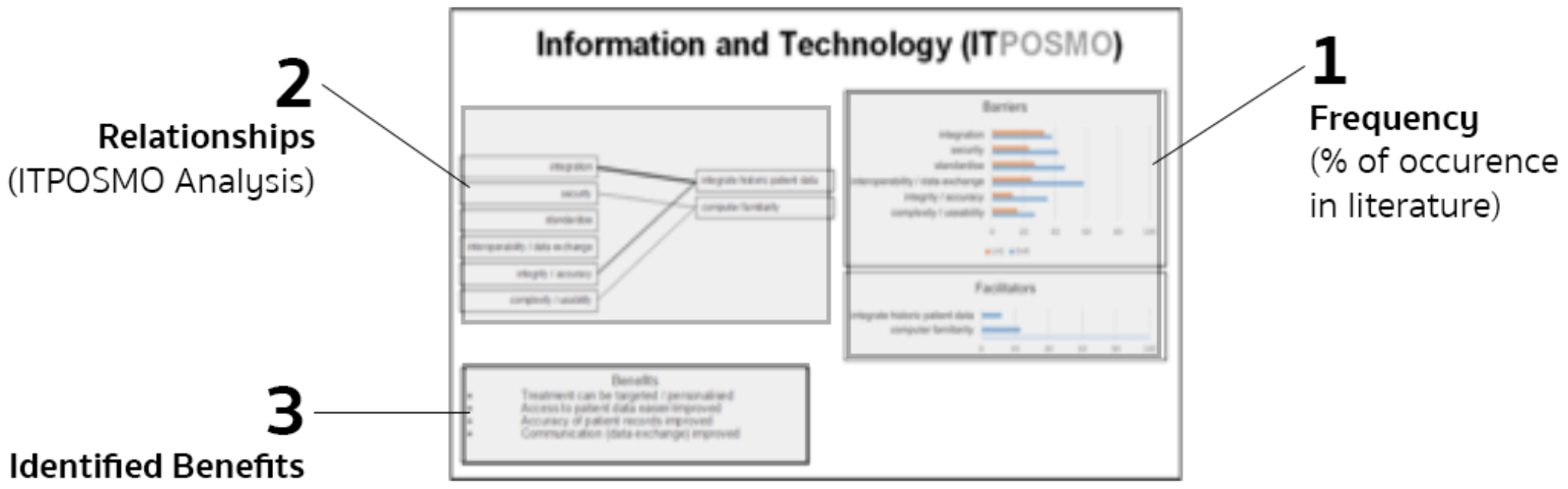

Figure 17: ITPOSMO-BBF diagram structure

Each of the four ITPOSMO-BBF diagrams presented in the Results chapter provide key data, including the percentage of EHR and LHS literature identifying a particular barrier or facilitator, the contextual relationships observed in authors' discussions of barriers and facilitators, and the corresponding benefits ascribed to resolving the barriers or engaging with the facilitators. The frequency of attention drawn by authors to the relationship between a barrier and a facilitator is shown using a weighted line. Figure 18 identifies the relationship between the number of authors discussing a relationship and the thickness or weight of the line identifying that relationship in the ITPOSMO-BBF diagrams.

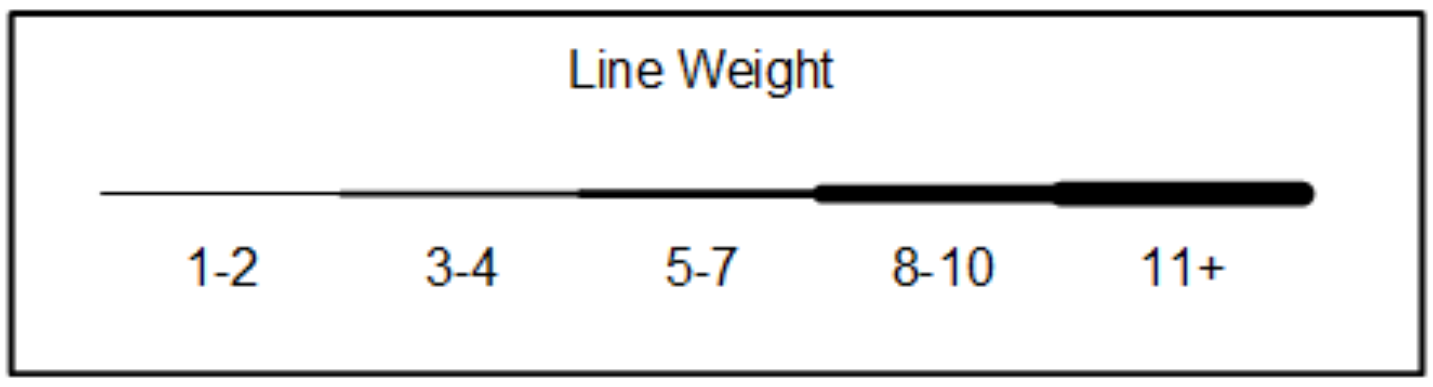

Figure 18: Line weight and number of authors

While ITPOSMO was developed as a retrospective analysis of projects that have already been completed, ITPOSMO-BBF can be used with data on barriers and facilitators to understand, plan for and mitigate potential barriers prior to a new EHR or LHS implementation.

Except for the fourth and seventh themes from Table 11 and Table 13, which were found to fall equally across the boundaries of two aspects, Figure 19 shows that each of the key themes could be located within a single ITPOSMO aspect. Themes are distributed outwards from the centre based on the strength of their heterogeneity from other aspects – that is, the more a theme identified within a particular aspect and apart from other aspects, the further out from the middle it is located.

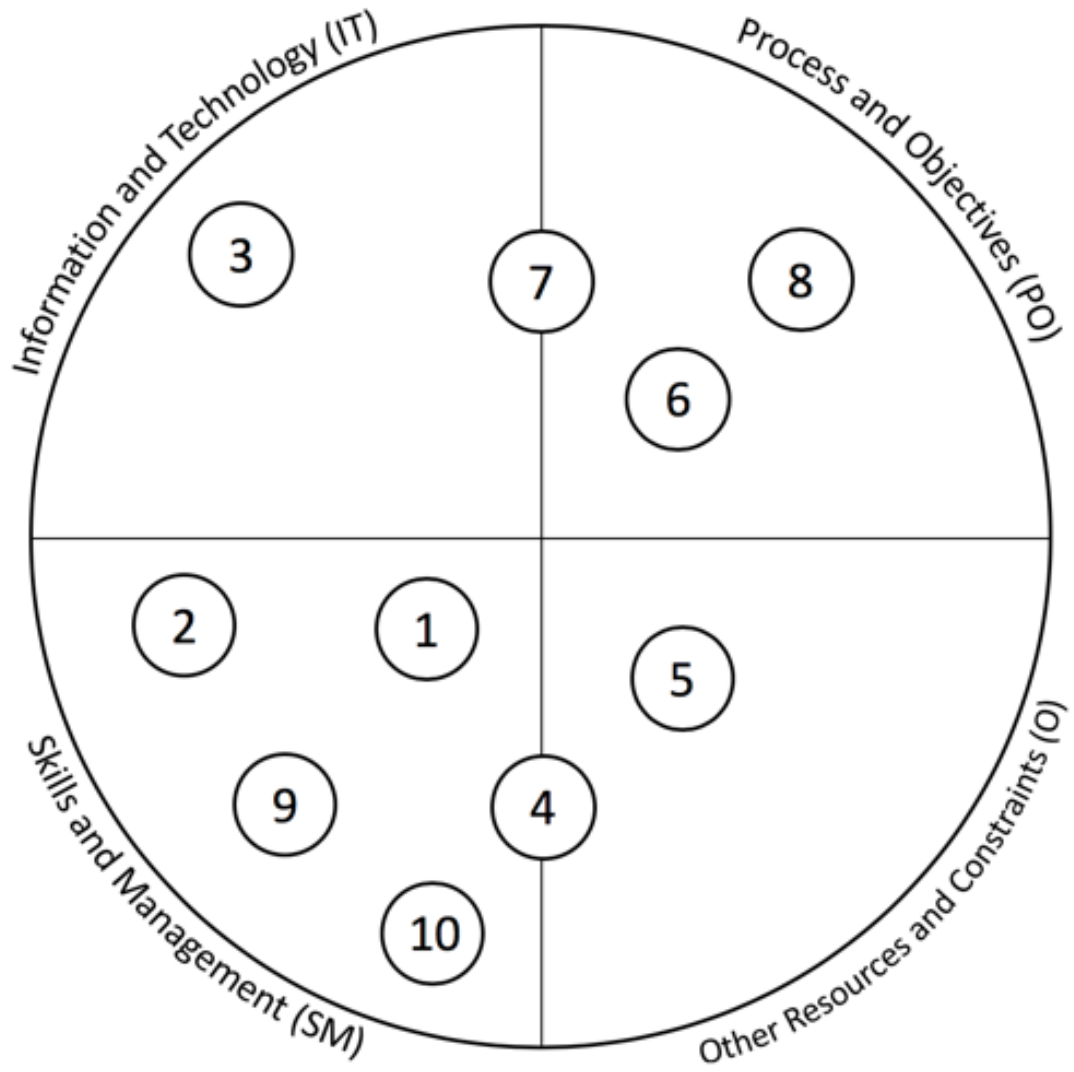

Figure 19: Linking the key themes to ITPOSMO aspects

### 6.4.1 Barriers

Every author presenting an HIS implementation spoke of barriers or challenges. Some presented as *barriers to entry*, such as the high cost to implement and support EHR and other health information systems (HIS) [234, 241, 237] or issues resulting from complicated and inconsistent legislation [240, 235, 263]. Others presented as *barriers to success*, including the need for data and systems standardisation [237, 238, 263], interoperability [242, 264, 263] or integration [242, 243]. Finally, there were *barriers to organisational culture* [455] such as clinical users and patients who expressed reticence or negative attitudes towards the use of computing systems and resistance to potential changes in workflow that it was claimed would disrupt the flow of patient care [241-243] or impact time management [264, 238, 398], and opposition to the requirement for staff development and training in information systems use [234, 241, 243]. Issues of data accuracy and integrity which could have lasting impact on patient care and health outcomes were also claimed to arise from using health information technology [237, 238, 263, 398]. Concerns were expressed that systems producing, storing, exchanging or amalgamating health data could negatively impact patient privacy and confidentiality [234, 242, 262, 264], with adequate data security remaining as an unresolved primary challenge [154, 234, 238].

Appendix A provides detail of the individual authors and frequency with which each barrier was discussed. Figure 20 presents the EHR barriers grouped within the referential thematic context in which each author discussed them. Barriers naturally presented from the literature in the context of four archetypal themes (shown in blue): (1) Human Factors issues; (2) Technological/ IT Issues; (3) Policy issues, and; (4) Finance issues. Sub-themes (shown in black) were identified that further contextualised small groups of identified barriers. Authors gave greater attention to human factors barriers affecting staff and clinicians. These barriers including clinician resistance, the need for training, support and user champions, and the potential for disruption to patient care and increased clinical workload. Further analysis was performed to separately identify *who* authors described as being primarily affected by some of the more frequently identified barriers. Figure 21 displays these lines of effect in red. While most of the attention in the literature is being directed towards barriers arising from staff and clinicians, this shows that it may be patients who are more frequently impacted.

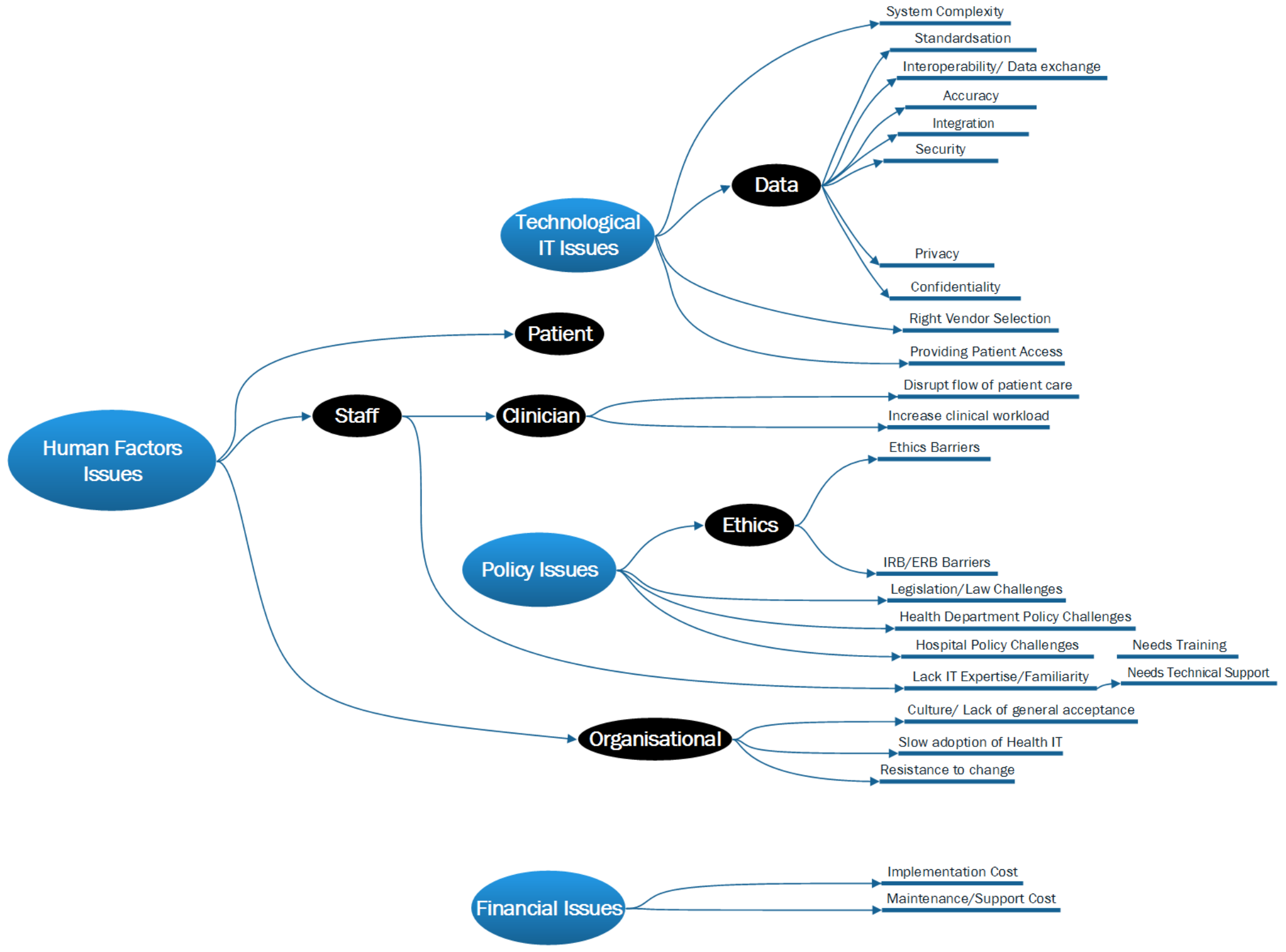

Figure 20: EHR Barriers

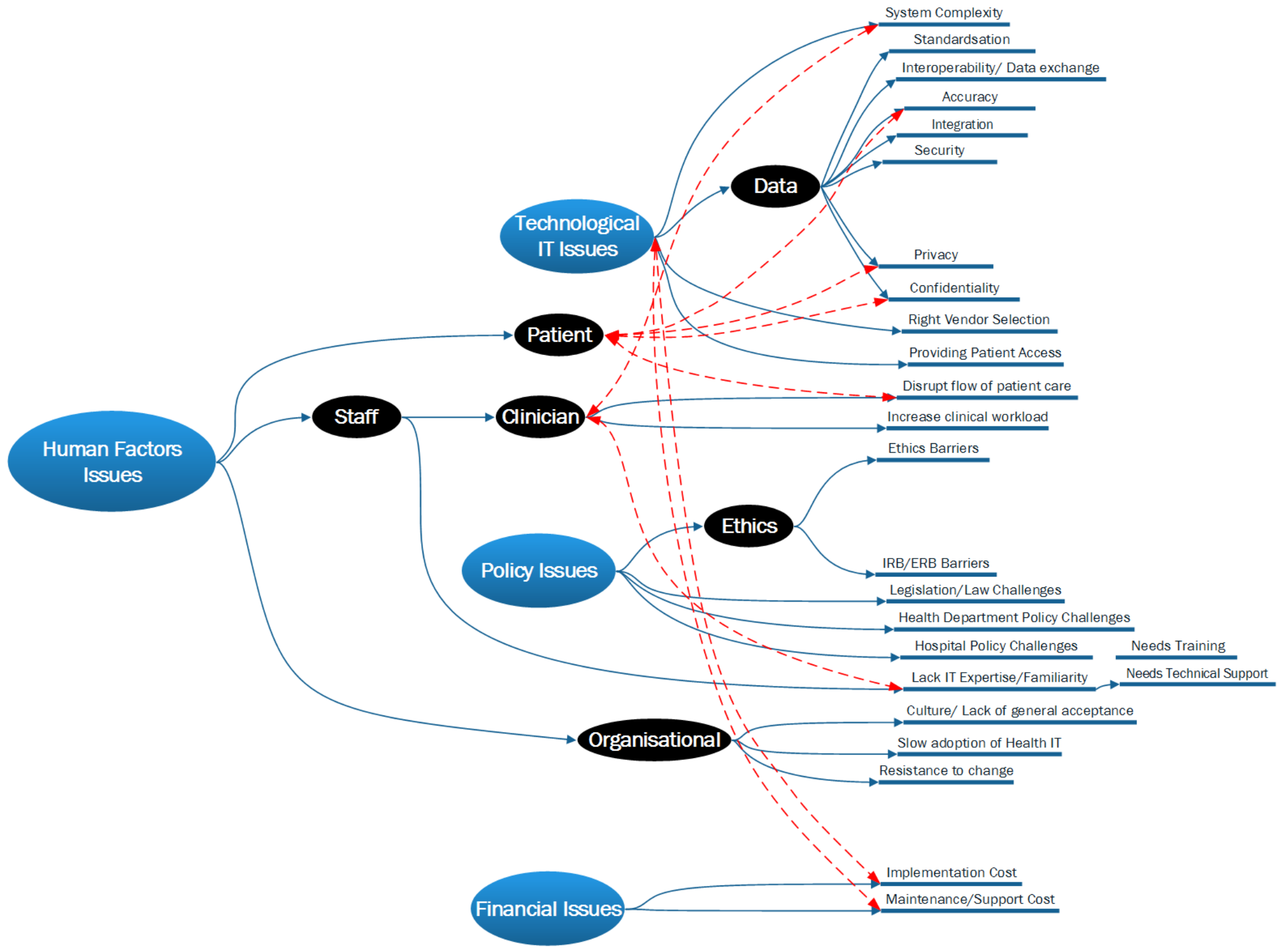

Figure 21: EHR Barriers indicating ‘*barrier directly affects*’

Finally, Figure 22 indicates how the identified barriers distribute within the four ITPOSMO aspects that will be used later in this chapter to contrast barriers and facilitators.

### 6.4.2 Facilitators of EHR Adoption

The literature demonstrates that patience [243], committed leadership [242, 262], systematically planned incremental implementations [234, 262] all have considerably positive effects. Early collaboration with clinical users [234, 242], provision of both general computer and EHR-specific training [234, 243, 237], and engaging user champions to drive acceptance and reduce user frustration [404, 234, 239] were identified as substantial human factor facilitators that counter resistance. Resistance is considered significant enough that some medical schools have recommended or even mandated specific training for doctors in the implementation and use of health information systems [456-458]. Appendix A shows that a total of fourteen facilitators were identified. These were described by authors as falling within eight of the ten key themes as shown in Table 11.

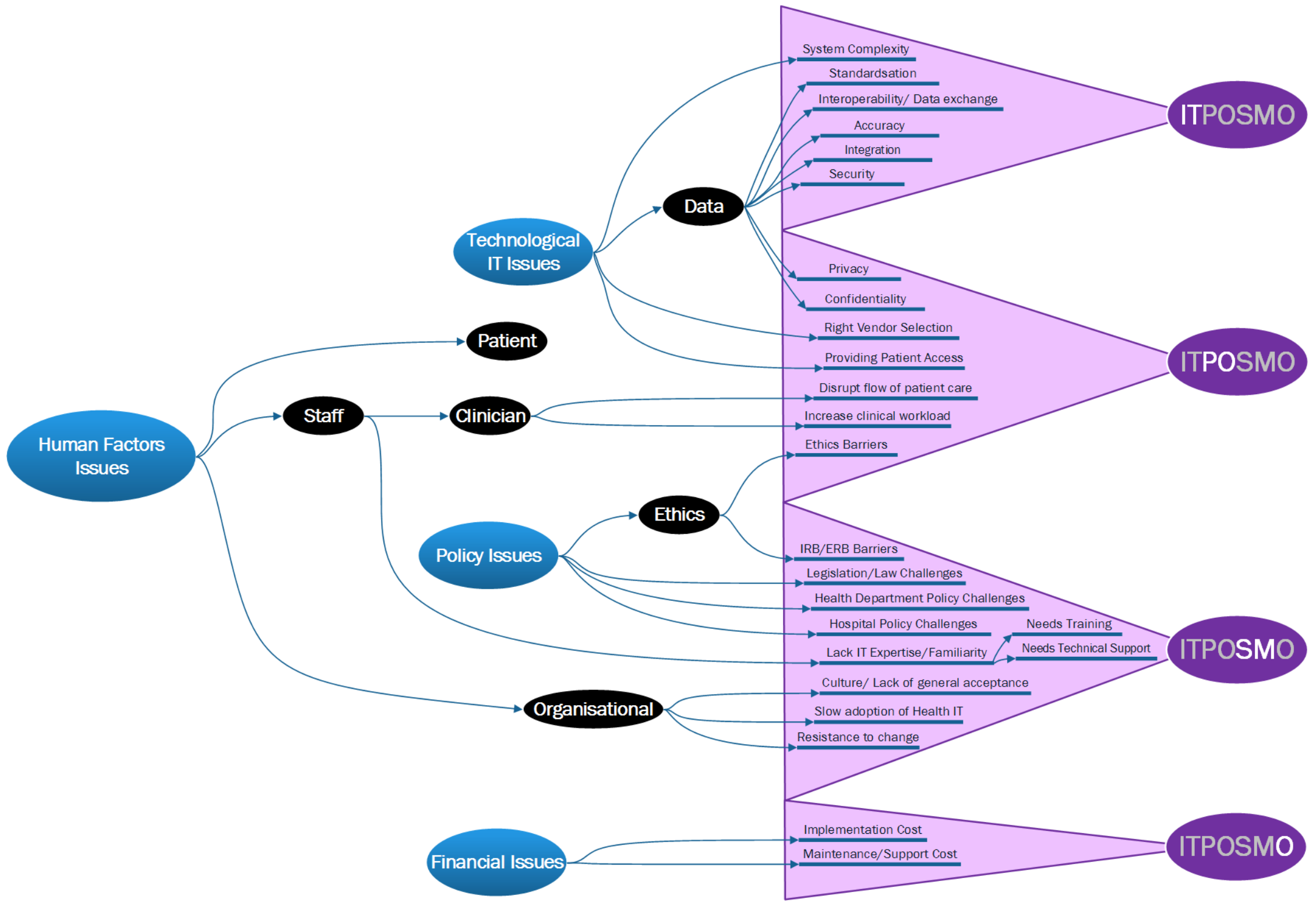

Figure 22: EHR Barriers within the ITPOSMO aspects

### 6.4.3 Benefits

Organisations contemplating HIS implementation do so with some intention of realising benefits [459]. While it could be argued that any benefit improves the health of patients, even indirectly, this research found that benefits fall broadly into two categories. First, there are those that have a direct positive effect on health outcomes for patients. These include any that increase patient safety [234, 263], reduce harm from treatment or medication errors [234, 238], or improve the overall quality of healthcare [241, 237, 238]. Second, there are those seeking to improve some metric of healthcare delivery: increasing efficiency and accountability [234, 241, 237], or reducing waste or overconsumption of resources, which accordingly increases the overall capacity of healthcare systems [234, 241, 262].

### 6.4.4 ITPOSMO-BBF: Information and Technology

A number of Information and Technology barriers were described by authors in Table 11 and Table 13. These were most often issues resulting from the stand-alone and bespoke nature of health systems, coupled with a lack of ability to combine systems or data in any simple, inexpensive or meaningful way. Little effort has been expended in devising facilitators to resolving these barriers, even though there are important benefits that could be realised. Figure 23 shows that the most frequently discussed facilitator that authors considered would realise a number of the listed benefits was the seemingly simple act of integrating historic patient data so that any new system presented a complete picture of the patient.

### 6.4.5 ITPOSMO-BBF: Process and Objectives

Many of the Process and Objectives (PO) barriers might reasonably appear to fall within the remit of ethicists. The small number remaining were raised by clinicians who work closest with patients, namely nurses and general practitioners. Many of the potential benefits authors felt would result from resolving the PO barriers would appear to deal directly with issues of patient safety and confidence with health services, yet surprisingly no single facilitator was directly attributable to resolving one of the PO barriers and realising the benefits presented in Figure 24.

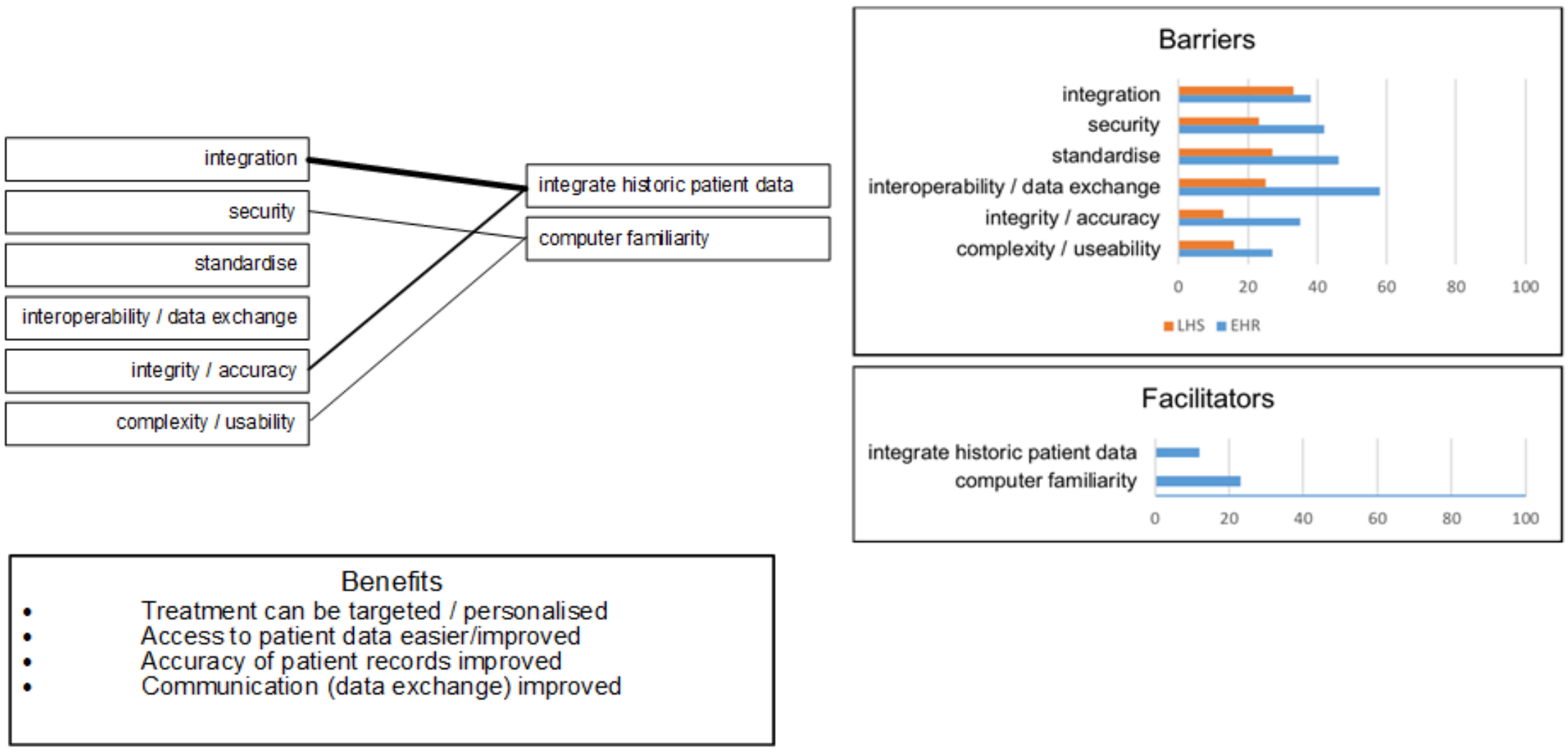

Figure 23: Information and Technology (ITPOSMO-BBF)

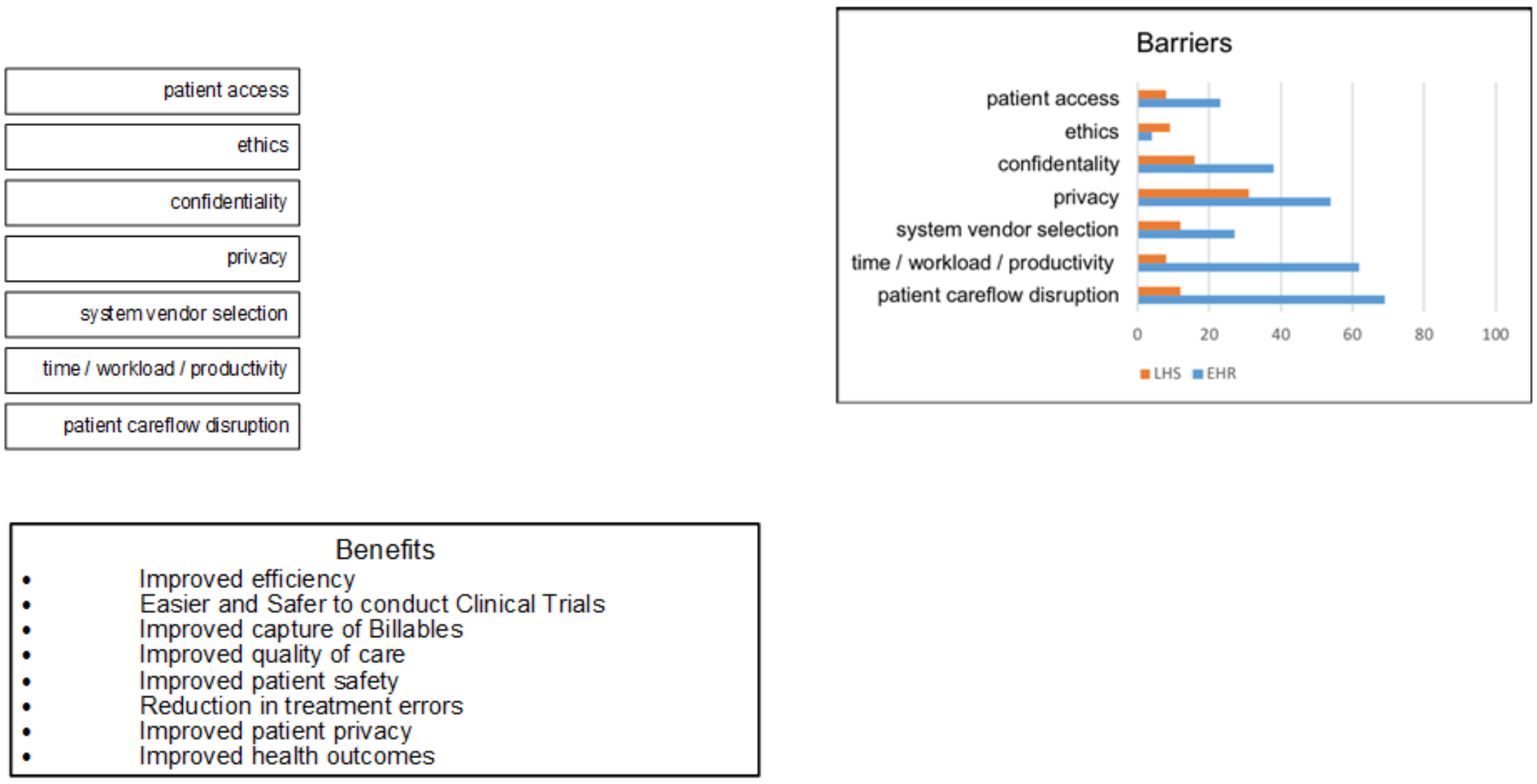

Figure 24: Process and Objectives (ITPOSMO-BBF)

### 6.4.6 ITPOSMO-BBF: Skills and Management

Most effort aimed at facilitating EHR has gone towards resolving ‘human factor’ barriers. This, in spite of the fact that authors only made reference to one benefit arising from facilitators in the Skills and Management (SM) element of ITPOSMO, as shown in Figure 25. Even in the case of technical support and training, these were described by authors in the context of developing skills and managing staff resistance, yet no single author reported that any of these facilitators was actually reducing staff resistance to technology or improving adoption rates for EHR. While there is strong interest directed towards resolving adoption issues, the facilitators presently being employed do not appear to have substantially resolved these issues, as the EHR adoption problem persists.

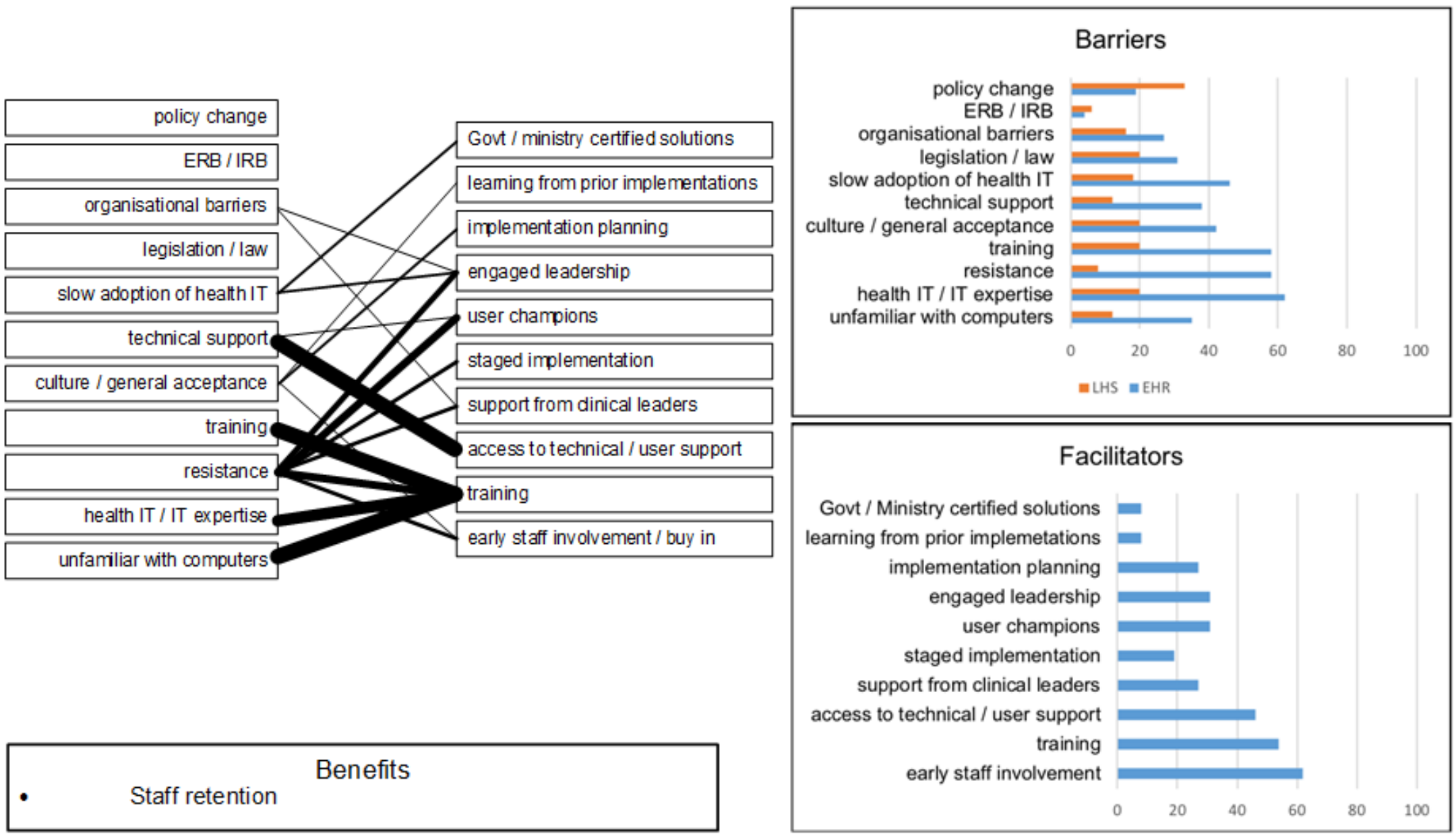

Figure 25: Skills and Management (ITPOSMO-BBF)

(ERB/IRB: Ethics Review Board/Institutional Review Board)

### 6.4.7 ITPOSMO-BBF: Other Resources and Constraints

The other resources and constraints (O) element shown in Figure 10 reviews those attributes not falling within the first three ITPOSMO elements, including finance, maintenance, and user and systems support. While financial incentives that had been enshrined in the laws of countries like the U.S. was discussed by more than half of all EHR papers, only three mentioned the potential for penalties for non-adoption contained in the same legislation to be a facilitator. Note however, that none of the three led to the belief that the threat of penalties had helped an implementation of HIS.

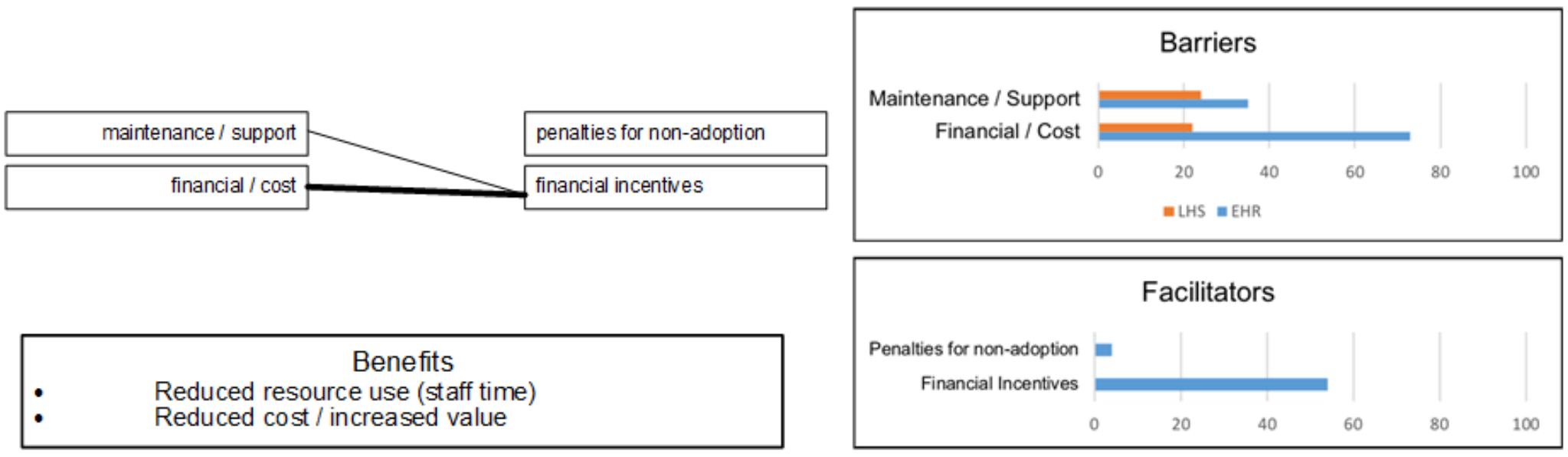

Figure 26: Other Resources and Constraints (ITPOSMO-BBF)

## 6.5 DISCUSSION

### 6.5.1 EHR in the context of HIS Adoption

A strong undertone was observed in the EHR literature, consistent with the idea that strong barriers still impede acceptance of EHR even when they are now largely ubiquitous at all levels of healthcare service delivery. One novel causal factor receiving attention is *digital disruption*. Digital disruption is a catch-all term for a range of related issues, described as *the changes facilitated by the introduction of digital technologies that occur at a pace and magnitude that disrupt established ways of value creation, social interactions, doing business and, more generally, our way of thinking* [460]. One group in Australia have attributed the failure of more than half of all EHR system implementations to poor understanding and management of digital disruption, failure to understand and manage disruption to clinical workflows, the anxiety this engenders in staff, staff dissatisfaction, and the concerns for the quality and safety of care being delivered during the digital transformation [461]. Elements of digital disruption are seen in almost all the barriers identified in this work. Facilitators, such as those which stipulate early staff involvement, staff training, and user championing have been promoted for many years as mitigants for these barriers. If the issues raised by [461] as elements of digital disruption are still evident, it is possibly because the selected facilitators do not adequately deal with the barriers identified, or they were not successfully employed by the authors during their hospital's implementation project. Policymakers and clinicians still struggle with barriers that only serve to limit widespread acceptance and adoption of HIS [462-464].

### 6.5.2 Analysis of LHS Themes

Several themes emerged during this work which are discussed in this section, which also includes topical analysis to investigate their effect on the LHS domain.

***Patient Focused:*** While clinicians argue they always practiced patient-focused medicine, normal clinical practice follows population medicine-based EBM [465]. Precision Medicine (PM) extends diagnostic practices with profiling techniques and therapies tailored to the individual [466, 467]. *Patient focus* is a key dimension that LHS improve [86]. PM approaches can be retrospective, as in CER, and prospective, when genotyping for treatment selection [466, 467]. Patient-centred care encourages data use in optimising care for individuals [422]. Aggregated patient records enable LHS to identify cohorts similar to the patient [468]. [76]. LHS is an efficient tool for integrating PM into practice. As the clinician enters attributes about the patient, the LHS refines a cohort of prior similar patients. It assesses the treatments they received to recommend one most likely to produce an optimal outcome.

***Technology Usage:*** The focus for health IT has shifted from issues of adoption to identifying how to best use technology to improve healthcare delivery and outcomes [442]. This shift is significant in creating LHS, and elevates issues in EMR/EHR interoperability, data standardisation and quality that must be resolved if LHS are to be truly practical and ubiquitous [469, 470, 442]. Health IT's ability to improve healthcare service delivery quality and efficiency is recognised [471, 472]. Enabling necessary data flow and integration of data sources are key abilities technology can deliver, representing core requirements to enabling LHS [473, 420]. Integration of learning into technology is observed in every LHS solution reviewed. Technology is fundamental to LHS. As EBM evolves from paper-based roots, clinicians and healthcare providers will realise benefits from coupling technology to Learning Health Organisations, thus realising LHS [469, 474].

***Decision Support:*** Healthcare providers evolved from considering health IT as a billing and documentation facilitator, to contemplating its active participation and capabilities to answer complex questions in care delivery [475]. LHS bring opportunities for improving speed and efficiency of clinical decision support [473, 476, 475]. LHS solutions are context-sensitive, incorporating cohort identification and risk modelling in real time to identify interventions for improving individual patient outcomes [432, 477]. LHS have potential to rapidly perform retrospective comparative effectiveness trials, evaluating treatment options against each other where they have been provided to similar patients [153, 477]. In contrast with Randomised Clinical Trials (RCTs), LHS is considered safer, and engenders greater confidence in accuracy of the treatment choice [153, 477].

***Application of LHS:*** Clinical epidemiology is an example of learning healthcare. EBM evolved from clinical epidemiology: statistically identifying the optimal treatment which becomes best practice for that condition [478, 479]. Conversely, the focus for PM is selecting from the available interventions the treatment that will best serve the individual. The primary driver towards *population medicine* was economic: maximising benefit while minimising cost, harm and waste [478]. While meant for benefiting the individual patient, population medicine has disadvantages in that the individual's best interests may conflict with those of the population and it is difficult to reconcile the two [478, 422]. Individual patients may be denied higher priced precision interventions in favour of lower cost population-optimised interventions [478]. The Heimdall framework demonstrates that LHS focus healthcare using population medicine and EBM directly onto the presenting patient. While EBM selects one treatment for all patients, LHS produces a cohort with attributes similar to the presenting patient, identifying the treatment most likely to be effective for this individual patient. LHS in this way delivers PM.

***LHS Barriers and Further Observations:*** Most authors discuss barriers to implementation. The most common are: cost [404, 480, 288]; *data interoperability* and *standardisation* [481, 392, 482]; *poor data* quality and *integrity* [257, 483, 435]; *informed consent* and *ethics review complications* [484, 485, 400]; *privacy* and *security issue*s [418, 481]; and *slow technology adoption* [481, 421, 486]. These issues are discussed in the same context as they were for adopting EHR/EMR. This suggests LHS are inheriting challenges from the EHR/EMR on which they depend.

### 6.5.3 ITPOSMO-BBF

While the benefits of LHS build and significantly expand on those put forward for EHR, the barriers described for both were similar. This tends towards confirming LHS are inheriting unresolved challenges from EHR. For this reason, this research also investigated the factors identified as facilitating EHR implementation to assess whether it is possible that these may aid in resolving LHS implementation challenges. This is discussed next.

### 6.5.4 Enabling LHS

The barriers identified by this study represent the substantive issues impeding implementation and adoption of LHS. However, few authors are asking the right questions, such as: how can health departments achieve subject matter expertise in all technology, legal, compliance and privacy aspects? Nor are they recognising that these issues must be resolved in order to achieve a secure data repository to support LHS [392, 84]. The literature contains abundant discussion of requirements or elements of a solution to one or more of the barriers, mostly revolving around calls for a new and common set of standards [392, 259]. Another key point is that while LHS are seen to have potential to significantly benefit the conduct of many types of clinical trial research, EHR were not discussed by any author as being beneficial in that same context. This, in spite of the fact that EHR are the constituent components for all LHS and are often where the patient data from clinical trials is recorded. However, we conclude that the absence of universal and effective LHS shows these barriers remain unresolved. While LHS have been developed that are intended to learn from evidence-based literatures, patents, genomics and other non-patient data, LHS can only be successful in their ultimate goal of delivering ubiquitous individualised or personalised healthcare when data from EHR are made available. Data sharing to create large-scale data warehouses will only occur when clinicians and patients can trust that methods and systems used are protecting their privacy. Even then, ethics review processes may still impede the realisation of knowledge that can come from LHS.

It became clear during this study that the majority of facilitation efforts are focused on human factors, and more specifically the mitigation of negative aspects arising from a general resistance to change. More facilitators fell within the Skills and Management aspect of ITPOSMO-BBF than any other: an aspect domain that primarily deals with staff, the skills they possess, whether these are sufficient to using the HIS being implemented, and the structure and style of management within the healthcare organisation [178, 179]. While many of these facilitators should lead towards a smoother and more successful implementation it is notable that while so much effort goes towards mitigation of barriers in this aspect domain, authors only ascribed the one benefit as arising from it: staff retention. This strongly suggests that the greatest mitigation effort results in the least amount of tangible benefit. Other areas described with benefits bringing more significant impact on patient safety, health outcomes and efficiency, such as those of the Process and Objectives aspects are left unresolved. Further research is needed to provide those implementing HIS with a more focused toolkit capable of mitigating a wider range of barriers and enabling delivery of the broadest possible benefits.

It is for those involved in developing HIS to actively participate in counteracting the barriers and changing negative perceptions. The barriers and issues for LHS identified in this research were largely similar to those previously ascribed to EHR, with addition of the key issue that *good quality EHRs are a necessity to enable LHS* [23]. Variations on Meaningful Use legislation seen in the USA, UK and Australia are aimed at supporting use of EHR in LHS, motivating expensive government-operated national solutions like Care.data (UK), Shared Care Records (NZ) and MyHealth (AUS). While the cost to implement and maintain standardised EHR repositories in support of LHS may seem substantial, the cost savings promoted as justification for engaging LHS are potentially many times more significant [487]. Many government, academic and private organisations are funding research into novel health technologies aimed at realising the benefits identified by this study. A key theme within the facilitating factors for EHRs is clinician involvement, whether it be through early involvement, or ongoing as HIS are integrated into the patient care environment. Many health technology implementations have lacked the input and involvement of appropriate stakeholder group members. Seeking input from all stakeholders who will impact and be impacted by the HIS is a significant factor in reducing resistance and increasing adoption of technology that has the potential to help many [488]. While there has been a call for clinicians and their training organisations to engage with technologists, those working in the technology sector must be similarly called to seek clinician involvement [488, 489]. Clinicians and technologists must work as co-investigators and leaders in the research and implementation of health technologies. Engaging each other as leaders and stakeholders to influence HIS design and implementation [489, 490].

While this thesis starts from the premise that comparative review of the EHR implementation literature can provide a framework for analysis of LHS implementation, with the potential to increase the number of successful implementations, future work to extend this might also include comparison through analysis using other established quality improvement and change management frameworks. Such analysis was outside the scope of this particular work.

## 6.6 SUMMARY

EHR lack standardisation in many areas including content, structure and process. This lack of standardisation was clear in several key themes arising out of our investigation of the barriers, benefits and facilitators of EHR implementations. This is not only true for the processes and forms that are the structure of EHR, and which EHR inherit from CCPS, but especially true where it concerns the content, or data, that the EHR is used to collect, and the inability of most EHR software products to integrate and interoperate with other HIS. Prior to this research no similar effort had been made to collectively identify the barriers, benefits and facilitators of LHS. Comparative analysis of these for EHR and LHS produced two significant findings: (1) LHS are inheriting many barriers from EHR; and, (2) much of the work supporting EHR and LHS is targeted at human factors barriers, and given there was no suggestion that facilitators in one domain created benefits in another, little effort has gone towards facilitators that would resolve barriers leading to improved healthcare service efficiency, patient safety and health outcomes.

LHS were conceived to capitalise on the large datasets of EHR held by many healthcare provider networks and state and federal health departments. While most potential LHS have been proposed or tested by researchers and healthcare services, it is possible that the lack of a standard terminology, nomenclature and definitions may have restrained many authors from identifying their health-AI/ML solutions as LHS. The proposed taxonomy can help researchers to standardise naming and presentation of LHS solutions, and was significant in developing the Heimdall Framework that identifies within the healthcare organisation the standard application for each LHS type. This supports the contention raised in the introduction of this chapter that as a concept becomes better documented and understood, standardisation can occur.

# Chapter 7: LAGOS Framework to guide integration of new healthcare tools

When we have developed and implemented a new tool, and that tool has become established, and even after we have re-evaluated and standardised that tool, effectively integrating it with the other tools and processes currently in use can still prove exceedingly difficult [491-493]. In these situations it is common to investigate a framework that can inform and assist integration [492, 493]. This thesis has presented a number of new tools. However, on reflection, an additional framework is necessary to inform integration of these tools into healthcare practice. This chapter presents LAGOS, a pathway-based framework that uses the established *levels of medical practice* to guide integration and application of each of the primarily clinically-focused tools of this thesis.

Section 7.1 introduces Pathway Theory and the LAGOS framework and describes the construction of each of LAGOS' five pathways. Section 7.2 presents an instructional case study in the application of LAGOS in three key areas of contemporary health research: (1) the creation of realistic synthetic data for secondary uses; (2) patient risk and clinical decision modelling; and, (3) approaches to empowering patients as active participants in their own healthcare.

## 7.1 LAGOS: BRINGING CCPS, EHR AND LHS TOGETHER

### 7.1.1 Pathway Theory

The use of a pathway theory and analytical framework helps highlight underlying values and fundamental relationships between otherwise fragmented concepts [494]. Different to the clinical or treatment pathways common to medicine and nursing, an application of pathway theory is designed around a consistent value or belief set [494]. In this case, we use the established levels of medical practice application [63, 495, 496] which are described on the Application Pathway. We start with this application pathway and use it as the basis for ordering all other pathways, bringing together the domains of LHS and CCPS and arranging them on the basis of how each function in the delivery of population, evidence-based or precision medicine.

### 7.1.2 The LAGOS Framework

LAGOS is an acronym drawn from the first initial of each of the five pathways shown in Figure 27. These pathways, described below, are Learning Health Systems, Applications, Guidance, Operational and Systems. They focus and converge on the individual patient presenting to the clinician, and broadly define the domain map covered by the LAGOS Framework. Each pathway radiates from the most general, or broadest application of that pathway's scope, toward the centre, which represents the most specific application. Thus, within LAGOS, as the viewer moves along each pathway toward the patient in the centre, the focus of elements at each layer shift from a population-based focus that seeks interventions or approaches applicable to an entire population, to a precision medicine focus that can identify the most appropriate and curative intervention for the individual patient.

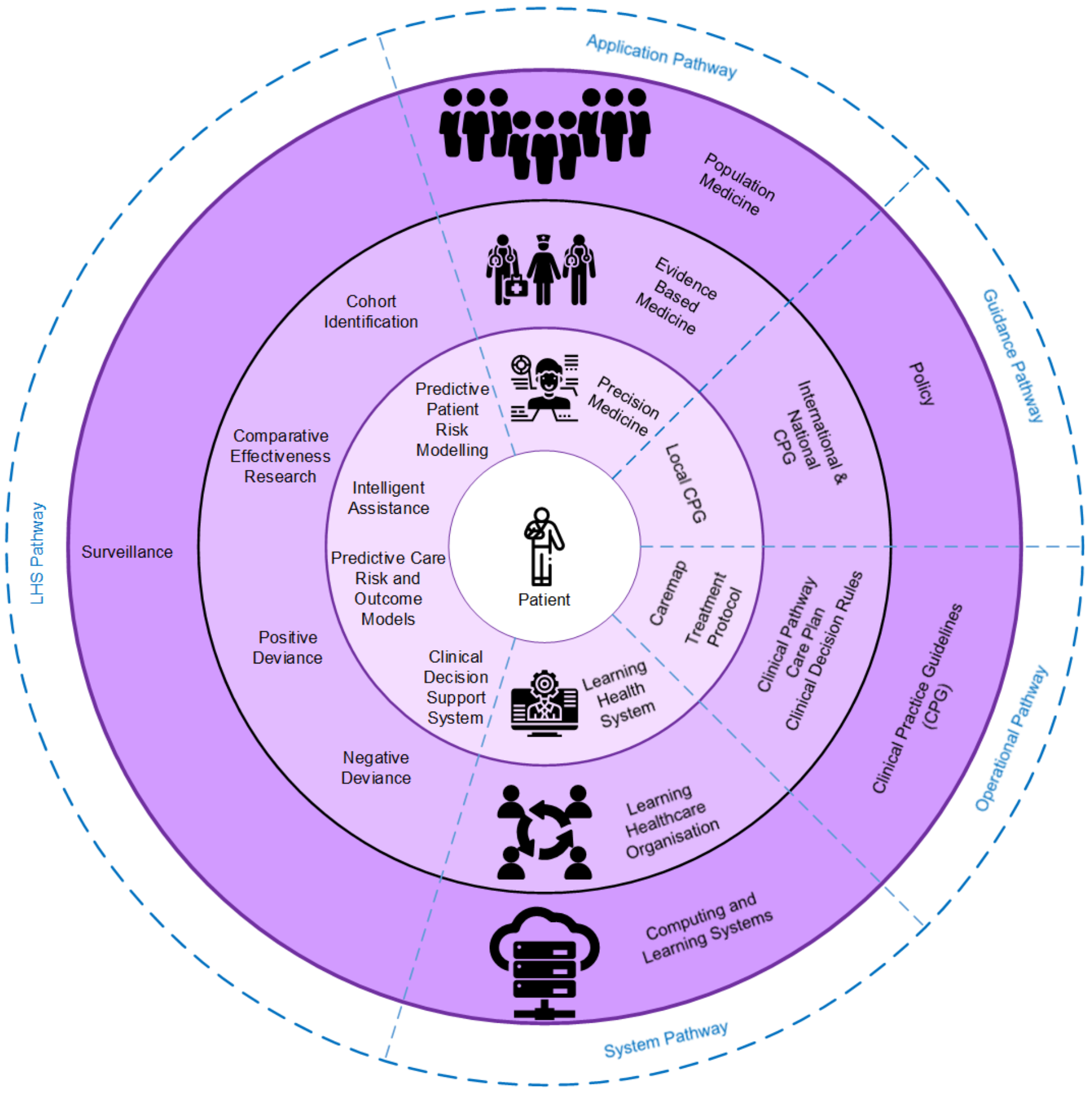

Figure 27: The LAGOS Framework

### *The Application Pathway*

The pathway of medical practice and its ongoing shift from population medicine through evidence-based medicine (EBM) and on towards precision medicine, is best described in the title of Horowitz *et al*'s [495] paper *From Evidence-Based Medicine to Medicine-Based Evidence*. Population medicine effectively promotes those activities that will improve general health for an entire population [497], and may be impacted and influenced by policy and financial concerns [497] that are not always informed by clear or convincing scientific evidence [278, 279]. EBM focuses on informing clinicians' practice with scientifically-proven current best treatment options for a particular condition. This has sometimes been characterised as a one-size-fits-all approach to medicine [63, 496]. EBM identifies the treatment or intervention that cures the largest number of patients with a given medical condition. Finally, precision medicine is an approach to customising medical treatment, basing the treatment decision on patient-specific factors [63]. Just as EBM is the scientific basis for and epidemiological application of population medicine [498, 495], precision medicine is seen to be the natural scientific evolution of EBM [63].

### *The System Pathway*

The system pathway supports our need for technology and learning in the provision and improvement of healthcare. In the outer arc is the computing technology on which everything else, including patient EHR, operate. On the next level the *learning healthcare organisation* employ learning approaches, research and trial approaches and stored EHRs in their goal of identifying new evidence-based knowledge and treatments. Proximal to the patient on this pathway are the LHS which are those systems presenting customised treatments for individual patients.

### *The LHS Pathway*

The complete LHS taxonomy shown in Figure 14 is represented in the LHS pathway of LAGOS. At the outer edge is *Surveillance*, which operates as an automated alert process within Health IT systems that: (a) monitors the entire population's EHR for diagnosis or clinical coding of a range of communicable disease; and, (b) can also be programmed to monitor for adverse treatment outcomes or drug reactions. Those LHS types at the second layer primarily work with or on the basis of EBM, or are used by clinicians in review of their, or other clinicians, treatment outcomes. This includes *Cohort Identification* which is most often used within learning health organisations to identify groups of patients who share one or more similar characteristics of interest. In the smallest circle are those LHS proximal to the patient which can be engaged by the clinician in the direct conduct of patient care. These models support personalised clinical decision making and predict risks and outcomes for an individual patient receiving the clinician-selected treatment. It is these LHS that meet the true definition of *precision medicine*.

***The Guidance and Operational Pathways***

CCPS are documents that define healthcare policy and procedure. They are arranged in the hierarchy shown previously in Figure 10 which describes both their distance from the patient and inherent operational nature. For example, policy is furthest from the individual patient in that it is set by governments to guide health services for entire populations. Policy also possesses the quality of being the least operational as they are the most general and least likely to be grounded on scientific evidence. CCPS may also be described based on their inherent nature; that is, whether their primary intention is only to provide guidance, or be operational and therefore engaged in clinical practice. Policy, and the levels of CPG discussed in Section 5.1.2, operate primarily in the guidance space and are presented on the *Guidance Pathway* based on their proximity to the patient. Policy is population-based while local CPGs are closest to the individual patient. There is overlap between the guidance and operational pathways in that local CPGs can also be seen with an operational content. On the *Operational Pathway*, local CPGs are observed at the population end as they would be applied to the general population of the health service or district who are diagnosed with that particular condition. Care plans, clinical pathways and clinical decision rules are found centrally along the pathway as they primarily focus on the diagnosis or condition and present with templated, general or recommended interventions applicable to the entire cohort. Caremaps and treatment protocols that have options that are selected based on individual patient need or treatment response, are accordingly are found closest to the patient.

## 7.2 LAGOS: CASE STUDY

### 7.2.1 LHS in the context of generating the Realistic Synthetic Electronic Health Record

Accessing EHR for secondary use purposes such as non-clinical research, patient or disease modelling or AI training presents with challenges: (i) the challenge of achieving ethics approval for access to collections of EHR; (ii) a practical difficulty when consent is required from each individual patient; and (iii) over-reliance on anonymisation as a privacy-protection mechanism can reduce or remove important contextual detail. The CoMSER Realistic Synthetic Electronic Health Record (RS-EHR) [269] and ATEN Realism in Synthetic Data [499] projects operate following the approach described in Figure 28 and focus on satisfying the need for access to EHR for secondary uses relying on a privacy-preserving knowledge-intensive method to generate locally realistic, but not real, synthetic EHR without needing access to the real EHR.

The relationship between LHS and RS-EHR can be two-way. LHS can help to provide the aggregated statistical data and knowledge described as rules and relationships that exist in EHR datasets. In this way, RS-EHR generation need never be exposed to real EHR during the definition or generation of synthetic EHR. Conversely, LHS can be built, trained and validated by projects like PAMBAYESIAN using collections of RS-EHR, prior to being productionised to work from real patient data for patients and clinicians.

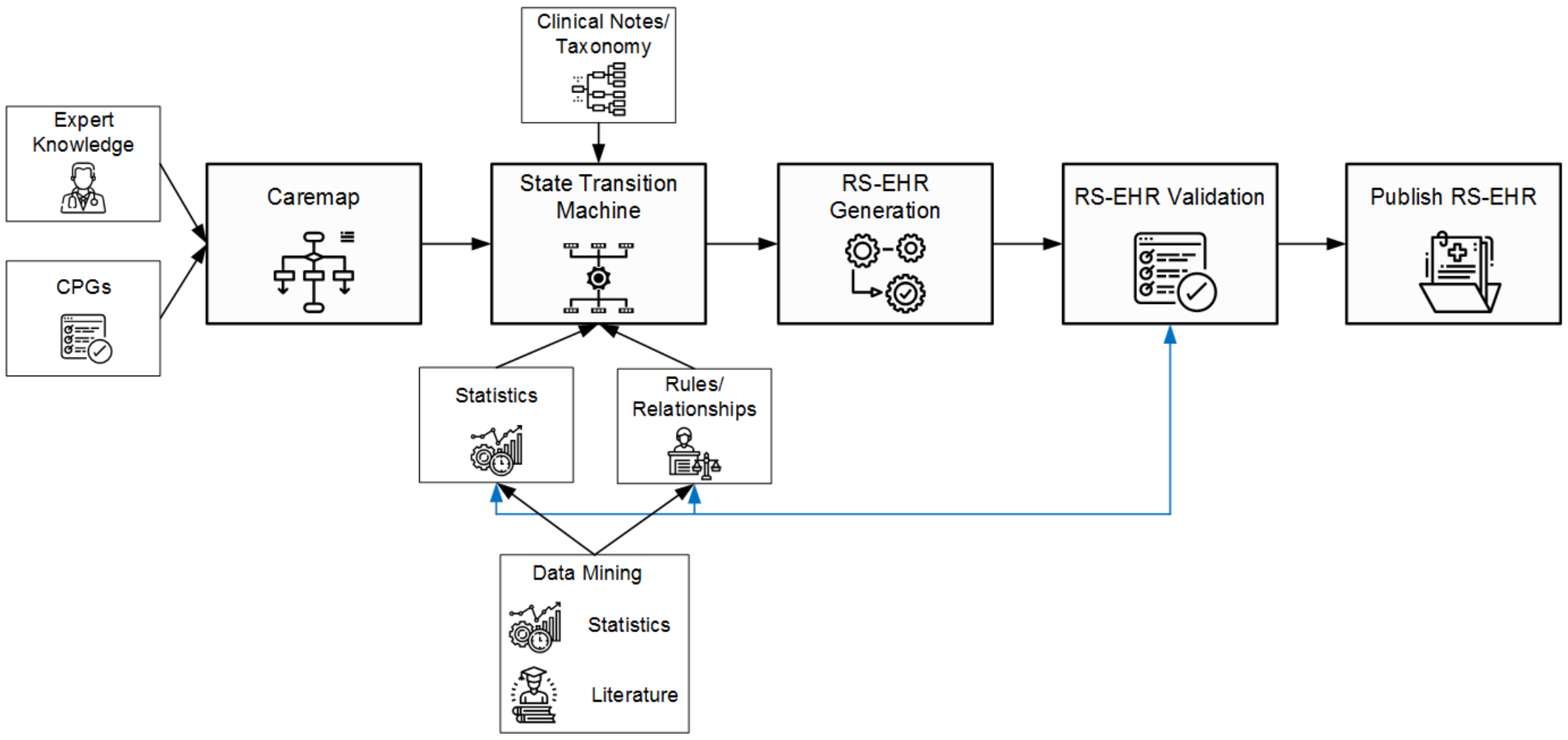

Figure 28: The ATEN Approach to RS-EHR Generation [499]

The LHS paradigm allowed us to fully exploit the routinely collected data from the healthcare system. This made development of knowledge-intensive methods for generating synthetic EHR successful, making it easy to create collections of realistic synthetic EHR for use in secondary uses where privacy concerns prevent release of real data. Further, development of knowledge-intensive models enables prediction of patient risk for particular negative outcomes and the recommending of appropriate and more effective treatments based on patient characteristics, history and current symptomatology possible.

To fulfil RS-EHR's aims the following LHS types from the LHS pathway, which apply to the levels of medical practice from the application pathway (in brackets) are needed:

1. *CI - learning evide*nce (EBM) and operating within the context of the *Learning Healthcare Organisation* level of the *System Pathway* to identify a prescribed cohort of patients with similar health conditions or characteristics such as demographics and symptomatology consistent with the disease to be modelled and generated;
2. *PD* and *ND* - *learning evidence* (EBM) and operating within the context of the *Learning Healthcare Organisation level* of the *System Pathway*; of commonly used treatments, both effective and ineffective to ensure synthetic patients receive realistic treatments and outcomes.
3. *PPRM* - *specific to patient* (Precision Medicine) and operating within the *Learning Health System* level of the *System Pathway* to identify patterns and model risk factors consistent with adverse events;
4. *CDSS* - *specific to patient* (Precision Medicine) and operating within the *Learning Health Syst*em level of the *System Pathway* to identify characteristics of synthetic patients that make them compatible for generation of specific disease or treatment outcomes;

### 7.2.2 LHS in the context of Patient Risk and Decision Modelling

There are numerous intelligent systems that may be capable of supporting clinical decision making for diagnosis, prognosis or treatment selection. Bayesian networks (BNs) are one such system [500]. BNs model uncertainty and allow the user to update prior belief, such as when assessing the probability for presence of a medical condition in light of new evidence (additional symptoms, risk factors and test results). However, the process of building these intelligent systems for chronic conditions is not yet fully explored and understood. The chronic condition presents as a particularly challenging clinical scenario as the patient's condition must be monitored for extended periods during which many decisions may be undertaken. Ideally, doctors and nurses should be able to monitor patients without the resource-intensive, expense and inconvenience of clinic visits, except when such visits are necessary. Current clinical records and care processes do not easily receive, integrate or enable patients in the home to collect and transmit self-monitoring data from inexpensive sensor-based devices like the Apple watch and continuous glucose monitors.

PAMBAYESIAN [367] is developing a new framework for distributed probabilistic decision-support systems. As shown in Figure 29, PAMBAYESIAN combines patient data with clinical expertise and patient input, for use in developing intelligent systems. The novelty of this framework is the use of "conventional" EHR (e.g. blood tests, imaging results) combined with near real-time continuous data from local sensors for learning and providing new knowledge. This allows for autonomy in a collaborative decision-making environment that includes clinicians and patients, enabling avoidance of unnecessary visits to a clinic or hospital. Referring to the *Thresholds* box in the top right corner of the figure, once the patient's condition crosses the diagnostic threshold (in green), the clinician prescribes the treatment (in yellow) and treatment review (in red) thresholds. The patient self-monitors the parameters of their condition and enters these into the LHS application. If assessment and prediction of their condition rises above the treatment threshold, the patient receives treatment, be it medication or otherwise. If it rises above the treatment review threshold, the clinician is alerted that the patient requires review so that an appointment can be offered.

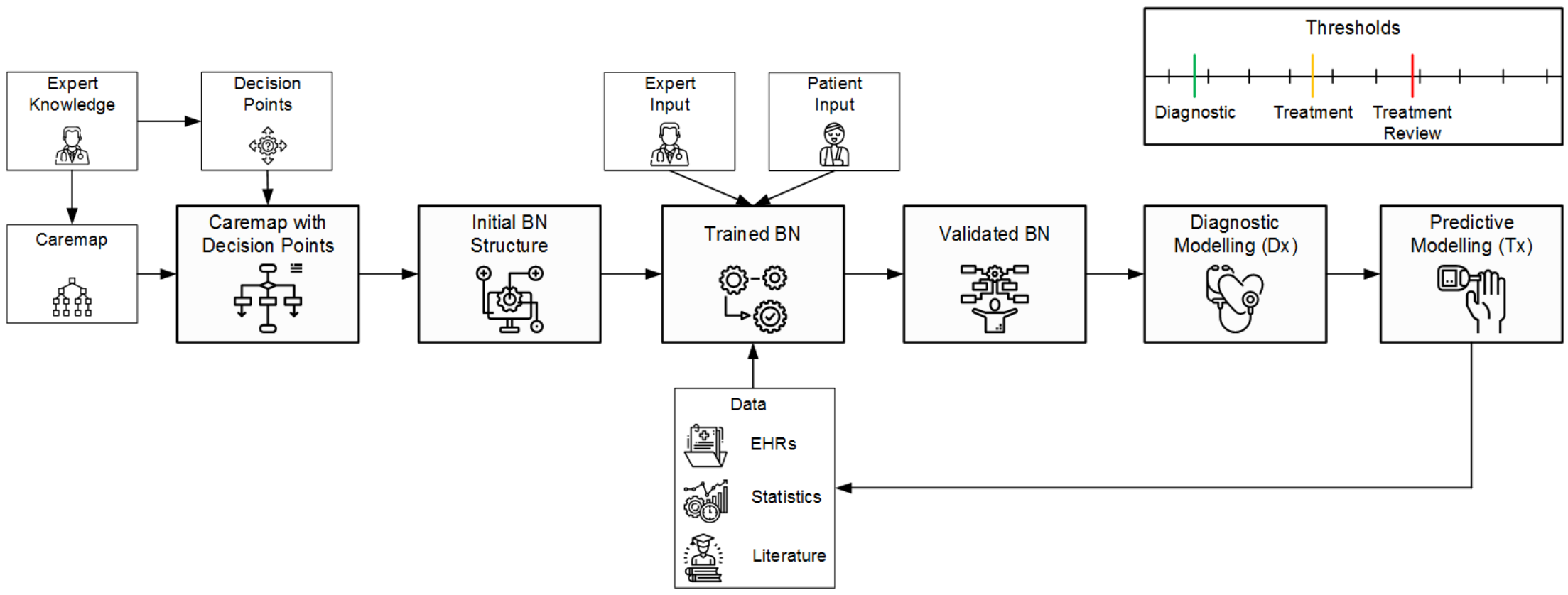

Figure 29: The PAMBAYESIAN Project as an LHS

To fulfil PAMBAYESIAN's aims the following LHS types from the LHS pathway, which apply to the levels of medical practice from the application pathway (in brackets) are needed:

1. CI - learning evidence (EBM) and operating within the context of the Learning Healthcare Organisation of the System Pathway to identify patients with similar demographic and clinical characteristics;
2. A CDSS - specific to patient (Precision Medicine) to collect and analyse daily data and operating within the Learning Health System level of the System Pathway to provide relevant patient feedback;
3. A PPRM - specific to patient (Precision Medicine) and operating within the Learning Health System level of the System Pathway to predict potential future adverse events, for example, where treatment may fail to improve the patient's experience.

### 7.2.3 LHS in the context of Empowering Patient Participation in Healthcare

Despite advances in modern medicine, many chronic conditions have generally proven incurable. The daily life of patients with chronic conditions is highly affected by disease progression; over time disease symptoms exacerbate until they overwhelm the patient. Patients must constantly evaluate their condition, making day-to-day decisions regarding care and relying on advice from their treating clinicians to guide those decisions. Again, despite medical advances, access to healthcare remains a significant issue for all patients. Regular appointments with doctors or nurses are time consuming, expensive, inconvenient and in many cases cannot be scheduled to coincide with times when the worst symptomatology may present.

PAMBAYESIAN aims to empower patients to undertake day-to-day self-care within boundaries; diagnostic, treatment and treatment review thresholds that are defined by the patient's clinician. As shown on the right side of Figure 30, home health monitoring devices and applications will be used to gather patient symptoms, measurements and reports about their condition, and with BN intelligence will tailor clinical knowledge and generate patient advices. In this way, PAMBAYESIAN will promote continuous monitoring of the patient's condition while supporting patient self-management and engagement of timely interventions. PAMBAYESIAN also promotes a more effective and efficient interaction model between patients and clinicians whereby expensive and time-consuming clinic visits need only occur when a patient's monitoring shows that their symptomatology has escalated and surpassed the treatment review threshold as discussed in the previous section.

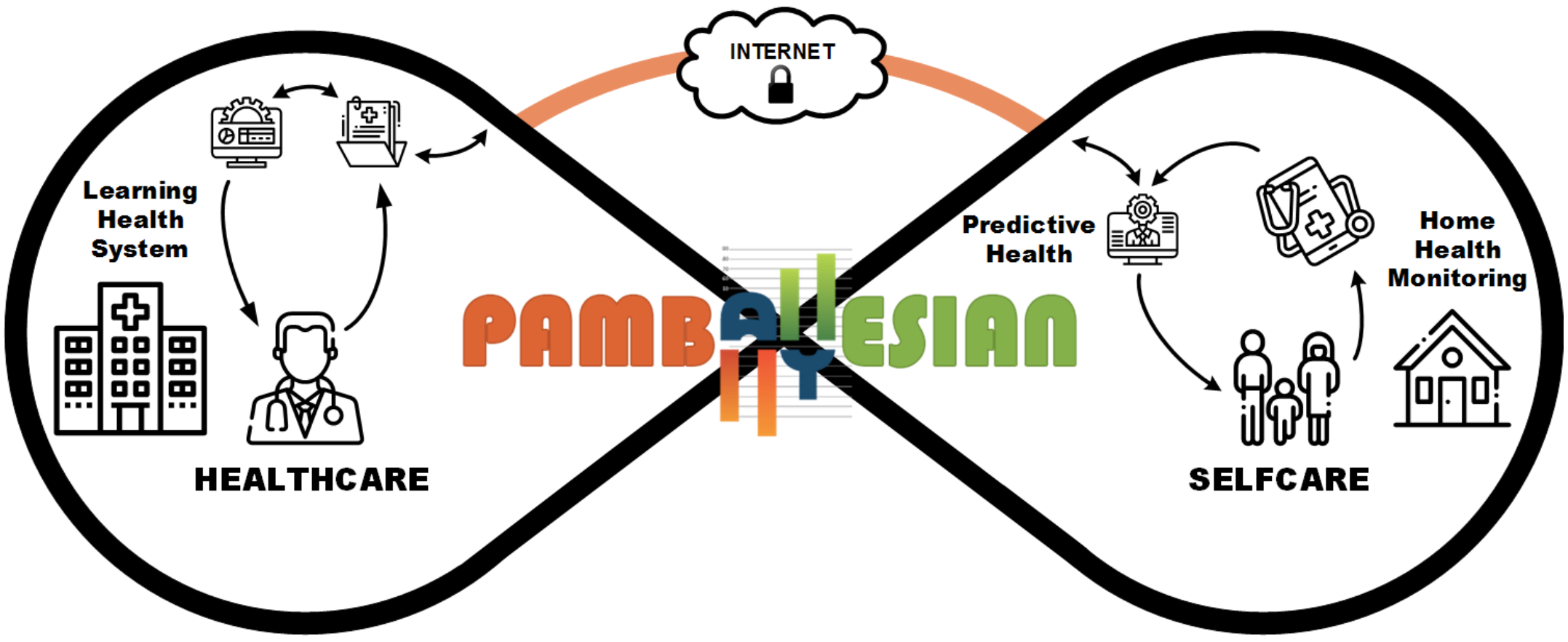

Figure 30: Using PAMBAYESIAN to Promote Patient Empowerment

To fulfil PAMBAYESIAN's aims the following LHS types from the LHS pathway, which apply to the levels of medical practice from the application pathway (in brackets) are needed:

1. CI - learning evidence (EBM) and operating within the context of the Learning Healthcare Organisation of the System Pathway to identify patients with similar demographic and clinical characteristics;

2. A CDSS - specific to patient (Precision Medicine) to collect and analyse daily data and operating within the Learning Health System level of the System Pathway to provide relevant patient feedback;
3. A PPRM - specific to patient (Precision Medicine) and operating within the Learning Health System level of the System Pathway to predict and identify potential future adverse events.

## 7.3 SUMMARY

Medical practice has focused on health and wellness recommendations for the entire population (population medicine), and identifying the treatment that positively benefits the largest number of patients with a particular condition (evidence-based medicine). However, clinicians have recently realised that with the added insight of modern genomics and data science, these approaches are no longer sufficient. LHS are a recent concept with little more than a decade of research but limited exposure. LHS have the potential to completely change the way medicine is practiced by guiding treatment selection on characteristics of the individual patient (precision medicine) instead of focusing on the disease (EBM). This chapter set out with the aim of unifying LHS with the domains of medicine, clinical care, health informatics and decision science. The LAGOS framework successfully incorporates these domains within five pathways, each with three levels corresponding to how the elements within that level relate to the three application levels of medicine: population medicine, evidence-based medicine and precision medicine. This chapter also introduced the work of the EPSRC-funded PAMBAYESIAN project and shows aspects of PAMBAYESIAN as a case study demonstrating how such health informatics and decision science projects are and should be regarded as LHS. LAGOS is further grounded through application to three LHS examples, demonstrating that it describes the interaction between LHS, CCPS and medical practice. We believe that LAGOS can help those standardising healthcare services by identifying related elements on the other pathways. This can help when seeking the appropriate system or documentation on which to base an LHS, or an LHS that can help deliver a particular level of medical practice.

# Chapter 8: TaSC: Towards a Standard for Caremaps

This Chapter presents a standardised method, model and notation for caremap content, structure and development. Section 8.1 presents TaSC, the standardised model for caremap structure and content and a development lifecycle approach to standardising the process by which caremaps are developed. Section 8.3 discusses the background and basis for clinical decisions, and extends the TaSC model with decision points (DP) that graphically represent clinical decisions. Section 8.4 summarises and concludes the chapter. The TaSC model is evaluated later in Chapter 9.

## 8.1 TASC: STANDARDISING CAREMAPS

A search using the three terms observed across clinical documentation literature: ‘caremap’, ‘CareMap’, and ‘care map’ was conducted using a range of databases. This search was supplemented using a citation search of the selected caremap papers. Once duplicates, papers not based in the nursing, medical or healthcare domains and those papers using the term ‘care map’ in other contexts were removed a core pool of 1215 papers remained for this review. Initially, each paper was reviewed using CA and TA to identify and classify terminology, construction and content elements, and to infer the development process.

### 8.1.1 Data Visualisation

Graphically modelling the process of patient care for a given medical condition or hospitalisation event is not new, and approaches have been proposed utilising a variety of presentation styles, including: UML process modelling to represent the ongoing clinical management of a chronic condition [501]; business process modelling notation (BPMN) to visually map the treatment flow encapsulated in clinical pathways [502]; and, influence diagrams to model the structure of complex clinical problems, identifying decisions to be made, the sequence in which those decisions may arise, the information available to make the decision and the probability of uncertain events [503]. Caremaps model patient care, and while their presentation style and content has changed markedly since conception in the 1980's, contemporary caremaps present as one example of an underdeveloped information visualisation approach currently used in clinical medicine [39]. Existing caremap visual modelling systems, in addition to lacking standardisation, generally lack comprehensive representation of all types of clinical decision points within the caremap.

Graphical modelling, termed Information visualisation, is the study of transforming data, information and knowledge into visual representations that can more easily convey meaningful patterns and trends hidden within large and otherwise abstract datasets [504-507]. Depending on the style or information presentation approach used, information visualisation can provide the reader with an ability to immediately comprehend vast datasets and perceive emergent properties, patterns, new insights, causation, and problems within the data and the methodology used to collect it [505-507]. A 2011 report by the US Institute of Medicine (IoM) described information visualisation in clinical medicine as underdeveloped when compared and contrasted with other scientific disciplines [508].

### 8.1.2 Structure

As discussed in Table 3, caremaps were transformed through the years from holistic and wordy texts [330, 318, 200] to illustrative graphs [16, 317]. This transition is shown in Figure 31, where the traditional caremap on the left requires completion of a large number of text-based fields, while the contemporary caremap on the right presents as a flow diagram of the care process for the clinician to follow.

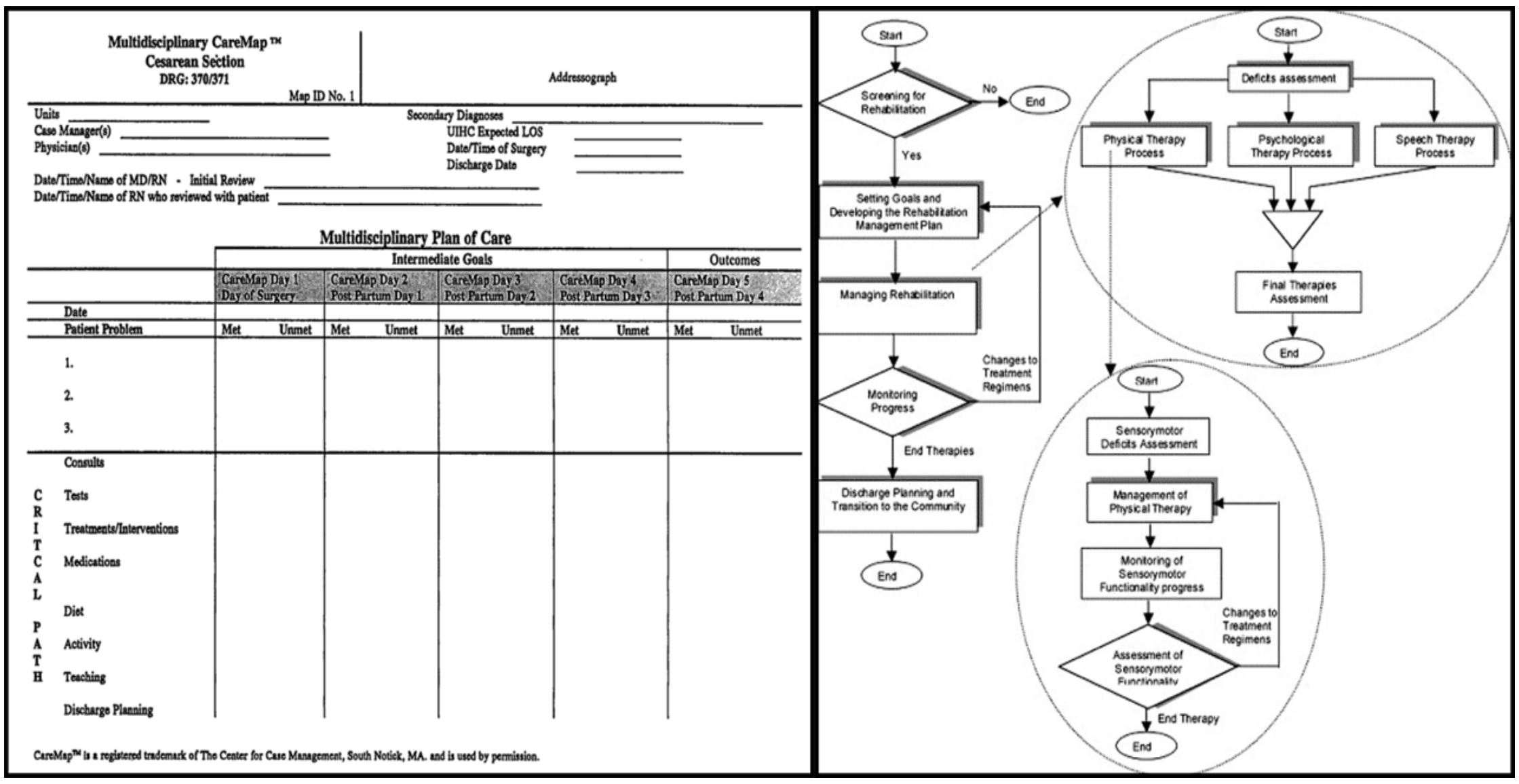

Figure 31: Traditional text-based [330] and Contemporary flow-based [509] caremap comparison

The vast majority of contemporary caremaps present either: (a) monochromatic like the right-hand example in Figure 31 [332, 320, 510], or; (b) as a full colour flow diagram like those shown in Figure 32 [511, 323]. The flow diagram represents a well-known and long established process modelling tool [512].

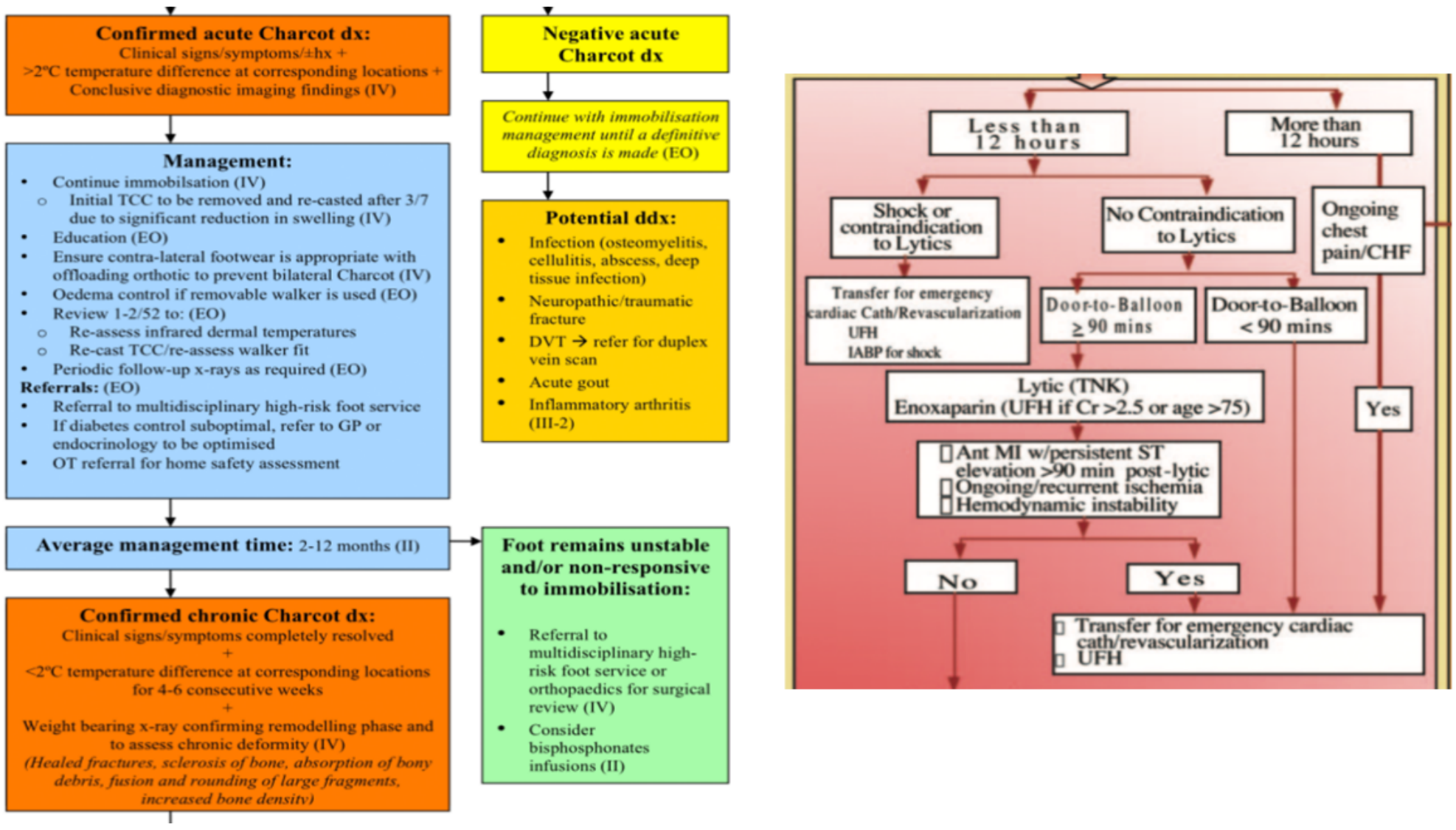

Figure 32: Colour caremap examples from [511] left and [323] right

In general, each of these caremaps has its own set of boxes and notations, with the most common box presenting as a rectangle representing a care process usually described as an *activity*. Contemporary caremaps contain a set of activities that represent medical care processes; however, the literature shows that there is no consistency in the way an activity is represented. Different shapes such as rectangular boxes with rounded [205] or square [16, 325] corners, or even arrows [513] have been used. In some cases, activities that lead to multiple mutually exclusive pathways are represented by a diamond [201, 514, 510]. The flow from one activity to another is sometimes illustrated with arrows [332, 317] or simple lines [320, 201]. The literature lacks a clear description as to whether the caremap should present with clear entry and exit points. For some, neither is present [317, 325], while for others these points are an implicit [201, 514] or explicit [509] part of the diagram. Finally, most of the contemporary caremaps reviewed in this review contain multiple pathways and are often presented as multi-level flow charts [16, 509, 510].

Contemporary caremaps lack a consistent structure. A problem that is amplified by the absence of an established formal representational model for caremap elements. To begin resolving these problems the entity relationship diagram (ERD) in Figure 33 was prepared to describe the relationship between structural elements of the caremap.

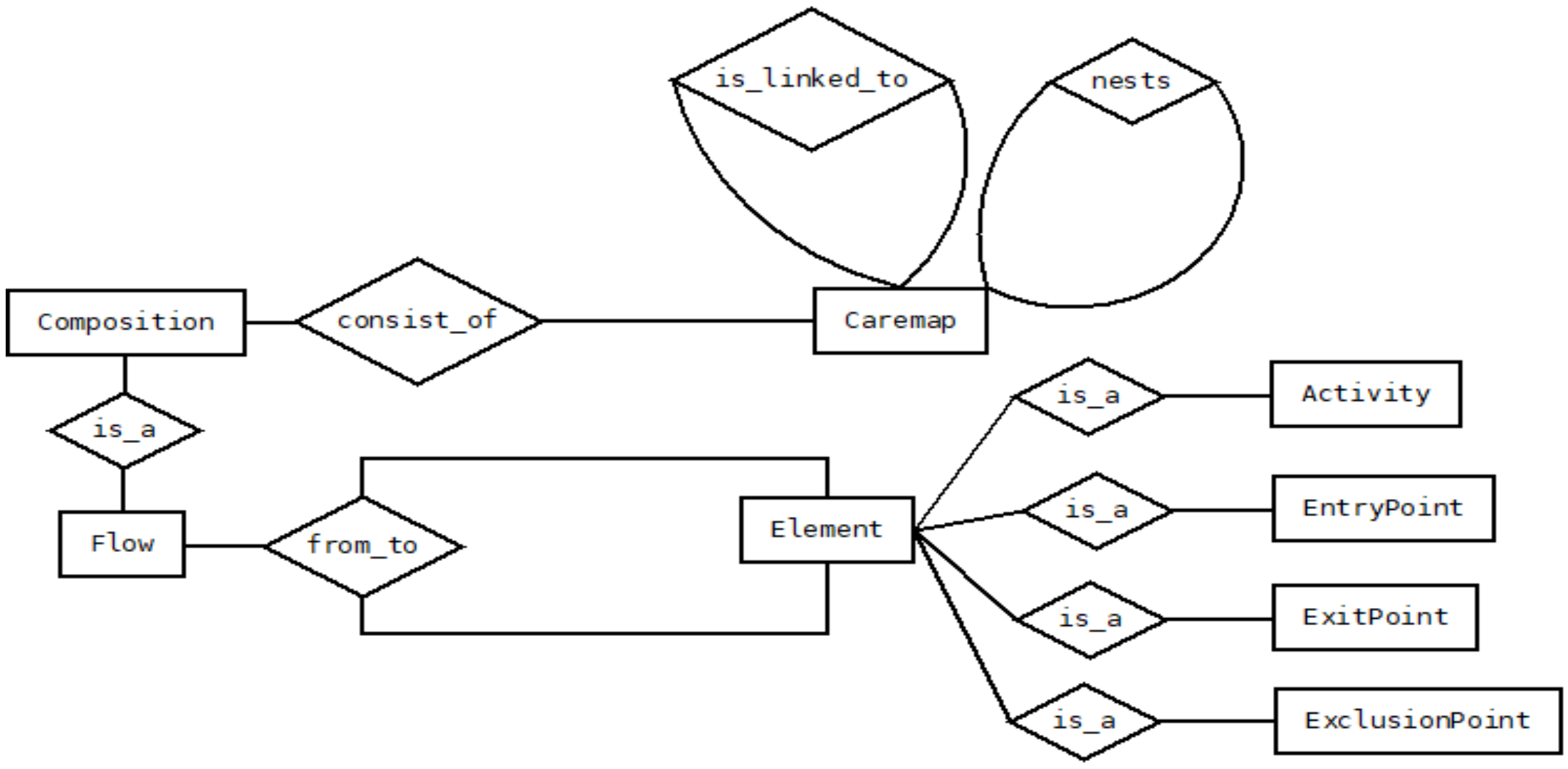

Figure 33: Entity relationship diagram for the caremap

Guided by the ERD and adapting the activity diagram representation from Unified Modelling Language (UML) version 1.0, which itself was an adaptation of Harel's statechart [515, 516], Table 14 presents the representational notation for caremap elements, while Figure 34 presents the standardised structural model for the caremap.

Table 14: Representational notation for structural elements of caremap

| | Element | Description | Notation |
|---|---|---|---|
| 1 | *Entry point* | Beginning of the caremap | 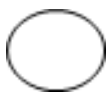 |
| 2 | *Exit point* | End of the caremap | 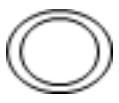 |
| 3 | *Exclusion point* | Exclusion from the caremap, as the patient does not belong to the targeted population | 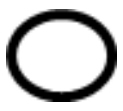 |
| 4 | *Activity* | A care or medical intervention that is associated with a medical content type *(see Table X in next section)* | 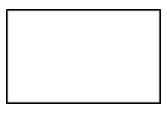 |
| 5 | *Nested Activity* | An activity that has an underlying caremap | 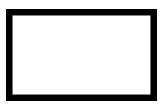 |
| 6 | *Flow* | Transition from one activity to another | 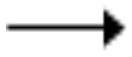 |
| 7 | *Multiple pathways* | Flow from an antecedent activity to a number of successors from which the clinician must choose the most appropriate ongoing path | 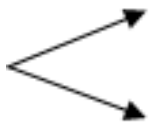 |
| 8 | *Nested caremap connection* | Connection between an activity and its nested caremap | 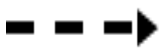 |
| 9 | *Multi-level caremap connection* | Connection between a series of linked caremaps | 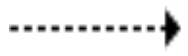 |

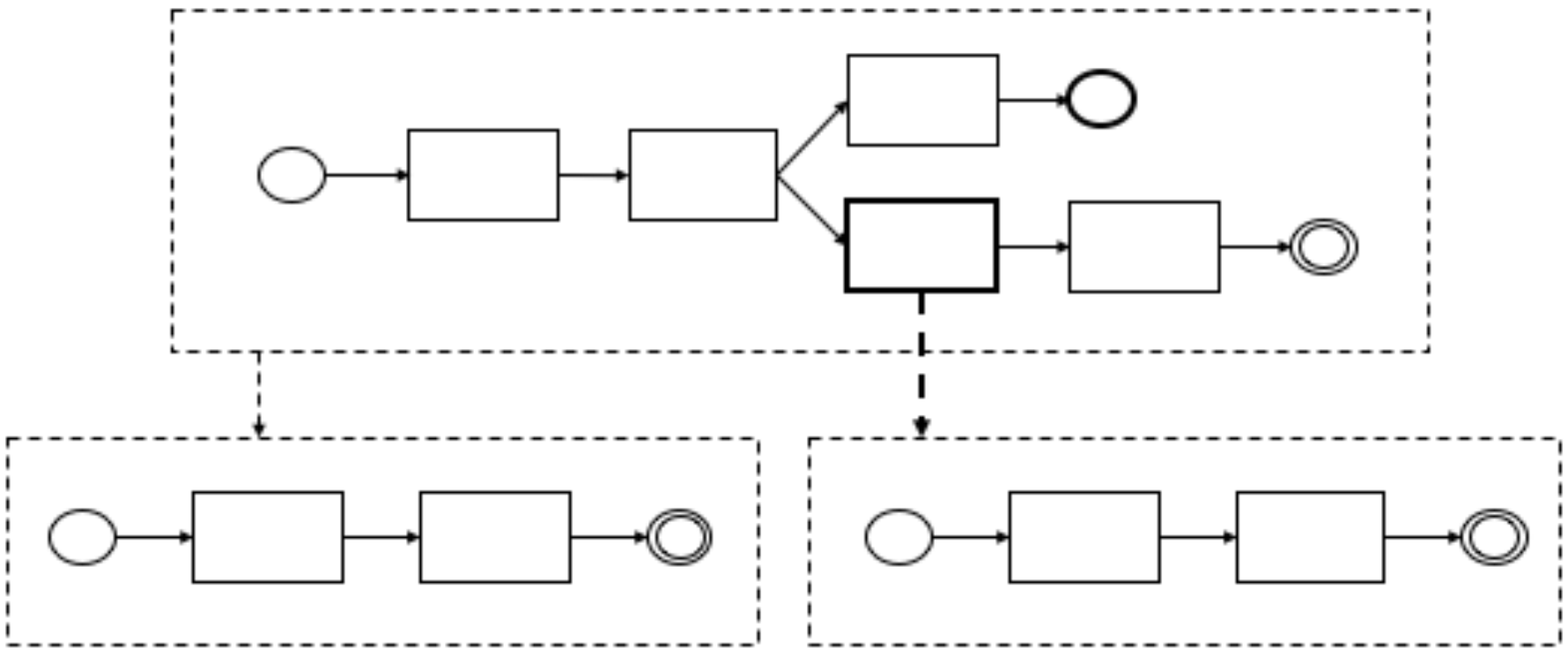

Figure 34: Structural model for caremaps

### 8.1.3 Content

Each node in the caremap represents a specific medical activity. *Diagnosis*, *initial* and *ongoing treatment/monitoring* are three archetypal medical activities most often represented in caremaps [310, 205, 514]. It is common for the caremap to contain a set of targeted outcomes, that is, points at which the patient exits the caremap (exit points) [16, 509, 325] with activities along a time scale, described either as a duration or inferred from the step-by-step ordered nature of the care process [323, 514, 510]. Finally, an explanation associated with the nodes (activities) or edges (arrows or paths) may be provided [317, 511, 325, 510]. This explanation serves to better describe the care being provided during the particular activity (node) or provides justification for the clinical decision to prefer one path (edge) over an alternative path.

The three archetypal contemporary caremap activity types relate to a set of specifically ordered medical activities that may also collect related patient health data, as described in Table 15. The caremap content model demonstrating how the three archetypes relate, and the temporal nature of the activities is shown in Figure 35. The three types are represented on different caremap levels, while described medical activities are shown as the components of the caremap. This proposed standard content model represents the information that should be captured by the caremap.

Table 15: Content type, activities and information captured in the caremap

| **Content Type** | **Activity** | **Data/Information Captured** |
|---|---|---|
| Diagnosis | Review patient records | Demographics<br>Medical history |
| | Collect patient history | Family history<br>Comorbidities |
| | Ask personal, lifestyle questions | Habits (risk factors) |
| | Clinical examination | Signs<br>Symptoms |
| | Targeted exam | Diagnostic test results |
| | Disease assessment | Diagnosis |
| Treatment | Set goals | Expected Outcomes |
| | Consider different interventions | Possible treatments |
| | Consider potential complications | Variances from expected outcomes |
| | Write prescription | Selected treatment<br>Treatment details |
| Monitoring | Review patient records | Previous test results<br>Previous symptoms |
| | Clinical exam | Signs/Symptoms |
| | Targeted exam | Diagnostic test results |
| | Evaluate goals | Progression |

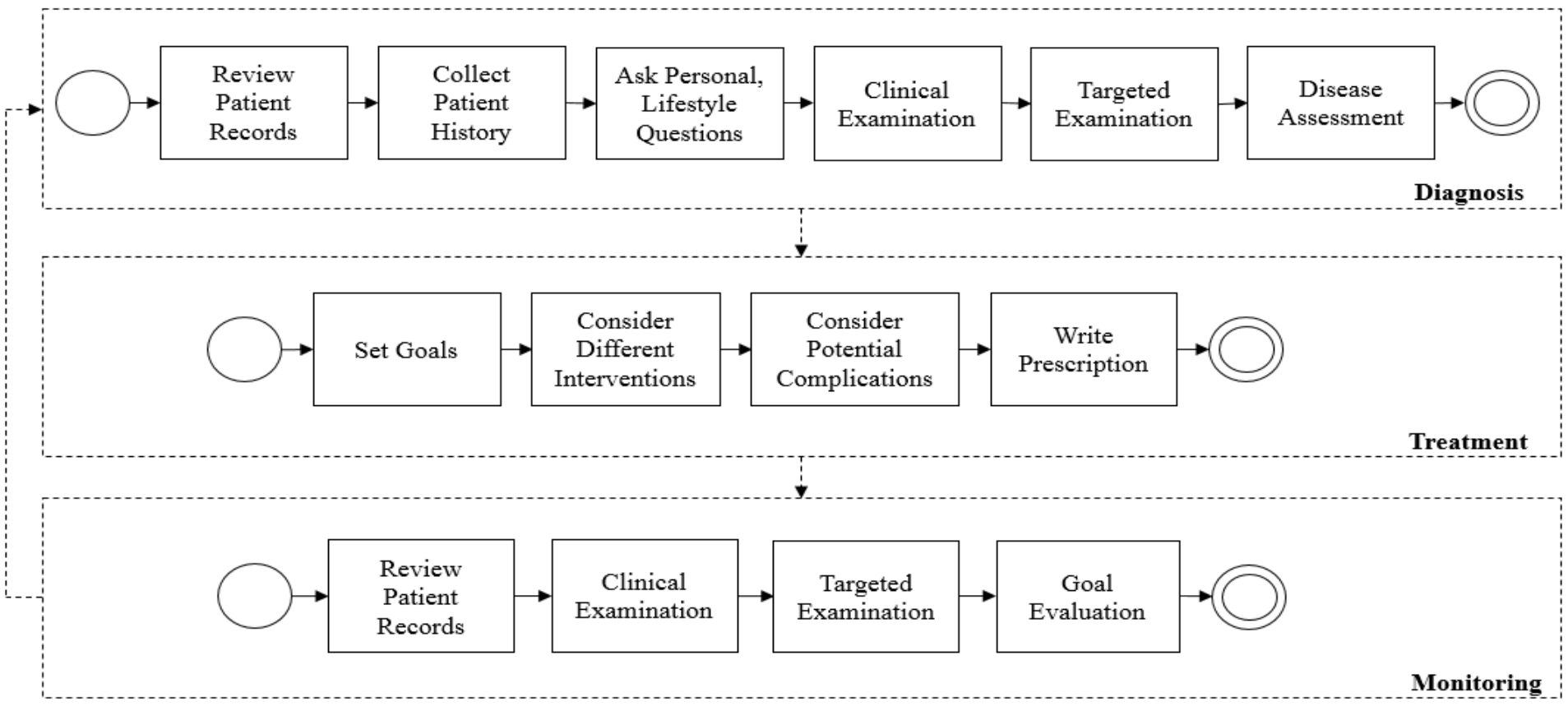

Figure 35: Content model templates for caremaps

## 8.2 DEVELOPMENT PROCESS

Despite the importance of standardising clinical processes, standardising the development process for contemporary caremaps has received very little attention in the literature. This review found that only one from every six caremap papers in the literature pool provides any detail regarding the development process. Some (*n* = *5*) discussed their development process with some degree of deliberateness [125, 310, 308, 517, 205]. From the others, the development process could only be inferred [320, 319, 509, 325].

Appendix C lists the caremap development steps described in the literature, and which authors described each step. The development steps have been clustered into three primary groups: (a) those undertaken before actual caremap development commenced; (b) those undertaken during development and refinement of the caremap, and; (c) those that come after the caremap has been refined and approved for implementation. The steps described have also been clustered into groups of fourteen *like steps*, those that appear similar or that may represent part of the same process step. The table also shows the order in which the particular author described undertaking each step, and from this the overall number of steps described or inferred from each.

The proposed caremap development process presents with six distinct phases, as shown in Figure 36. During the initial preparation phase the conceptual framework is decided, the multidisciplinary team assembled, and training should be conducted covering the overall caremap development process. The next phase clarifies current practice for the condition or patient cohort and seeks to identify any existing or anticipated variance issues that may exist. A review of the available evidence is the final step prior to development of the draft caremap. Once developed, the caremap should be evaluated and with consensus, implemented into practice. As Figure 36 shows, caremap development should operate as a lifecycle learning process. As new variance or evidence are identified, the process continues and the caremap is re-evaluated and updated as necessary.

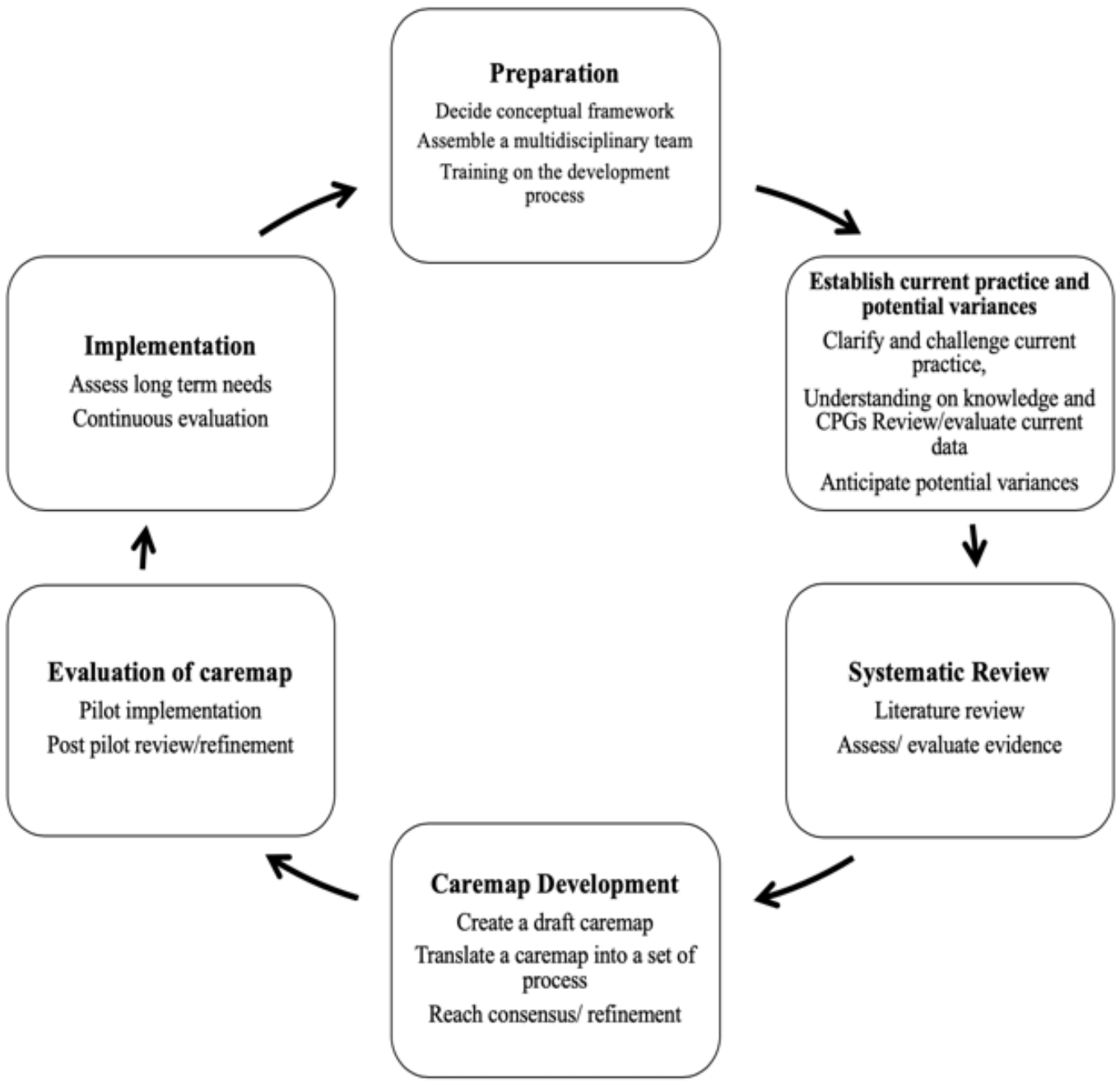

Figure 36: Caremap development lifecycle

## 8.3 EXTENDING TASC WITH CLINICAL DECISION POINTS

The standard in TaSC for nodes within the caremap is that they represent activities conducted in the performance of patient care. However, nodes are often also seen to represent within their perspective one or more latent clinical decisions, described in this paper as decision points (DP), that the clinician may consider as part of selecting the treatment path for the individual patient. The existence of these latent DPs within caremaps, and the lack of an approach to identifying and representing them has already been noted [39]. This section explores these issues and presents an extension to TaSC for identifying and representing DPs along with decision criterion.

The term *allopathic medicine* was coined by homeopathist Samuel Hahnemann in the 1800's to describe with his own personal degree of disdain what he saw as the growing practice of science-based medicine: a practice that as early as the 1850's was already being described as *modern medicine* [518, 519]. For as long as doctors have practiced modern medicine they have sought approaches, even graphical ones, for making quick, accurate and confident clinical decisions [520-523]. The diverse clinical presentation of disease, ever-increasing variety of potential treatment options, constantly changing opinions regarding those treatment options, and unlimited scope for patient response to treatment all complicate the clinical decision-making process [524, 525]. Patient care involves making a series of clinical decisions to determine the next treatment activity after consideration of all information available for this patient up to that point [526]. The more accurate, detailed and current that information can be, the more reliable the resulting clinical decision is and the closer the clinician comes to providing precision, rather than just population or evidence-based medicine [527, 528].

Medical decision-making is inevitably performed under conditions of uncertainty. Where diagnostic tests are used to guide decisions, differences are often observed between the diagnostic and treatment threshold values being applied for a particular test result [529, 530]. The diagnostic threshold assists the clinician to estimate the probability that the patient has the disease, and evaluate whether further tests are required [530]. The treatment threshold is the point where the probability of disease is so high that treatment can no longer be withheld [529, 530]. Application of these threshold models in clinical practice guidelines (CPGs) and care pathways is often represented similarly to the if/then/else statements common to software programming. However, graphical presentation in a caremap requires multiple potential paths, some leading to different treatment options and others toward continued diagnostic evaluation. In every case, all outcome options for a diagnostic test must be represented.

While not specifically describing an intention to do so, prior examples have attempted in part to address the need for contextualising the inherent, or latent, DPs in caremaps. Some provided the clinician with a graphical cue representing where a clinical decision needs to be made, while failing to provide the justification or necessary criteria that would help them to easily identify the best path for their patient, for example: Figure 37 from [509] which identifies decision nodes as a diamond but fails to provide detail to assist the clinician to know why a particular path must be followed.

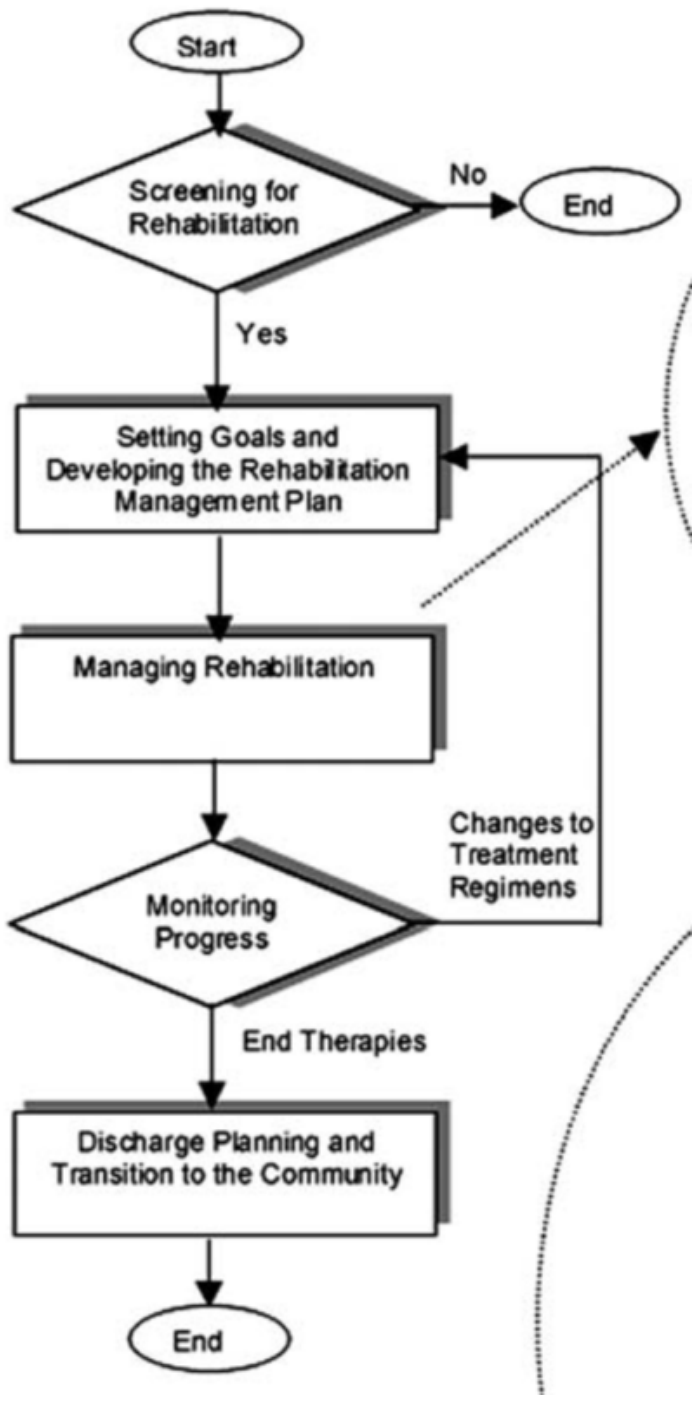

Figure 37: Example from [509]

Others present quite granular justification for the threshold values or criteria to be considered when making clinical decisions, but they lack any visual cue for simple differentiation of DPs from standard activity nodes, such as the examples shown in Figure 38 from [531, 511, 323].

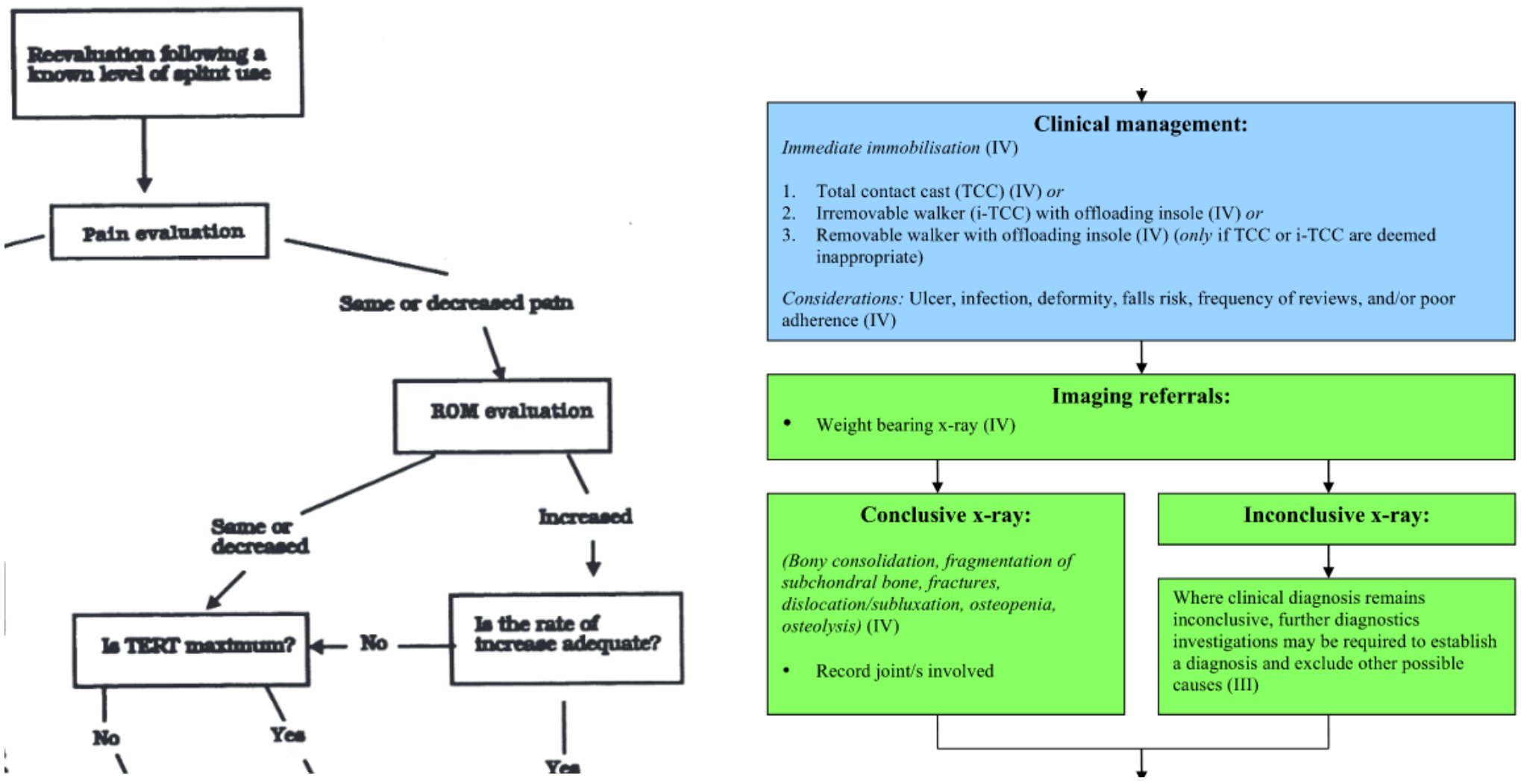

Figure 38: Examples from [528] (left) and [508] (right)

This work proposes that the way forward for integrating DPs into the TaSC caremap model is a combination of both approaches: (1) a visual cue for clinicians to easily identify when a clinical decision needs to be made combined with (2) easy to identify criteria that enables easy selection of the appropriate treatment path based on the accumulated knowledge regarding the current patient. We suggest that caremaps are more useful when the latent decisions that exist within the caremap, along with the criteria to support clinical decision making, are presented graphically as a standard component of the caremap model.

### 8.3.1 Clinical Decisions

There are many clinical decisions that might inhabit a particular caremap node. For example, a treatment activity may require the clinician to consider whether aseptic technique is required, which dressing to use or the selection of a clinical resource to assist during treatment. The majority of these decisions do not impact directly on the flow of care or the pathway of the patient within the caremap. In identifying DPs to be defined within the caremap we are concerned with those decisions that have an impact on the path to be taken by the patient: DPs that are critical to patient flow.

Clinical decisions that may give rise to DPs in a caremap result from six aspects of clinical work identified by [532] as follows:

***Clinical Evidence:*** The identification and selection of clinical evidence from clinical trials and clinical practice guidelines for use in the creation of tools like caremaps necessitates decisions regarding how to gather the right clinical findings properly and interpret them soundly.

***Diagnosis:*** During diagnosis decisions are made regarding the selection and interpretation of diagnostic tests.

***Prognosis:*** Prognosis requires decisions of how to anticipate a given patient's likely course.

***Therapy:*** Therapy decisions consider how to select treatments that do more good than harm.

***Prevention:*** Screening and reducing a patient's risk for disease are prevention decisions.

***Education:*** Consideration of how to teach the clinician, patient or patient's family what is needed fall within the remit of education decisions.

### 8.3.2 Extending TaSC with Clinical Decisions

***Structure:*** Caremaps with DPs present as flow diagrams. The e-TaSC entity relationship model (ERM) that describes the relationship among structural elements of a caremap with DPs is shown in Figure 39. The extended TaSC structural elements and notation are presented in Table 16. The elements are inspired by the standardised pictorial elements seen in UML and hard state chart notations. The additional elements related to extending TaSC with DPs are shaded in grey. The standardised structural model of the caremap with DPs can be drawn from the updated content model shown in Figure 40.

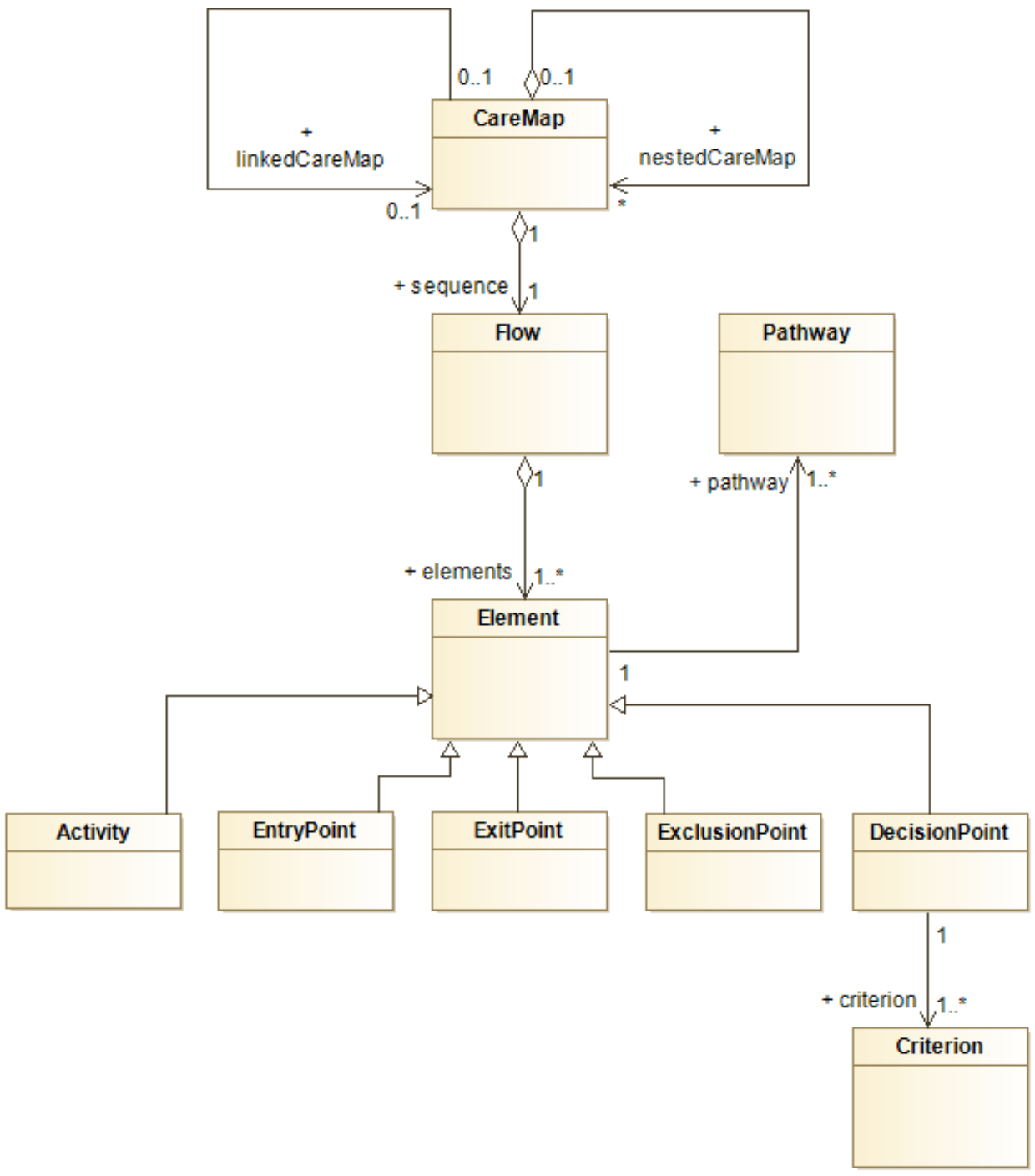

Figure 39: The TaSC ERM for the caremap with decision points

***Content:*** The same three main content types are captured in the extended TaSC model; diagnosis, treatment, and management/monitoring. As shown in Table 17, these broad content types are related to a set of specific medical activities and DPs. The additional DPs related to e-TaSC are shaded with grey. The supporting exemplar content model is shown in Figure 40. Each content type represents a different caremap level, while the activities and decisions are components of the caremap.

Table 16: The extended TaSC representative notation

| | **Element** | **Description** | **Notation** |
|---|---|---|---|
| 1 | *Entry point* | Beginning of the caremap | 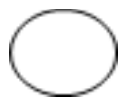 |
| 2 | *Exit point* | End of the caremap | 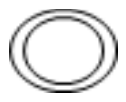 |
| 3 | *Exclusion point* | Exclusion from the caremap, as the patient does not belong to the targeted population | 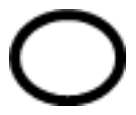 |
| 4 | *Activity* | A care or medical intervention that is associated with a medical content type *(see Table X in next section)* | 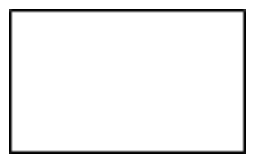 |
| 5 | *Nested Activity* | An activity that has an underlying caremap | 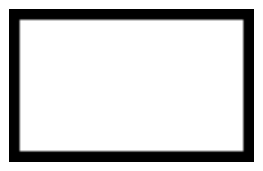 |
| 6 | *Decision* | A cognitive process of selecting a course of action that is associated with a medical content type *(see Table X in next section)* | 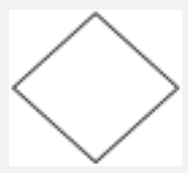 |
| 7 | *Nested Decision* | A decision that has an underlying caremap | 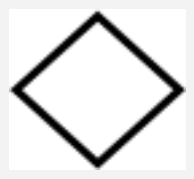 |
| 8 | *Flow* | Transition from one activity to another along the pathway |  |
| 9 | *Multiple pathways* | Flow from an antecedent activity to a number of successors from which a decision point arises | 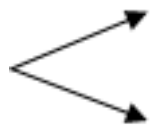 |
| 10 | *Decision Criterion* | Conditional values used to identify the path to be taken based on the clinical decision being made |   |
| 11 | *Nested caremap connection* | Connection between an activity and its nested caremap | 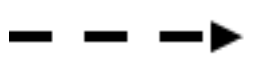 |
| 12 | *Multi-level caremap connection* | Connection between a series of linked caremaps | 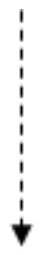 |

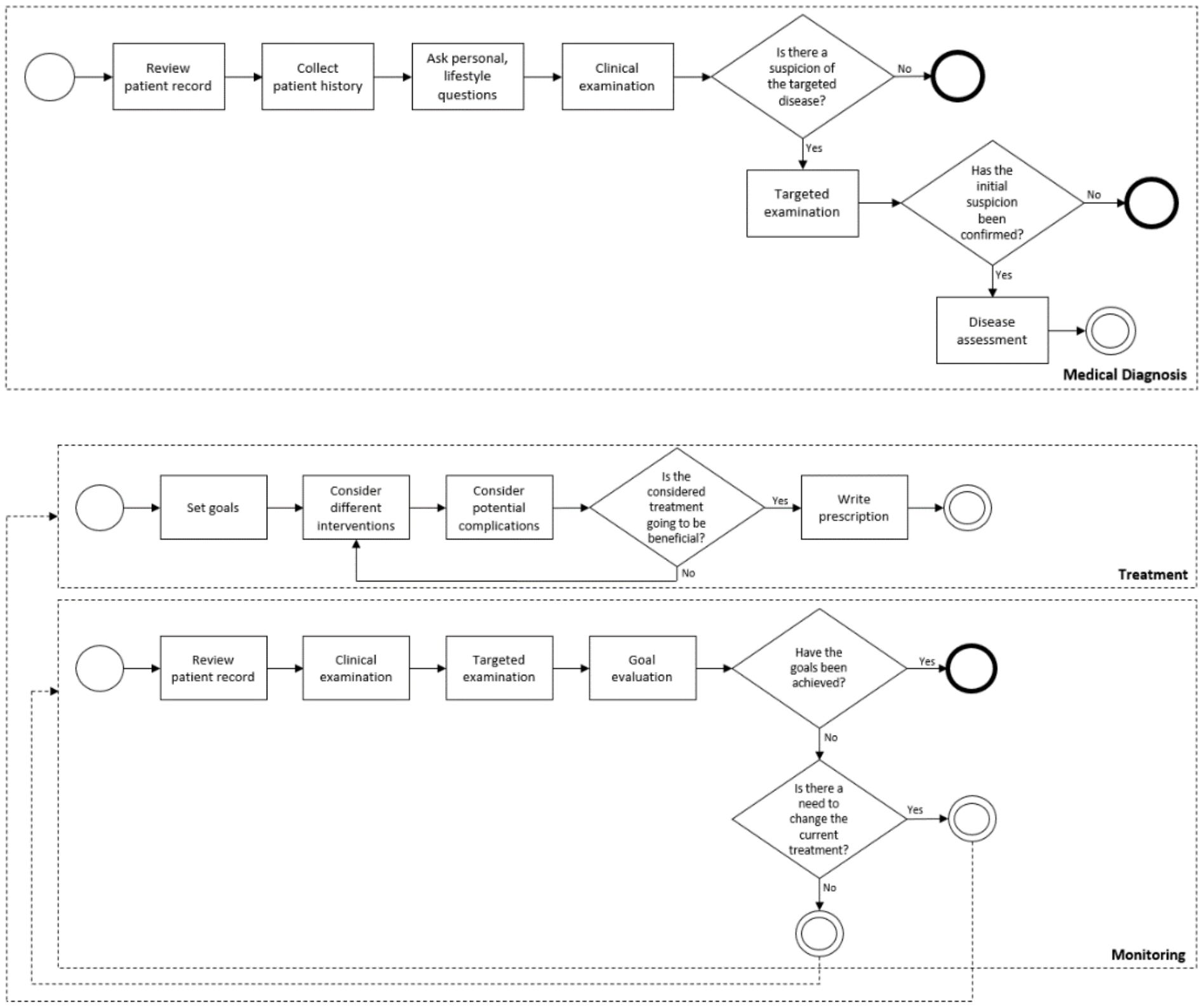

Figure 40: The updated TaSC content model for caremaps with decision points

Table 17: Caremap content

| **Content Type** | **Activity** ***(associated with Content Type)*** | **Decision** ***(associated with Content Type)*** |
|---|---|---|
| Diagnosis | Review patient records | Is there a suspicion of the targeted disease? |
| | Collect patient history | |
| | Ask personal, lifestyle questions | |
| | Clinical examination | |
| | Targeted examination | Has the initial suspicion been confirmed? |
| | Disease assessment | |
| Treatment | Set goals | Is the considered treatment going to be beneficial? |
| | Consider different interventions | |
| | Consider potential complications | |
| | Write prescription | |
| Monitoring | Review patient records | Have the goals been achieved?<br>Is there a need to change the current treatment? |
| | Clinical examination | |
| | Targeted examination | |
| | Evaluative goals | |

## 8.4 SUMMARY

This chapter presented a solution for standardising caremap structure and content, and an approach for caremap development distilled directly from analysis of the CCPS literature. During the course of refining and evaluating TaSC the presence of DPs undescribed in the original caremaps specification and that can assist clinicians in identify the appropriate treatment path for patients, was realised. As a result, an extension to TaSC was developed to incorporate a standard approach to describing DPs relevant to path selection, and based around the six aspects of clinical work.

If used consistently, the methods presented in this chapter will bring standardisation to caremaps and ensure that as clinical staff move between busy units in a tertiary care setting, they are not distracted from the patient in effort to understand the care flow model. Every caremap would be familiar and time can be given over to treating their patient, not trying to understand the document. Using TaSC and its extension as the example, standardisation is evaluated in the following chapter.

# Chapter 9: Evaluating TaSC: Studies in Caremap Development

This chapter presents four case studies directly applying the TaSC standardised structure and development approach for caremaps proposed in Section 8.1. The first, a caremap for the labour and birth event, presents a simple and direct example that occurred prior to development of the TaSC approach. However, in development of the second case study in Gestational Diabetes Mellitus it was discovered that caremaps become more useful when the latent decisions that exist as divergences in the flow diagram, along with the decision criteria that identify the path to be taken, are presented. It was during development of this case study that the DPs extension presented in Section 8.3.2 was formalised. The third case study in Rheumatoid Arthritis was the first to be developed using the complete TaSC model. Finally, the fourth case study in trauma care proved to be the most complicated of the three, and a fine test case for more complex evaluation of caremaps generally, as well as of the refined TaSC approach. Each case study is evaluated quantitatively through recording and comparative reflection of a number of metrics, including the number of clinicians and information scientists involved and the total hours of each resource that were required to complete the caremap. As the fourth case study was also to play a structural part in the work presented in the clinical fellow's own research, the opportunity presented for it to also be qualitatively assessed using a short convenience survey of a small number of experienced trauma clinicians. The resulting *caremaps with decision points* were seen to become an invaluable tool in the development of Bayesian Networks (BNs) for use as clinical decision support learning health systems.

## 9.1 STUDY I: THE LABOUR AND BIRTH CAREMAP

The labour and birth process represents an excellent example for a first-pass evaluation case study to assess the development process for caremaps. Labour and birth have easily defined start and end points, limited temporal variance, and a small number of easily identified treatment paths.

### 9.1.1 Inputs

Inputs for the labour and birth caremap were: (a) clinical practice guidelines for intrapartum care at Middlemore Hospital, Counties Manukau District Health Board of New Zealand (NZ); (b) clinical expertise and consensus from midwives and obstetricians who practice in that facility; and, (c) publicly available incidence and treatment statistics for that facility published annually by the Ministry of Health in NZ.

### 9.1.2 Development

An iterative development process was used wherein we created an initial version of the caremap based on the clinical practice guideline (CPG) and evidence derived from the treatment statistics. The initial caremap was revised and refined during a number of sessions with the clinicians. The resulting labour and birth caremap based on the Middlemore Hospital CPG is shown in Figure 41.

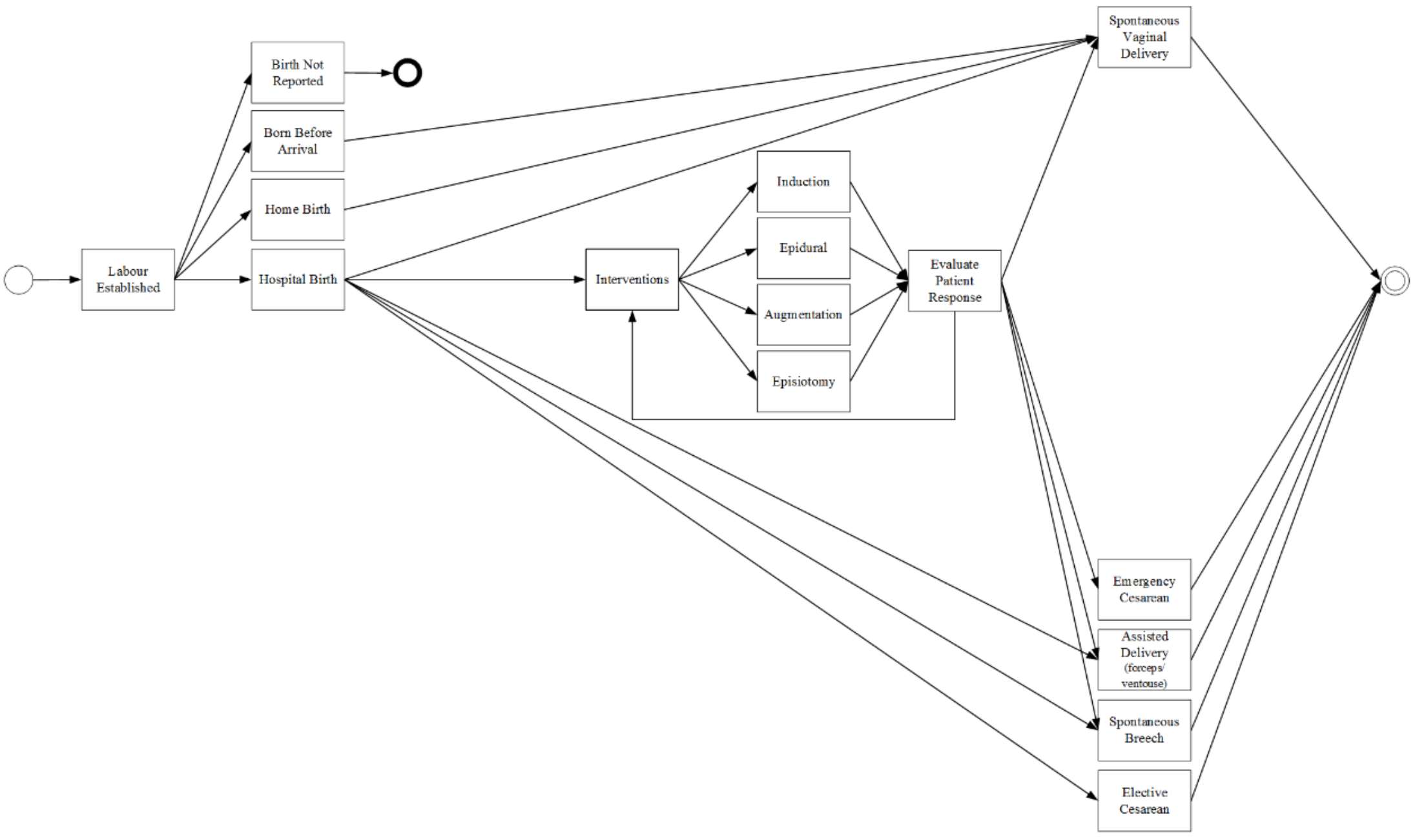

Figure 41: Labour and Birth Caremap

### 9.1.3 Validation

The Ministry of Health in NZ annually publish aggregated and anonymised maternity and newborn data and statistics for each of the country's birthing units and hospitals. These statistics are presented as a set of contingency tables whereby the possible birthing outcomes and clinical interventions are correlated with a whole range of demographic and clinical variables (maternal age at birth, ethnicity, deprivation, maternal BMI, obstetric interventions and so on). Using the 2014 data, we calculated the most likely treatment paths and care experiences for all birthing mothers at the Middlemore Hospital birthing unit. A state transition machine was developed and computerised, and realistic synthetic electronic health records (RS-EHR) for all 8,731 mothers were synthetically generated [269]. The synthetic treatment paths for each woman were digitally compared against the caremap in Figure 41 to ensure a valid path solution could be resolved for every recorded birth. In this way we demonstrated that the caremap is representative of the entire range of patient presentations and treatment options as undertaken by clinicians and experienced by delivering mothers.

### 9.1.4 Evaluation

Evaluation metrics for development of the labour and birth caremap are provided in Table 18.

Table 18: Evaluation Metrics for Caremap Development

| Metric | No. | Description |
|---|---|---|
| Clinicians | 3 | Senior Practicing Obstetric Clinician<br>Two Senior Midwives |
| Hours with Clinicians | 30 | Over 24 weeks |
| Hours without Clinicians | 60 | Over 24 weeks |
| Iterations | 7 | |

## 9.2 STUDY II: THE GESTATIONAL DIABETES MELLITUS CAREMAPS

As part of PAMBAYESIAN [367] we are creating a Bayesian Network (BN) model [533] to predict treatment needs for individual mothers with gestational diabetes mellitus (GDM). The process initially required three caremaps, for: (1) the midwifery booking visit; (2) GDM diagnosis; and, (3) clinical management of the patient's condition. Later, a fourth *sequela* caremap was developed for postnatal assessment and prediction of the likelihood of: (a) the mother going on to develop Type 2 Diabetes[5] (T2D) [534]; and, (b) the child going on to develop some form of diabetes[6], either Type 1 Diabetes (T1D) or early-onset T2D [535, 536].

---

[5] Women diagnosed with GDM have a seven-times greater risk of developing T2D postnatally (Lie, Hayes, Lewis-Barned et al, 2013).

[6] The prevalence of impaired glucose tolerance (IGT) is as high as 20% in some studies of children of GDM mothers (Plagemann, Harder, Kohler et al, 1997).

### 9.2.1 Background

GDM occurs in 2-25% of pregnancies [537, 538] and, depending on the diagnostic criteria used, rates across the United Kingdom (UK) may be as high as 17% [539, 538]. While the original definition for GDM was based on maternal risk for developing diabetes postpartum, newer glucose criteria have been developed based on risk of maternal and neonatal complications [540, 541]. While a number of international standards provide diagnostic thresholds for GDM, in 2015 the National Institute for Health and Care Excellence (NICE) published an updated guidance for diabetes in pregnancy [542]. The Barts Health Trust (BHT) CPG used in development of the GDM caremaps was based on this 2015 NICE guideline.

### 9.2.2 Inputs

Inputs for the gestational diabetes caremaps were: (a) a clinical practice guideline from Barts and The London Hospital in East London that is currently in use for the care of women with diabetes in pregnancy; and, (b) clinical expertise and consensus from midwives and diabetologists of the same facility.

As part of the work undertaken on the PAMBAYESIAN project's GDM case study, this author assisted another PhD student to undertake an AGREE II protocol review undertaken to assess the quality of CPGs that would contribute to PAMBAYESIAN's CDSS [167]. AGREE II is a widely accepted and validated tool for assessing the methodological quality of CPGs [543]; however, it should be noted that it does not assess the implementation of that guideline. The AGREE II instrument [544] comprises 23 items arranged in 6 domains: (1) scope and purpose, (2) stakeholder involvement, (3) rigor of development, (4) clarity and presentation, (5) applicability, and (6) editorial independence. Responses are scored on a Likert scale from 1 to 7 (1 = strongly disagree, 7 = strongly agree). The AGREE II study's primary focus, therefore, is less on the *focus* and more on the *form* of the guideline: who wrote the guideline, conflicts of interest, the processes used for evaluation of evidence, the quality and referencing of that evidence, and so on. The AGREE II committee cite these as primary issues that affect the quality and reliability of CPGs and their effect on the care delivered in hospitals. It has been observed that low CPG usage rates coupled with increased clinical resource wastage conflict with the CPGs primary purpose to such degree that issues of who wrote or funded its development and the processes they relied upon to select and evaluate evidence may not actually be impacting on the quality of care to any significant degree [545, 546]. It does not matter how well a guideline scores on the AGREE II protocol, AGREE studies are silent as to how well the guideline is being applied in clinical practice.

Rather than seeking guidelines or guideline reviews from the literature, which in many cases only consider ideal guidelines from national or international non-clinical organisations (as well as those from medical associations, professional colleges, health insurance providers, government health departments, and so on), this guideline review sought a representative selection of local CPGs developed by, or required to be used by, clinicians in English-speaking western hospitals. The criteria for inclusion were that the document: (1) was explicitly identified as a guideline; (2) was produced by or for the hospital in an English-speaking country; (3) included diagnostic criteria and recommendations concerning gestational diabetes; (4) demonstrated some evaluation or inclusion of evidence; and, (5) was easily accessible over the internet to health service consumers. In addition, it was required that: (6) two CPGs from each country were identified. Where more than two from any one country met these requirements, those most recently updated were used. Given that all Canadian hospitals searched (n=11) referred to the same nationally developed Canadian Diabetes Association (CDA) CPG, its singular inclusion was allowed in order to assess the reason for its clinical popularity. The resulting CPG collection included two from each of AUS, NZ and the UK, and the single CDA CPG from Canada. These are listed in Table 19.

All of the CPGs were assessed as having some deficiencies, most notably a lack of user involvement, assessment of resource implications and conflicts of interests, and an across the board lack of external review. It was positively noted that none of the CPG significantly deviated from recommendations of the national or international advising bodies on which their recommendations were based. The UK local CPGs broadly adhered to NICE recommendations, with minor refinements, or tweaks, in the BHT one which the CPG states is due to a patient population predominately classified as high-risk.

Table 19: Reviewed Guidelines and URLs

| | | Author Organisation | Year | Title |
|---|---|---|---|---|
| 1 | AUS | Royal Women's Hospital (RWH) | 2017 | Management of Gestational Diabetes |
| 2 | AUS | King Edward Memorial Hospital (KEMH) | 2017 | Diabetes in Pregnancy |
| 3 | NZ | Auckland DHB (ADHB) | 2013 | Diabetes in Pregnancy |
| 4 | NZ | Hutt Valley DHB (HVDHB) | 2015 | Diabetes: Pre-existing and Gestational |
| 5 | CA | Canadian Diabetic Assoc. (CDA) | 2013 | Diabetes and Pregnancy |
| 6 | UK | Nottingham University Hospital (NUH) | 2016 | Management of pregnant women with diabetes |
| 7 | UK | Barts Health Trust (BHT) | 2015 | Diabetes - Pregnancy, Labour and Puerperium |

| | URL |
|---|---|
| 1 | https://bit.ly/2OcMFFJ |
| 2 | https://bit.ly/2N1cAeJ |
| 3 | https://bit.ly/2Iemf0J |
| 4 | https://bit.ly/2xBuakU |
| 5 | https://bit.ly/2IdXoud |
| 6 | http://bit.do/exsYa |
| 7 | http://bit.do/exsXy |

While the PAMBAYESIAN AGREE II study rated the Canadian CPG best overall [167], it lacked usability for application in daily care. Development of a practical care pathway and caremap would be necessary in order for midwives and obstetricians to consistently apply it in practice. The second-highest ranked CPG was from NZ ADHB. It provided simple and clear clinical pathway diagrams, but its use of URL links to other hospital documentation meant those sections could only be followed and understood by someone working within the hospital computer network. And while it provided detailed instructions for post-natal follow-up, it failed to provide appropriately detailed recommendations for foetal surveillance during the pregnancy. The UK CPGs were produced to be generally consistent with the recommendations of the NICE guidelines. As the NICE guidelines are already considered rigorous in their investigation and assessment of evidence, the UK guidelines scores were not impacted so much by the lack of evidence, but by the presentation, style and lack of adherence to the format prescribed by AGREE II and other CPG development protocols.

The primary goal of this AGREE II assessment was to comparatively evaluate adequacy and appropriateness of the BHT guideline that would be used in production of the caremaps presented later in this section. The AGREE II outcomes showed that while it was due for updating, the BHT CPG was still of sufficient quality and recency to be highly suitable for this purpose [167].

### 9.2.3 Development

An iterative development process was used wherein the health informatician, decision scientist and midwifery fellow all worked together to deliver an initial version of the caremap based on CPG and clinical expertise. The initial caremap was revised and refined during a small number of sessions with the midwife and clinicians. As an example of the output of this process, Figure 42 presents the resulting caremap for clinical management of the patient diagnosed with GDM.

### 9.2.4 Extending the Caremaps

While using the caremaps to develop of BNs to support diagnostic and treatment decisions for GDM we found that the process was significantly easier and more efficient when the latent decisions embedded in each caremap were identified and included in the caremap. The GDM caremaps were redeveloped as *caremaps with decision points*. Figure 43 shows the resulting caremap with decision points for the midwifery booking visit, Figure 44 shows diagnosing GDM, and Figure 45 shows management of the diagnosed patient.

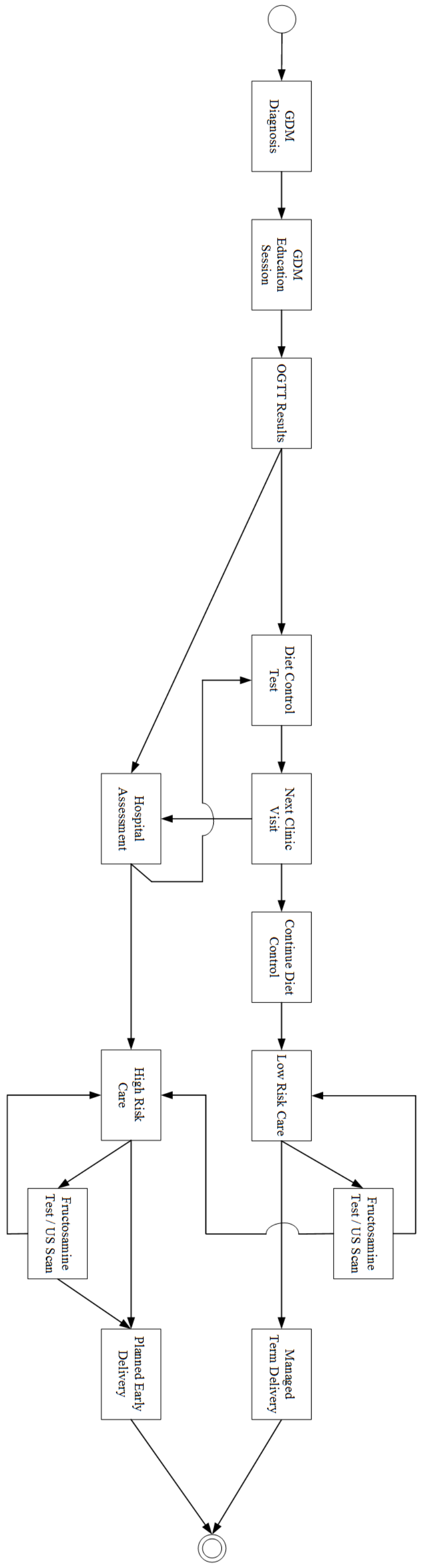

Figure 42: GDM Management Caremap

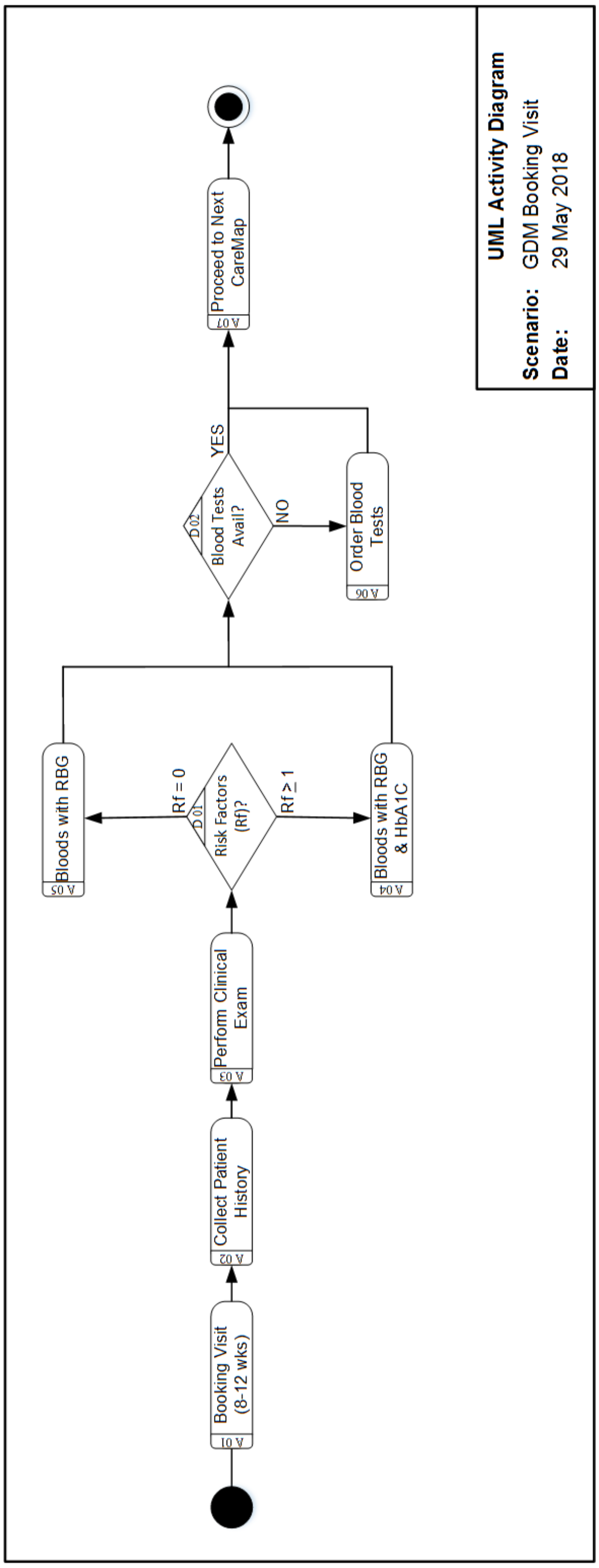

Figure 43: GDM Booking Visit Caremap with Decision Points

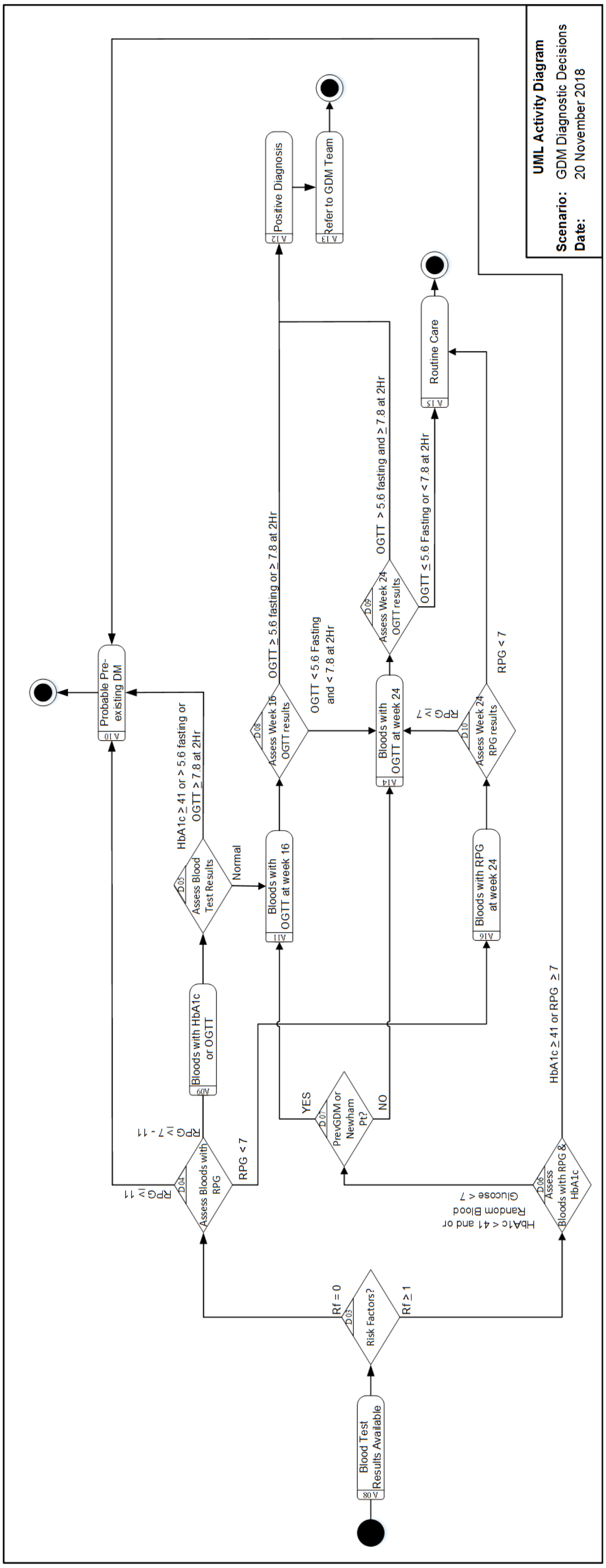

Figure 44: GDM Diagnostic Caremap with Decision Points

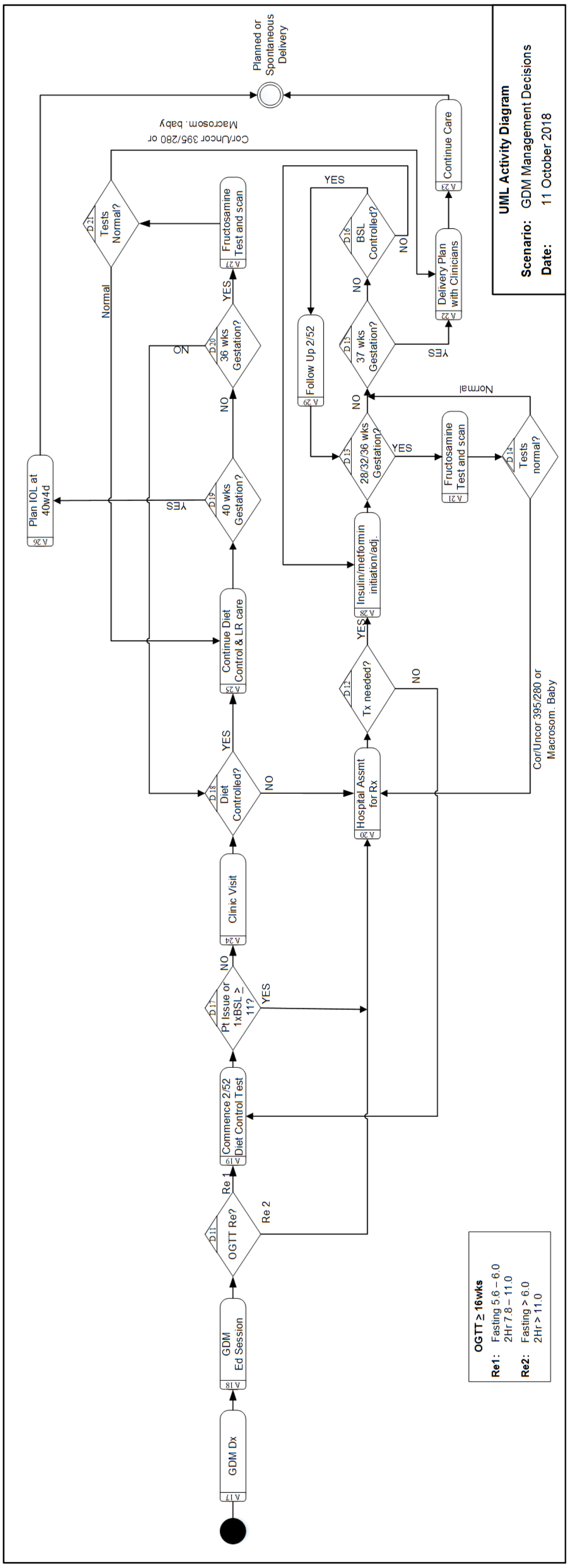

Figure 45: GDM Management Caremap with Decision Points

### 9.2.5 Validation

Validation was performed through consultation seeking consensus from three participating diabetologists with tertiary care experience treating obstetric patients under the CPGs used in development of the caremaps. The caremaps shown in Figures 43-45 are those which expert assessment found to be correctly representative of the overall care process for these patients.

### 9.2.6 Evaluation

Evaluation metrics for development of the gestational diabetes caremaps are provided in Table 20.

Table 20: Evaluation Metrics for Caremap Development

| Metric | No. | Description |
|---|---|---|
| Clinicians | 2 | Senior Practicing Clinician<br>Clinical Research Midwife |
| Hours with Clinicians | 14 | Over 16 weeks |
| Hours without Clinicians | 27 | Over 16 weeks |
| Iterations | 4 | |

## 9.3 STUDY III: THE RHEUMATOID ARTHRITIS CAREMAPS

PAMBAYESIAN's second, but this thesis' third case study focuses on the diagnosis, initial treatment and ongoing management of Rheumatoid Arthritis (RA).

### 9.3.1 Background

RA is a chronic disease and presently the most common autoimmune inflammatory disorder in adults [547, 548]. The condition is characterised by progressive damage to the articular cartilage at the end of the bones, where they come together to form joints [547]. RA has a gradual but significantly detrimental effect on the patient's ability to perform normal activities of daily living (ADLs), reduces quality of life (QoL), and increases mortality [548]. While there is significant year-on-year development of new knowledge and treatments for the condition, a cure has thus far eluded clinicians [547, 548]. RA was therefore a suitable case study condition for PAMBAYESIAN as: (a) it is a long-term chronic condition; (b) has a large number of pharmaceutical treatment strategies that include steroids, non-steroidal anti-inflammatories, disease modifying antirheumatic drugs (DMARDs), biologics, and vaccines; (c) it has patients with a large number of therapeutic, non-therapeutic and psychosocial needs that must be addressed in a community setting; and, (d) there is considerable day-to-day variability in the condition that must be modelled and understood in order for any potential intervention to be beneficial.

### 9.3.2 Inputs

Inputs for the RA caremaps were: (a) a current clinical practice guideline from Barts and the London Hospital in East London; (b) literature, including systematic reviews, for the condition and common treatment options published during the last three years; and, (c) clinical expertise and consensus from a senior rheumatologist and her practicing clinical research fellow.

### 9.3.3 Development

While development of the RA caremaps was also an iterative process, where the process differed from the GDM case study was that the information scientist performed considerably more self-directed research and used this to develop much larger portions of the caremap that were presented and reviewed regularly with the clinicians. Figure 46 presents the diagnostic caremap, Figure 47 presents the initial management caremap, and Figure 48 provides the approach for ongoing management of RA.

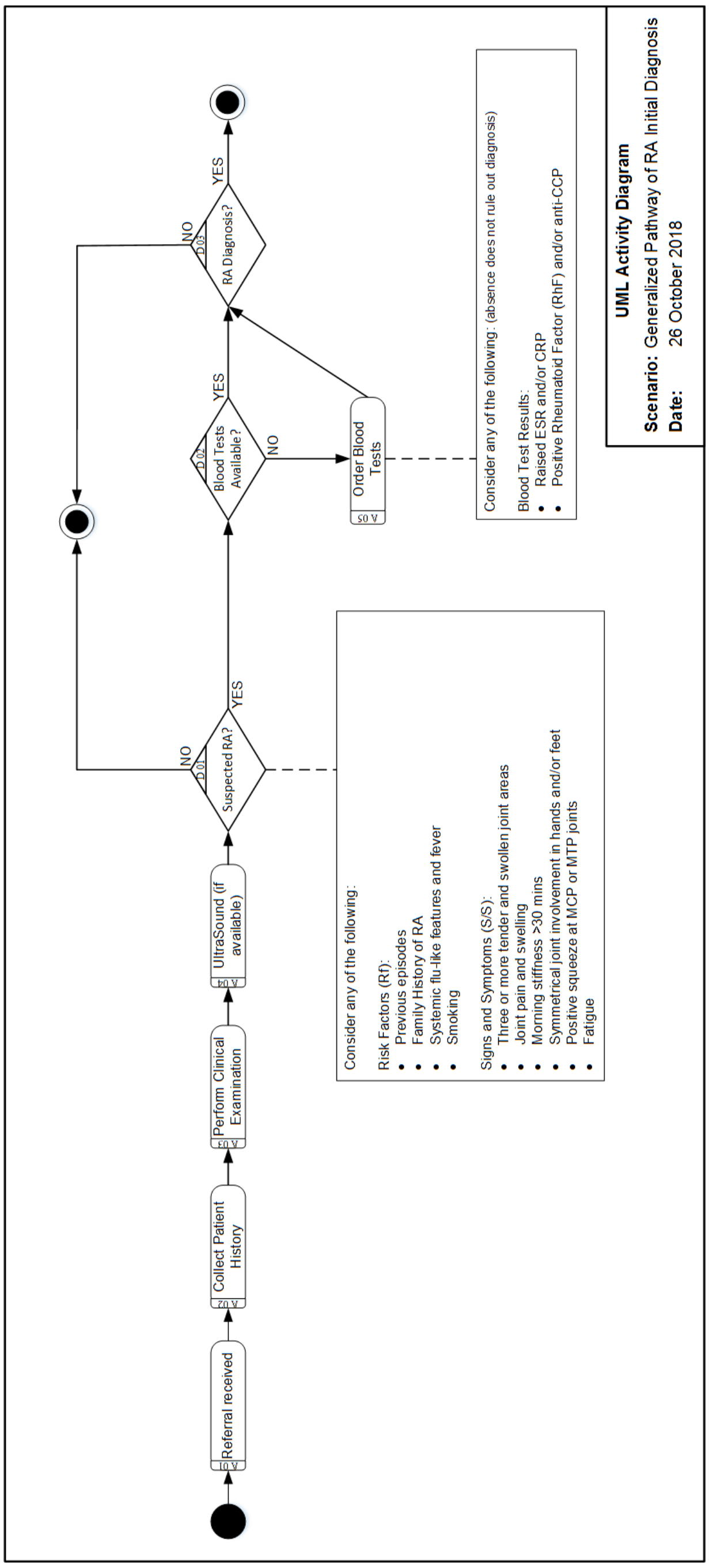

Figure 46: RA Diagnosis Caremap with Decision Points

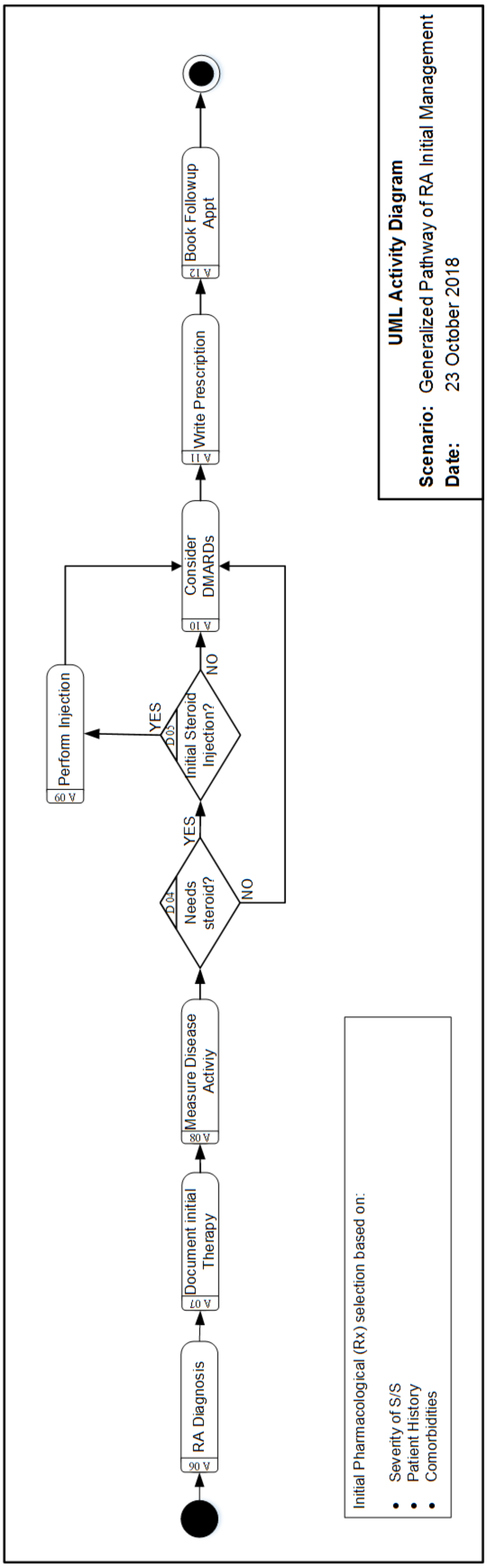

Figure 47: RA Initial Management Caremap with Decision Points

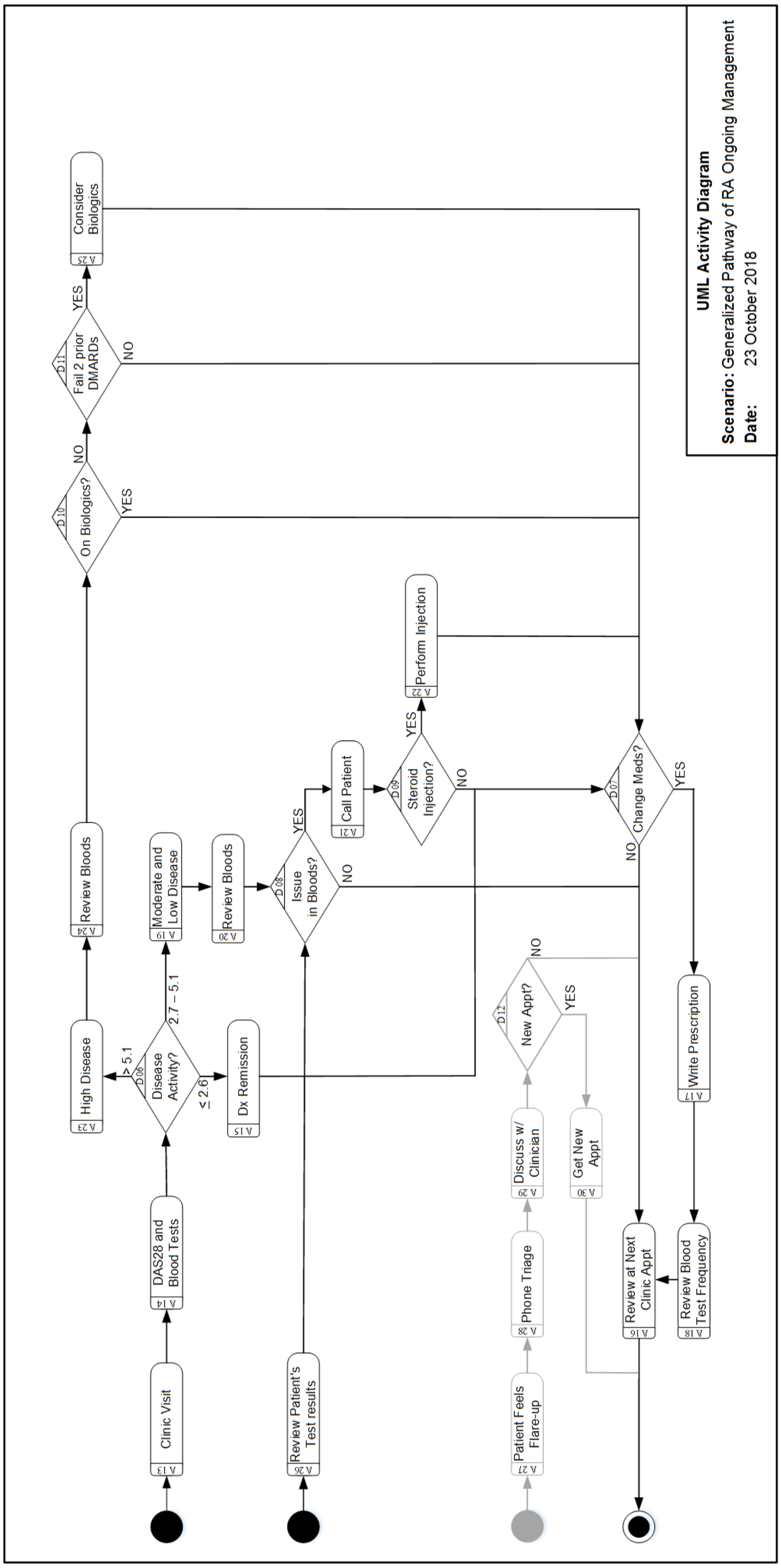

Figure 48: RA Ongoing Management Caremap with Decision Points

### 9.3.4 Validation

As with the GDM caremap, validation was achieved through consultation and consensus of the participating treating clinicians.

### 9.3.5 Evaluation

Evaluation metrics for development of the RA caremaps are provided in Table 21.

Table 21: Evaluation Metrics for Caremap Development

| Metric | No. | Description |
|---|---|---|
| Clinicians | 2 | Senior Practicing Clinician<br>Clinical Research Fellow |
| Hours with Clinicians | 9 | Over 16 weeks |
| Hours without Clinicians | 23 | Over 16 weeks |
| Iterations | 3 | |

## 9.4 STUDY IV: THE HEMS TRAUMA ASSESSMENT CAREMAPS

The Helicopter Emergency Medical Service (HEMS) is a physician-led and well-established component of trauma systems in most first-world countries [549, 550]. Helicopters are capable of transporting the major trauma patient significantly faster than ground-based services, and while costly, this mode of transport is seen to have a significant impact on reducing mortality [551-553]. HEMS clinicians have a history of seeking algorithms to make critical life-and-death medical decisions (such as whether or not to amputate) in trauma care situations: these range from simple score-based systems like MESS [554] which is based on observable symptoms and patient responses, through to complex CDSS built on medical AI or Bayesian Networks [555-558]. HEMS crews are literally at the bleeding edge of critical care practice, and as a result, are often on the forefront of any new research that could make an appreciable difference in outcomes for their patients.

### 9.4.1 Background

When HEMS clinicians arrive on-scene a range of triage and treatment processes developed during the last several decades and refined into an ever-growing collection of mnemonic terms are engaged. Mnemonics are a memory-aid learning strategy: catchy phrases to prompt recall of a process or subject [559, 560]. Over the years simple mnemonics like ABC, which stands for *airway*, *breathing* and *circulation* [561], have been extended and enhanced by a variety of first responder and trauma care organisations. St John Ambulance first responder manuals added *danger* and *response* to create DRABC [562]. Military medics preferred the addition of a prefix for *catastrophic haemorrhage* [561], and later appended *disability* and *exposure*, resulting in the more comprehensive cABCDE [563]. Given that any effort to sequence prehospital care must faithfully report the activities of clinicians, and those activities are guided directly by these mnemonics, this case study set out to develop caremaps based on a number of prehospital care mnemonics.

### 9.4.2 Inputs

Inputs for the trauma care maps included: (a) a clinical reference textbook prescribed for prehospital emergency medical training [564]; (b) clinical practice guidelines issued by state or federal health authorities and intended for use by paramedics and ambulance personnel; (c) current literature on prehospital emergency care; and, (d) clinical expertise from a practicing trauma fellow.

### 9.4.3 Development

The trauma fellow determined which current prehospital care plans could be targeted. Two primary targets most commonly used in HEMS practice were resolved from his ongoing survey and interviews with prehospital emergency care clinicians. They were: (i) SCREAMER – *Scene survey*, *Communicate*, *Read the scene*, *Everyone accounted for*, *Assess patients*, *Method of extraction*, *Evacuation route* and *Right facility*; and, (ii) cABCDE – *Catastrophic haemorrhage*, *Airway*, *Breathing*, *Circulation*, *Disability* and *Environment and Exposure*.

A textbook was located that described SCREAMER [564]. According to several clinicians interviewed by the trauma fellow, the nomenclature used in SCREAMER had been influenced by its authors' experiences attending motor vehicle traffic accidents. Encouraged by the clinical experts, the caremap adapted from SCREAMER uses adapted language intended to ensure application in a much wider range of prehospital emergency care scenarios. The same textbook [564], along with a range of clinical literature [565-567], described the cABCDE primary survey approach. An initial overview caremap was created for each, and was refined during short consultations with the trauma fellow.

The complete SCREAMER caremap can be viewed at:

http://www.mclachlandigital.com/screamer.png

The complete cABCDE caremap can be viewed at:

http://www.mclachlandigital.com/cabcde.png

### 9.4.4 Validation

Initial review and validation was conducted with the trauma fellow and an experienced HEMS clinician. Once they agreed on the content of each caremap, extended validation was performed by the trauma fellow through a process of sending out the caremaps and inviting review by a select number of the prehospital emergency care clinicians who had been interviewed prior to and during the caremaps creation. Some minor modifications and 'fine tuning' was performed on the basis of these reviews that included the addition of clinical factors and symptomatology for a number of decision nodes, and a process loop for triage and treatment in situations involving multiple casualties.

### 9.4.5 Evaluation

Evaluation metrics for development of the trauma caremaps are provided in Table 22.

Table 22: Evaluation Metrics for Caremap Development

| Metric | No. | Description |
|---|---|---|
| Clinicians | 1 | Senior Practicing Trauma Clinician |
| Hours with Clinician | 7 | Over 6 weeks |
| Hours without Clinicians | 25 | Over 10 weeks |
| Iterations | 8 | |

## 9.5 TASC EVALUATION AND DISCUSSION

### 9.5.1 Caremaps Resource Evaluation

One approach to evaluating the effect of standardisation on caremaps is quantitative: we assess the time taken to develop the caremap and achieve consensus, and the overall cost of development. Table 23 brings together the metrics recorded for each case study, as well as the number of individual caremaps or caremap segments produced by each case study.

Table 23: Evaluation Metrics Combined

| Case Study | Clinicians | Clinician Hrs | InfSc Hrs | Clinician Wks | InfSc Wks | Iterations | Total Caremaps |
|---|---|---|---|---|---|---|---|
| 1 | 3 | 30 | 60 | 24 | 24 | 7 | 1 |
| 2 | 2 | 14 | 27 | 16 | 16 | 4 | 3 |
| 3 | 2 | 9 | 23 | 16 | 16 | 3 | 3 |
| 4 | 1 | 7 | 25 | 6 | 10 | 9 | 12 |

The initial caremap standard, TaSC, was developed during the period between case studies one and two. As TaSC was refined and familiarity increased, and even as the standard was extended with DPs, the time taken to deliver each individual caremap reduced. Table 24 shows that each component of the final case study's two large amalgamated caremaps was developed with a per-caremap labour time of just 2.7 hours. Similarly, when compared to the first case study, the final case study's cost-per-caremap reduced by almost 2700%.

Table 24: Individual Caremap Cost and Labour

| Case Study | Clinician Rate/ Hr (£) | Clinician Hrs | Clinician Total (£) | InfSci Rate/ Hr (£) | InfSci Hrs | InfSci Total (£) | Total (£) | Total £ Per Caremap | Hrs Per Caremap |
|---|---|---|---|---|---|---|---|---|---|
| 1 | 84 | 30 | 2520.00 | 65 | 60 | 3900.00 | 6420.00 | 6420.00 | 90.0 |
| 2 | 104 | 14 | 1456.00 | 80 | 27 | 2160.00 | 3616.00 | 1205.33 | 13.7 |
| 3 | 104 | 9 | 936.00 | 80 | 23 | 1840.00 | 2776.00 | 925.33 | 10.7 |
| 4 | 123 | 7 | 861.00 | 80 | 25 | 2000.00 | 2861.00 | 238.42 | 2.7 |

### 9.5.2 Trauma Caremaps Evaluation Survey

Modelling of clinical work flow processes, sometimes confusingly described as *patient flow*, is common, especially for emergency care, and serves an initial step in development of simulations that can assist clinicians and provider organisations to address problems [568-570]. While developing the trauma caremaps, the trauma fellow visited the offices of Kent Surrey Sussex Air Ambulance (KSSAA) and observed that their staff were engaged in an ostensibly similar work flow modelling process using post-it notes arranged in sequence across a large whiteboard to represent activity nodes. Their process flow focused on the clinician role: on ordering activities the HEMS clinician undertakes to prepare for, travel to, attend and care for the trauma patient. They included nodes for activities like dressing in their flight suits and checking and refilling medical equipment packs. In contrast, the caremaps presented in this thesis focus on care of the patient: on identifying the health issues, treatment and clinical response of the patient from arrival at the patient's location through to transport of the patient to hospital. We preferred to focus on patient care because, while planning and training to get to the patient expediently is no doubt beneficial, there are only limited gains to be made here. We believe there are greater improvements to be found in understanding the processes and procedures that clinicians undertake in evaluating the patient's condition and providing first-line care during what is described as *the golden hour*.

Another approach to evaluating TaSC is qualitative: evaluation through responses to a convenience survey on the accuracy characteristics of the delivered caremaps and their decision points. The mnemonic caremaps from the fourth case study were evaluated through the operation of a survey instrument using a forced-choice Likert scale. The survey questions posed to clinicians are found in Table 25.

Table 25: Questionnaire for evaluating Trauma Caremaps

| Q | Survey Response Prompt<br>After viewing the caremap… | Aspect Evaluated |
|---|---|---|
| 1 | I find that the caremap faithfully reproduces the necessary structure represented in the mnemonic. | Accuracy of caremap structure to mnemonic. |
| 2 | I find that the caremap path progression is similar to the path a clinician would take in applying the mnemonic during assessment and treatment of the trauma patient. | Accuracy of caremap path progression to actual treatment. |
| 3 | I find that the placement and purpose of decision points within the caremap are similar to the decisions a clinician would make during each stage of the caremap. | Accuracy of decision points in placement and purpose. |
| 4 | I find that the criterion associated with the paths away from each decision point are similar to the diagnostic and treatment thresholds and other criteria clinicians would expect to find in the CPG used in the caremap's development. | Accuracy of decision point criterion to the CPG. |
| 5 | I find that the caremap structure and path progression, and the placement and purpose and criterion for decision points when read together has neither conflicts nor inconsistencies as would be expected in the actual trauma patient treatment process. | Accuracy and realism for the entire caremap with decision points and hence for the entire clinical logic flow. |
| 6 | I find the caremap to be simpler and easier to use in practice than the CPGs and clinical literature its development was based on. | Accuracy of claim that the caremap has utility in clinical practice |

Participating in the survey were seven self-identifying experienced emergency and trauma clinicians. None of the survey respondents had any involved in development of the trauma caremaps. Similarly. none were exposed to the caremaps prior to completing the survey. In each case the clinician examined the SCREAMER and cABCDE trauma caremaps in answering whether the caremap possessed qualities similar to the CPGs, mnemonic sequence and actual care process flow. The results of the survey demonstrates that the caremap structure and path, when examined independently, are considered by clinicians to have accuracy 88% (Q1) and 76% (Q2) respectively, while the placement and purpose and criterion used to describe decision points were 86% (Q3) and 95% (Q4) accurate. The caremaps were assessed overall to be 93% accurate when all elements were examined jointly, and 95% (Q6) easier to use than the clinical documentation they were based on. Hence, this survey indicates that, for caremaps developed using the extended TaSC approach, practicing clinicians found a high degree of accuracy when examining the caremaps both independently and jointly. While the conclusion cannot be generalised due to the limited number of participants, it does provide support for the potential clinical utility of using the extended TaSC approach to deliver accurate caremaps. The survey response legend and individual responses are provided in Appendix E.

### 9.5.3 TaSC Benefits, Limitations and Significance

The TaSC approach is promising for efficient and standardised production of caremaps with high clinical accuracy. An important benefit is that TaSC is generic: it can be applied to any medical condition or practice. Extending TaSC with DPs increased the caremap's utility, providing clinicians with a visual prompt for when significant clinical decisions must be made, as well as the evidence-based criteria to support selection of the appropriate treatment path for the current patient.

The TaSC caremap is as robust, applicable and accurate as the CPG, medical literature and expert guidance allows. Different local health districts develop their own CPGs with diagnostic and treatment thresholds customised for the local population. One issue for application is that even when the caremap with DPs is based on a national CPG and evidence-based literature, the aspect of local CPGs and clinical expertise may colour the caremap and DP criteria with local influence that limits the resulting caremap's general applicability.

Of significance to TaSC is visual simplicity and standard appearance. We propose that if the range of caremaps appended to CPGs and in use within a facility were to be standardised using TaSC, clinicians could engage with TaSC caremaps in clinical practice without the effort that adapting to different presentation styles and notations as they move between units within the hospital must require. The presence of decision criteria acts as a simple CDSS, prompting the clinician with the path to be taken and next treatment activity required while still being entirely grounded in the CPG and evidence-based medical literature they would normally refer to. TaSC may have the potential to expedite treatment, improve treatment consistency and save time for both the patient and clinician. All of which reduces healthcare cost and resource consumption and improves the patient's quality of life.

## 9.6 SUMMARY

This chapter has described four case studies in the development of caremaps. While the first case study provides a pre-standardisation baseline example, the latter three apply TaSC in assessment of the aims and objectives of standardisation of clinical documentation. The latter three were also developed within the context of the PAMBAYESIAN project: used as a formal approach to elicit expert information and validate the LHS designer's understanding of clinical information prior to use in development of Bayesian-based CDSS LHS.

Quantitative assessment showed that TaSC reduced the resources, significantly time and cost, required to produce the clinical caremap. Qualitative assessment with experienced clinicians found that the structure, care path, and DP in caremaps developed using TaSC were highly accurate and significantly easier to use than the CPGs the caremap was based on. TaSC-based Caremaps imbued with DP can be applied in clinical care as one type of LHS: a CDSS – prompting the clinical decision-making processes with criteria for consideration and using the existing evidence to direct treatment.

Developing caremaps with clinical experts along with the responses received from clinicians during the survey supports the research hypothesis that through standardisation of clinical documentation, we can support development and implementation of LHS.

# Chapter 10: Discussion and Conclusions

The hypothesis for this research was *that by understanding the ongoing barriers that have inhibited health IT implementation, it is possible to develop models to achieve the standardisation necessary for development and implementation of LHS with a new approach to standardising CCPS.*

Rather than approaching the hypothesis from the perspective of LHS alone, we began by investigating the primary source which EHR, the building blocks of LHS, were designed to replace: clinical documentation. While many healthcare organisations focused standardisation efforts on the stored data format or data entry method, these approaches have failed to deal with the original lack of standardisation at the source. Clinical documentation lacked standardisation. This is a flaw inherited by EHR that has been a significant barrier to adoption. The thesis has shown that understanding the barriers enabled development of models that can help integrate new technology tools into clinical practice, and that standardisation of the caremap in particular provides insight into a new approach that could truly lead to implementation and adoption of LHS.

This chapter summarises why the thesis has provided strong support for the research hypothesis. It reviews the content and contributions and relates them back to the objectives defined in Chapter 1. It also describes how the chapters and contributions relate to the academic literature drawn from this thesis. The majority of these papers have, as at the time of submission, been peer reviewed and are, or are in the process of being, published. Chapter 3 described in detail the common methodologies applied throughout this research. Within each chapter the application of these methods is described, along with any methodology unique to the needs of that chapter's research. Examples include the development and application of ITPOSMO-BBF in Chapter 6, and the use of Pathway Theory in Chapter 7.

## 10.1 REVIEW OF RESEARCH OBJECTIVES

### Objective 1: The Lack of Standardisation

In chapter 4 we saw that research established a lack of awareness as to what constituted, and how to classify solutions as, LHS. It was also observed that only around half of the self-identified LHS literature reported on a developed solution, and of those, none were found to be in clinical use. In order to investigate standardisation as a causal issue or potential mitigant, it was first necessary to understand the fundamental elements of LHS, being EHR. We investigated their purpose and the role standardisation had, or should have had historically, in development of the clinical documentation underpinning EHR. In chapter 2 we established that EHR were originally intended to be digital replacements for much of the clinical documentation commonly used in healthcare. A review of literature in chapter 4 identified a lack of standardisation in the CCPS which EHR were developed to replace. This discovery underpinned the rest of the research.

### Objective 2: The relationships between CCPS, EHR and LHS

This objective had four parts: (a) qualification of the relationship between CCPS, EHR and the more recent invention, LHS; (b) the role for standardisation in this relationship; (c) the role, or application, of each in clinical practice; and, (d) investigation of a unifying model that integrates CCPS, EHR and LHS and describes their relationship to the levels of medical application and treatment of individual patients.

The direct relationship between CCPS and EHR was established in literature which showed that an important motivation for the development of EHR had been to create CCPS and capture patient data in digital form. Chapter 4 showed that the direct relationship between EHR and LHS was also already well understood as EHR are widely described as the basic building block necessary for LHS, which by definition seek to expose new knowledge from large datasets of EHR. The literature review in chapter 4 was the first comprehensive literature review of EHR and LHS. It also established that interdependent relationships and similar issues, suggesting inheritance of issues, exist between CCPS, EHR and LHS. As the relationships between CCPS, EHR and LHS were established, resolving the first branch of Objective 2.

Determining the role standardisation plays in the relationships between CCPS, EHR and LHS first required investigation of the degree of standardisation currently present in each: a lack of standardisation in all was established in chapter 4. This lack of standardisation could also have been established through the gathering and analysis of CCPS and EHR implementation documentation directly from a range of hospitals. However, this additional step was unnecessary as much of the collected literature reproduced either extracts from or entire copies of exemplar clinical documents. From these it was possible to establish the vast differences in structure and content between examples of the same document type, supporting this thesis' contention that standardisation was an issue.

Resolving the remaining two branches of Objective 2 occurred contemporaneously, with each becoming apparent as the LAGOS framework presented in chapter 7 was developed. Investigation and identification of the relationships between CCPS, EHR and LHS clarified their connection to corresponding levels of medical practice.

**Objective 3: Standardisation as A basis for Learning Health Systems**

Meeting this objective required us to investigate, develop and evaluate an approach to standardising CCPS. Given the range of CCPS identified in chapter 5, it became clear that seeking to standardise all types would far exceed the thesis scope. It was therefore necessary to identify a single CCPS type that could serve as an example. Caremaps were chosen and an approach to their standardisation was developed and evaluated through several case studies.

During work on the second case study presented in chapter 8 it became apparent that caremaps not only contain latent DP, but that identification and rendering of those DP would improve clinical understanding and overall utility of the resulting caremap. Six aspects of clinical practice from which clinical decisions arise had previously been identified [532]. Different approaches were observed for presenting criterion for clinical decisions on caremaps. However, activities identifiable as DPs were not always evident and no previous standard describing DP identification and presentation in clinical documentation of a graphical form was identified.

## 10.2 CONTRIBUTIONS

The novel contributions of this thesis can be summarised as follows:

- Identification and characterisation for each type of CCPS identified from the literature, along with a taxonomy for CCPS incorporating a hierarchy and describing the relationship and heritance between each. The taxonomy and characterisation are applied through a case study examination of clinical documents for Type 2 Diabetes.
- A taxonomy identifying the nine types of LHS, with validity demonstrated through classification of the self-identified LHS solutions identified during the literature review.
- A unified framework, Heimdall, that describes how the nine types of LHS situate within the learning healthcare organisation and how LHS focus clinical practice toward delivery of precision medicine.
- A new approach, ITPOSMO-BBF, that extends on the established ITPOSMO tool used in IT implementation gap and failure analysis, for comparing and contrasting the benefits, barriers and facilitating factors. A case study that applies ITPOSMO-BBF to EHR and LHS implementations.
- A unifying framework, LAGOS, using pathway theory for aligning the domains of medicine, clinical care, health informatics and decision science. These are incorporated within five pathways, each with three levels corresponding to how the elements of these domains relate to the three *applications of medical practice*: population, evidence-based and precision medicine.
- A new approach, TaSC, for standardising the structure, content and development of caremaps. TaSC is used as a demonstration that standardisation of clinical documents can have both a beneficial effect on the application and use of the clinical document, and development and engagement with LHS.
- A case study that applies the standardisation approach, TaSC, to a selection of caremaps developed and used in conjunction with clinical expertise to develop CDSS for the EPSRC-funded PAMBAYESIAN project. The results demonstrated great potential for clinical benefit as the standardised approach efficiently produced caremaps that clinicians found to be highly accurate and, due to their standard appearance, easy to engage with.

## 10.3 FUTURE DIRECTIONS

This thesis presents the TaSC approach and demonstrates it is repeatable. However, the method described is limited in application because it requires information scientists to lead the process and information technology tools that are non-intuitive for most clinicians. Future work should create online tools for developing caremaps that can be used in clinical research and patient care. Such tools would need to provide a user interface and intuitive approach that would allow clinicians to self-create and validate caremaps adhering to TaSC.

The TaSC approach presented in this thesis draws on CPGs and other clinical documentation to develop accurate caremaps. Caremaps have previously been used to create *state transition machines* (STMs) that, with the addition of incidence and treatment statistics, can be used to generate synthetic data described as the *realistic synthetic electronic health record* (RS-EHR). Populating the caremap is time consuming, as it requires locating and aggregating large sets of statistics for each node and arc of the caremap. These statistics describe the likelihood that a given treatment will be applied to the patient, the probability that this treatment will resolve the patient's condition, and the likely path along the caremap that the clinician will select to progress treatment. Future work should investigate an accelerated solution to locate, aggregate and populate caremap nodes with knowledge from RCTs, systematic reviews and publicly available incidence and treatment statistics. Such a solution would represent a significant precision medicine LHS that could be used to position the patient within the caremap, and efficiently fine tune DPs in near real-time to demographic, risk and treatment response factors relevant to the individual patient.

The work described in this thesis developed an approach to standardising one clinical document type and discussed its application in the creation of Bayesian-based CDSS; this is one type of LHS. While accuracy and engagement of the TaSC approach was assessed non-clinically with a small number of clinicians, future work in this area should involve assessment in the conduct of pre-clinical and clinical trials that evaluate the research outputs of projects that have used standardised caremaps as the foundation for developing LHS, like the EPSRC-funded PAMBAYESIAN. Such a study has been proposed under the title *Accelerating Development of Improved Clinical Decision Support Applications for Diabetes* (ADiCAD). This funding application was successful in its initial evaluation with Barts and The London Charity (BLC), and was invited to proceed to a full funding submission that the charity will evaluate late in 2019.

## 10.4 CONCLUSION

The research and frameworks presented in this thesis have supported the research hypothesis by showing that, through understanding the barriers inhibiting health IT implementation, it is possible to develop models to achieve standardisation, mitigate the barriers, and increase acceptance of LHS.

LHS could revolutionise healthcare through the potential of precision medicine to deliver advanced decision-making and predictive solutions; assisting health systems to reduce resource expenditure and improve patient experience and outcomes. However, like EHR before them, healthcare providers and clinicians have been slow to adopt LHS.

A lack of awareness persists within the LHS domain which inhibits creation of a critical mass of researchers. While the number of proposed and published LHS solutions continues to rise, very few LHS have been used in clinical practice. The public failings of systems considered to be LHS, including IBM Watson and Google DeepMind, have made clinicians and patients wary of LHS solutions. Many also continue to express concern over potential privacy implications arising from the *secondary use* of personal medical data. LHS can only ever be as good as the EHR they data-mine for new knowledge, and the *Triple-I* issues of data *integrity*, *integration* and *interoperability* are another significant factor that presently restrain our ability to accurately develop new knowledge from EHR.

While standardisation has been a successful approach used in a wide range of business applications, its application in healthcare has not always been accepted nor has it been entirely successful. Much of the focus of standardisation has been on the management of healthcare services and the approach to clinical practice through evidence-based medicine. However, directing standardisation towards those items that affect and record patient care, namely the clinical documentation, may be one way in which hospital management and clinicians can bring about the advantages of standardisation, successfully implement and engage with learning health systems, and realise the benefits precision medicine has to offer.

# Bibliography

1. OECD. (2016). Health Expenditure and Financing. Retrieved from: OECD Publishing: Paris. https://stats.oecd.org/Index.aspx?DataSetCode=SHA
2. Dieleman, J., T. Templin, N. Sadat, P. Reidy, A. Chapin, K. Foreman, A. Haakenstad, T. Evans, C. Murray, and C. Kurowski. (2016). National spending on health by source for 184 countries between 2013 and 2040. The Lancet, 387, 2521-2535.
3. Emergo. (2016). Worldwide Spending on Healthcare. Retrieved from https://www.emergobyul.com/resources/worldwide-health-expenditures
4. Delloitte. (2017). 2017 global health care sector outlook. Retrieved from: London, UK. https://www2.deloitte.com/content/dam/Deloitte/global/Documents/Life-Sciences-Health-Care/gx-lshc-2017-health-care-outlook-infographic.pdf
5. Bramwell, C., R. Don, I. Porter, H. Lloyd, U. Kadak, M. Rijken and J. Valderas. (2016). Caring for people with multiple chronic conditions in the United Kingdom: Policy and practices with a focus on England and Scotland. Retrieved from: London, UK. http://www.icare4eu.org/pdf/Country_Factsheet_United_Kingdom_ICARE4EU.pdf
6. NICE. (2015). Costing Statement: Implementing the NICE guideline on social care of older people with social care needs and multiple long-term conditions. Retrieved from: London, UK. https://www.nice.org.uk/guidance/ng22/resources/costing-statement-pdf-554559661
7. Campbell, D. (2018). NHS £20bn boost risks being spent to pay off debts, experts warn The Guardian. Retrieved from https://www.theguardian.com/society/2018/oct/21/nhs-20bn-cash-risks-paying-off-debts
8. Owen, J. (2015). Crippling PFI deals leave Britain £222bn in debt. The Independent. Retrieved from https://www.independent.co.uk/money/loans-credit/crippling-pfi-deals-leave-britain-222bn-in-debt-10170214.html
9. Anani, N., M. Mazya, R. Chen, T. Moreira, O. Bill, N. Ahmed, N. Wahlgren and S. Koch. (2017). Applying openEHR's guideline definition language to the SITS international stroke treatment registry: a European retrospective observational study. BMC medical informatics and decision making, 17(1).
10. Franz, B. and J. Murphy. (2015). Electronic medical records and the technological imperative: the retrieval of dialogue in community-based primary care. Perspectives in biology and medicine, 58(4), 480-492.
11. Edberg, D. and W. J. (2018). Healthcare Transformation: The Electronic Health Record. In (Ed.), Behavioral Medicine and Integrated Care. Springer.
12. Lougheed, M., N. Thomas, N. Wasilewski, A. Morra and J. Minard. (2018). Use of SNOMED CT® and LOINC® to standardize terminology for primary care asthma electronic health records. Journal of Asthma, 55(6), 629-639.
13. Silow-Carroll, S., J. Edwards and D. Rodin. (2012). Using electronic health records to improve quality and efficiency: the experiences of leading hospitals. Retrieved from:

14. Dinev, T., V. Albano, H. Xu, A. D'Atri and P. Hart. (2016). Individuals' Attitudes Towards Electronic Health Records: A Privacy Calculus Perspective. In (Ed.), Advances in Healthcare Informatics and Analytics. Springer Publishing.
15. Poissant, L., J. Pereira, R. Tamblyn and Y. Kawasumi. (2005). The impact of electronic health records on time efficiency of physicians and nurses: a systematic review. Journal of the American Medical Informatics Association, 12(5), 505-516.
16. Chu, S. and B. Cesnik. (1998). Improving Clinical Pathway Design: Lessons Learned from a Computerised Prototype. International Journal of Medical Informatics, 51.
17. Gemmel, P., D. Vandaele and W. Tambeur. (2008). Hospital Process Orientation (HPO): The development of a measurement tool. Total Quality Management & Business Excellence, 19(12), 1207-1217.
18. Panz, V., F. Raal, J. Paiker, R. Immelman and H. Miles. (2005). Performance of the CardioChek (TM) PA and Cholestech LDX (R) point-of-care analysers compared to clinical diagnostic laboratory methods for measurement of lipids: Cardiovascular topic. Cardiovascular Journal of South Africa, 16(2), 112-116.
19. Zander, K. (1992). Quantifying, managing and improving quality Part 1: How CareMaps link C.Q.I. to the patient. The New Definition, 1-3.
20. DiabetesUK. (2014). NHS missing chances to improve diabetes care and save money as well as lives. Retrieved from website: https://www.diabetes.org.uk/about_us/news/nhs-missing-chances-to-improve-diabetes-care-and-save-money-as-well-as-lives
21. IoM. (2013). Best care at lower cost: The path to continuously learning health care in America. Retrieved from: Washington, USA.
22. Pym, H. (2017). Could the NHS save money by getting it right first time? Retrieved from website: https://www.bbc.co.uk/news/health-40825987
23. McLachlan, S., H. Potts, K. Dube, D. Buchanan, S. Lean, O. Johnson, B. Daley, T. Gallagher, W. Marsh, and N. Fenton. (2018). The Heimdall Framework for supporting characterisation of learning health systems. BCS Journal of Innovation in Health Informatics, 25(2).
24. English, M., G. Irimu, A. Agweyu, D. Gathara, J. Oliwa, P. Ayieko and C. Forrest. (2016). Building learning health systems to accelerate research and improve outcomes of clinical care in low- and middle-income countries. PLOS One, 13(4).
25. Mullins, C., L. Wingate, H. Edwards, T. Tofade and A. Wutoh. (2018). Transitioning from learning healthcare systems to learning healthcare communities. Journal of Comparative Effectiveness Research, 7(6), 603-614.
26. Nwaru, B., C. Friedman, J. Halamka and A. Sheikh. (2017). Can learning health systems help organisations deliver personalised care? BMC Medicine, 15(177).
27. Beaupre, L. (2005). Effectiveness of a caremap for treatment of elderly patients with hip fracture. (Doctor of Philosophy), University of Alberta, Alberta, CA. Retrieved from https://elibrary.ru/item.asp?id=9041885
28. Flynn, R. (2002). Clinical governance and governmentality. Health, Risk and Society, 4(2), 155-173.
29. Hubner, U., D. Flemming, K. Heitmann, F. Oemig, S. Thun, A. Dickerson and M. Veenstra. (2010). The need for standardised documents in continuity of care: Results of standardising the eNurse summary. Studies in Health Technology and Informatics, 160(2), 1169-1173.

30. Morris, A. (2002). Decision support and safety of clinical environments. BMJ Quality and Safety in Healthcare, 11, 69-75.

31. Rycroft-Malone, J., M. Fontenla, K. Seers and D. Bick. (2009). Protocol-based care: The standardisation of decision-making? Journal of Clinical Nursing, 18(10), 1490-1500.

32. Trowbridge, R. (Weingarten, S.). Critical Pathways. In J. Eisenberg and D. Kamerow (Ed.), Making healthcare safer - A critical anlysis of patient safety practices. Rockville, MD: AHRQ Publications.

33. Donnelly, K. (2006). SNOMED-CT: The advanced terminology and coding system for eHealth. Studies in health technology and informatics, 121.

34. NHS. (2018). Quality and Outcomes Framework (QOF), enhanced services and core contract extraction specifications. Retrieved from https://digital.nhs.uk/data-and-information/data-collections-and-data-sets/data-collections/quality-and-outcomes-framework-qof

35. NHS. (2019). Commission for Quality and Innovation. Retrieved from https://www.england.nhs.uk/nhs-standard-contract/cquin/

36. NICE. (2018). Standards and Indicators. Retrieved from https://www.nice.org.uk/standards-and-indicators

37. Ilott, I., A. Booth, J. Rick and M. Patterson. (2010). How do nurses, midwives and health visitors contribute to protocol-based care? A synthesis of the UK literature. International Journal of Nursing Studies, 47(6), 770-780.

38. McDonald, A., K. Frazer and D. Cowley. (2013). Caseload management: An approach to making community needs visible. British Journal of Community Nursing, 18(3), 140-147.

39. McLachlan, S., E. Kyrimi, K. Dube and N. Fenton. (2019). Clinical caremap development: How can caremaps standardise care when they are not standardised? Paper presented at the 12th International Joint Conference on Biomedical Systems and Technologies (BIOSTEC 2019), volume 5: HEALTHINF, Prague, Czech Republic.

40. Mitchell, S., A. Plunkett and J. Dale. (2014). Use of formal advance care planning documents: A national survey of UK Paediatric Intensive Care units Archives of disease in childhood, 99(4), 327-330.

41. Scott, I., G. Mitchell, E. Reymond and M. Daly. (2013). Difficult but necessary conversations: The case for advanced care planning. Medical Journal of Australia, 199(10), 662-666.

42. Wilson, M. (2002). Making nursing visible? Gender, technology and the care plan as script. Information Technology and People, 15(2), 139-158.

43. Harrison, J., J. Young, M. Price, P. Butow and M. Solomon. (2009). What are the unmet supportive care needs of people with cancer? A systematic review. Supportive care in cancer, 17(8), 1117-1128.

44. NHS. (2018). Funding and Efficiency. Retrieved from https://www.england.nhs.uk/five-year-forward-view/next-steps-on-the-nhs-five-year-forward-view/funding-and-efficiency/

45. Martin, G., D. Kocman, T. Stephens, C. Peden and R. Pearse. (2017). Pathways to professionalism? Quality improvement, care pathways and the interplay of standardisation and clinical anatomy. Sociology of Health and Illness, 39(8), 1314-1329.

46. Wears, R. (2015). Standardisation and its discontents. Cognition, Technology and Work, 17(1), 89-94.
47. Campbell, D. (2017). NHS nurses are too busy to care for patients properly, research shows. The Guardian. Retrieved from https://www.theguardian.com/society/2017/sep/29/nhs-nurses-are-too-busy-to-care-for-patients-properly-research-shows
48. CQC. (2017). The state of care in NHS acute hospitals: 2014-2016. Retrieved from: London, UK. https://www.cqc.org.uk/sites/default/files/20170302_stateofhospitals_web_0.pdf
49. Bozic, K., J. Maselli, P. Pekow, P. Lindenauer, T. Vail and A. Auerbach. (2010). The influence of procedure volumes and standardization of care on quality and efficiency in total joint replacement surgery. Journal of Bone and Joint Surgery, 92(16), 2643-2652.
50. Hashimoto, H., N. Ikegami, K. Shibuya, N. Izumida, H. Noguchi, H. Yasunaga, H. Miyata, J. Acuin, and M. Reich. (2011). Cost containment and quality of care in Japan: is there a trade-off? The Lancet, 378, 1174-82.
51. Balakrishnan, M., A. Raghavan and G. Suresh. (2017). Eliminating undesirable variation in neonatal practice: balancing standardization and customization. Clinics in Perinatology, 44(3), 529-540.
52. Maeng, D., X. Yan, T. Graf and G. Steele. (2016). Value of primary care diabetes management: long-term cost impacts. American Journal of Managed Care, 22(3), e88-e94.
53. Taylor, C., J. Hepworth, P. Buerhaus, R. Dittus and T. Speroff. (2007). Effect of crew resource management on diabetes care and patient outcomes in an inner-city primary care clinic. BMJ Quality and Safety, 16(4), 244-247.
54. Blind, K. (2009). Standardisation: A catalyst for innovation. Rotterdam, Netherlands: Rotterdam School of Management, Erasmus University.
55. Edum-Fotwe, F., A. Gibb and M. Benford-Miller. (2004). Reconciling construction innovation and standardisation on major projects. Engineering, Construction and Architectural Management, 11(5), 366-372.
56. Featherston, C., J. Ho, L. Brevignion-Dodin and E. O'Sullivan. (2015). Mediating and catalysing innovation: A framework for anticipating the standardisation needs of emerging technology. Technovation, 48, 25-40.
57. Williams, R., I. Graham, K. Jakobs and K. Lyytinen. (2011). China and Global ICT standardisation and innovation. Technology Analysis & Strategic Management, 23(7), 715-724.
58. Blind, K. (2016). The impact of standardisation and standards on innovation. In J. Edler (Ed.), Handbook of Innovation Policy Impact. Edward Elgar Publishing.
59. IoM. (2007). The Learning Healthcare System: Workshop Summary. Retrieved from: https://www.ncbi.nlm.nih.gov/pubmed/21452449
60. IoM. (2011). Digital Infrastructure for the Learning Health System: The foundation for continuous improvement in health and health care. Retrieved from: Washington DC. https://www.ncbi.nlm.nih.gov/pubmed/22379651
61. Friedman, C., N. Allee, B. Delaney, A. Flynn, J. Silverstein, K. Sullivan and K. Young. (2017). The science of Learning Health Systems: Foundations for a new journal. Learning Health Systems, 1(1).

62. Krumholz, H. (2014). Big data and new knowledge in medicine: The thinking, training and tools needed for a learning health system. Health Affairs, 33(7), 1163-1170.
63. Beckmann, J. and D. Lew. (2016). Reconciling evidence-based medicine and precision medicine in the era of big data: Challenges and opportunities. Genome Medicine, 8(1).
64. Sitapati, A., H. Kim, B. Berkovich, R. Marmor, S. Singh, R. El-Kareh and L. Ohno-Machado. (2017). Integrated precision medicine: the role of electronic health records in delivering personalized treatment. Wiley Interdisciplinary Reviews: Systems Biology and Medicine, 9(3), e1378.
65. IBM. (2016). IBM Watson for Oncology. Retrieved from https://www.ibm.com/us-en/marketplace/ibm-watson-for-oncology
66. Treanor, M. (2018). How Watson technology can help with addiction care. IBM Client Success Field Notes. Retrieved from website: https://www.ibm.com/blogs/client-voices/watson-technology-addiction-care/
67. Chen, A. (2018). IBM's Watson gave unsafe recommendations for treating cancer. The Verge. Retrieved from website: https://www.theverge.com/2018/7/26/17619382/ibms-watson-cancer-ai-healthcare-science
68. Hernandez, D. and T. Greenwald. (2018). IBM has a Watson dilemma. The Wall Street Journal. Retrieved from website: https://www.wsj.com/articles/ibm-bet-billions-that-watson-could-improve-cancer-treatment-it-hasnt-worked-1533961147
69. Iyer, P. and N. Camp. (1995). Nursing Documentation: A Nursing Process Approach (2nd edn. ed.). St Louis, MO, USA: Mosby Year-book.
70. Kuhn, T., P. Basch, M. Barr and T. Yackel. (2015). Clinical Documentation in the 21st Century: Executive Summary of a Policy Position Paper From the American College of Physicians. Annals of Internal Medicine, 162(4), 301-303.
71. Tornvall, E. and S. Wilhelmsson. (2008). Nursing documentation for communicating and evaluating care. Journal of Clinical Nursing, 17, 2116-2124.
72. Freire, V., M. Lopes, G. Keenan and K. Lopez. (2018). Nursing students' diagnostic accuracy using a computer-based clinical scenario simulation. Nursing Education Today, 71, 240-246.
73. Johansson, M., E. Pilhammar and A. Willman. Nurses' clinical reasoning concerning management of peripheral venous cannulae. Journal of Clinical Nursing, 18, 3366-3375.
74. Bossen, C., L. Jensen and F. Udsen. (2013). Evaluation of a comprehensive EHR based on the DeLone and McLean model for IS success: Approach, results, and success factors. International Journal of Medical Informatics, 82(10), 940-953.
75. Kopanitsa, G., Z. Tsvetkova and H. Veseli. (2012). Analysis of metrics for the usability evaluation of EHR management systems. In J. Mantas (Ed.), Quality of Life through Quality of Information. European federation for Medical Informatics and IOS Press.
76. Seymour, T., D. Frantsvog and T. Graeber. (2012). Electronic Health records (EHR). American Journal of Health Sciences, 3(3), 201-210.
77. Zandieh, S., K. Yoon-Flannery, G. Kuperman, D. Langsam, D. Hyman and R. Kaushal. (2008). Challenges to EHR Implementation in Electronic- Versus

Paper-based Office Practices. Journal of General Internal Medicine, 23(6), 755-761.
78. Smoyer, W., P. Embi and S. Moffatt-Bruce. (2018). Creating Local Learning Health Systems: Think Globally, Act Locally. JAMA, 316(23), 2481-2482.
79. Fischbach, F. (1991). Documenting Care Communication, the Nursing Process and Documentation Standards. Philadelphia, PA, USA: F.A. Davis.
80. Fontenot, S. (2013). The Affordable Care Act and electronic health care records. Does Today's.
81. Peleg, M., D. Beimel, D. D'ori and Y. Denekamp. (2008). Situation-Based Access Control: Privacy management via modeling of patient data access scenarios. Journal of Biomedical Informatics, 41(6), 1028-1040.
82. Hayrinen, K., K. Saranto and P. Nykanen. (2008). Definition, structure, content, use and impacts of electronic health records: A review of the research literature. International Journal of Medical Informatics, 77(5).
83. Miller, R. and I. Sim. (2004). Physicians' use of electronic medical records: barriers and solutions. Health Affairs, 23(2), 116-126.
84. Friedman, C., J. Rubin, J. Brown, M. Buntin, M. Corn, L. Etheredge and C. Gunter. (2015). Toward a science of learning systems: A research agenda for the high-functioning learning health system. Journal of American Medical Informatics Association (JAMIA), 22, 43-50.
85. Abernethy, A., L. Etheredge, P. Ganz, P. Wallace, R. German, C. Neti, P. Back and S. Murphy. (2010). Rapid-learning System for Cancer Care. Journal of Clinical Oncology, 28(27).
86. Foley, T. and L. Vale. (2017). What role for learning health systems in quality improvement within healthcare providers? Learning Health Systems, e10025.
87. Friedman, C., A. Wong and D. Blumenthal. (2010). Achieving a nationwide learning health system. Science Transitional Medicine, 2(57), 1-3.
88. Christian, C., M. Gustafson, E. Roth, T. Sheridan, T. Gandhi, K. Dwyer, M. Zinner and M. Dierks. (2006). A prospective study of patient safety in the operating room. Surgery, 139(2), 159-173.
89. Valenstein, P. (2008). Formatting pathology reports: Applying four design principles to improve communication and patient safety. Archives of Pathology and Laboratory Medicine 1(84).
90. Agaimy, A., N. Vassos, B. Markl, N. Meidenbauer, J. Kohler, J. Spatz and W. Hohenberger. (2013). Anorectal gastrointestinal stromal tumors: a retrospective multicenter analysis of 15 cases emphasizing their high local recurrence rate and the need for standardized therapeutic approach. International journal of colorectal disease 28(8), 1057-1064.
91. Hryb, K., R. Kirkhart and R. Talbert. (2007). A call for standardized definition of dual diagnosis. Psychiatry, 4(9).
92. Jarden, R. and S. Quirke. (2010). Improving safety and documentation in intrahospital transport: development of an intrahospital transport tool for critically ill patients. Intensive and Critical Care Nursing, 26(2), 101-107.
93. Munyisia, E., P. Yu and D. Hailey. (2011). The changes in caregivers' perceptions about the quality of information and benefits of nursing documentation associated with the introduction of an electronic documentation system in a nursing home. International journal of medical informatics 80(2), 116-126.

94. Stanton, T., J. Latimer, C. Maher and M. Hancock. (2011). A modified Delphi approach to standardize low back pain recurrence terminology. European spine journal, 20(5), 744-752.
95. Langguth, B., T. Kleinjung and M. Landgrebe. (2011). Tinnitus: the complexity of standardization. Evaluation & the health professions, 34(4), 429-433.
96. Nystedt, K., J. Hill, A. Mitchell and F. Goodwin. (2005). The standardization of radiation skin care in British Columbia: a collaborative approach. Oncology nursing forum, 32(6).
97. Blair, W. and B. Smith. (2012). Nursing documentation: Frameworks and barriers. Contemporary Nurse, 41(2), 160-168.
98. Laszlo, G. (2006). Standardisation of lung function testing: helpful guidance from the ATS/ERS Task Force. BMJ Thorax, 61, 744-746.
99. Mahler, M. (2019). Lack of standardisation of ANA and implications for drug development and precision medicine. Annals of the rheumatic diseases, 78(5), e33.
100. Armbruster, D. and R. Miller. (2007). The Joint Committee for Traceability in Laboratory Medicine (JCTLM): A Global Approach to Promote the Standardisation of Clinical Laboratory Test Results. Clinical Biochemistry Review, 28(3), 105-114.
101. Little, R. and C. Rohlfing. (2009). HbA1c Standardization: Background, Progress and Current Issues. Laboratory Medicine, 40(6), 368-373.
102. Harper, E. (2014). Can big data transform electronic health records into learning health systems? Nursing Informatics, 201, 470-475.
103. Crandall, R. and J. Graham. (1989). The effect of fuel economy standards on automobile safety. Journal of Law and Economics, 32.
104. Curey, R., M. Ash, L. Thielman and C. Barker. (2004). Proposed IEEE inertial systems terminology standard and other inertial sensor standards. Paper presented at the PLANS 2004. Position Location and Navigation Symposium, Monterey, CA.
105. Jones, K. (2003). Miscommunication between pilots and Air Traffic Control. Language Problems and Language Planning, 27(3), 233-248.
106. Griffith, R., H. Griffiths and S. Jordan. (2003). Administration of medicines part 1: the law and nursing. Nursing Standard, 18(2), 47-56.
107. Hoag, S., H. Ramachandruni and R. Shangraw. (1997). Failure of Prescription Prenatal Vitamin Products to Meet USP Standards for Folic Acid Dissolution: Six of nine vitamin products tested failed to meet USP standards for folic acid dissolution. Journal of the American Pharmaceutical Association, 37(4), 397-400.
108. Milstien, J., A. Costa, S. Jadhav and R. Dhere. (2009). Reaching international GMP standards for vaccine production: challenges for developing countries. Expert review of Vaccines, 8(5), 559-566.
109. de Vries, H. (1997). Standardisation - What's in a name? Terminolog, 4(1), 55-83.
110. ISO/IEC. (2004). Standardisation and Related Activities - General Vocabulary. Retrieved from: Geneva.
111. OECD. (2005). Oslo Manual: Guidelines for collecting and interpreting innovation data. Retrieved from: Paris.
112. Baumol, W. (2002). Entrepreneurship, innovation and growth: The David-Goliath symbiosis. Journal of Entrepreneurial Finance, 7(2), 1-10.

113. Allen, R. and R. Sriram. (2000). The role of standards in innovation. Technological Forecasting and Social Change, 64(2-3), 171-181.
114. Egyedi, T. and M. Sherif. (2008). Standards' dynamics through an innovation lens: Next generation ethernet networks. Paper presented at the First ITU-T Kaleidoscope Academic Conference-Innovations in NGN: Future Network and Services.
115. Reddy, N., S. Cort and D. Lambert. (1989). Industrywide technical product standards. R & D Management, 19(1).
116. Cargill, C. (1995). A five-segment model for standardization. In B. Kahin and J. Abbate (Ed.), Standards policy for information infrastructure. Cambridge, MA: MIT Press.
117. Belleflamme, P. (2002). Coordination on formal vs. de facto standards: a dynamic approach. European Journal of political economy, 18(1), 153-176.
118. Tether, B., C. Hipp and I. Miles. (2001). Standardisation and particularisation in services: Evidence from Germany. Research Policy, 30, 1115-1138.
119. Kosawatz, J. (2017). Rekindling the Spark. Mechanical Engineering Magazine Selected Articles, 139(11), 28-33.
120. Mazda. (2019). Performance: Skyactiv-X. Retrieved from https://www.mazda.com.au/imagination-drives-us/performance---skyactiv-x/
121. Mizuno, H. (2017). Nissan gasoline engine strategy for higher thermal efficiency. Combustion Engines, 169(2), 141-145.
122. Chrisafis, A. and A. Vaughan. (2017). France to ban sales of petrol and diesel cars by 2040. The Guardian. Retrieved from https://www.theguardian.com/business/2017/jul/06/france-ban-petrol-diesel-cars-2040-emmanuel-macron-volvo
123. Coren, M. (2018). Nine countries say they'll ban internal combustion engines. So far, it's just words. Retrieved from https://qz.com/1341155/nine-countries-say-they-will-ban-internal-combustion-engines-none-have-a-law-to-do-so/
124. Corbett, E. (2016). Standardised care vs. personalisation: Can they coexist? Quality and Process Improvement, Health Catalyst. Retrieved from https://www.healthcatalyst.com/standardized-care-vs-personalization-can-they-coexist
125. Giffin, M. and R. Giffin. (1994). Market Memo: Critical Pathways Produce Tangible Results. Health Care Strategy Management, 12(7).
126. Rotter, T., J. Kugler, R. Koch, H. Gothe, S. Twork, J. van Oostrum and E. Steyerberg. (2008). A systematic review and meta-analysis of the effects of clinical pathways on length of stay, hospital costs and patient outcomes. BMC Health Services Research, 8(265).
127. Zarzuela, S., N. Ruttan-Sims, L. Nagatakiya and K. DeMerchant. (2015). Defining standardisation in healthcare. Mohawk Shared Services, 15.
128. Keyhani, S., R. Falk, E. Howell, T. Bishop and D. Korenstein. (2014). Overuse and systems of care: A systematic review. Medical Care, 51(6).
129. Martin, E. (2014). Eliminating waste in healthcare. ASQ Healthcare Update, 14.
130. Binks, C. (2017). Standardising the delivery of oral health care practice in hospitals. Nursing Times, 113(11), 18-21.
131. Leotsakos, A., H. Zheng, R. Croteau, J. Leob, H. Sherman, C. Hoffman and M. Duguid. (2014). Standardisation in patient safety: The WHO High 5's Project. International Journal for Quality in Health Care, 26(2), 109-116.

132. Freeman, K., R. Field and G. Perkins. (2015). Variations in local trust Do Not Attempt Cardiopulmonary Resuscitation (DNACPR) policies: A review of 48 English healthcare trusts. BMJ Open, 5.
133. De Pietro, C. and I. Francetic. (2018). E-Health in Switzerland: The laborious adoption of the federal law on electronic health records (EHR) and health information exchange (HIE) networks. Health Policy, 122(2), 69-74.
134. Piwek, L., S. Andrews and A. Joinson. (2016). The rise of consumer health wearables: Promises and barriers. PloS Medicine, 13(2), e1001953.
135. Shiffman, R., P. Shekelle, J. Overhage, J. Slutsky, J. Grimshaw and A. Deshpande. (2003). Standardised reporting of clinical practice guidelines: A proposal from the Conference on Guideline Standardisation. Annals of Internal Medicine, 139(6), 493-498.
136. Timmermans, S. and A. Angell. (2001). Evidence-based medicine, clinical uncertainty, and learning to doctor. Journal of health and social behavior, 342-359.
137. Audige, L., S. Goldhahn, M. Daigl, J. Goldhahn, M. Blauth and B. Hanson. (2014). How to document and report orthopedic complications in clinicalstudies? A proposal for standardization. Archives of orthopaedic and trauma surgery, 134(2), 269-275.
138. Carlesso, L., J. MacDermid and L. Santaguida. (2010). Standardization of adverse event terminology and reporting in orthopaedic physical therapy: application to the cervical spine. journal of orthopaedic & sports physical therapy, 40(8), 455-463.
139. Krumm, R., A. Semjonow, J. Tio, H. Duhme, T. Burkle, Haier, J. and B. Breil. (2014). The need for harmonized structured documentation and chances of secondary use–Results of a systematic analysis with automated form comparison for prostate and breast cancer. Journal of biomedical informatics, 51, 86-99.
140. Sankaran, S. (2001). Methodology for an organisational action research thesis. Graduate College of Management. Southern Cross University. Retrieved from https://opus.lib.uts.edu.au/bitstream/10453/9830/1/2006009109.pdf
141. Walker, D. (1997). Choosing an appropriate research methodology Construction Management and Economics, 15, 149-159.
142. Howe, K. (1992). Getting over the quantitative–qualitative debate. American Journal of Education, 100, 236-256.
143. Kothari, C. (2004). Research Methodology: Methods and Techniques. New Delhi, India: New Age International.
144. Onwuegbuzie, A. and N. Leech. (2007). On Becoming a Pragmatic Researcher: The Importance of Combining Quantitative and Qualitative Research Methodologies. International Journal of Social Research Methodology, 8(5), 375-387.
145. Davis, K., N. Drey and D. Gould. (2009). What are scoping studies? A review of the nursing literature. International Journal of Nursing Studies, 46(10), 1386-1400.
146. Arksey, H. and L. O'Malley. (2005). Scoping Studies: Towards a methodological framework. International Journal of Social Research Methodology, 8(1), 19-32.
147. Greenhalgh, T., G. Robert, F. Macfarlane, P. Bate, O. Kyriakidou and R. Peacock. (2005). Storylines of research in diffusion of innovation: A meta-narrative approach to systematic review. Social Science and Medicine, 61, 417-430.
148. Greenhalgh, T., G. Wong, G. Westhorp and R. Pawson. (2011). Protocol - realist and meta-narrative evidence synthesis: Evolving standards (RAMESES). BMC Medical Research Methodology, 11(115).
149. Vaismoradi, M., H. Turunen and T. Bondas. (2013). Content analysis and thematic analysis: Implications for conducting a qualitative descriptive study. Nursing and Health Sciences, 15(3), 398-405.
150. Joffe, H. and L. Yardley. (2004). Content and Thematic Analysis. Research Methods for Clinical and Health Psychology, 56.
151. Fereday, J. and E. Muir-Cochrane. (2006). Demonstrating rigor using thematic analysis: A hybrid approach of inductive and deductive coding and theme development. International Journal of Qualitative Methods, 5(1), 80-92.
152. Vaismoradi, M., J. Jones, H. Turunen and S. Snelgrove. (2016). Theme development in qualitative content analysis and thematic analysis. Journal of Nursing Education and Practice, 6(5), 100-110.
153. Angus, D. (2015). Fusing randomised trials with big data: The key to self-learning health care systems. Journal of the American Medical Association (JAMA), 314(8), 767-768.
154. Boonstra, A., A. Versluis and J. Vos. (2014). Implementing electronic health records in hospitals: A systematic literature review. BMC Health Services Research, 14(370).
155. Goldberg, D., A. Kuzel, L. Feng, J. DeShazo and L. Love. (2012). EHRs in Primary Care Practices: Benefits, Challenges and Successful Strategies. American Journal of Managed Care, 18(2), e48-e54.
156. Basole, R., M. Braunstein and J. Sun. (2015). Data and analytics challenges for a learning healthcare system. ACM Journal of Data and Information Quality, 6(2).
157. Bhandari, R., A. Feinstein, S. Huestis, E. Krane, A. Dunn, L. Cohen, M. Kao, B. Darnall, and S. Mackey. (2017). Pediatric-Collaborative Health Outcomes Information Registry (Peds-CHOIR): A learning health system to guide pediatric pain research and treatment. PAIN, 157(9), 2033-2044.
158. Crawford, L., G. Matczak, E. Moore, R. Haydar and P. Coderre. (2017). Patient-centered drug development and the Learning Health System. Learning Health Systems, e10027, 1-7.
159. Daniel, C., D. Ouagne, E. Sadou, N. Paris, S. Hussain, M. Jaulent and D. Kalra. (2015). Cross border semantic interoperability for learning health systems: The EHR4CR semantic resources and services. Learning Health Systems, e10014.
160. Eisenhardt, K. (1989). Building theories from case study research. Academy of Management Review, 14(4).
161. Lee, A. (1989). A scientific method for MIS case studies. MIS Quarterly, March 1989, 33-50.
162. Trellis, W. (1987). Application of a case study methodology. The Qualitative Report, 3(3).
163. Cable, G. (1994). Integrating case study and survey research methods: An example in information systems. European Journal of Information Systems, 3(2).
164. Yin, R. (2011). Applications of Case Study Research: SAGE.
165. Benbasat, I., D. Goldstein and M. Mead. (1987). The case research strategy in studies of Information Systems. MIS Quarterly, 11(3), 369-386.

166. Baxter, P. and S. Jack. (2008). Qualitative case study methodology: Study design and implementation for novice researchers. The Qualitative Report, 13(4).
167. Daley, B., G. Hitman, E. Fenton and S. McLachlan. (2019). Assessment of the Methodological Quality of Local Clinical Practice Guidelines on the Identification and Management of Gestational Diabetes Manuscript accepted for publication in BMJ Open.
168. Bero, L., R. Grilli, J. Grimshaw, E. Harvey, A. Oxman and M. Thomson. (1998). Closing the gap between research and practice: an overview of systematic reviews of interventions to promote the implementation of research findings. The Cochrane Effective Practice and Organization of Care Review Group. BMJ (Clinical Research Ed.), 317(7156), 465-468.
169. Cook, J., D. Nuccitelli, S. Green, M. Richardson, B. Winkler, R. Painting, R. Way, P. Jacobs, and A. Skuce. (2013). Quantifying the consensus on anthropogenic global warming in the scientific literature. Environmental Research Letters, 8(2).
170. Ganter, B. and R. Willie. (1997). Applied lattice theory: Formal Concept Analysis. In G. Gratzer (Ed.), General Latice Theory. Birkhauser.
171. Stumme, G. (2009). Formal Concept Analysis. In (Ed.), Handbook on Ontologies. Berlin: Springer.
172. Stumme, G., R. Taouil, Y. Bastide, N. Pasquier and L. Lakhal. (2002). Computing iceberg concept lattices with TITANIC. Data & knowledge engineering, 42(2), 189-222.
173. Ammenwerth, E., S. Graber, G. Herrmann, T. Burkle and J. Konig. (2003). Evaluation of health information systems—problems and challenges. International journal of medical informatics, 71(2), 125-135.
174. Kitson, A., J. Rycroft-Malone, G. Harvey, B. McCormack, K. Seers and A. Titchen. (2008). Evaluating the successful implementation of evidence into practice using the PARiHS framework: theoretical and practical challenges. Implementation Science, 3(1).
175. Pagliari, C. (2007). Design and evaluation in eHealth: challenges and implications for an interdisciplinary field. Journal of medical Internet research, 9(2), e15.
176. Pucihar, A., K. Bogataj and M. Wimmer. (2007). Gap Analysis Methodology for Identifying Future Ict Related eGovernment Research Topics-Case of" Ontology and Semantic Web" in the Context of eGovernment. Paper presented at the BLED 2007.
177. Elkadi, H. (2013). Success and failure factors for e-government projects: A case from Egypt. Egyptian Infomatics Journal, 14(2), 165-173.
178. Heeks, D., D. Mundy and A. Salazar. (1999). Why Healthcare Information Systems Succeed or Fail. Retrieved from: Manchester, UK.
179. Syamsuddin, I. (2016). Novel gap analysis framework for cloud health information systems. Journal of Theoretical and Applied Information Technology, 87(3), 415-421.
180. Rugchatjaroen, K. (2015). Success of Electronic Government Project in Bangkok Metropolis: An ITPOSMO approach. International Journal of Social Science and Humanity, 5(9), 783-787.
181. Maarten, H. (2016). A modern Socrates discourse in a local e-government setting. Future of E-government: learning from the Past, 4(3), 1-12.

182. Comreid, L. (1996). Cost analysis: Initiation of HBMC and First CareMap. Nursing Economics, 14(1), 34-39.
183. Wall, D. and M. Proyect. (1997). Critical Pathway Implementation Guide. Chicago, USA: Precept Press.
184. Donabedian, A. (1988). The quality of care: How can it be assessed? Journal of the Americal Medical Association, 260(12), 1743-1748.
185. Rubin, H., P. Pronovost and G. Diette. (2001). The advantages and disadvantages of process-based measures of health care quality. International Journal for Quality in Health Care, 13(6), 469-474.
186. Coates, E., G. Fuller, I. Wrench, M. Wilson, T. Stephens and D. Hind. (2016). A quality improvement clinical pathway for enhanced recovery after elective caesarean section: Results of a consensus exercise and survey. Faculty of Medicine, Dentistry and Health. University of Sheffield.
187. Deneckere, S., M. Euwema, P. Van Herck, C. Lodewijckx, M. Panella, W. Seremus and K. Vanhaecht. (2012). Care pathways lead to better teamwork: Results of a systematic review. Social science & medicine, 75(2), 264-268.
188. Jun, G., J. Ward, Z. Morris and J. Clarckson. (2009). Health care process modelling: Which method when? International Journal for Quality in Health Care, 21(3), 214-224.
189. Neuhauser, D., L. Provost and B. Bergman. (2011). The meaning of variation to healthcare managers, clinical and health service researchers, and individual patients. BMJ Quality & Safety, 20, i36-i40.
190. Pappas, P., C. Kauffman, D. Andes, C. Clancy, K. Marr, L. Ostrosky-Zeichner, A. Reboli, M. Schuster, J. Vazquez, T. Walsh, and T. Zaoutis. (2015). Clinical practice guideline for the management of candidiasis: 2016 update by the Infectious Diseases Society of America. Clinical Infectious Diseases, 62(4), e1-e50.
191. Seidman, M., R. Gurgel, S. Lin, S. Schwartz, F. Baroody, J. Bonner and S. Ishman. (2015). Clinical practice guideline: Allergic rhinitis. Otolaryngology - Head and Neck Surgery, 152, s1-s43.
192. Beskow, L., C. Dombeck, C. Thompson, J. Watson-Ormond and K. Weinfurt. (2015). Informed consent for bio-banking: consensus-based guidelines for adequate comprehension. Genetics in Medicine, 17(3).
193. Hennaut, E., H. Duong, B. Chiodini, B. Adams, K. Lolin, S. Blumenthal and K. Ismaili. (2018). Prospective cohort study investigating the safety and efficacy of ambulatory treatment with oral cefuroxime-axetil in febrile children with urinary tract infection. Frontiers in Pediatrics, 6.
194. RHW. (2019). Local Operating Procedures: A-Z Links. Retrieved from https://www.seslhd.health.nsw.gov.au/royal-hospital-for-women/policies-and-publications
195. Holmes, J., P. Sokolove, W. Brant and N. Kuppermann. (2002). A clinical decision rule for identifying children with thoracic injuries after blunt torso trauma. Annals of Emergency Medicine, 39(5), 492-499.
196. Downey, L., F. Cluzeau, K. Chalkidou, M. Morton, R. Sadanandan and S. Bauhoff. (2017). Clinical pathways, claims data and measuring quality/measuring quality of care: The authors reply. Health Affairs, 36(2), 382a.
197. Handelsman, Y., Z. Bloomgarden, G. Grunberger, G. Umpierrez, R. Zimmerman, T. Bailey and J. Davidson. (2015). American Association of Endocrinologists and American College of Endocrinology: Clinical practice guidelines for developing a diabetes mellitus comprehensive care plan - 2015. Endocrine Practice, 21, 1-87.
198. Salmonson, H., G. Sjoberg and J. Brogren. (2018). The standard treatment protocol for paracetamol poisoning may be inadequate following overdose with modified release formulation: A pharmacokinetic and clinical analysis of 53 cases. Clinical Toxicology, 56(1), 63-68.
199. Campbell, H., R. Hotchkiss, N. Bradshaw and M. Porteous. (1998). Integrated Care Pathways. British Medical Journal (BMJ), 316(7125).
200. Holocek, R. and S. Sellards. (1997). Use of Detailed Clinical Pathway for Bone Marrow Transplant Patients. Journal of Pediatric Oncology Nursing, 14(4).
201. Li, W., K. Liu, H. Yang and C. Yu. (2014). Integrated clinical pathway management for medical quality improvement based on a semiotically inspired systems architecture. European Journal of Information Systems, 23(4), 400-471.
202. O'Neill, E. and N. Dluhy. (2000). Utility of structured caer approaches in education and clinical practice. Nursing Outlook, 48(3), 132-135.
203. Kehlet, H. (2011). Fast-track Surgery: An update on physiological care principles to enhance recovery. Langenbeck's Archives of Surgery, 396(5), 585-590.
204. Solsky, I., A. Edelstein, M. Brodman, R. Kaleya, M. Rosenblatt, C. Santana, D. Feldman, P. Kischak, D. Somerville, S. Mudiraj, and I. Leitman. (2016). Perioperative care map improves compliance with best practices for the morbidly obese. Surgery, 160(6), 1682-1688.
205. Thompson, G., E. deForest and R. Eccles. (2011). Ensuring diagnostic accuracy in pediatric emergency medicine. Clinical Pediatric Emergency Medicine, 12(2), 121-132.
206. Yazbeck, A. (2014). Reengineering of business functions of the hospital when implementing care pathways. (Ph.D), University of Ljubljana. Retrieved from
207. Bumgarner, S. and M. Evans. (1999). Clinical Care Map for the ambulatory lapropscopic cholecystectomy patient. Journal of PeriAnaesthesia Nursing 14(1), 12-16.
208. Cholock, L. (2001). Care Map for the Management of Asthma in a Pediatric Population in a Primary Health Care Setting. (Master of Nursing), University of Manitoba, Winnipeg, Manitoba. Retrieved from
209. Lorenzi, N., A. Kouroubali and D. Detmer. (2009). How to successfully select and implement electronic health records (EHR) in small ambulatory practice settings. BMC Medical Informatics & Decision Making, 9(15).
210. Tripathi, M. (2012). EHR evolution: policy and legislation forces changing the EHR. Journal of AHIMA, 83(10), 24-29.
211. Shulman, L., R. Miller, E. Ambinder, P. Yu and J. Cox. (2008). Principles of Safe Practice Using an Oncology EHR System for Chemotherapy Ordering, Preparation, and Administration: Part 1 of 2. Journal of Oncology Practice, 4(4).
212. Okun, S., M. Williams, J. Sensmeier, M. Troseth, D. Conrad, M. Sugrue, L. Johnson, J. Murphy, and B. Frink. (2011). Enhancing the visibility of Nusring through standardised nursing language in the Electronic Health Record. Paper presented at the Summer Institute in Nursing Informatics (SINI) 2011 Part IV, University of Maryland, Baltimore.
213. Blobel, B. (2010). Standardized and flexible health data management with an archetype driven EHR system (EHRflex). Paper presented at the Seamless

Care, Safe Care: The Challenges of Interoperability and Patient Safety in Health Care: Proceedings of the EFMI Special Topic Conference, Reykjavik, Iceland
214. Wang, L., M. Miller, M. Schmitt and F. Wen. (2013). Assessing readability formula differences with written health informatics materials: Application, results and recommendations. Research in Social and Administrative Pharmacy, 9(5), 503-516.
215. Ungan, M. (2006). Standardization through process documentation. Business Process Managment Journal, 12(2), 135-148.
216. Mann, R. and J. Williams. (2003). Standards in Medical Record Keeping. Clinical Medicine, 3, 329-332.
217. Meidani, Z., F. Sadoughi, M. Maleki, S. Tofighi and A. Marani. (2012). Organization's Quality Maturity as a Vehicle for EHR Success. Journal of Medical Systems, 36(3), 1229-1234.
218. Casey, J., B. Schwartz, W. Stewart and N. Adler. (2016). Using Electronic Health Records for population health research: A review of methods and applications. Annual Review of Public Health, 37.
219. Johnson, O.A., H.S. Fraser, J.C. Wyatt and J.D. Walley. (2014). Electronic health records in the UK and USA. The Lancet, 384(9947).
220. Scott, P., P. Curley, P. Williams, I. Linehan and S. Shaba. (2016). Measuring the operational impact of digitized hospital records: A mixed methods study. BMC Medical Informatics and Decision Making, 16(143).
221. Trotman, J., J. Trinh, Y. Kwan, J. Estell, J. Fletcher, K. Archer, K. Lee, K. Foo, J. Curnow, A. Bianchi, L. Wignall, E. Verner, R. Gasiorowski, E. Siedlecka, and I. Cunningham. (2017). Formalising multidisciplinary peer review: Developing a haematological malignancy-specific electronic proforma and standard operating procedure to facilitate procedural efficiency and evidence-based clinical practice. Internal Medicine Journal, 47(5).
222. Greenhalgh, T., K. Stramer, T. Bratan, E. Byrne, J. Russell and H. Potts. (2010). Adoption and non-adoption of a shared electronic summary record in England: A mixed-method case study. British Medical Journal, 340(c3111).
223. HOCCOPA. (2013). The dismantled National Programme for IT in the NHS Retrieved from: London, UK. https://publications.parliament.uk/pa/cm201314/cmselect/cmpubacc/294/294.pdf
224. Miller, A., B. Moon, S. Anders, R. Walden, S. Brown and D. Montella. (2015). Integrating computerised clinical decision support systems into clinical work: A meta-synthesis of qualitative research. International Journal of Medical Informatics, 84.
225. Miriovsky, B., L. Shulman and A. Abernethy. (2012). Importance of health information technology, electronic health records, and continuously agregating data to comparative effectiveness research and learning healthcare. Journal of Clinical Oncology, 30(34), 4243-4248.
226. Koppel, R. and D. Kreda. (2010). Healthcare IT Usability and Sustainability for Clinical Needs: Challenges of design, workflow, and contractual relations. Study of Health Technology and Informatics, 157, 7-14.
227. Greenhalgh, T., H. Potts, G. Wong, P. Bark and D. Swinglehurst. (2009). Tensions and paradoxes in Electronic Patient Record research: A systematic literature review using the meta-narrative method. The Millbank Quarterly, 87(4), 729-788.

228. Sheikh, A., T. Cornford, N. Barber, A. Avery, A. Takian, V. Lichtner, D. Petrakaki, S. Crowe, K. Marsden, and K. Cresswell. (2011). Implementation and adoption of nationwide electronic health records in secondary care in England: final qualitative results from prospective national evaluation in “early adopter” hospitals. British Medical Journal, 343.
229. Topaz, M., C. Ronquillo, L. Peltonen, L. Pruinelli, R. Sarmiento, M. Badger, S. Ali, A. Lewis, M. Georgsson, E. Jeon, J. Tayaben, D. Alhurwail, and Y. Lee. (2016). Nurse Informaticians Report Low Satisfaction and Multi-level Concerns with Electronic Health Records: Results from an International Survey. Paper presented at the AMIA Annual Symposium Proceedings.
230. Imison, C., S. Castle-Clarke, R. Watson and N. Edwards. (2016). Delivering the benefits of Digital Health Care. Retrieved from:
231. Liyanage, H., A. Correa, S. Liaw, C. Kuziemsky, A. Terry and S. de Lusignan. (2015). Does informatics enable or inhibit the delivery of patient-centred, coordinated, and quality-assured care: A delphi study. IMIA Yearbok of Medical Informatics, 10(1).
232. Protti, D. (2015). Missed connections: The adoption of Information Technology in Canadian healthcare. Retrieved from:
233. Tolar, M. and E. Balka. (2012). Caring for individual patients and beyond: Enhancing care through secondary use of data in a general practice setting. International Journal of Medical Informatics, 81.
234. Cherry, B., M. Carter, D. Owen and C. Lockhart. (2008). Factors affecting electronic health record adoption in long-term care facilities. Journal for Healthcare Quality, 30(2).
235. Houser, S. and L. Johnson. (2008). Perceptions regarding electronic health record implementation among health information management professionals in Alabama: A statewide survey and analysis. Perspectives in Health Information Management, 5(6).
236. Jawhari, B., D. Ludwick, L. Keenan, D. Zakus and R. Hayward. (2016). Benefits and challenges of EMR implementations in low resource settings: a state-of-the-art review. BMC Medical Informatics and Decision Making, 16(116).
237. Nguyen, L., E. Bellucci and L. Nguyen. (2014). Electronic health records implementation: An evaluation of information system impact and contingency factors. International Journal of Medical Informatics, 83.
238. Schumaker, R. and K. Reganti. (2014). Implementation of Electronic Health Record (EHR) system in the heathcare industry. International Journal of Privacy and Health Information Management, 2(2), 57-71.
239. Terry, A., C. Thorpe, G. Giles, J. Brown, S. Harris, G. Reid, A. Thind and M. Stewart. (2008). Implementing Electronic Health Records. Canadian Family Physician, 54.
240. Chao, W., H. Hu, C. Ung and Y. Cai. (2013). Benefits and challenges of electronic health record system on stakeholders: A qualitative study of outpatient physicians. Journal of Medical Systems, 37.
241. Jamoom, E., V. Patel, M. Furukawa and J. King. (2014). EHR adopters vs. non-adopters: Impacts of, barriers to, and federal initiatives for EHR adoption. Healthcare Amsterdam, 2(1).
242. Kaye, R., E. Kokia, V. Shalev, D. Idar and D. Chinitz. (2010). Barriers and success factors in health information technology: A practitioners perspective. Journal of Management and Marketing in Healthcare, 3(2).

243. Lluch, M. (2011). Healthcare professionals' organisational barriers to health information technologies: A literature review. International Journal of Medical Informatics, 80.
244. Bellack, J. (2016). Creating a continuously learning health system through technology: A call to action. Journal of Nursing Education, 55(1), 3-5.
245. Yoder-Wise, P. (2015). The continuously learning health system: Recommendations for the Josiah Macy Jr. Foundation. Journal of Continuing Education in Nursing, 46(9), 379-380.
246. Bernstein, J., C. Friedman, P. Jacobsen and J. Rubin. (2015). Ensuring public health's future in a national-scale learning health system. American Journal of Preventative Medicine, 48(4), 480-487.
247. Brody, H. and F. Miller. (2013). The Research-Clinical Practice Distinction, Learning Health System and Relationships. Hastings Centre Report, 43(5), 41-47.
248. Faden, R., N. Kass, S. Goodman, P. Pronovost, S. Tunis and T. Beuchamp. (2013). An ethics framework for a Learning Health Care System: A departure from traditional research ethics and clinical ethics. Retrieved from:
249. Forrest, C., P. Margolis, M. Seid and R. Colletti. (2014). PEDSnet: How a prototype pediatric learning health system is being expanded into a national network. Health Affairs, 33(7), 1171-1177.
250. Codman, E., W. Mayo and J. Clark. (1913). Standardization of Hospitals: Report of the Committee Appointed by the Clinical Congress of Surgeons of North America. Retrieved from:
251. Nightingale, F. (1863). Notes on Hospitals, 3rd Ed. London: Longman, Roberts & Green.
252. Flores, M., G. Glusman, K. Brogaard, N. Price and L. Hood. (2013). P4 Medicine: How systems medicine will transform the healthcare sector and society. Personalised Medicine, 10(6), 565-576.
253. Horne, R., J. Bell, J. Montgomery, M. Ravn and J. Tooke. (2015). A new social contract for medical innovation. The Lancet, 385(9974), 1153-1154.
254. Cano, I., M. Lluch-Ariet, D. Gomez-Cabrero, D. Maier, S. LKalko, M. Cascante, J. Tegner, F. Miralles, D. Herrera, and J. Roca. (2014). Biomedical research in a Digital Health Framework. Journal of Translational Medicine, 12(Suppl 2).
255. Chen-Shie, H. and Y. Chang. (2016). Incremental Learning for Alzheimer's Disease on Medical Cloud Service Environment. DEStech Transactions on Materials Science and Engineering.
256. Miralles, F., D. Gomez-Cabrero, M. Lluch-Ariet, J. Tegner, M. Cascante and J. Roca. (2014). Predictive medicine: outcomes, challenges and opportunities in the Synergy-COPD project. Journal of Translational Medicine, 12(2).
257. Deeny, S. and A. Steventon. (2015). Making sense of the shadows: Priorities for creating a learning healthcare system based on routinely collected data. BMJ Quality Safety, 24, 505-515.
258. Everson, J. (2016). The implications and impact of 3 approaches to health information exchange: Community, enterprise, and vendor-mediated health information exchange. Learning Health Systems, e10021, 1-9.
259. Etheredge, L. (2007). A rapid-learning health system. Health Affairs, 26(2), 107-118.
260. Curcin, V. (2017). Embedding data provenance into the Learning Health System to facilitate reproducible research. Learning Health Systems, e10019.

261. Faden, R., N. Kass, D. Whicher, W. Stewart and S. Tunis. (2013). Ethics and Informed Consent for Comparative Effectiveness Research with prospective electronic clinical data. Medical Care, 51(8).
262. Ludwick, D. and J. Doucette. (2009). Adopting electronic medical records in primary care: Lessons learned from health information systems implementation experience in seven countries. International Journal of Medical Informatics, 78.
263. Tang, P., J. Ash, D. Bates, M. Overhage and D. Sands. (2006). Personal health records: definitions, benefits and strategies for overcoming barriers to adoption. Journal of the American Medical Informatics Association, 13(2).
264. McGinn, C., S. Genier, J. Duplantie, N. Shaw, C. Sicotte, L. Mathieu, Y. Leduc, F. Legare, and M. Gagnon. (2011). Comparison of user groups' perspectives of barriers and facilitators to implementing electronic health records: A systematic review. BMC Medicine, 9(46).
265. Greene, S., R. Reid and E. Larson. (2012). Implementing the learning health system: From concept to action. Annals of Internal Medicine, 157, 207-210.
266. Jenicek, M. (1997). Epidemiology, Evidence-Based Medicine and Evidence-Based Public Health. Journal of Eidemiology, 7(4), 187-197.
267. Haynes, B. and A. Haines. (1998). Barriers and bridges to evidence based clinical practice. British Medical Journal, 317, 273-276.
268. Lowes, L., G. Noritz, A. Newmeyer, P. Embi, H. Yin and W. Smoyer. (2016). 'Learn From Every Patient': Implementation and early results of a Learning Health System. Developmental Medicine and Child Neurology, 59, 183-191.
269. McLachlan, S., K. Dube and T. Gallagher. (2016). Using CareMaps and health statistics for generating the realistic synthetic Electronic Healthcare Record. Paper presented at the International Conference on Healthcare Informatics (ICHI'16), Chicago, USA.
270. Pletcher, M., Lo, B., & Grady, D. (2014). Informed consent in randomised quality improvement trials: A clinical barrier for learning health systems. Journal of the American Medical Association (JAMA), 174(5), 668-670.
271. Schneiderman, B., C. Plaisant and B. Hesse. (2013). Improving healthcare with interactive visualization. Computer 46(5), 58-66.
272. Victores, A., K. Coggins and M. Takashima. (2015). Electronic health records and resident workflow: A time-motion study of otolaryngology residents. The Laryngoscope 125(3), 594-598.
273. Andreu-Perez, J., D. Leff, H. Ip and G. Yang. (2015). From wearable sensors to smart implants—toward pervasive and personalized healthcare. EEE Transactions on Biomedical Engineering, 62(12), 2750-2762.
274. McLachlan, S., K. Dube, D. Buchanan, S. Lean, O. Johnson, H.H. Potts, T. Gallagher, W. Marsh, and N. Fenton. (2018). Learning Health Systems: The Research Community Awareness Challenge. BCS Journal of Innovation in Health Informatics, 25(1).
275. IEEE. (2017, 23-26 Aug). Proceedings of the 5th International Conference on Healthcare Informatics. Paper presented at the, Park City, Utah, USA.
276. Fung-Kee-Fung, M., D. Maziak, J. Pantarotto, J. Smylie, L. Taylor, T. Timlin, T. Cacciotti and S. Madore. (2018). Regional process redesign of lung cancer care: a learning health system pilot project. Current Oncology, 25(1).
277. Ling, S. and P. McGann. (2018). Changing Healthcare Service Delivery to Improve Health Outcomes For Older Adults: Opportunities Not to Be Missed. Journal of the American Geriatrics Society, 66(2), 235-238.

278. Agrawal, P. and J. Kosowsky. (2009). Clinical practice guidelines in the emergency department. Emergency Medicine Clinics, 27(4), 555-567.
279. Koon, A., B. Hawkins and S. Mayhew. (2016). Framing and the health policy process: a scoping review. Health policy planning, 31(6), 801-816.
280. Allsop, J. (2016). Health policy and the NHS: towards 2000: Routledge.
281. De Vos, P. and P. Van der Stuyft. (2015). Sociopolitical determinants of international health policy. International Journal of Health Services, 45(2), 363-377.
282. Gimbel, S., P. Kohler, P. Mitchell and A. Emami. (2017). Creating academic structures to promote nursing's role in global health policy. International nursing review, 64(1), 117-125.
283. Storm, I. (2016). Towards a HiAP cycle: Health in All Policies as a practice-based improvement process. (PhD), University of Amsterdam. Retrieved from
284. Benita, C., S. Katherine, K. Anita, J. Marlene, L. Suzanne, M. Kathy, W. Caroline, A. Deborah, R. Jordan, and S. Rosana. (2018). Original qualitative research-Indicators to guide health equity work in local public health agencies: a locally driven collaborative project in Ontario. Health Promotion Chronic Disease Prevention in Canada, 38(7-8), 277.
285. Nomura, S., H. Sakamoto, S. Glenn, Y. Tsugawa, S. Abe, M. Rahman, J. Brown, S. Ezoe, C. Fitzmaurice, and T. Inokuchi. (2017). Population health and regional variations of disease burden in Japan, 1990–2015: a systematic subnational analysis for the Global Burden of Disease Study 2015. The Lancet, 390(10101), 1521-1538.
286. Starfield, B. and L. Shi. (2002). Policy relevant determinants of health: an international perspective. Health policy, 60(3), 201-218.
287. Ducharme, J. (2005). Clinical guidelines and policies: can they improve emergency department pain management? The Journal of Law, Medicine and Ethics, 33(4), 783-790.
288. LeFevre, M. (2017). From authority- to evidence-based medicine: Are clinical practice guidelines moving us forwards or backwards? Annals of Family Medicine, 15(5).
289. Matz, P., R. Meagher, T. Lamer, W. Tontz Jr, T. Annaswamy, R. Cassidy, C. Cho, P. Dougherty, J. Easa, and D. Enix. (2016). Guideline summary review: an evidence-based clinical guideline for the diagnosis and treatment of degenerative lumbar spondylolisthesis. The Spine Journal, 16(3), 439-448.
290. Rodrigues, G., H. Choy, J. Bradley, K. Rosenzweig, J. Bogart, W. Curran Jr, E. Gore, C. Langer, A. Louie, and S. Lutz. (2015). Definitive radiation therapy in locally advanced non-small cell lung cancer: Executive summary of an American Society for Radiation Oncology (ASTRO) evidence-based clinical practice guideline. Practical radiation oncology, 5(3), 141-148.
291. Shekelle, P., M. Aronson and J. Melin. (2016). Overview of clinical practice guidelines. UpToDate, 28.
292. Bernis, L. (2007). Obstetric fistula: guiding principles for clinical management and programme development, a new WHO guideline. International Journal of Gynecology & Obstetrics, 99(S1).
293. World Health Organisation. (2013). Diagnostic criteria and classification of hyperglycaemia first detected in pregnancy. Retrieved from: http://www.who.int/diabetes/publications/Hyperglycaemia_In_Pregnancy/en/

294. Srivastava, A., A. Brewer, E. Mauser-Bunschoten, N. Key, S. Kitchen, A. Llinas, C. Ludlam, J. Mahlangu, K. Mulder, and M. Poon. (2013). Guidelines for the management of hemophilia. Haemophilia, 19(1), e1-e47.
295. ESC. (2018). European Society of Cardiology: Guidelines for the diagnosis and management of syncope. Eur Heart J., 39(21), 1883-1948. doi:https://doi.org/10./eurheartj/ehy037
296. DHE. (2016). About National Clinical Guidelines. Danish Health Authority. Retrieved from https://www.sst.dk/en/national-clinical-guidelines/about-national-clinical-guidelines
297. NCEC. (2014). Framework for Endorsement of National Clinical Guidelines July 2014. Retrieved from: Dublin, Ireland.
298. Woolf, S., R. Grol, A. Hutchinson, M. Eccles and J. Grimshaw. (1999). Potential benefits, limitations, and harms of clinical guidelines. BMJ, 318(7182), 527-530.
299. Bosse, G., J. Breuer and C. Spies. (2006). The resistance to changing guidelines–what are the challenges and how to meet them. Best Practice Research Clinical Anaesthesiology, 20(3), 379-395.
300. Dodek, P., N. Cahill and D. Heyland. (2010). The relationship between organizational culture and implementation of clinical practice guidelines: a narrative review. Journal of Parenteral Enteral Nutrition, 34(6), 669-674.
301. Harris, L. (2009). Tort reform as carrot-and-stick. Harv. J. on Legis., 46, 163.
302. Cheng, D. (1998). Fast track cardiac surgery pathways early extubation, process of care, and cost containment. Anesthesiology: The Journal of the American Society of Anesthesiologists, 88(6), 1429-1433.
303. Stephen, A. and D. Berger. (2003). Shortened length of stay and hospital cost reduction with implementation of an accelerated clinical care pathway after elective colon resection. Surgery, 133(3), 277-282.
304. De Bleser, L., R. Depreitere, K. Waele, K. Vanhaecht, J. Vlayen and W. Sermeus. (2006). Defining pathways. Journal of nursing management, 14(7), 553-563.
305. Goode, C. (1995). Evaluation of patient and staff outcomes with hospital-based managed care. (PhD), University of Iowa. Retrieved from
306. Van Herck, P., K. Vanhaecht and W. Sermeus. (2004). Effects of clinical pathways: do they work? Journal of Integrated Care Pathways, 8(3), 95-105.
307. Dagliati, A., L. Sacchi, A. Zambelli, V. Tibollo, L. Pavesi, J. Holmes and R. Bellazzi. (2017). Temporal electronic phenotyping by mining careflows of breast cancer patients. Journal of biomedical informatics, 66, 136-147.
308. Hydo, B. (1995). Designing an effective clinical pathway for stroke. American Journal of Nursing, 95(3), 44-50.
309. Jones, A. (2003). Perceptions on the development of a care pathway for people diagnosed with schizophrenia on acute psychiatric units. Journal of psychiatric mental health nursing, 10(6), 669-677.
310. Huang, B., P. Zhu and C. Wu. (2012). Customer-centered Careflow Modeling Based on Guidelines. Journal of Medical Systems, 36.
311. Marr, J. and B. Reid. (1992). Implementing managed care and case management: The neuroscience experience. Journal of Neuroscience Nursing, 24(5), 281-285.
312. Ogilvie-Harris, D., D. Botsford and R. Hawker. (1993). Elderly patients with hip fractures: Improving outcomes with the use of care maps with high-quality medical and nursing protocols. Journal of Orthopaedic Trauma, 7(5), 428-437.

313. Wilson, D. (1995). Effect of managed care on selected outcomes of hospitalised surgical patients. (Master of Nursing), University of Alberta, Alberta, Canada. Retrieved from
314. Hampton, D. (1993). Implementing a managed care framework through care maps. Journal of Nursing Administration, 23(5), 21-27.
315. Zander, K. (2002). Integrated care pathways: Eleven international trends. Journal of Integrated Care Pathways, 6, 101-107.
316. Blegen, M., R. Reiter, C. Goode and R. Murphy. (1995). Outcomes of hospital-based managed care: A multivariate analysis of cost and quality. Managed Care: Obstetrics and Gynaecology, 86(5), 809-814.
317. Houltram, B. and M. Scanlan. (2004). Care Maps: Atypical Antipsychotics. Nursing Standard, 18(36), 42-44.
318. Gordon, D. (1996). Critical Pathways: A Road to Institutionalizing Pain Management. Journal of Pain and Symptom Management, 11(4).
319. Hill, M. (1998). The Development of Care Management Systems to Achieve Clinical Integration. Advanced Practice Nursing Quarterly, 4(1).
320. Dickinson, C., M. Noud, R. Triggs, L. Turner and S. Wilson. (2000). The Antenatal Ward Care Delivery Map: a team model approach. Australian Health Review, 23(3), 68-77.
321. Philie, P. (2001). Management of blood-borne fluid exposures with a rapid treatment prophylactic caremap: One hospital's four-year experience. Journal of Emergency Nursing, 27(5), 440-449.
322. Griffith, D., D. Hampton, M. Switzer and J. Daniels. (1996). Facilitating the recovery of open-heart surgery patients through quality improvement efforts and CareMAP implementation. American Journal of Critical Care, 5(5), 346-352.
323. Saint-Jacques, H., V. Burroughs, J. Watkowska, M. Valcarcel, P. Moreno and M. Maw. (2005). Acute coronary syndrome critical pathway: Chest PAIN caremap. A qualitative research study provider-level intervention. Critical Pathways in Cardiology, 4(3), 145-604.
324. Mackay, D., M. Myles, C. Spooner, H. Lari, L. Tyler, S. Blitz, A. Senthilselvan and B. Rowe. (2007). Changing the process of care and practice in acute asthma in the emergency department: Experience with an asthma care map in a regional hospital. Canadian Journal of Emergency Medicine 9(5), 353-365.
325. Royall, D., P. Brauer, L. Bjorklund, O. O'Young, A. Tremblay, K. Jeejeebhoy, D. Heyland, R. Dhaliwah, D. Klein, and D. Mutch. (2014). Development of a dietary management care map for metabolic syndrome. Canadian Journal of Dietetic Practice and Research, 75(3), 132-139.
326. Jones, A. and I. Norman. (1998). Managed mental health care: Problems and possibilities. Journal of Psychiatric and Mental Health Nursing, 5, 21-31.
327. Schwoebel, A. (1998). Care Mapping: A common sense approach. Indian Journal of Pediatrics, 65(2), 257-264.
328. Williams, K., R. Ford, I. Bishop, D. Loiterton and J. Hickey. (2007). Realism and selectivity in data-driven visualisations: A process for developing viewer-oriented landscape surrogates. Landscape and Urban Planning, 81.
329. Feigin, J. (1996). Medical Care Management. Allergy and Asthma Proc., 17(6), 359-361.
330. Goode, C. and M. Blegen. (1993). Developing a caremap for patients with a cesaerian birth: A multidisciplinary process. Journal of Perinatal and Neonatal Nursing, 7(2), 40-49.

331. DeJesse, P., C. Bland, O. Fuller and J. Macbride. (1995). Managed care: A view from the inside. Medical Marketing and Media, 5.
332. Chan, E., J. Russell, W. Williams, G. Van Arsdell, J. Coles and B. McCrindle. (2005). Postoperative chylothorax after cardiothoracic surgery in children. The Annals of Thoracic Surgery, 80(5), 1864-1870.
333. Reading, S., A. Snaghi, B. Huda, J. Saravanamutha, G. Toms, F. Braggins, B. Dawlatly, R. Davison, M. Hannon, R. Lu, and D. McEneaney. (2015). Maternity Services: Diabetes - Pregnancy, Labour and Peurperium. Barts Health NHS Trust. London, UK.
334. RWH. (2010). Bladder Management - Intrapartum and Postpartum. The Royal Womens Hospital. Melbourne, AUS.
335. TCHaW. (2010). Abdominal wall defects in neonates: Initial, pre and post-operative management Practice Guideline. The Children's Hospital at Westmead. Westmead, NSW, AUS.
336. Thompson, D., H. Berger, D. Feig, R. Gagnon, T. Kader, E. Keely, S. Kozak, E. Ryan, M. Sermer, and C. Vinokuroff. (2013). Clinical Practice Guidelines: Diabetes and Pregnancy. Canadian Journal of Diabetes, 37, S168-S183.
337. deForest, E. and G. Thompson. (2012). Advanced nursing directives: Integrating validated clinical scoring systems into nursing care in the pediatric emergency department. Nursing Research and Practice.
338. Sackman, J. and L. Citrin. (2014). Cracking down on the cost outliers. Healthcare Finance Management, 68(3), 58-63.
339. Ebell, M. (2010). AHRQ White Paper: use of clinical decision rules for point-of-care decision support. Medical Decision Making, 30(6), 712-721.
340. Lockie, F., S. Dalton, E. Oakley and F. Babl. (2013). Triggers for head computed tomography following paediatric head injury: comparison of physicians' reported practice and clinical decision rules. Emergency Medicine Australasia, 25(1), 75-82.
341. Wang, H., T. Zhou, Y. Zhang and J. Li. (2015). Research and Development of Semantics-based Sharable Clinical Pathway Systems. Journal of Medical Systems, 2015(39).
342. McGinn, T., G. Guyatt, P. Wyer, C. Naylor, I. Stiell and W. Richardson. (2000). Users' guides to the medical literature: XXII: how to use articles about clinical decision rules. JAMA, 284(1), 79-84.
343. Goode, C. (1993). Evaluation of patient and staff outcomes with hospital based managed care. (PhD), University of Iowa. Retrieved from
344. Tastan, S., S. Hatipoglu, E. Iyigun and S. Kilic. (2012). Implementation of a clinical pathway in breast cancer patients undergoing breast surgery. European Journal of Oncology Nursing, 16(4), 368-374.
345. Morreale, M. (1997). Evaluation for a care-map for community acquired pneumonia hospital patients. (Masters), Queens University, BC, Canada. Retrieved from
346. Sternberg, S. (2007). Influence of stroke clinical pathway on documentation. (Master of Science in Nursing), Clemson University. Retrieved from
347. Nielsen, J., M. Sally, R. Mullins, M. Slater, T. Groat, X. Gao, J. de la Cruz, M. Ellis, M. Schreiber, and D. Malinoski. (2017). Bicarbonate and mannitol treatment for traumatic rhabdomyolysis revisited. The American Journal of Surgery, 213(1), 73-79.

348. So, L., A. Lau, L. Yam, T. Cheung, E. Poon, R. Yung and K. Yuen. (2003). Development of a standard treatment protocol for severe acute respiratory syndrome. The Lancet, 361(9369), 1615-1617.
349. Sunde, K., M. Pytte, D. Jacobsen, A. Mangschau, L. Jensen, C. Smedsrud, T. Draegni and P. Steen. (2007). Implementation of a standardised treatment protocol for post resuscitation care after out-of-hospital cardiac arrest. Resuscitation, 73(1), 29-39.
350. Miñambres, E., J. Pérez-Villares, M. Chico-Fernández, A. Zabalegui, J. Dueñas-Jurado, M. Misis, F. Mosteiro, G. Rodriguez-Caravaca, and E. Coll. (2015). Lung donor treatment protocol in brain dead-donors: a multicenter study. The Journal of Heart Lung Transplantation, 34(6), 773-780.
351. Government., U. (2018). Health Matters: Preventing Type 2 Diabetes. Retrieved from: London, UK. https://www.gov.uk/government/publications/health-matters-preventing-type-2-diabetes/health-matters-preventing-type-2-diabetes
352. NICE. (2018). National Strategy and Policy to prevent Type 2 Diabetes. Retrieved from: https://pathways.nice.org.uk/pathways/preventing-type-2-diabetes/national-strategy-and-policy-to-prevent-type-2-diabetes
353. NHS. (2014). NHS England: Action for Diabetes. Retrieved from: https://www.england.nhs.uk/rightcare/wp-content/uploads/sites/40/2016/08/act-for-diabetes-31-01.pdf
354. Abhilash, M., M. Paul, M. Varghese and R. Nair. (2011). Effect of long term intake of aspartame on antioxidant defense status in liver. Food and Chemical Toxicology, 49(6), 1203-1207.
355. Kleine, L., J. Whitfield and A. Boynton. (1986). The glucocorticoiddexamethasone and the tumour-promoting artificial sweetener saccharin stimulate protein kinase C from T51B rat liver cells. Biochemical and biophysical research communications, 135(1), 33-40.
356. Stepien, M., T. D'uarte-Salles, V. Fedirko, A. Trichopoulou, P. Lagiou, C. Bamia and G. Fagherazzi. (2016). Consumption of soft drinks and juices and risk of liver and biliary tract cancers in European cohort. European Journal of Nutrition, 55(1), 7-20.
357. IDF. (2017). IDF Clinical Practice Recommendations for Managing Type 2 Diabetes in Primary Care. Retrieved from: https://www.idf.org/e-library/guidelines/128-idf-clinical-practice-recommendations-for-managing-type-2-diabetes-in-primary-care.html
358. Davies, M., D. D'Alessio, J. Fradkin, W. Kernan, C. Mathieu, G. Mingrone and J. Buse. (2018). Management of hyperglycaemia in type 2 diabetes, 2018. A consensus report by the American Diabetes Association (ADA) and the European Association for the Study of Diabetes (EASD). Diabetologia, 61(12), 2461-2498.
359. NICE. (2017). Type 2 Diabetes in adults: Management. Retrieved from: https://www.nice.org.uk/guidance/ng28
360. SIGN. (2017). SIGN 154: Pharmacological management of glycemic control in people with Type 2 Diabetes. Retrieved from: https://www.sign.ac.uk/assets/sign154.pdf
361. BHT. (2015). Barts Health NHS Trust: Diabetes – Pregnancy, Labour and Puerperium. Retrieved from:
362. QMUL-CEG. (2017). Type 2 Diabetes: Reducing hypoglycaemia. Retrieved from:

https://www.qmul.ac.uk/blizard/ceg/media/blizard/images/documents/ceg-documents/Type-2-diabetes-Reducing-hypoglycaemia,-Jan-2017.pdf

363. NHS. (2016). NHS London Clinical Networks – Clinical Management: Optimal Pathway. How to deliver Type 1 diabetes services. Retrieved from: http://www.londonscn.nhs.uk/wp-content/uploads/2016/09/dia-t1-commissioning-pk-bk2-092016.pdf

364. HUH. (2014). Homerton University Hospital: New Patient Clinic Pathway. Retrieved from: Homerton, UK. http://www.homerton.nhs.uk/our-services/services-a-z/d/diabetes/clinics/new-patient-clinic-pathway/

365. NHS. (2016). NHS Mid Essex: My Personal Diabetes Health Care Plan and Record. Retrieved from: Essex, UK. https://midessexccg.nhs.uk/livewell/information-guides/diabetes/1525-diabetes-handbook/file

366. NHS. (2015). NHS Dudley: My personal diabetes handheld record and care plan. Retrieved from: Dudley, UK. http://www.mysurgerywebsite.co.uk/website/M83041/files/Diabetes-My-Personal-Diabetes-Handheld-Record-and-Care-Plan.pdf

367. PamBayesian. (2018). The PamBayesian Project. Retrieved from http://www.pambayesian.org

368. Pambayesian. (2018). Gestational Diabetes Booking Visit Caremap. Retrieved from www.pambayesian.org/wp-content/uploads/2019/05/GDM-Booking-Visit.pdf

369. Pambayesian. (2019). Gestational Diabetes Diagnostic Decisions Caremap. Retrieved from www.pambayesian.org/wp-content/uploads/2019/05/GDM-Diagnostic-Decisions.pdf

370. Pambayesian. (2019). Gestational Diabetes Management Decisions Caremap. Retrieved from www.pambayesian.org/wp-content/uploads/2019/05/GDM-Management-Decisions.pdf

371. Bilici, E., G. Despotou and T. Arvanitis. (2018). The use of computer-interpretable clinical guidelines to manage care complexities of patients with multimorbid conditions: A review. Digital Health, 4. doi:10.1177/2055207618804927

372. De Clercq, P., K. Kaiser and A. Hasman. (2008). Computer-Interpretable Guideline formalisms. Studies in health technology and informatics, 139, 22-43.

373. McLachlan, S. (2019). The largest single barrier to Learning Health Systems. A presentation at the 2nd Annual Pambayesian Workshop Event. Queen Mary University of London. London, UK.

374. Johnson, L., K. Edward and J. Giandinoto. (2018). A systematic literature review of accuracy in nursing care plans and using standardised nursing language. Collegian, 25(3), 355-361.

375. Finch, C., G. Valuri and J. Ozanne-Smith. (1999). Injury surveillance during medical coverage of sporting events-development and testing of a standardised data collection form. Journal of Science and Medicine in Sport, 2(1), 42-56.

376. Guerlain, S., K. LeBeau, M. Thompson, C. Donnelly, H. McClelland, S. Syverud and J. Calland. (2001). The effect of a standardized data collection form on the examination and diagnosis of patients with abdominal pain. Paper presented at the The Human Factors and Ergonomics Society Annual Meeting Los Angeles, California.

377. Pedersen, A., E. Mikkelsen, D. Cronin-Fenton, N. Kristensen, T. Pham, L. Pedersen and I. Petersen. (2017). Missing data and multiple imputation in clinical epidemiological research. Clinical Epidemiology, 9, 157.
378. Buchert, A. and G. Butler. (2016). Clinical Pathways: Driving high-reliability and high-value care. Pediatric Clinics of North America, 63, 317-328.
379. Kinsman, L., T. Rotter, E. James, P. Snow and J. Willis. (2010). What is a Clinical Pathway? Development of a definition to inform the Debate. BMC Medicine, 8(31).
380. Brennan, P., M. Brands, L. Caldwell, F. Fonseca, N. Turley, S. Foley and S. Rahimi. (2018). Surgical specimin handover from the operating theatre to laboratory: Can we improve patient safety by learning from aviation and other high-risk organisations? Journal of Oral Pathology & Medicine, 47(2), 117-120.
381. Gu, X., H. Liu and K. Itoh. (2018). Inter-department patient handoff quality and its contributing factors in chinese hospitals. Cognition, Technology and Work, 1(11).
382. Kripalani, S., F. LeFevre, C. Phillips, M. Williams, P. Basaviah and D. Baker. (2007). Deficits in communication and information transfer between hospital-based and primary care physicians: Implications for patient safety and continuity of care. Journal of the Americal Medical Association, 297(8), 831-841.
383. Leape, I. (2018). The preventability of medical injury. In M. Bogner (Ed.), Human Error in Medicine. CRC Press.
384. Naveh, E., T. Katz-Navon and Z. Stern. (2006). Readiness to report medical treatment errors: The effects of safety procedures, safety information and priority of safety. Medical Care, 44(2), 117-123.
385. Richter, J., A. McAlearny and M. Pennell. (2015). Evaluating the effect of safety culture on error reporting: A comparison of managerial and staff perspectives. American Journal of Medical Quality, 30(6), 550-558.
386. Yelland, I., B. Kahan, E. Dent, K. Lee, M. Voysey, A. Forbes and J. Cook. (2018). Prevalence and reporting of recruitment, randomisation and treatment errors in clinical trials: A systematic review. Clinical Trials, 15(3), 278-285.
387. McLachlan, S., K. Dube, O. Johnson, D. Buchanan, H. Potts, T. Gallagher and N. Fenton. (2019). A framework for analysing learning health systems: Are we removing the most impactful barriers? Learning Health Systems. doi:10.1002/lrh2.10189
388. Gold, M., M. Hossain and A. Mangum. (2015). Consumer engagement in Health IT: Distinguishing rhetoric from reality. eGEMS, 3(1).
389. Koppel, R., J. Metlay, A. Cohen, B. Abaluck, A. Localio, S. Kimmel and B. Strom. (2005). Role of Computerized Physician Order Entry Systems in facilitating Medication Errors. Journal of the American Medical Association (JAMA), 293(10).
390. Menachemi, N. and T. Collum. (2011). Benefits and drawbacks of electronic health record systems. Risk Management and Healthcare Policy, 4, 47-55.
391. Corrigan, D., Munnelly, G., Kazienko, P., Kajdanowicz, T., Soler, J., Mahmoud, S., Porat, T., Kostopoulou, O., Curcin, V., & Delaney, B. (2017). Requirements and validation of a prototype learning health system for clinical diagnosis. Learning Health Systems, e10026.
392. Davidson, A. (2015). Creating Value: Unifying silos into public health business intelligence. eGEMS, 2(4).

393. Furukawa, M., J. King, V. Patel, C. Hsiao, J. Adler-Milstein and A. Jha. (2014). Despite substantial progress in EHR adoption, Health Information Exchange and patient engagement remain low in office settings. Health Affairs, 33(9), 1672-1679.
394. Khurshid, A. (2017). A tale of two cities: Developing health information platforms for a learning health system in Austin and in New Orleans. Learning Health Systems, e10017.
395. Mason, A. and A. Barton. (2013). The emergence of a Learning Healthcare System. Clinical Nurse Specialist, 27(1).
396. Morain, S., N. Kass and C. Grossman. (2016). What alows a health care system to become a learning health cae system: Results from interviews with health system leaders. Learning Health Systems, 1(1).
397. Weng, C., Y. Li, S. Berhe, M. Boland, J. Gao, G. Hruby, R. Steinman, C. Lopez-Jiminez, L. Busacca, G. Hripcsak, S. Bakken, and J. Bigger. (2013). An integrated model for patient care and clinical trials (IMPACT) to support clinical research visit scheduling workflow for future learning systems. Journal of Biomedical Informatics, 46, 642-652.
398. Tout, D., K. Leeds, E. Murphy, U. Srarkar, C. Lyles, T. Mekonnen and A. Chen. (2015). Facilitators and Barriers to implementing electronic referral and/or consultation systems: A qualitative study of 16 health organisations. BMC Health Services Research, 15.
399. Rumbold, J. and B. Pierscionek. (2017). A critique of the regulation of data science in healthcare research in the European Union. BMC Medical Ethics, 18(27).
400. Marsolo, K., P. Margolis, C. Forrest and R. Colletti. (2015). A digital architecture for a network-based learning health system - Integrating chronic care management, quality improvement and research. eGEMS, 3(1).
401. Mirnezami, R., J. Nicholson and A. Darzi. (2012). Preparing for precision medicine. New England J. of Medicine, 366(6).
402. Winden, T., E. Chen and G. Melton. (2016). Representing Residence, Living Situation, and Living Conditions: An Evaluation of Terminologies, Standards, Guidelines, and Measures/Surveys. Paper presented at the 2016 Conference of the American Medical Informatics Association.
403. Jha, A., C. DesRoches, E. Campbell, K. Donelan, S. Rao, T. Ferris, A. Shields, S. Rosenbaum, and D. Blumenthal. (2009). Use of electronic records in U.S. Hospitals. New England J. of Medicine, 360(16).
404. Boonstra, A. and M. Broekhuis. (2010). Barriers to the acceptance of electronic medical records by physicians from systematic review to taxonomy and interventions. BMC Health Services Research, 10(1).
405. Musen, M., M. Blackford and R. Greenes. (2014). Clinical decision-support systems. In (Ed.), Biomedical Informatics. London: Springer.
406. Turley, C. (2016). Leveraging a statewide clinical data warehouse to expand boundaries of the Learning Health System. eGEMS, 4(1).
407. Pare, G., L. Raymond, A. Ortiz de Guinea, P. Poba-Nzaou, M. Trudel, J. Marsan and T. Micheneau. (2014). Barriers to organizational adoption of EMR systems in family physician practices: A mixed-methods study in Canada. International Journal of Medical Informatics, 83, 548-558.
408. Zandeih, S., K. Yoon-Flannery, G. Kuperman, D. Langsam, D. Hyman and R. Kaushal. (2008). Challenges to EHR Implementation in Electronic- Versus

Paper-based Office Practices. Journal of General Internal Medicine, 23(6), 755-761.
409. Adler-Milstein, J., G. Daniel, C. Grossman, C. Mulvany, R. Nelson, E. Pan, V. Rohrbach and J. Perlin. (2014). Return on Information: A Standard Model for Assessing Institutional Return on Electronic Health Records. Retrieved from:
410. Scheuner, M., H. de Vries, B. Kim, R. Meili, S. Olmstead and S. Teleki. (2009). Are electronic health records ready for genomic medicine? Genetics in Medicine, 11(7).
411. Forrest, C., W. Crandall, C. Bailey, P. Zhang, M. Joffe, R. Colletti and J. Adler. (2014). Effectiveness of Anti-TFNa for Crohn Disease: Research in a pediatric learning health system. Pediatrics, 134, 37-44.
412. Bradley, E., L. Curry, S. Ramanadhan, L. Rowe, I. Nembhard and H. Krumholz. (2009). Research In Action: Using positive deviance to improve quality of health care. BMC Implementation Science, 4(25).
413. Lewis, G., H. Kirkham and R. Vaithianathan. (2013). How health systems could avert 'triple fail' events that are harmful, are costly, and result in poor patient satisfaction. Health Affairs, 32(4), 669-676.
414. Foley, T. and F. Fairmichael. (2015). Site visit to Geisinger Health Systems. The Learning Healthcare Project. Retrieved from http://www.learninghealthcareproject.org/publication/6/63/site-visit-to-geisinger-health-system
415. Garg, A., N. Adhikari, M. Rosas-Arellano, P. Devereaux, J. Sam and R. Haynes. (2005). Effects of computerised clinical decision suport systems on practitioner performance and patient outcomes. Journal of the American Medical Association (JAMA), 293(10).
416. Kawamoto, K., C. Houlihan, A. Balas and D. Lobacj. (2005). Improving clinical practice using clinical decision support systems: A systematic review of trials to identify features critical to success. British Medical Journal, 330, 765-768.
417. Wyatt, J. and D. Spiegelhalter. (1991). Field trials of medical decision-aids: potential problems and solutions. Paper presented at the Fifteenth Annual Symposium on Computer Applications in Medical Care., Washington, DC.
418. Abernethy, A., A. Ahmad, S. Zafar, J. Wheeler, J. Barsky-Reese and H. Lyerly. (2010). Electronic patient-reported data capture as a foundation of rapid learning cancer care. Medical Care, 48(6), S32-S38.
419. Ye, Y., M. Wamukoya, A. Ezeh, J. Emina and O. Sankoh. (2012). Health and Demographic Surveillance Systems: A step towards full civil registration and vital statistics in sub-Saharan Africa? BMC Public Health, 12.
420. Ethier, J. (2016). Integrating resources for translational research: A unified approach for learning health systems. (Doctorate), Universite Pierre et Marie Curie, Paris. Retrieved from
421. Feeley, T., G. Sledge, L. Levit and P. Ganz. (2014). Improving the quality of cancer care in America through health information technology. Journal of the American Medical Informatics Association, 21, 772-775.
422. Jameson, J. and D. Longo. (2015). Precision Medicine - Personalized, Problematic and Promising. New England J. of Medicine, 372(23).
423. Braithwaite, S., & Stine, N. (2013). Health-weighted composite quality metrics offer promise to improve health outcomes in a learning health system. eGEMS, 1(2).

424. Azar, J., N. Adams and M. Boustani. (2015). The Indiana University Center for healthcare innovation and implementation science: Bridging healthcare research and delivery to build a learning healthcare system. Zeitschrift für Evidenz, Fortbildung und Qualität im Gesundheitswesen, 109(2), 138-143.
425. Wysham, N., Howie, L., Patel, K., & Cameron, C. (2016). Development and refinement of a learning health systems training program. eGEMS, 4(1).
426. Coiera, E. (2017). The forgetting health system. Learning Health Systems, 2. doi:10.1002/lrh2.10023
427. Cresswell, K., H. Mozaffar, L. Lee, R. Williams and A. Sheikh. (2016). Safety risks associated with the lack of integration and interfacing of hospital health information technologies: A qualitative study of hospital electronic prescribing systems in England. British Medical Journal of Quality and Safety, 0, 1-12.
428. Kuchinke, W., C. Ohmann, R. Verheij, B. van Veen and B. Delaney. (2016). Development towards a learning health system - Experiences with the privacy protection model of the TRANSFoRm project. In (Ed.), Data Protection on the Move. Netherlands: Springer.
429. Schneeweiss, S. (2014). Learning from Big Health Care Data. New England J. of Medicine, 370(23), 2161-2163.
430. Steinwachs, D. (2015). Transforming public health systems: Using data to drive organizational capacity for quality improvement and efficiency. eGEMS, 2(4).
431. Doebbeling, B. and M. Flanagan. (2011). Emerging Perspectives on Transforming the Healthcare System: Redesign Strategies and a Call for Needed Research. Medical Care, 49(12), s59-s64.
432. Etheredge, L. (2014). Rapid learning: A breakthrough Agenda. Health Affairs, 33(7).
433. Gardner, W. (2015). Policy capacity in the learning healthcare system. International Journal of Health Policy Management, 4(12).
434. Lee, S., Kelley, M., Cho, M., Kraft, S., James, C., Constantine, M., Meyer, A., Diekema, D., Capron, A., Wilfond, B., & Magnus, D. (2016). Adrift in the gray zone: IRB perspectives on research in the Learning Health System. AJOB Empirical Bioethics, 7(2), 125-134.
435. Wilk, A. and J. Platt. (2016). Measuring Physicians' Trust: A scoping review with implications for public policy. Social Science and Medicine, 165, 75-81.
436. Krumholz, H., S. Terry and J. Waldstreicher. (2016). Data acquisition, curation and use for a continuously learning health system. Journal of the American Medical Association (JAMA), 316(16), 1669-1670.
437. Wallace, P., N. Shah, T. Dennen, P. Bleicher and W. Crown. (2014). Optum Labs: Building a novel node in the Learning Health Care System. Health Affairs, 33(7), 1187-1194.
438. Wiley. (2017). Learning Health Systems. Retrieved from http://onlinelibrary.wiley.com/journal/10.1002/(ISSN)2379-6146
439. Wong, G., T. Greenhalgh, G. Westhorp, J. Buckingham and R. Pawson. (2013). RAMESES publication standards: Realist syntheses. BMC Medicine, 11(21).
440. Mandl, K., I. Kohane, D. McFadden, G. Weber, M. Natter, J. Mandel, S. Schneeweiss, S. Weiler, J. Klann, J. Bickel, W. Adams, Y. Ge, X. Zhou, J. Perkins, K. Marsolo, and S. Murphy. (2014). Scalable collaborative infrastruture for a learning healthcare system (SCILHS): Architecture. Journal of the American Medical Informatics Association, 21, 615-620.

441. Rouse, W., Johns, M., & Pepe, K. (2017). Learning in the Health Care Enterprise. Learning Health Systems, e10024.
442. Yu, P., D. Artz and J. Warner. (2014). Electronic health records (EHRs): Supporting ASCO's vision of cancer care. Electronic Health Records, ASCO Educational Book(2014).
443. Yu, P., Hoffman, M., & Hayes, D. (2015). Biomarkers and Oncology. Archives of Pathology and Laboratory Medicine, 139.
444. Kelley, M., James, C., Alessi, S., Korngiebel, D., Wijangco, I., Rosenthal, E., Joffe, S., Cho, M., Wilfond, B., & Lee, S. (2015). Patient perspectives on the learning health system: The importance of trust and shared decision making. American Journal of Bioethics, 15(9), 4-17.
445. McNolty, L. and R. Payne. (2015). Relying on Trust for Research on Medical Practice in Learning Health Systems. The American Journal of Bioethics, 15(9), 30-32.
446. Roth, M., J. Rubin, K. Omollo, C. Friedman and J. Seagull. (2016). The Learning Health-System: A New Frontier for Human Factors. Paper presented at the 2016 International Symposium on Human Factors and Ergonomics in Health Care: Improving the Outcomes.
447. Friedman, C. and M. Rigby. (2013). Conceptualising and creating a global learning health system. International Journal of Medical Informatics, 82, e63-e71.
448. Kwon, S., M. Florence, P. Grigas, M. Horton, K. Hovarth, M. Johnson, G. Jurkovich, W. Klamp, K. Peterson, T. Quigley, W. Raum, T. Rogers, R. Thirlby, E. Farrokhi, and D. Flum. (2011). Creating a Learning Healthcare System in surgery: Washington State's Surgical Care and Outcomes Assessment Program (SCOAP) at 5 Years. Surgery, 151(2).
449. Amin, W., Tsui, F., Borromeo, C., Chuang, C., Espino, J., Ford, D., Hwang, W., Kapoor, W., Lehman, H., Martich, G., Morton, S., Paranjape, A., Shirey, W., Sorenson, A., Becich, M, & Hess, R. (2014). PaTH: Towards a learning health system in the Mid-Atlantic region. Journal of American Medical Informatics Association (JAMIA), 21, 633-636.
450. Kumar, S., Hanss, T., Johnson, L., Freidman, C., Donaldson, K., Tyus, J., Omollo, K., & Rubin, J. (2015). Leveraging contextual inquiry methods to empower patients in a Learning Health System. Paper presented at the 48th International Conference on System Sciences, Hawaii.
451. Rubin, J. (2017). Patient empowerment and the Learning Health System. Learning Health Systems, e10030.
452. Rubin, J. and C. Friedman. (2014). Weaving together a healthcare improvement tapestry: Learning Health System brings together health IT data stakeholders to share knowledge and improve health. Journal of AHIMA, 85(5), 38-43.
453. Forrest, C., P. Margolis, C. Bailey, K. Marsolo, M. Del-Beccaro, J. Finkelstein, D. Milov, V. Vieland, B. Wolf, F. Yu, and M. Kahn. (2014). PEDSnet: A national pediatric learning health system. Journal of the American Medical Informatics Association, 21, 602-606.
454. Hassan, H., E. Shehab and P. J. (2010). Toward full public E-Service environment in developing countries. World Academic Science Engineering and Technology, 4(6), 618-622.
455. Cresswell, K. and A. Sheikh. (2013). Organizational issues in the implementation and adoption of health information technology innovations: an interpretative review. International journal of medical informatics, 82(5), e73-e86.
456. Lea, A., D. Pearson, S. Clamp, O. Johnson and R. Jones. (2008). Undergraduate Learning: Using the Electronic Medical Record within Medical Undergraduate Education. Education For Primary Care, 19(6), 656-659.
457. MelbUni. (2017). Graduate Certificate in Health Informatics and Digital Health. Retrieved from https://mdhs-study.unimelb.edu.au/degrees/graduate-certificate-in-health-informatics-and-digital-health/overview
458. UTAS. (2017). Introduction to Health Informatics. Retrieved from http://www.utas.edu.au/courses/chm/units/crh500-introduction-to-health-informatics
459. Sugarman, J. and R. Calif. (2014). Ethics and Regulatory Complexities for Pragmatic CLinical Trials. Journal of the Americal Medical Association, 311(23).
460. Reimer, K. and J. Hamann. (2015). Digital Disruption Intermediaries. Retrieved from https://ses.library.usyd.edu.au/bitstream/2123/12761/7/ADTL_Digital%20Disruptive%20Intermediaries-final.pdf
461. Sullivan, C. and A. Staib. (2017). Digital disruption 'syndromes' in a hospital: Important considerations for the quality and safety of patient care during rapid digital transformations. Australian Health Review, (May 2018).
462. Kowitlawakul, Y., S. Chan, J. Pulcini and W. Wang. (2016). Factors influencing nursing students' acceptance of electronic health records for nursing education (EHRNE) software program. Nurse Education Today, 35(1), 189-194.
463. Nicholls, J., H. Potts, B. Coleman and D. Patterson. (2015). Legal and professional implications of shared care: A case study in oral anticoagulation stroke prevention therapy. BMC Health Services Research, 15(93).
464. Rathert, C., T. Porter, T. Mittler and M. Fleig-Palmer. (2017). Seven years after Meaningful Use: Physicians' and nurses' experiences with electronic health records. Health Care Management Review, 2017.
465. Hughes, K., E. Ambinder, G. Hess, P. Yu, E. Bernstam and M. Routbort. (2017). Identifying health information technology needs of oncologists to facilitate the adoption of genomic medicine: Recommendations from the 2016 American society of clinical oncology omics and precision oncology workshop. Journal of Clinical Oncology, 35.
466. Mimezami, R., J. Nicholson and A. Darzi. (2012). Preparing for Precision Medicine. New England J. of Medicine, 366(6), 489-91.
467. Sacristan, J. (2013). Patient-centred medicine and patient-oriented research: Improving health outcomes for individual patients. BMC Medical Informatics & Decision Making, 13(6).
468. Harwood, J., C. Butler and A. Page. (2016). Patient-centred care and population health: Establishing their role in orthapaedic practice. The Journal of Bone and Joint Surgery, 98(10).
469. Cresswell, K., D. Bates and A. Sheikh. (2017). Ten key considerations for the successful optimisation of large-scale health information technology. Journal of the American Medical Informatics Association, 24(1).
470. Rioth, M., J. Warner, B. Savani and M. Jagasia. (2016). Next generation long term transplant clinics: Improving resource utilisation and the quality of care through health information technology. Bone Marrow Transplantation, 51(1).

471. Chaudhry, B., J. Wang, M. Maglione, W. Mojica, E. Roth and P. Shekelle. (2006). Systematic review: impact of health information technology on quality, efficiency and cost of medical care. Annals of Internal Medicine, 144(10).
472. Pritchard, D., F. Moeckel, M. Villa, L. housman, C. McCarthy and H. McLeod. (2017). Strategies for integrating personalised medicine into healthcare practice. Personalised Medicine, 14(2).
473. Buntin, M., M. Burke, M. Hoaglin and D. Blumenthal. (2011). The benefits of health information technology: A review of the recent literature shows predominately positive results. Health Affairs, 30(3).
474. Yu, P. (2015). Knowledge bases, clinical decision support systems, and rapid learning in oncology. Journal of Oncology Practice, 11(2).
475. Zalon, M. (2016). Technology and continuously learning health systems. The Journal of Continuing Education in Nursing, 47(6).
476. Chorpita, B., E. Daleiden and A. Bernstein. (2016). At the intersection of Health Information Technology and Decision Support: Measurement feedback systems... and beyond. Administration and Policy in Mental Health and Mental Health Services Research, 43(3).
477. Fihn, S., J. Francis, C. Clancy, C. Nielson, K. Nelson, J. Rumsfeld, T. Cullen, J. Bates, and G. Graham. (2014). Insights from advanced analytics at the Veterans Affairs Health Administration. Health Affairs, 33(7).
478. Dreyer, N. (2011). Making observational studies count: Shaping the future of comparitive effectiveness research. Epidemiology, 22(3), 295-297.
479. Muir Gray, J. (2013). The shift in personalised and population medicine. The Lancet, 382(9888).
480. Horvath, K. and M. Johnson. (2011). Creating a Learning Healthcare System in surgery: Washington State's surgical care and outcomes assessment program (SCOAP) at five years Surgery, 151(2), 146-52.
481. AMA. (2016). Shared Electronic Medical Records. Retrieved from https://myhealthrecord.gov.au/internet/mhr/publishing.nsf/content/home
482. Hillstead, R., J. Bigelow, A. Bower, F. Girosi, R. Meili, R. Scoville and R. Taylor. (2005). Can electronic medical record systems transform health care? Potential health benefits, savings and costs. Health Affairs, 24(5), 1103-1117.
483. Denaxas, S., C. Friedman, A. Geissbuhler, H. Hemingway, D. Kalra, M. Kimura and J. Wyatt. (2015). Discussion of "combining health data uses to ignite health system learning". Methods of Information in Medicine, 54(6), 488-499.
484. Faden, R., T. Beauchamp and N. Kass. (2014). Informed consent, comparative effectiveness, and learning health care. New England Medical Journal, 370(8).
485. Lean, S., H. Guesgen, I. Hunter and A. Tretiakov. (2016). Working toward automated coding in general practice. Massey University. Palmerston North. Retrieved from http://goo.gl/PLk1Si
486. Nishimura, A., J. Carey, P. Erwin, J. Tilburn, M. Murad and J. McCormick. (2013). Improving understanding in the research informed consent process: a systematic review of 54 interventions tested in randomized control trials. BMC Medical Ethics, 14(28).
487. Morain, S. and N. Kass. (2016). Ethics issues arising in the transition to Learning Health Care Systems: Results from interviews with leaders from 25 health systems. eGEMS, 4(2).
488. Chey, W. and B. Speigel. (2016). The Digital Doctor: How Technologies Enhance Health Care. Helio Gastroenterology, February 2016.

489. Clemensen, J., S. Larsen, M. Kyng and M. Kirkevold. (2007). Participatory Design in Health Sciences: Using Cooperative Experimental Methods in Developing Health Services and Computer Technology. Qualitative Health Research, 17(1).
490. Dickson, H. and C. Ham. (2008). Engaging doctors in leadership: Review of the literature. Retrieved from:
491. Boulos, M., I. Maramba and S. Wheeler. (2006). Wikis, blogs and podcasts: a new generation of Web-based tools for virtual collaborative clinical practice and education. BMC Medical Education, 6(41).
492. Leyland, M., D. Hunter and J. Dietrich. (2009). Integrating Change Management into Clinical Health Information Technology Project Practice. Paper presented at the 2009 World Congress on Privacy, Security, Trust and the Management of e-Business.
493. May, C., F. Mair, T. Finch, A. MacFarlane, C. Dowrick, S. Treweek, T. Rapley, L. Ballini, B. Ong, A. Rogers, E. Murray, G. Elwyn, F. Legare, J. Gunn, and V. Montori. (2009). Development of a theory of implementation and integration: Normalization Process Theory. Implementation Science, 4(29).
494. Sperling, D. (1984). Assessment of technological choices using a pathway methodology. Transportation Research, 18(4), 343-353.
495. Horowitz, R., A. Hayes-Conroy, R. Carichio and B. Singer. (2017). From evidence-based medicine to medicine-based evidence. The American Journal of Medicine, 130(11), 1246-1250.
496. Masic, I., M. Miokovic and B. Muhamedagic. (2006). Evidence-Based medicine: New approaches and challenges. Acta Informatica Medica, 16(4).
497. Lavigne, J., J. Brown and G. Matzke. (2017). Population health and medicine: Policy and financial drivers. American Journal of Health-System Pharmacy, 74(18), 1413-1421.
498. Gray, M. and W. Ricciardi. (2010). From public health to population medicine: The contribution of public health to health care services. European Journal of Public Health, 20(4), 366-367.
499. McLachlan, S., K. Dube, T. Gallagher, B. Daley and J. Walonoski. (2018). The ATEN framework for creating the realistic synthetic electronic health record. Paper presented at the 11th International Joint Conference on Biomedical Engineering Systems and Technologies (BIOSTEC 2018), Madiera, Portugal.
500. Fenton, N. and M. Neil. (2010). Comparing risks of alternative medical diagnosis using Bayesian arguments. Journal of Biomedical Informatics, 43, 485-495.
501. Knape, T., L. Hederman, V. Wade, M. Gargan, C. Harris and Y. Rahman. (2003). A uml approach to process modelling of clinical practice guidelines for enactment. Studies in health technology and informatics, 635-640.
502. Scheuelein, H., F. Rauchfuss, Y. Dittmar, R. Molle, T. Lehmann, N. Pienkos and U. Settmacher. (2012). New methods for clinical pathways—business process modeling notation (BPMN) and tangible business process modeling (t. BPM). Langenbeck's archives of surgery, 397(5), 755-761.
503. Gomez, M., C. Bielza, J. Fernandez del Pozo and S. Rios-Insua. (2007). A graphical decision-theoretic model for neonatal jaundice. Medical Decision Making, 27(3), 250-265.
504. Liu, S., W. Cui, Y. Wu and M. Liu. (2014). A survey on information visualization: recent advances and challenges. The Visual Computer, 30(12), 1373-1393.

505. Moere, A., M. Tomitsch, C. Wimmer, B. Cristoph and T. Grechenig. (2012). Evaluating the effect of style in information visualization. IEEE transactions on visualization and computer graphics, 18(12), 2739-2748.
506. Ware, C. (2004). Information Visualisation: Perception for Design, 2nd Ed. San Francisco, USA: Morgan Kaufmann Publishers for Elsevier.
507. Wood, J., T. Isenberg, J. Dykes, N. Boukhelifa and A. Slingsby. (2012). Sketchy rendering for information visualization. IEEE Transactions on Visualization and Computer Graphics, 18(12), 2749-2758.
508. IoM. (2011). Health IT and Patient Safety: Building Safer Systems for Better Care. Retrieved from: https://www.ncbi.nlm.nih.gov/pubmed/24600741
509. Panzarasa, S., S. Madde, S. Quaglini, C. Pistarini and M. Stefanelli. (2002). Evidence-based careflow management systems: the case of post-stroke rehabilitation. Journal of biomedical informatics, 35(2), 123-139.
510. Ye, Y., Jiang, Z., Diao, X. & Du, G. (2009). Knowledge Based hybrid variance handling for patient care workflows based on Clinical Pathways. Paper presented at the IEEE/INFORMS International Conference on Service Operations, Logistics and Informatics.
511. Milne, T., J. Rogers, E. Kinnear, H. Martin, P. Lazzarini, T. Quinton and F. Boyle. (2013). Developing an evidence-based clinical pathway for the assessment, diagnosis and management of acute Charcot Neuro-Arthropathy: a systematic review. Journal of foot and ankle research, 6(1).
512. Gilbreth, F. and L. Gilbreth. (1921). Process Charts. USA: American Society of Mechanical Engineers.
513. Gopalakrishna, G., M. Langendam, R. Scholten, P. Bussuyt and M. Leeflang. (2016). Defining the Clinical Pathway in Cochrane Diagnostic Test Accuracy Reviews. BMC Medical Research Methodology, 16(153).
514. van de Klundert, J., P. Gorissen and S. Zeemering. (2010). Measuring Clinical Pathway Adherence. Journal of Biomedical Informatics, 43.
515. Dumas, M. and A. Hofstede. (2001). UML Activity Diagrams as a Workflow Specification Language. Heidelberg, Germany: Springer.
516. Harel, D. (1987). Statecharts: A visual formalism for complex systems. Science of computer programming, 8, 231-274.
517. Lodewijckx, C., M. Decramer, W. Sermeus, M. Panella, S. Deneckere and K. Vanhaecht. (2012). Eight-step method to build the clinical content of an evidence-based care pathway: the case for COPD exacerbation. Trials, 13(1).
518. Weatherall, M. (1996). Making medicine scientific: empiricism, rationality, and quackery in mid-Victorian Britain. Social History of Medicine, 9(2), 175-194.
519. Whorton, J. (2004). Nature cures: The history of alternative medicine in America. USA: Oxford University Press on Demand.
520. Barker, L. (1916). The teaching of clinical medicine. Science, 43(1119), 799-810.
521. Croskerry, P. (2013). From mindless to mindful practice—cognitive bias and clinical decision making. New England Journal of Medicine, 368(26), 2445-2448.
522. Marewski, J. and G. Gigerenzer. (2012). Heuristic decision making in medicine. Dialogues in clinical neuroscience, 14(1).
523. Marshall, H. (1920). Why Relief of Joint Troubles Is Easy or Difficult. The Boston Medical and Surgical Journal, 183(15), 430-438.

524. Czaja, A. (2010). Difficult treatment decisions in autoimmune hepatitis. World journal of gastroenterology, 16(8).
525. Heesen, C., A. Solari, A. Giordano, J. Kasper and S. Kopke. (2011). Decisions on multiple sclerosis immunotherapy: new treatment complexities urge patient engagement. Journal of the neurological sciences, 306(1-2), 192-197.
526. Zhang, B., A. Tsiatis, E. Laber and M. Davidian. (2013). Robust estimation of optimal dynamic treatment regimens for sequential treatment decisions. Biometrika, 100(3), 681-694.
527. Ciarlegio, A., E. Petkova, R. Ogden and T. Tarpey. (2015). Treatment decisions based on scalar and functional baseline covariates. Biometrics, 71(4), 884-894.
528. Dancey, J., P. Bedard, N. Onetto and T. Hudson. (2012). The genetic basis for cancer treatment decisions. Cell, 148(3), 409-420.
529. Djulbegovic, B., S. Elqayam, T. Reljic, I. Hozo, B. Miladinovic and A. Tsalatsanis. (2014). How do physicians decide to treat: an empirical evaluation of the threshold model. BMC medical informatics and decision making, 14(1).
530. Pauker, S. and J. Kassirer. (1980). The threshold approach to clinical decision making. New England J. of Medicine, 302(20), 1109-1117.
531. McClure, P., L. Blackburn and C. Dusold. (1994). The use of splints in the treatment of joint stiffness: biologic rationale and an algorithm for making clinical decisions. Physical Therapy, 74(12), 1101-1107.
532. Richardson, W., M. Wilson, J. Nishikawa and R. Hayward. (1995). The well-built clinical question: a key to evidence-based decisions. ACP J Club, 123(3), A12-13.
533. Fenton, N. and M. Neil. (2012). Risk Assessment and Decision Analysis with Bayesian Networks: CRC Press.
534. Lie, M., L. Hayes, N. Lewis-Barned, C. May, M. White and R. Bell. (2013). Preventing Type 2 diabetes after gestational diabetes: women's experiences and implications for diabetes prevention interventions. Diabetic Medicine, 30, 986-993.
535. Plagemann, A., T. Harder, R. Kohloff, W. Rohde and G. Dorner. (1997). Glucose tolerance and insulin secretion in children of mothers with pregestational IDDM or gestational diabetes. Diabetologia, 40, 1094-1100.
536. Sellers, E., H. Dean, L. Shafer, P. Martens, W. Phillips-Beck, M. Heaman, H. Prior, A. Dart, J. McGavock, M. Morris, A. Torshizi, S. Ludwig, and G. Shen. (2016). Exposure to Gestational Diabetes Mellitus: Impact on the Development of Early-Onset Type 2 Diabetes in Canadian First Nations and Non–First Nations Offspring. Diabetes Care, 39(12), 2240-2246.
537. Greenhalgh, T., M. Clinch, N. Afsar, Y. Choudhury, R. Sudra, D. Campbell-Richards, A. Claydon, G.A. Hitman, P. Hanson, and S. Finer. (2015). Socio-cultural influences on the behaviour of South Asian women with diabetes in pregnancy: qualitative study using a multi-level theoretical approach. BMC Medicine, 13(1), 120. doi:10.1186/s12916-015-0360-1
538. Zhu, Y. and C. Zhang. (2016). Prevalence of Gestational Diabetes and Risk of Progression to Type 2 Diabetes: a Global Perspective. Current Diabetes Reports, 16(1).
539. Farrar, D., M. Simmonds, S. Griffin, A. Duarte, D. Lawlor, M. Sculpher, L. Fairley, S. Golder, D. Tuffnell, M. Bland, F. Dunne, D. Whitelaw, W. Wright, and T. Sheldon. (2016). Chapter 4: Prevalence of gestational diabetes in the UK and Republic of Ireland: A systematic review In (Ed.), Health Technology Assessment.
540. HAPO Study Cooperative Research Group. (2008). Hyperglycemia and adverse pregnancy outcomes. New England Journal of Medicine, 358(19), 1991-2002.
541. International Association of Diabetes and Pregnancy Study Groups. (2010). Recommendations on the Diagnosis and Classification of Hyperglycemia in Pregnancy. Diabetes Care, 33(3), 676.
542. NICE. (2015). Diabetes in pregnancy: Management of diabetes and its complications from preconception to the postnatal period (NICE guideline NG3). Retrieved from: London, UK.
543. Siering, U., M. Eikermann, E. Hausner, W. Hoffman-Eber and E. Neugebauer. (2013). Appraisal Tools for Clinical Practice Guidelines: A Systematic Review. PLOS ONE, 8(12).
544. Brouwers, M., M. Kho, G. Browman, F. Cluzeau, G. Feder, B. Fervers, S. Hanna and J. Makarski. (2010). AGREE II: Advancing guideline development, reporting and evaluation in healthcare. Canadian Medical Association Journal, 182, E839-842.
545. Peterson, P. and J. Rumsfeld. (2011). The evolving story of guidelines and health care: Does being NICE help? Annals of Internal Medicine, 155(4).
546. Shalom, E., Y. Shahar, Y. Parmet and E. Lunenfeld. (2015). A multiple-scenario assessment of the effect of a continuous-care, guideline-based decision support system on clinicians' compliance to clinical guidelines. . International Journal of Medical Informatics, 84(4), 248-262.
547. McInnes, I. and G. Schett. (2017). Pathogenetic insights from the treatment of rheumatoid arthritis. The Lancet, 389(10086), 2328-2337.
548. Singh, J., K. Saag, S. Bridges, E. Akl, R. Bannuru, M. Sullivan, E. Vaysbrot, C. McNaughton, and M. Osani. (2015). 2015 American College of Rheumatology Guideline for the Treatment of Rheumatoid Arthritis. Arthritis & rheumatology, 68(1), 1-26.
549. Crewdson, K., D. Lockey and G. Davies. (2007). Outcome from paediatric cardiac arrest associated with trauma. Resuscitation, 75, 29-34.
550. Taylor, C., S. Jan, K. Curtis, A. Tzannes, Q. Li, C. Palmer, C. Dickson and J. Myburgh. (2012). The cost-effectiveness of physician staffed Helicopter Emergency Medical Service (HEMS) transport to a major trauma centre in NSW, Australia. Injury, Int. J. Care Injured, 43, 1843-1849.
551. Andruszkow, H., R. Lefering, M. Frink, P. Mommsen, C. Zeckey, K. Rahe and F. Hildebrand. (2013). Survival benefit of helicopter emergency medical services compared to ground emergency medical services in traumatized patients. Critical Care, 17(3).
552. Boyd, C., K. Corse and R. Campbell. (2006). Emergency Interhospital Transport of the Major Trauma Patient: Air versus Ground. The Journal of Trauma, 29(6), 789-794.
553. Galvagno, S., R. Sikorski, J. Hirshon, D. Floccare, C. Stephens, D. Beecher and S. Thomas. (2015). Helicopter emergency medical services for adults with major trauma. Cochrane Database of Systematic Reviews, 12.
554. Helfet, D., T. Howey, R. Sanders and K. Johansen. (1990). Limb salvage versus amputation. Preliminary results of the Mangled Extremity Severity Score. Clinical Orthopaedics and Related Research, 256, 80-86.

555. Black, J., M. Ward and D. Lockey. (2004). Appropriate use of helicopters to transport trauma patients from incident scene to hospital in the United Kingdom: an algorithm. Emergency Medicine Journal, 21, 355-361.
556. Coats, T., S. Keogh, H. Clark and M. Neal. (2001). Prehospital Resuscitative Thoracotomy for Cardiac Arrest after Penetrating Trauma: Rationale and Case Series. Journal of Trauma, Injury, Infection and Critical Care, 50(4).
557. Kleber, C., M. Giesecke, T. Lindner, N. Haas and C. Buschmann. (2014). Requirement for a structured algorithm in cardiac arrest following major trauma: epidemiology, management errors, and preventability of traumatic deaths in Berlin. Resuscitation, 85(3), 405-410.
558. Sherren, P., C. Reid, K. Habig and B. Burns. (2013). Algorithm for the resuscitation of traumatic cardiac arrest patients in a physician-staffed helicopter emergency medical service. Critical Care, 17(2), P281.
559. Bellezza, F. (1981). Mnemonic devices: Classification, characteristics, and criteria. Review of Educational Research, 51(2), 247-275.
560. Starmer, A., N. Spector, R. Srivastava, A. Allen, C. Landrigan and T. Sectish. (2012). I-pass, a mnemonic to standardize verbal handoffs. Pediatrics, 129(2), 201-204.
561. Nutbeam, T. and M. Boylan. (2013). The primary survey. In T. Nutbeam and M. Boylan (Ed.), ABC of Prehospital Emergency Medicine. John Wiley and Sons.
562. SJA. (2015). The Primary Survey. Retrieved from http://www.sja.org.uk/sja/first-aid-advice/what-to-do-as-a-first-aider/the-primary-survey.aspx
563. Pountney, A. and D. McDonagh. (2014). Thorax and Abdominal Injuries on the Field. Sports Injuries: Prevention, Diagnosis, Treatment and Rehabilitation,, 1-18.
564. Nutbeam, T. and M. Boylan. (2013). ABC of Prehospital Emergency Medicine. West Sussex, UK: Wiley Blackwell: BMJ Books.
565. Ellerton, J. (2006). Casualty care in mountain rescue, 2nd Ed: reeds Ltd.
566. Greaves, I., K. Porter and J. Garner. (2009). Trauma Core Manual, 2nd Ed. London, UK: Hodder Arnold.
567. McGrath, A. and D. Whiting. (2015). Recognising and assessing blunt abdominal trauma. Emergency Nurse, 22(10).
568. Djohan, V., L. Churilov and J. Wassertheil. (2002). Business integration in an acute emergency department in Australia: a clinical process modelling perspective. Paper presented at the The 6th Pacific Asia Conference on Information Systems, Tokyo, Japan.
569. Onggo, B. (2009). Towards a unified conceptual model representation: a case study in healthcare. Journal of Simulation, 3(1), 40-49.
570. Wang, J., J. Li and P. Howard. (2013). A system model of work flow in the patient room of hospital emergency department. Health care management science, 16(4), 341-351.

# Appendices

## Appendix A
## EHR: Scoping Review and Analysis

### EHR Challenges: Barriers, Benefits and Facilitators

A meta-analysis of 26 scoping or meta-analysis reviews

**Barriers Mentioned**

| Author/Year | Title | Financial/Cost | Maintenance and Support Costs | Slow Adoption | Lack of HIT/IT Expertise | Culture / General Acceptance | Organisational | Technological/ IT / Adoption Issues | Interoperability/Data Exchange | Integration | Data Standardisation | Data Accuracy | Staff Training | Adequate Technical Support | Unfamiliarity woth Computers | Staff Resistance to Change | Privacy | Security | Confidentiality | Legislation/Law/Legal | Policy | Ethics | IRB/ERB | Selection of the Right System/Vendor | System Complexity/ usability | Disrupt the flow of Pt Care | Increased Clinician Time / Workload | Providing Patient Access |
|---|---|---|---|---|---|---|---|---|---|---|---|---|---|---|---|---|---|---|---|---|---|---|---|---|---|---|---|---|
| | | 19 | 9 | 11 | 16 | 11 | 7 | 18 | 15 | 10 | 12 | 9 | 15 | 10 | 9 | 15 | 14 | 11 | 10 | 8 | 5 | 1 | 1 | 7 | 7 | 18 | 16 | 6 |
| Harle et al 2012 | Overcoming challenges to achieving meaningful Use | X | | X | | X | X | X | X | | | | | | | | | | | | | | | | | | | X |
| Cherry et al 2008 | Factors affecting EHR adoption in long-term | X | | | X | X | | X | X | | X | | X | | X | X | X | X | X | X | | | | X | X | X | X | |
| Jamoom et al 2014 | EHR Adopters vs non-adopters | X | X | | | | | | | | | | X | X | | | | | | | X | | | | | X | X | |
| Haughom et al 2011 | Overcoming Barriers to EHR | | | X | | X | | X | X | | | | | | | | | X | X | | | | | | X | X | | |
| Kaye et al 2010 | Barriers and Success factors in HIT | X | | X | X | X | | X | X | X | X | | X | X | | X | X | | X | X | | | | X | | X | X | |
| Lluch 2011 | Healthcare Professionals' organisational barriers | X | | | X | X | X | | X | X | X | | X | X | X | | | | | X | | | | | | X | X | X |
| Gagnon et al 2009 | Users Perspectives of Barriers and Facilitators | | | X | | | X | | | X | | | | | | | | | | | | | | | | | | X |
| McGinn et al 2011 | Comparison of Users' groups perspectives of barrier | X | X | | X | X | | X | X | | X | | | | X | X | X | X | X | | | | | X | X | | X | |
| Zandeih et al 2008 | Challenges to EHR implementation | | | | | | | | | | | | X | X | X | X | X | X | X | | | | | | | X | X | |
| Weng et al 2012 | Using EHRs to integrate research with patient care | | | | | | | | | | X | X | | | | | | | | X | | | X | | | | X | X |
| Tang et al 2006 | Personal Health records: Definitions benefits | | | X | | | X | X | X | X | X | X | X | | | | X | X | | X | | | | | | X | | |
| Hersh 2004 | Health Care IT: Progress and Barriers | X | | | X | | | X | X | | | | X | | | | X | X | X | | | | | | | X | | |
| Akanbi et al 2014 | Use of EHR in sub-saharan africa | X | X | | X | | | X | | | | | | | | | | | | | | | | | | | X | |
| Tout et al 2015 | Facilitators and barriers to implementing electronic | | X | | X | X | | | X | X | | X | | | | X | | | | | | | | | | X | X | |
| Terry et al 2008 | Implementing EHR | | | X | | | | | | | | | X | X | X | | | | | | | | | | | X | X | |
| Schumaker et al 2016 | Implementing EHR in the healthcare industry | X | X | | X | | | X | X | X | X | X | X | | | X | X | X | X | | | | | X | | X | X | X |
| Nguyen et al 2014 | EHR implementation: An evaluation | X | X | X | X | X | X | X | X | X | X | X | X | X | | X | X | X | X | | | | | | X | X | X | X |
| Ludwick et al 2009 | Adopting EMR in primary care | X | X | | X | X | | X | | X | X | | X | | X | X | X | | | | | | | | | X | X | |
| Kumar et al 2010 | Overcoming barriers to EMR | X | | | | X | | X | X | X | X | | X | | | X | X | | | X | X | | | | X | | | |
| Jawhari et al 2016 | Benefits and challenges of EMR implementations | X | | | | | | X | | | | X | | | | X | | | | | X | X | | X | | X | X | |
| Jha et al 2009 | Use of EHR in US hospitals | X | X | X | X | | | | X | | | | | | | X | | | | | X | | | | | | | |
| Heisey-Grove et al 2014 | Challenges to EHR Adoption | X | | | X | X | | X | X | | | X | X | X | | X | X | X | | | | | | X | | X | | |
| Golberg et al 2012 | EHRs in primary care practices | X | | X | X | | | X | | | X | | | X | X | X | | | | | X | | | | X | X | X | |
| Boonstra et al 2014 | Implementing EHR in Hospitals | X | | | X | | X | X | | | | X | X | X | X | X | X | X | X | | | | | X | X | X | | |
| Houser et al 2008 | Perceptions regarding EHR Implementation | X | | X | X | | X | X | | | | | X | | | X | X | | | X | | | | | | | | |
| Chao et al 2013 | Benefits and challenges of EHR system | X | X | X | X | | | X | X | X | X | X | | | X | | X | X | X | X | | | | | | X | X | |

**Benefits Claimed / Factors that Facilitate / Country**

| Author/Year | Improved Quality of Care | Efficiency | Accuracy of Documentation | Easier Access to Pt Info | Reduced Costs | Reduced Resources (Staff Time) | Reduction in errors | Improved capture of billables | Improved Staff Retention | Increased/ Enhanced Confidentiality | Improved Communication/Exchange | Increased Patient Safety | Comprehensive Implementation Planning | Learning from Prior Implementations | Staged Implementation Processes | staff involvement and early buy-in | Facilitation of computer familiarity | User Champions | Staff Training | Access to Technical and User Support | Strong/ Engaged Leadership/Management | Support From Clinical Leaders | Penalties for non-adoption | Financial Incentives/ Funding | Govt/Health Ministry Certified Solutions | Integrating Historic Data | USA | UK | Israel | Canada | Australia | Netherlands |
|---|---|---|---|---|---|---|---|---|---|---|---|---|---|---|---|---|---|---|---|---|---|---|---|---|---|---|---|---|---|---|---|---|
| | 16 | 17 | 7 | 17 | 9 | 10 | 7 | 7 | 1 | 2 | 11 | 8 | 7 | 2 | 5 | 16 | 6 | 8 | 14 | 12 | 8 | 7 | 1 | 14 | 2 | 3 | 16 | 1 | 1 | | | |
| Harle et al 2012 | | | | | | | | | | | | | | | | | | | | | | | | | | | X | | | | | |
| Cherry et al 2008 | X | X | X | X | X | X | X | X | X | | | X | X | X | X | X | X | X | X | X | X | X | | | | X | X | | | | | |
| Jamoom et al 2014 | X | X | | X | X | X | | | | X | | | | | | | | | | X | | | X | X | X | | X | | | | | |
| Haughom et al 2011 | X | X | | X | | | | | | | | | | | X | X | | | | | | X | | X | | | X | | | | | |
| Kaye et al 2010 | X | | | | X | | | | | | | | | | | X | | | | | X | | | X | | | | | X | | | |
| Lluch 2011 | X | | | | | | | X | | | | | | | | X | | X | X | X | | X | | X | | | | X | | | | |
| Gagnon et al 2009 | | | | | | | | | | | | | | | | | | | | | | | | | | | | | | X | | |
| McGinn et al 2011 | | | | X | | | | | | | X | | | | | X | X | | X | | X | X | | | X | | | | | X | | |
| Zandeih et al 2008 | | X | | X | | | | X | | | X | | | | | | | X | X | X | | | | | | | X | | | | | |
| Weng et al 2012 | | | | | | | | | | | | | | | | | | | | | | | | | | | X | | | | | |
| Tang et al 2006 | X | X | | X | X | | | | | | X | X | | | | X | | | | | | | | X | | | X | | | | | |
| Hersh 2004 | X | | | | | | | | | | | | | | | | | | | | | | | | | | X | | | | | |
| Akanbi et al 2014 | X | X | | X | X | X | | | | | | X | | | | | | | | | | | | | | | | | | | | |
| Tout et al 2015 | | X | | X | | X | | | | | | | X | | | X | | | | | X | | | X | | X | X | | | | | |
| Terry et al 2008 | X | X | | X | | | | | | | | X | | | X | X | | X | X | X | | | | | | | | | | X | | |
| Schumaker et al 2016 | X | X | X | X | | | X | | | X | X | X | X | | X | X | X | | X | X | X | | | X | | X | X | | | | | |
| Nguyen et al 2014 | X | X | X | X | | X | X | | | | X | X | X | | | X | | X | X | X | | | | | | | | | | | X | |
| Ludwick et al 2009 | | | | | | | | | | | X | | | | X | X | | X | X | | X | X | | X | | | | | | X | | |
| Kumar et al 2010 | | X | X | X | X | X | X | X | | | X | | | | | X | | | X | | | | | | | | X | | | | | |
| Jawhari et al 2016 | X | X | X | X | | X | X | X | | | X | X | X | | | X | X | X | X | | | X | | X | | | | | | X | | |
| Jha et al 2009 | X | X | | | | | | | | | | | | | | | | | | X | | | | X | | | X | | | | | |
| Heisey-Grove et al 2014 | | | | | | | | | | | | | | | | X | | | X | | | | | X | | | X | | | | | |
| Golberg et al 2012 | X | X | X | X | | | | | | | X | | X | | | X | X | | X | X | X | | | X | | | X | | | | | |
| Boonstra et al 2014 | | X | | X | X | | | | | | | | X | X | | X | | X | X | X | X | X | | X | | | | | | | | X |
| Houser et al 2008 | X | X | X | X | X | X | X | X | | | X | | | | | | | | | X | | | | X | | | X | | | | | |
| Chao et al 2013 | X | X | | X | X | X | X | | | | X | X | | | | | X | | X | X | | | | | | | | | | | | |

# Appendix B
# LHS: Benefits and Barriers

**Benefits**

| Year | Lead Author | Improve quality of care | improve clinical outcomes | improve patient quality of life | Reduce variance (good/bad/fragmented practice) | Reduction in Ineffective Treatment/Care | Shared Decision Making (between different disciplines) | Empowering Patients | Identify and apply evidence (EBM) | Estimation of costs | reduction of costs / resource utilisation | Reduced hospital stay (LOS) | Identify/improve education/training (C=clinician, P=patient, B=both) | Enhanced collaboration with patients | Improved satisfaction of Care team/Patients (C, P or Both) | Increased efficiency | Improved clinical job performance | Improved communication between clinicians (sharing info) | Incorporate skills from other profs/sectors | Reduce documentation | Enhance accountability | Auditing | Shield against liability |
|---|---|---|---|---|---|---|---|---|---|---|---|---|---|---|---|---|---|---|---|---|---|---|---|
| 75 | TOTALS | 45 | 30 | 22 | 44 | 6 | 2 | 5 | 26 | 5 | 38 | 26 | 18 | 26 | 14 | 30 | 7 | 17 | 29 | 6 | 0 | 5 | 3 |
| 1993 | Goode | X | | | X | | | X | | | X | X | P | X | | X | | | | | | | |
| 1993 | Goode | X | X | X | X | | | | | | X | X | | X | X | X | X | X | X | | | | |
| 1994 | Lumsdon | | X | | X | | | | | | | | | | | | | | X | | | | |
| 1994 | Giffin | X | X | | X | X | | | | | X | X | P | X | P | X | | X | X | | | | |
| 1995 | Hydo | X | | X | X | | | | X | | X | X | | X | | X | | X | X | | | | |
| 1996 | Gordon | X | X | X | | | | X | | X | X | | | | | | | X | | | | | |
| 1996 | Comried | X | X | | X | | | | | | X | | P | X | P | X | | | X | | | | |
| 1997 | Holecek | X | | | X | | | | | | X | | X | | | X | | | X | X | | | |
| 1997 | Chu | X | X | | X | | | | X | | X | X | | | | X | | | | | | | |
| 1997 | Wilson | | | | X | | | | | | | X | | | | | | | | | | | |
| 1998 | Hill | | X | X | X | | | | X | | X | X | | | P | X | | X | X | | | | |
| 1999 | Matula | | X | X | X | | | | | | X | X | | X | | | | | X | | | | |
| 1999 | Denton | X | X | | X | | | | | | | | | | | | | X | | | | | |
| 1999 | Bumgarner | X | | | | | | | | | X | X | P | X | P | | | | X | X | | | |
| 2000 | Dickinson | X | X | | X | | | X | X | | | | X | X | X | X | X | | X | | | X | |
| 2001 | Philie | X | | | | X | | | | | | | | | | X | | | | | | | |
| 2001 | Sherman | X | | | X | | | | X | | X | X | | | | X | | | X | | | | |
| 2001 | Lanziere | | X | X | | | | | X | | X | X | X | | | | | X | | | | | |
| 2001 | Cholock | X | X | | X | | | | X | | X | X | X | X | | X | | X | X | | | | |
| 2001 | Quaglini | | | | | | | | X | | | | | | | | | | | | | | |
| 2002 | Micik | X | | X | | | | | | | | | | | | X | | | | | | | |
| 2002 | Panzarasa | X | | | X | | | | X | | | | | | | X | X | X | | | | | |
| 2003 | Mingmei | X | X | X | X | | | | | | X | X | | X | X | X | X | X | X | | | | |
| 2003 | Jones | X | | | | | | | X | | X | X | | X | | | | | X | | | | |
| 2004 | Beaupre | X | X | X | X | | | | | | | X | | | | X | | | X | | | | |
| 2004 | Herck | | | X | X | | X | | | | X | X | | | C | X | | X | | | | | |
| 2004 | Houltram | X | | X | X | | | | X | | | | B | X | | | | | X | | | | |
| 2004 | Houltram | X | | X | | | | | X | | | | B | X | | | | | | | | | |
| 2004 | Houltram | X | | X | | | | | X | | | | | | | | | | | | | | |
| 2004 | Repasky | X | X | X | X | | | | | | | | | | | | | | | | | X | |
| 2006 | Lawson | | | X | | | | | | | X | | | X | | | | | | | | | |
| 2006 | Tay | | | | X | | | | | X | | | | | | | | | X | X | | | |

**Barriers/Challenges**

| Year | Lead Author | Lack of or poorly graded evidence | Potential Bias/Conflict of Interest | Cultural barriers within org. | Complexity | Lack of resources | Potential conflicts with other guidelines/protocols | Removes individual decision-making ("cookbook medicine") | Fails to account for patient variance | Difficulty in involving all stakeholders | Development process: time, commitment, and resources | Create more paperwork | Clinician disagrees with recommendations (deviates) | Unpredictable clinical course | Refusal to adopt/use (resistance) | Lack of awareness for existence of artefact | Integration into the existing information system | Paper-based artefacts cause inefficiencies in care processes |
|---|---|---|---|---|---|---|---|---|---|---|---|---|---|---|---|---|---|---|
| 75 | TOTALS | 1 | 1 | 1 | 3 | 1 | 1 | 8 | 3 | 1 | 9 | 1 | 2 | 1 | 4 | 1 | 1 | 1 |
| 1993 | Goode | | | | | | | | | | | | | | | | | |
| 1993 | Goode | | | | | | | | | | | | | | | | | |
| 1994 | Lumsdon | | | | | | | | | | | | | | | | | |
| 1994 | Giffin | | | | | | | X | X | | X | | | | | | X | |
| 1995 | Hydo | | | | | | | | | | | | | X | | | | |
| 1996 | Gordon | | | | | | | | | | | | | | | | | |
| 1996 | Comried | | | | | | | | | | | | | | | | | |
| 1997 | Holecek | | | | | | | | | | | | | | | | | |
| 1997 | Chu | | | | | | | | | | | | | | | | | |
| 1997 | Wilson | | | | | | | | | | | | | | | | | |
| 1998 | Hill | | | | | | | | | | | | | | | | | |
| 1999 | Matula | | | | | | | | | | | | | | | | | |
| 1999 | Denton | | | | | | | | | X | | | | | | | | |
| 1999 | Bumgarner | | | | | | | | | | | | | | | | | |
| 2000 | Dickinson | | | | | | | | | | X | | | | | | | |
| 2001 | Philie | | | | | | | | | | | | | | | | | |
| 2001 | Sherman | | | | | | | | | | | | | | | | | |
| 2001 | Lanziere | | | | | | | | | | | | | | | | | |
| 2001 | Cholock | | | | | | | | | | | | | | | | | |
| 2001 | Quaglini | | | | | | | | | | | | | | | | | |
| 2002 | Micik | | | | | | | | | | | | | | | | | |
| 2002 | Panzarasa | | | | | | | | | | | | | | | | | |
| 2003 | Mingmei | | | | | | | | | | | | | | | | | |
| 2003 | Jones | | | | | | | | | | | | | | | | | |
| 2004 | Beaupre | | | | | | | | | | | | | | | | | |
| 2004 | Herck | | | | | | | | | | | | | | | | | |
| 2004 | Houltram | | | | | | | | | | | | | | | | | |
| 2004 | Houltram | | | | | | | | | | | | | | | | | |
| 2004 | Houltram | | | | | | | | | | | | | | | | | |
| 2004 | Repasky | | | | | | | | | | | X | | | X | | | |
| 2006 | Lawson | | | | | | | | | | | | | | | | | |
| 2006 | Tay | | | | X | | | X | | | X | | | | | | | |

| | | | | | | | | | | | | | | | | | | | | | | | | | | | | | | | | | | | | | | | | | |
|---|---|---|---|---|---|---|---|---|---|---|---|---|---|---|---|---|---|---|---|---|---|---|---|---|---|---|---|---|---|---|---|---|---|---|---|---|---|---|---|---|---|
| 2007 | Sternberg |  | X |  | X |  |  |  | X |  | X |  |  |  |  |  |  |  | X |  |  |  |  |  |  |  |  |  |  |  |  |  |  |  |  |  |  |  |  |  |  |
| 2007 | Kwan | X |  |  | X |  |  |  |  |  | X | X | P | X |  |  |  | X | X | X |  |  |  |  |  |  |  |  |  |  | X | X |  | X |  |  |  |  |  |  |  |
| 2007 | Mackey | X | X |  | X | X |  |  | X |  |  |  |  |  |  |  | X |  |  |  |  |  |  |  |  |  |  |  |  |  |  |  |  |  |  |  |  |  |  |  |  |
| 2008 | Rotter | X |  |  |  |  |  |  |  |  | X | X |  |  |  |  |  |  |  |  |  |  |  |  |  |  |  |  |  |  | X |  |  | X |  |  |  |  |  |  |  |
| 2008 | Osborn |  |  |  |  |  |  |  |  |  |  |  |  | X | P |  |  |  |  |  |  |  |  |  |  |  |  |  |  |  |  |  |  | X |  |  |  |  |  |  |  |
| 2008 | Gemmel |  |  |  |  |  |  |  |  |  | X |  |  |  |  | X |  |  | X |  |  |  |  |  |  |  |  |  |  |  |  |  |  |  |  |  |  |  |  |  |  |
| 2009 | Lougheed | X |  |  |  |  |  |  | X |  |  |  |  |  |  |  |  |  |  |  |  |  |  |  |  |  |  |  |  |  |  |  |  |  |  |  |  |  |  |  |  |
| 2010 | Una | X | X | X | X |  |  |  |  |  | X |  |  |  |  |  |  |  | X |  |  |  |  |  |  |  |  |  |  |  |  |  |  |  |  |  |  |  |  |  |  |
| 2010 | Kinsman |  |  |  |  |  |  |  |  |  |  |  |  |  |  |  |  |  |  |  |  |  |  |  |  |  |  |  |  |  |  |  |  |  |  |  |  |  |  |  |  |
| 2010 | van de Klundert | X | X | X | X |  |  |  | X |  | X |  |  |  |  | X |  |  |  |  |  |  |  |  |  |  |  |  |  |  | X |  |  |  |  |  |  |  |  |  |  |
| 2011 | Ryu |  | X |  |  | X |  |  |  |  | X | X |  |  |  |  |  |  |  |  |  |  |  |  |  |  |  |  |  |  |  |  |  |  |  |  |  |  |  |  |  |
| 2012 | Lodewijckx |  |  |  | X |  |  |  | X |  |  |  |  | X |  |  |  |  | X |  |  |  |  |  |  |  |  |  |  |  |  |  |  | X |  |  |  |  |  |  |  |
| 2012 | Huang | X |  |  | X |  |  |  | X |  |  | X | B |  |  |  |  |  | X |  |  |  |  |  |  |  |  |  |  |  |  |  |  |  |  |  |  |  |  |  |  |
| 2012 | Tastan |  |  | X |  |  |  |  |  | X | X | X |  | X | B |  | X | X | X |  |  |  |  |  |  |  |  |  |  |  |  |  |  |  |  |  |  |  |  |  |  |
| 2012 | MDD |  |  |  |  | X |  |  | X | X |  |  |  |  |  |  |  |  |  |  |  |  |  |  |  |  |  |  |  |  |  |  |  |  |  |  |  |  |  |  |  |
| 2012 | Thompson | X | X |  | X |  |  |  | X |  |  |  | P | X |  | X |  |  | X |  |  |  |  |  |  |  |  |  |  |  | X |  |  |  |  |  |  |  |  |  |  |
| 2012 | Huang (2) | X |  | X |  |  |  |  |  |  | X |  |  | X |  | X |  |  |  |  |  |  |  |  |  |  |  |  |  |  |  |  |  |  |  |  |  |  |  |  |  |
| 2013 | Milne |  | X | X |  |  |  |  |  |  |  |  |  |  |  |  |  |  |  |  |  |  |  |  |  |  |  |  |  |  |  |  |  |  |  |  |  |  |  |  |  |
| 2013 | Morrow | X |  |  |  |  |  |  |  |  |  |  | P |  |  | X |  |  |  |  |  |  |  |  |  |  |  |  |  |  |  |  |  |  |  |  |  |  |  |  |  |
| 2014 | Carter |  |  |  | X |  |  |  |  |  |  |  |  | X |  |  |  |  | X |  |  |  |  |  |  |  |  |  |  |  |  |  |  |  |  |  |  |  |  |  |  |
| 2014 | Royall | X |  | X |  |  |  |  | X |  |  |  |  |  |  |  |  |  |  |  |  |  |  |  |  |  |  |  |  |  |  |  |  |  |  |  |  |  |  |  |  |
| 2015 | Wang | X |  |  |  |  |  |  |  | X | X |  |  |  |  | X |  |  |  |  |  |  |  |  |  |  |  |  |  |  |  |  |  | X |  |  |  |  |  |  |  |
| 2016 | Buchert | X | X |  | X |  |  | X |  |  | X |  | P | X | P | X |  |  |  |  |  |  |  |  |  |  |  |  |  |  |  |  |  |  |  |  |  |  |  |  |  |
| 2016 | Gopalakrishna |  |  |  |  |  |  |  |  |  |  |  |  |  |  |  |  |  |  |  |  |  |  |  |  |  |  |  |  |  |  |  |  |  |  |  |  |  |  |  |  |
| 2016 | Gordon | X | X | X | X |  |  |  | X |  | X | X |  | X | P | X |  | X | X | X |  |  |  |  |  |  |  |  |  |  |  |  |  |  |  |  |  |  |  |  |  |
| 2016 | Solsky |  |  |  | X |  |  |  | X |  |  |  |  |  |  |  |  |  |  |  |  |  |  |  |  |  |  |  |  |  |  |  |  |  |  |  |  |  |  |  |  |
| 2017 | Hussain | X |  |  | X |  |  |  |  |  | X | X | P | X |  |  |  | X | X | X |  |  |  |  |  |  |  |  |  |  |  |  |  |  |  |  |  |  |  |  |  |
| 2017 | Dagliati |  |  |  |  |  |  |  |  |  |  |  |  |  |  |  |  |  |  |  |  |  |  |  |  |  |  |  |  |  |  |  |  |  |  |  |  |  |  |  |  |
| 2005 | Ducharme |  |  |  |  |  |  |  |  |  |  |  |  |  |  |  |  |  |  |  |  |  | X |  | X | X |  | X |  | X | X | X |  |  |  | X |  | X |  |  |  |
| 1997 | Ramos |  |  |  |  |  |  |  |  |  | X | X |  |  |  | X |  |  |  |  |  |  |  |  |  |  |  |  |  |  |  |  |  |  |  |  |  | X |  |  |  |
| 2003 | Powell |  |  |  |  |  |  |  |  |  |  |  |  |  |  |  |  |  |  |  |  |  |  |  |  |  | X | X | X |  |  |  |  | X |  | X |  | X | X |  |  |
| 2006 | Vanhaecht |  |  |  |  |  |  |  |  |  |  |  |  |  |  |  |  |  |  |  |  |  |  |  |  |  |  |  |  |  |  |  |  |  |  |  |  |  |  |  |  |
| 2012 | Schrijvers | X | X |  | X |  | X | X | X |  | X | X |  | X | B | X |  | X |  |  |  | X |  |  |  |  |  |  |  |  | X |  |  |  |  |  |  |  |  |  |  |
| 2014 | Li | X | X |  | X |  |  |  |  |  | X |  |  |  |  | X | X |  |  |  |  | X |  |  |  |  |  |  |  |  |  |  |  |  |  |  |  |  |  |  | X |
| 2012 | Yang | X |  |  | X |  |  |  |  |  |  |  |  |  |  |  |  |  |  |  |  |  |  |  |  |  |  |  |  |  |  |  |  |  |  |  |  |  |  |  |  |
| 2006 | De Bleser | X | X |  | X |  |  |  |  |  | X |  | B |  | P | X |  | X |  |  |  | X | X |  |  |  |  |  |  |  |  |  |  |  |  |  |  |  |  |  |  |
| 2007 | Hurley | X |  |  | X |  |  |  |  |  | X |  |  |  |  |  |  |  |  |  |  |  |  |  |  |  |  |  |  |  |  |  |  |  |  |  |  |  |  |  |  |
| 2012 | Rosenfeld |  | X |  | X | X |  |  | X |  | X |  |  |  |  |  |  |  |  |  |  |  |  |  |  |  |  |  |  |  |  |  |  |  |  |  |  |  |  |  |  |
| 1993 | Hampton |  |  |  |  |  |  |  |  |  |  |  |  |  |  |  |  |  |  |  |  |  |  |  |  |  |  |  |  |  |  |  |  |  |  |  |  |  |  |  |  |
| 1994 | Lumsdon |  |  |  |  |  |  |  |  |  |  |  |  |  |  |  |  |  |  |  |  |  | X |  |  |  |  |  |  |  |  |  |  |  |  |  |  |  |  |  |  |
| 1992 | Marr | X | X |  | X |  |  |  |  |  | X |  | P | X |  |  |  |  | X |  |  |  |  |  |  |  |  |  |  |  |  |  |  |  |  |  |  |  |  |  |  |
| 1994 | Riley | X | X |  | X |  |  |  |  |  |  | X |  |  |  | X |  |  |  |  |  |  |  |  |  |  |  |  |  |  |  |  |  |  |  |  |  |  |  |  |  |
| 1993 | Ogilvie-Harris |  |  |  |  |  |  |  |  |  |  |  |  |  |  |  |  |  |  |  |  |  |  |  |  |  |  |  |  |  |  |  |  |  |  |  |  |  |  |  |  |
| 1995 | Blegen |  |  |  | X |  |  |  |  |  |  |  |  |  |  |  |  |  |  |  |  |  |  |  |  |  |  |  |  |  |  |  |  |  |  |  |  |  |  |  |  |
| 2009 | Ye |  |  |  |  |  |  |  |  |  |  |  |  |  |  |  |  |  |  |  |  |  |  |  |  |  |  |  |  |  |  |  |  |  |  |  |  |  |  |  |  |
| 2014 | Sackman |  |  |  |  |  |  |  |  |  |  |  |  |  |  |  |  |  |  |  |  |  |  |  |  |  |  |  |  |  |  |  |  |  |  |  |  |  |  |  |  |

# Appendix C
# Caremap: Development Process Analysis

| Phase | Step | Activity | 15 TOTALS | 1993 Goode | 1994 Giffin | 1995 Hydo | 1996 Comried | 1998 Hill | 2000 Dickinson | 2001 Quaglini | 2012 Lodewijckx | 2012 Huang (2) | 2014 Carter | 2014 Royall | 2005 Ducharme | 1997 Ramos | 2003 Powell | 2012 Rosenfeld | 1995 Blegen |
|---|---|---|---|---|---|---|---|---|---|---|---|---|---|---|---|---|---|---|---|
| BEFORE DEVELOPMENT | 1 | Identify Group of patients/condition/public health issue | 7 | | 1 | | | | | | 1 | | | | 1 | 1 | 1 | 1 | 2 |
| | | Assess Viability of patients/disease for improvement | 3 | | | | | | | | | | | | 2 | 2 | 2 | | |
| | | Decide conceptual framework | 1 | | | | | | 1 | | | | | | | | | | |
| | 2 | Assemble a multidisciplinary team | 8 | 1 | 3 | 2 | 2 | 1 | | | 2 | | | | | | | 2 | 1 |
| | 3 | Challenge traditional practice | 2 | 5 | | | | 4 | | | | | | | | | | | |
| | | Clarify current practice/ situation | 6 | | 2 | 1 | 1 | 2 | | | 3 | | | | | | | | 3 |
| | | Common understanding on CPGs and knowledge among experts | 2 | 2 | | | | | | | | 1 | | | | | | | |
| | | Review and evaluate current data (benchmarks) | 4 | 4 | | | | 3 | | | | | | | | | | 5 | 3 |
| | | Anticipate potential complications | 1 | | | | | 5 | | | | | | | | | | | |
| | | Identify key variances | 1 | | | | | 6 | | | | | | | | | | | |
| | 4 | Literature review | 2 | 3 | | | | | | | 4 | | | | | | | | |
| | | Identify patterns found in the literature | 1 | | | | | | | | | | | 1 | | | | | |
| | | International Delphi studt: content validity | 1 | | | | | | | | 5 | | | | | | | | |
| | 5 | Grading of evidence | 2 | | | | | | | | 6 | | | | | | | | 4 |
| | | Assess the evidence (strength, availability) | 3 | | | | | | | | | | | | 3 | | 3 | 4 | |
| | 6 | Train nurses on how to develop care maps | 1 | | | | 3 | | | | | | | | | | | | |
| DURING | 7 | Identify and manage Conflicts of Interest (CoI) | 1 | | | | | | | | | | | | | | | 3 | |
| | | Transform evidence into policy | 1 | | | | | | | | | | | | 4 | | | | |
| | | Add HPIP knowledge (care guidlines) | 1 | | | | | | | | | 2 | | | | | | | |
| | | Translate into computer interpretable care guidlines | 1 | | | | | | | | | 3 | | | | | | | |
| | | Make artefact Organization-Specific (i.e. revising different Dx standards) | 2 | | | | | | | | | 4 | | | | | 6,8 | | |
| | | Add patient specific knowledge (instances of different OSCs) | 1 | | | | | | | | | 5 | | | | | | | |
| | | CPGs are graphically represented through an authoring tool and a relational database | 1 | | | | | | | 1 | | | | | | | | | |
| | | CPGs are translated into the workflow process definition language | 1 | | | | | | | 2 | | | | | | | | | |
| | | WPDL code is used to build a Petri net -like structure | 1 | | | | | | | 3 | | | | | | | | | |
| | | Develop an ideal pathway | 3 | | | 3 | | | | | | | | | | | | 6 | 4 |
| | | Create discipline specific pathways | 2 | | 5 | 5 | | | | | | | | | | | | | |
| | | Create draft pathway based on experienced multidiciplinary team | 9 | 6 | 4 | | 4 | | 2 | | 7 | | 1 | 2 | | 3 | 4 | | |
| | 8 | detailed description of each intervention | 1 | | | | | | | | 8 | | | | | | | | |
| | | Translate into a set of process | 1 | | | | | | | | 9 | | | | | | | | |
| | 9 | Reach consensus on the content and flow | 3 | 7 | | 6 | | | | | | | | | | | | 7 | |
| | | Refine pathway using different multidisciplinary team from initial | 2 | | | | | | 3 | | | | 2 | | | | | | |
| | | Refine artefact using two or more voting rounds (Delphi Panel) | 2 | | | | | | | | | | | | | | 5 | 8 | |
| | | Refine pathway using external stakeholders | 1 | | | | | | | | | | 3 | | | | | | |
| | | Multi-disciplinary consultation prior to implementation | 1 | | | | | | | | | | | | | | 7 | | |
| | 10 | Training before implementation | 2 | | 6 | 7 | | | | | | | | | | | | | |
| | 11 | Piloting by multidisciplinary teams in different organisations | 1 | | | | | | | | 10 | | | | | | | | |
| | | Put pathway to the test | 1 | | | 8 | | | | | | | | | | | | | |
| | | Assess care map's usability principles | 1 | | | | | | | | | | | 3 | | | | | |
| | 12 | Identify problem areas based on the documented variances | 1 | | | 4 | | | | | | | | | | | | | |
| | | Modifications based on observed variations | 3 | | | 9 | 6 | | | | | 6 | | | | | | | |
| | | Period of public comment and review | 1 | | | | | | | | | | | | | | | 9 | |
| | | Assess long term information system needs | 1 | | 10 | | | | | | | | | | | | | | |
| | | Revise CPGs | 1 | | | | | | | | | 8 | | | | | | | |
| AFTER | 13 | Implementation of care map /guideline | 4 | 8 | 7 | | 5 | | | | | | | | | | | 10 | |
| | | Assessment, revision and full implementation | 2 | 9 | 8 | | | | | | | | | | | | | | |
| | 14 | IN-Practice (post implementation) evaluation and review | 1 | | | | | | | | | | | | | | 9 | | |
| | | Periodic review of variance reports | 1 | | 9 | | | | | | | | | | | | | | |

# Appendix D
# Clinical Documentation Terminology Use (1993-2016)

| Author (year) | Policy | Clinical Decision Rule | Case Management Plan | Clinical Practice Guideline | Caremap | Clinical Pathway | Care Pathway | Critical Pathway | Clinical/Treatment Protocol | Care Plan | Careflow |
|---|---|---|---|---|---|---|---|---|---|---|---|
| Goode (1993) | | | | | □ | | □ | □ | | | |
| Goode (1993) p5 | | | | | ● | | | ● | | | |
| Goode (1993) p17 | | | | □ | □ | | | | | | |
| Goode (1993) p215 | | | | | | | | ● | | ● | |
| Giffin (1994) | | | | ● | | | | ● | | | |
| Giffin (1994) | | | | | □ | | | □ | | | |
| Seaberg (1994) | | ● | | ● | | | | | | | |
| Hydo (1995) | | | | | | □ | □ | | | | |
| Hydo (1995) | | | ● | | | ● | | | | | |
| Hydo (1995) | | | | | | ● | | | | ● | |
| Stiell (1995) | | □ | | □ | | | | | | | |
| Gordon (1996) | | | | | □ | | | □ | | | |
| Gordon (1996) | | | | ● | | | | | ● | | |
| Gordon (1996) | | | | ● | | | | ● | | | |
| Holocek et al (1997) | | | □ | | □ | □ | | □ | | | |
| Morreale (1997) | | | | | □ | □ | | □ | | | |
| Morreale (1997) | | | | ● | | | | | | ● | |
| Morreale (1997) | | | | | □ | | | | | □ | |
| Hill (1998) | | | | | □ | □ | | | | | |
| Chu et al (1998) | | | | | ● | ● | | | | | |
| Chu et al (1998) | | | | | | ● | | | ● | | |
| Denton (1999) | | | □ | | □ | □ | □ | □ | | | |
| Dickinson (2000) | | | | | ● | ● | | | | | |
| Dickinson (2000) | | | | ● | ● | | | | | | |
| Philie (2001) | | | | | ● | | | | ● | | |
| Philie (2001) | | | | | □ | | | □ | | | |
| Quaglini (2001) | | | | ● | | | | | | | ● |
| Lanzieri (2001) | | | | | □ | | | □ | | | |
| Trowbridge et al (2001) | | | | ● | | ● | | | | | |
| Micik (2002) | | | | | | □ | | □ | | | |
| Panzarasa (2002) | | | | ● | | | | | ● | | ● |
| Powell (2003) | | | | ● | | | ● | | | | |
| Jones (2003) | ● | | | ● | | | | | | | |
| Jones (2003) | | | | ● | | | ● | | | | |
| Li (2004) | | | | ● | ● | | | | | | |
| Li (2004) | | | | | □ | □ | | | | □ | |
| Beaupre (2004) | | | | | □ | □ | | | | | |
| Beaupre (2004) | ● | | | | ● | | | | | | |
| Van Herck (2004) | | | | | □ | □ | □ | □ | | | |
| Agrawal et al (2009) | ● | ● | | ● | | ● | | ● | ● | | |
| Agrawal et al (2009) | | ● | | ● | | | | | | | |
| Brehaut (2005) | | ● | | ● | | | | | | | |
| Repasky (2005) | | | | | ☐ | | ☐ | | | | |
| Ducharme (2005) | ● | ● | | ● | | ● | | | ● | | |
| Ducharme (2005) | | ● | | ● | | | | | | | |
| Dunning (2006) | | ● | | ● | | | | | | | |
| De Blesser et al (2006) | | | | | ☐ | ☐ | ☐ | | | | |
| Vanhaecht (2006) | | | | | ☐ | ☐ | ☐ | ☐ | | | |
| Tay (2006) | | | | | ☐ | | ☐ | ☐ | | | |
| Gaddis (2007) | | ● | | ● | | ● | | | | | |
| Sternberg (2007) | | | | | | ● | | | | ● | |
| Sternberg (2007) | | | | | | ☐ | | | ☐ | | |
| Mackey (2007) | | | | ● | ● | | | | | | |
| Kwan (2007) | | | | | ☐ | ☐ | ☐ | ☐ | | | |
| Richman (2008) | | ☐ | | ☐ | | | | | | | |
| Rotter (2008) | | | | | ☐ | ☐ | ☐ | ☐ | ☐ | | |
| Rotter (2008) | | | | | | | | | ● | ● | |
| Rotter (2008) | | | | ● | | ● | | | | | |
| Lougheed (2008) | | | | | ● | ● | | | | | |
| Lougheed (2008) | | | | ● | | ● | | | | | |
| Gemmel et al (2008) | | | | | ☐ | ☐ | | ☐ | | | |
| Isern (2008) | | | | ● | | | | | | | ● |
| Ebell (2010) | | ● | | ● | | | | | | | |
| Nakanishi (2010) | | | | | | ● | | | | ● | |
| Kinsman et al (2010) | | | | ☐ | ☐ | ☐ | | | | | |
| Kinsman et al (2010) | | | | ● | | ● | | | | | |
| Kinsman et al (2010) | | | | | | ● | | | | ● | |
| Kinsman et al (2010) | | | | | | ● | | | ● | | |
| Lodewijckx (2012) | | | | ● | | | ● | | | | |
| Tastan et al (2012) | | | | | | ● | | | ● | | |
| Tastan et al (2012) | | | | | | ● | | | | ● | |
| Huang et al (2012) | | | | ● | | ● | | | | | |
| Huang et al (2012) | | | | ☐ | | | | | ☐ | | |
| Huang et al (2012) | | | | | | ● | | | | ● | |
| Lockie et al (2013) | | ● | | ● | | | | | | | |
| Jarayaj (2013) | | | | | | | ● | | ● | | |
| Jabbour (2013) | | | | ● | | ● | | | | | |
| Jabbour (2013) | | | | | | ● | | | ● | | |
| Morrow (2013) | | | | | ☐ | ☐ | | | | | |
| Yu (2014) | ● | | | ● | | | | | | | |
| Yu (2014) | | | | ● | | | | | | | ● |
| Li et al (2014) | | | | | | ● | | | | ● | |
| Royall (2014) | | | | ● | ● | | | | | | |
| Carter (2015) | | | | ● | | ● | | | | | |
| Carter (2015) | | | | | | ☐ | ☐ | | | | |
| Wang et al (2015) | | ● | | | | ● | | | | | |
| Wang et al (2015) | | | | ● | | ● | | | | | |
| Roopsawang (2015) | | | | ☐ | ☐ | ☐ | | | | | |
| Buchert et al (2016) | | | | ☐ | ☐ | ☐ | ☐ | ☐ | ☐ | | |
| Buchert et al (2016) | | | | ● | | ● | | | | | |
| Dagliati et al (2017) | | | | ● | | | | | ● | | ● |
| Dagliati et al (2017) | | | | | | ● | ● | | | | |
| Gopalakrishna (2016) | | | | ● | | ● | | | | | |
| Gopalakrishna (2016) | | | | | ☐ | ☐ | | | | | |
| Solsky et al (2016) | | | | ● | | | | | ● | | |
| Alonso-Coello (2016) | | ● | | ● | | | | | | | |

LEGEND:

● = Narrative provided by the authors identified that these items were **distinct/different**

☐ = Narrative provided by the authors identified these items as the **same or synonymous**

## References for Appendix D

Agrawal, P., & Kosowsky, J. M. (2009). Clinical practice guidelines in the emergency department. Emergency Medicine Clinics, 27(4), 555-567.

Alonso-Coello, P., Oxman, A. D., Moberg, J., Brignardello-Petersen, R., Akl, E. A., Davoli, M., ... & Guyatt, G. H. (2016). GRADE Evidence to Decision (EtD) frameworks: a systematic and transparent approach to making well informed healthcare choices. 2: Clinical practice guidelines. bmj, 353, i2089.

Beaupre, L. A. (2005). Effectiveness of a caremap for treatment of elderly patients with hip fracture. A thesis submitted in partial completion of the degree of Doctor of Philosophy, University of Alberta: Canada.

Brehaut, J. C., Stiell, I. G., Visentin, L., & Graham, I. D. (2005). Clinical decision rules "in the real world": how a widely disseminated rule is used in everyday practice. Academic emergency medicine, 12(10), 948-956.

Buchert, A., & Buttler, G. (2016). Clinical Pathways: Driving high-reliability and high-value care. Pediatric Clinical of North America, 63, pp 317-328.Chu, S., & Cesnik, B. (1998). Improving clinical pathway design: Lessons learned from a computerised prototype. Int. Journal of Medical Informatics, 51, 1-11.

Carter, P., Laurie, G. T., & Dixon-Woods, M. (2015). The social licence for research: why care. data ran into trouble. Journal of medical ethics, 41(5), 404-409.

Dagliati, A., Sacchi, L., Zambelli, A., Tibollo, V., Pavesi, L., Holmes, J., & Bellazzi, R. (2017). Temporal electronic phenotyping by mining careflows of breast cancer patients. Journal of Biomedical Informatics, 66, pp 136-147.De Bleser, L., Depreitere, R., De Waele, K., Vanhaecht, K., Vlayden, J., & Sermeus, W. (2006). Defining Pathways. Journal of Nursing Management, 14, pp 553-563.

Denton, M., Wentworth, S., Yellowlees, P., & Emmerson, B. (1999). Clinical pathways in mental health. Australasian Psychiatry, 7(2), 75-77.

Dickinson, C., & Noud, M. (2000). The antenatal ward care delivery map: a team model approach. Australian Health Review, 23(3), 68-77.

Ducharme, J. (2005). Clinical guidelines and policies: can they improve emergency department pain management? The Journal of Law, Medicine & Ethics, 33(4), 783-790.

Dunning, J., Daly, J. P., Lomas, J. P., Lecky, F., Batchelor, J., & Mackway-Jones, K. (2006). Derivation of the children's head injury algorithm for the prediction of important clinical events decision rule for head injury in children. Archives of disease in childhood, 91(11), 885-891.

Ebell, M. (2010). AHQR White Paper: Use of Clinical Decision Rules for Point-of-Care Decision Support. Medical Decision Support, Nov 2010, pp 712-721.

Gaddis, G. M., Greenwald, P., & Huckson, S. (2007). Toward improved implementation of evidence-based clinical algorithms: clinical practice guidelines, clinical decision rules, and clinical pathways. Academic Emergency Medicine, 14(11), 1015-1022.

Gemmel, P., Vandaele, D., & Tambeur, W. (2008). Hospital Process Orientation (HPO): The development of a measurement tool. Total Quality Management & Business Excellence, 19(12), 1207-1217.

Giffin, M., & Giffin, R. B. (1994). Critical pathways produce tangible results. Health Care Strategic Management, 12(7), 1-17.

Goode, C. (1993). Evaluation of patient and staff outcomes with hospital based managed care. Thesis in partial fulfillment of the degree of PhD, University of Iowa.

Gopalakrishna, G., Langendam, M. W., Scholten, R. J., Bossuyt, P. M., & Leeflang, M. M. (2016). Defining the clinical pathway in cochrane diagnostic test accuracy reviews. BMC medical research methodology, 16(1), 153.

Gordon, D. B. (1996). Critical pathways: a road to institutionalizing pain management. Journal of pain and symptom management, 11(4), 252-259.

Hill, M. (1998). The development of care management systems to achieve clinical integration. Advanced practice nursing quarterly, 4(1), 33-39.

Holocek, R., & Sellards, S. (1997). Use of a detailed clinical pathway for Bone Marrow Transplant patients. Journal of Pediatric Oncology Nursing, 14(4), 252-257.

Huang, B., Zhu, C., & Wu, C. (2012). Customer-centred careflow modelling based on guidelines. Journal of Medical Systems, 36, pp 3307-3319.

Hydo, B. (1995). Designing an effective clinical pathway for stroke. The American Journal of Nursing, 95(3), pp 44-51.

Isern, D., & Moreno, A. (2008). Computer-based execution of clinical guidelines: a review. International journal of medical informatics, 77(12), 787-808.

Jabbour, M., Curran, J., Scott, S. D., Guttman, A., Rotter, T., Ducharme, F. M., ... & Paprica, A. (2013). Best strategies to implement clinical pathways in an emergency department setting: study protocol for a cluster randomized controlled trial. Implementation science, 8(1), 55.

Jayaraj, R., Whitty, M., Thomas, M., Kavangh, D., Palmer, D., Thomson, V., ... & Nagel, T. (2013). Prevention of Alcohol-Related Crime and Trauma (PACT): brief interventions in routine care pathway–a study protocol. BMC public health, 13(1), 49.

Jones, A., (2003). Perceptions on the development of a care pathway for people diagnosed with schizophrenia on acute psychiatric wards. Journal of Psychiatric and Mental Health Nursing, 10, pp 669-677.

Kinsman, L., Rotter, T., James, E., Snow, P., & Willis, J. (2010). What is a clinical pathway? Development of a definition to inform the debate. BMC Medicine, 8(31).

Kwan, J. (2007). Care pathways for acute stroke care and stroke rehabilitation: from theory to evidence. Journal of Clinical neuroscience, 14(3), 189-200.

Lanzieri, M., Hobbs, R., Roy, P., & Richmond, R. (2001). Use of a clinical care map for the management of congestive heart failure in a community hospital. Congestive Heart Failure, 7(1), 37-42.

Li, M. (2004). Application of the Bayesian belief network model to evaluate variances in a clinical caremap: Radical prostatectomy case study (Doctoral dissertation, University of Ottawa (Canada)).

Li, W., Liu, K., Yang, H., & Yu, C. (2014). Integrated clinical pathway management for medical quality improvement – based on a semiotically inspired systems architecture. European Journal of Information Systems, 23(4), 400-417.

Lockie, F., Dalton, S., Oakley, E., Babl, F. (2013). Triggers for head computed tomography following paediatric head injury: Comparison of physicians' reported practice and clinical decision rules. Emergency Medicine Australasia, 25, pp 75-82.

Lodewijckx, C., Decramer, M., Sermeus, W., Panella, M., Deneckere, S., & Vanhaecht, K. (2012). Eight-step method to build the clinical content of an evidence-based care pathway: the case for COPD exacerbation. Trials, 13(1), 229.

Lougheed, M. D., Olajos-Clow, J., Szpiro, K., Moyse, P., Julien, B., Wang, M., & Day, A. G. (2009). Multicentre evaluation of an emergency department asthma care pathway for adults. Canadian Journal of Emergency Medicine, 11(3), 215-229.

Mackey, D., Myles, M., Spooner, C. H., Lari, H., Tyler, L., Blitz, S., ... & Rowe, B. H. (2007). Changing the process of care and practice in acute asthma in the emergency department: experience with an asthma care map in a regional hospital. Canadian Journal of Emergency Medicine, 9(5), 353-365.

Micik, S., & Borbasi, S. (2002). Effect of support programme to reduce stress in spouses whose partners 'fall off'clinical pathways post cardiac surgery. Australian Critical Care, 15(1), 33-40.

Morreale, M. J. (1997). Evaluation of a care map for community-acquired pneumonia hospital inpatients. Thesis in completion of the degree of Master of Science, Queens University, Kingston: Canada.

Morrow, J., McLachlan, H., Forster, D., Mary-Ann Davey, D. P. H., MEpi, G. D. S., & Post Grad Dip Health Ed, M. (2013). Redesigning postnatal care: exploring the views and experiences of midwives. Midwifery, 29(2), 159-166.

Nakanishi, M., Sawamura, K., Sato, S., Setoya, Y., & Anzia, N. (2010). Development of a clinical pathway for long-term inpatients with schizophrenia. Psychiatry and Clinical Neurosciences, 64, pp 99-103.

Panzarasa, S., Madde, S., Quaglini, S., Pistarini, C., & Stefanelli, M. (2002). Evidence-based careflow management systems: the case of post-stroke rehabilitation. Journal of biomedical informatics, 35(2), 123-139.

Philie, P. C. (2001). Management of bloodborne fluid exposures with a rapid treatment prophylactic caremap: One hospital's 4-year experience. Journal of Emergency Nursing, 27(5), 440-449.

Powell, C. V. (2003). How to implement change in clinical practice. Paediatric respiratory reviews, 4(4), 340-346.

Quaglini, S., Stefanelli, M., Lanzola, G., Caporusso, V., & Panzarasa, S. (2001). Flexible guideline-based patient careflow systems. Artificial intelligence in medicine, 22(1), 65-80.

Repasky, T. M. (2005). A frequently used and revised ED chest pain pathway. Journal of Emergency Nursing, 31(4), 368-370.

Richman, P. B., Vadeboncoeur, T. F., Chikani, V., Clark, L., & Bobrow, B. J. (2008). Independent Evaluation of an Out-of-hospital Termination of Resuscitation (TOR) Clinical Decision Rule. Academic Emergency Medicine, 15(6), 517-521.

Roopsawang, I., & Belza, B. (2015). When a Care Map is Not a True Predictor of Clinical Outcomes. Journal of gerontological nursing, 41(10), 5-7.

Rotter, T., Kugler, J., Koch, R., Gothe, H., Twork, S., van Oostrum, J. M., & Steyerberg, E. W. (2008). A systematic review and meta-analysis of the effects of clinical pathways on length of stay, hospital costs and patient outcomes. BMC health services research, 8(1), 265.

Royall, D., Brauer, P., Bjorklund, L., O'Young, O., Tremblay, A., Jeejeebhoy, K., ... & Mutch, D. M. (2014). Development of a dietary management care map for metabolic syndrome. Canadian Journal of Dietetic Practice and Research, 75(3), 132-139.

Seaberg, D. C., & Jackson, R. (1994). Clinical decision rule for knee radiographs. The American journal of emergency medicine, 12(5), 541-543.

Solsky, I., Edelstein, A., Brodman, M., Kaleya, R., Rosenblatt, M., Santana, C., . . . Shamamian, P. (2016). Perioperative care map improves compliance with best practices for the morbidly obese. Surgery, 160(6), 1682-1688.

Stiell, I. G., Greenberg, G. H., Wells, G. A., McKnight, R. D., Cwinn, A. A., Cacciotti, T., ... & Smith, N. A. (1995). Derivation of a decision rule for the use of radiography in acute knee injuries. Annals of emergency medicine, 26(4), 405-413.

Sternberg, S., (2007). Influence of Stroke Clinical Pathway on Documentation. Thesis presented in partial fulfillment of the degree of Master of Science in Nursing, Clemson University.

Tastan, S., Hatipoglu, S., Iyigun, E., & Kilic, S. (2012). Implementation of a clinical pathway in breast cancer patients undergoing breast surgery. European Journal of Oncology Nursing, 16, pp 368-374.

Tay, H. L., Raja Latifah, R. J., & Razak, I. A. (2006). Clinical pathways in primary dental care in Malaysia: Clinicians' knowledge, perceptions and barriers faced. Asia Pacific Journal of Public Health, 18(2), 33-41.

Trowbridge, R. and Weingarten, S. (2001), "Critical pathways", in Eisenberg, J. and Kamerow, D. (Eds), Making Health Care Safer – A Critical Analysis of Patient Safety Practices, Evident Report/Technology Assessment No. 43, AHRQ Publications, Rockville, MD, available at: www.ahcpr.gov/clinic/ptsafety

Vanhaecht, K., Witte, K. D., Depreitere, R., & Sermeus, W. (2006). Clinical pathway audit tools: a systematic review. Journal of nursing management, 14(7), 529-537.

Van Herck, P., Vanhaecht, K., & Sermeus, W. (2004). Effects of clinical pathways: do they work?. Journal of Integrated Care Pathways, 8(3), 95-105.

Wang, H., Zhou, T., Zhang, Y., Chen, L., & Li, J. (2015). Research and Development of Semantics-based sharable clinical pathway systems. Journal of Medical Systems, 39, p 73.

Yu, B. (2014). Enforcing Careflows and Treatment Consents in Electronic Medical Record Systems (Doctoral dissertation, George Mason University).

# Appendix E
# Trauma Caremap Survey

Table 26: Survey Response Legend

| Value | Text Description |
|---|---|
| 6 | Strongly Agree |
| 5 | Agree |
| 4 | Moderately Agree |
| 3 | Moderately Disagree |
| 2 | Disagree |
| 1 | Strongly Disagree |

Table 27: Survey Responses and Totals

| Clinician | Q1 | Q2 | Q3 | Q4 | Q5 | Q6 |
|---|---|---|---|---|---|---|
| 1 | 5 | 4 | 5 | 6 | 6 | 6 |
| 2 | 6 | 6 | 6 | 5 | 6 | 6 |
| 3 | 5 | 4 | 4 | 6 | 5 | 6 |
| 4 | 5 | 5 | 6 | 6 | 6 | 6 |
| 5 | 6 | 5 | 5 | 5 | 5 | 5 |
| 6 | 6 | 5 | 5 | 6 | 6 | 6 |
| 7 | 4 | 3 | 5 | 6 | 5 | 5 |
| Potential Score | 42 | 42 | 42 | 42 | 42 | 42 |
| Actual Score | 37 | 32 | 36 | 40 | 39 | 40 |
| Percentage | 88 | 76 | 86 | 95 | 93 | 95 |

# Appendix F
# Thesis Relationship to Published Works

During the conduct of this research, much of the contributions contained herein have been presented and/or published in the academic literature. Figure 49 identifies the relationship between contributions presented in this thesis, and their publication in the academic literature. A given paper's contributions may have been drawn from one or more chapters of the thesis. This often occurred because the research question and literature review that formed the basis of a published paper were contained in one chapter, and the new framework or research result are presented in another.

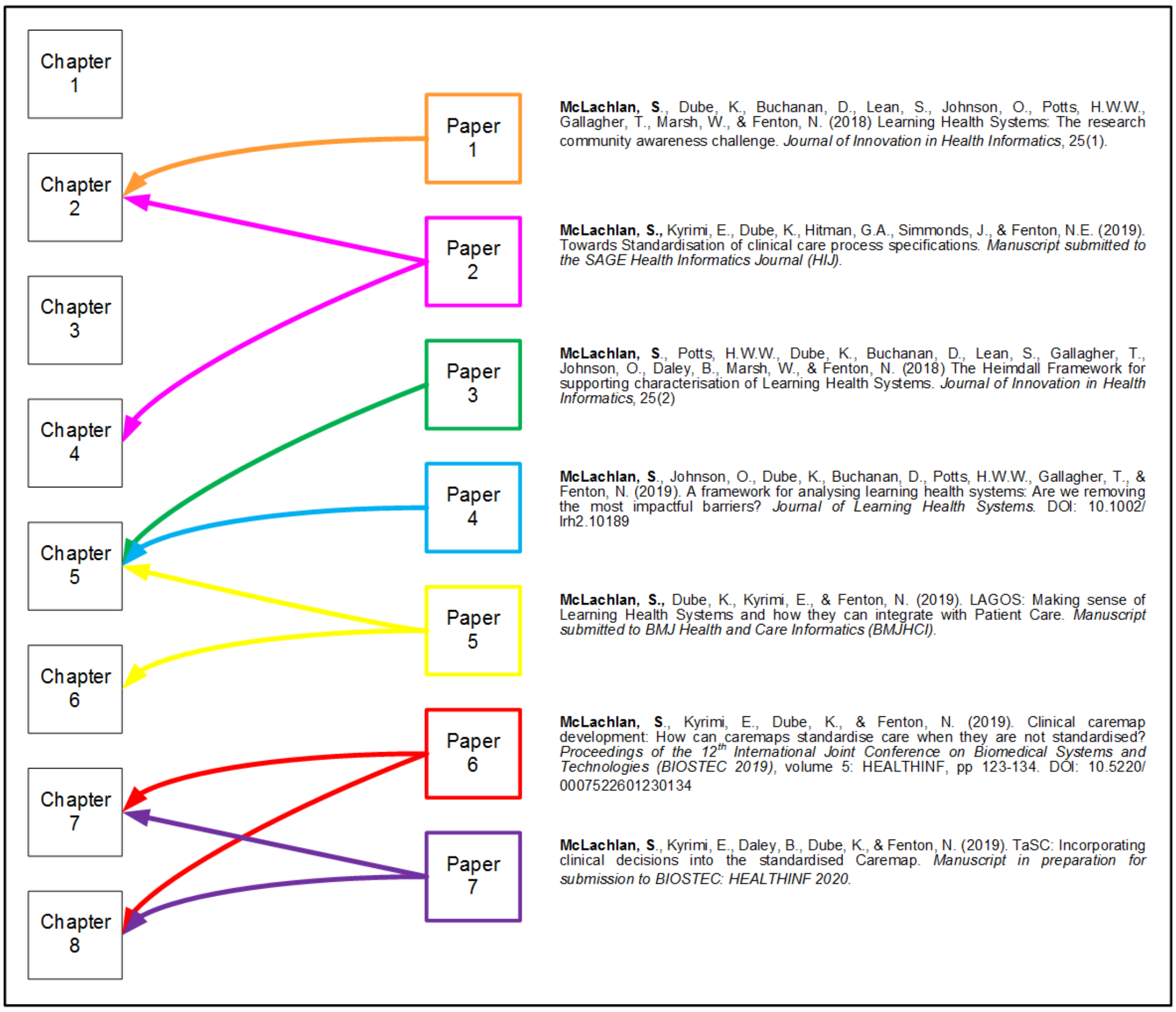

Figure 49: Papers to Chapters Relational Diagram

Paper 1: *The Community Awareness Challenge* identified that many authors were unaware that their health informatics solutions represented LHS, and that this, along with the lack of a complete taxonomy that would allow authors to identify their solutions within the LHS domain, was inhibiting development of a critical mass of researchers within the domain. This paper was published in the *Journal of Innovation in Health Informatics (JIHI)* in January 2018 and Google Scholar reports it has achieved five citations and currently has an Altmetric score of 7.

Paper 2: *The Taxonomy and Hierarchy for CCPS* reports on a literature review of CCPS nomenclature, types, content and usage. It presents the expanded De Bleser approach, a research methodology that was used to develop a taxonomy of the most common and necessary types of CCPS in contemporary use. It also presents an exemplar case study of CCPS for Type 2 Diabetes. This paper has been submitted to the Sage *Health Informatics Journal (HIJ)* and is currently in the review stage.

Paper 3: *The Heimdall Framework* used an extensive literature review to resolve a taxonomy for LHS, and presented the unifying Heimdall Framework showing how the nine LHS types integrate with healthcare delivery and the learning health organisation in order to focus medical practice towards precision medicine. The taxonomy was validated through classification of all presented or proposed solutions self-identified as LHS in the literature. Heimdall was published in the 25$^{th}$ issue of the *Journal of Innovation in Health Informatics (JIHI)* in 2018 and Google Scholar reports it has already garnered five citations while Researchgate shows an additional citation bringing the total to six. Heimdall currently has an Altmetric score of 15.

Paper 4: *LHS Barriers, Benefits and Facilitators* expands on the well-known ITPOSMO methodology for evaluating Information Technology projects, using it as the basis for ITPOSMO-BBF: a new approach to analysing barriers, benefits and facilitators for HIT implementations. ITPOSMO-BBF is applied to a comparison and analysis of barriers, benefits and facilitators identified during EHR and LHS implementations. The expansion on ITPOSMO presented in the paper was published with only minor changes by the *Journal of Learning Health Systems (LHS)* in early 2019 and in only a few months has already achieved an Altmetric score of 15.

Paper 5: *Integrating LHS with Patient Care* begins by providing a characterisation and conceptualisation for LHS and describing the LHS learning lifecycle. The paper goes on to present the LAGOS framework, which unifies LHS with the domains of medicine, clinical care, health informatics and decision science across five pathways. The relationships between each pathway are demonstrated and evaluated through three case studies related to the EPSRC-funded PAMBAYESIAN project. In October 2019 this paper was accepted for publication in the *British Medical Journal of Health and Care Informatics (BMJHCI)*.

Paper 6: *Towards a Standard for Caremaps* presents a standardised approach to caremap creation focusing on three key elements: structure, content and development process. The name of this approach has been abbreviated following the capitalisations in the italicised heading above, resolving to the acronym TaSC. The caremaps paper was presented at the *12$^{th}$ International Joint Conference on Biomedical Systems and Technologies (BIOSTEC)* in Prague during February 2019.

Paper 7: *Incorporating Clinical Decisions* provides a rational and description for resolving the latent decision points that inhabit activity nodes in the caremap.